\documentclass[acmsmall]{acmart}

\usepackage{tcolorbox}
\usepackage{float}
\usepackage{xcolor}
\usepackage[normalem]{ulem}
\usepackage{natbib}
\usepackage{hyperref}

\hypersetup{colorlinks,urlcolor=blue}

\makeatletter
\DeclareUrlCommand\ULurl@@{%
  \def\UrlLeft{\uline\bgroup}%
  \def\UrlRight{\egroup}}
\def\ULurl@#1{\hyper@linkurl{\ULurl@@{#1}}{#1}}
\DeclareRobustCommand*\ULurl{\hyper@normalise\ULurl@}

\usepackage{geometry}
\usepackage{multirow,bigstrut}
\usepackage{colortbl}
\usepackage{longtable}
\usepackage[multiple]{footmisc}
\usepackage{changepage}
\usepackage[printonlyused]{acronym} 

\newcolumntype{P}[1]{>{\centering\arraybackslash}p{#1}}

\definecolor{Gray1}{gray}{0.95}
\definecolor{Gray2}{gray}{0.89}

\AtBeginDocument{%
  \providecommand\BibTeX{{%
    \normalfont B\kern-0.5em{\scshape i\kern-0.25em b}\kern-0.8em\TeX}}}

\begin{document}

\title[Common Limitations of Image Processing Metrics: A Picture Story]{Common Limitations of Image Processing Metrics: \\A Picture Story}

\author{Annika Reinke}
\authornote{Shared first authors.}
\email{a.reinke@dkfz.de}
\affiliation{%
  \institution{German Cancer Research Center (DKFZ) Heidelberg, Division of Intelligent Medical Systems and HI Helmholtz Imaging}
  \country{Germany}
}
\affiliation{%
  \institution{Faculty of Mathematics and Computer Science, Heidelberg University}
  \country{Heidelberg, Germany}
}

\author{Minu D. Tizabi}
\authornotemark[2]
\affiliation{%
  \institution{German Cancer Research Center (DKFZ) Heidelberg, Division of Intelligent Medical Systems}
  \country{Germany}
}
\affiliation{%
  \institution{National Center for Tumor Diseases (NCT), NCT Heidelberg, a partnership between DKFZ and University Medical Center Heidelberg}
  \country{Germany}
}

\author{Carole H. Sudre}
\affiliation{%
 \institution{MRC Unit for Lifelong Health and Ageing at UCL and Centre for Medical Image Computing, Department of Computer Science, University College London}
 \country{London, UK}}
 \affiliation{%
 \institution{School of Biomedical Engineering and Imaging Science, King's College London}
 \country{London, UK}}
 
 \author{Matthias Eisenmann}
\affiliation{%
  \institution{German Cancer Research Center (DKFZ) Heidelberg, Division of Intelligent Medical Systems}
  \country{Germany}
}

\author{Tim Rädsch}
\affiliation{%
  \institution{German Cancer Research Center (DKFZ) Heidelberg, Division of Intelligent Medical Systems and HI Helmholtz Imaging}
  \country{Germany}
}

\author{Michael Baumgartner}
\affiliation{%
  \institution{German Cancer Research Center (DKFZ) Heidelberg, Division of Medical Image Computing}
  \country{Germany}
}

\author{Laura Acion}
\affiliation{%
  \institution{Instituto de Cálculo, CONICET -- Universidad de Buenos Aires}
  \country{Buenos Aires, Argentina}
}

\author{Michela Antonelli}
\affiliation{%
 \institution{School of Biomedical Engineering and Imaging Science, King's College London}
 \country{London, UK}}
\affiliation{%
 \institution{Centre for Medical Image Computing, University College London}
 \country{London, UK}}

\author{Tal Arbel}
\affiliation{%
  \institution{Centre for Intelligent Machines and MILA (Québec Artificial Intelligence Institute), McGill University}
  \country{Montréal, Canada}}

\author{Spyridon Bakas}
\affiliation{%
  \institution{Division of Computational Pathology, Dept of Pathology \& Laboratory Medicine, Indiana University School of Medicine, IU Health Information and Translational Sciences Building}
  \country{Indianapolis, USA}}
\affiliation{%
  \institution{Center for Biomedical Image Computing and Analytics (CBICA), University of Pennsylvania, Richards Medical Research Laboratories FL7}
  \country{Philadelphia, PA, USA}}
 
\author{Peter Bankhead}
\affiliation{%
  \institution{Institute of Genetics and Cancer, University of Edinburgh}
  \country{Edinburgh, UK}}

\author{Arriel Benis}
\affiliation{%
  \institution{Department of Digital Medical Technologies, Holon Institute of Technology}
  \country{Holon, Israel}
}
\affiliation{%
  \institution{European Federation for Medical Informatics}
  \country{Le Mont-sur-Lausanne, Switzerland}
}

\author{Matthew Blaschko}
\affiliation{%
  \institution{Center for Processing Speech and Images, Department of Electrical Engineering, KU Leuven}
  \country{Kasteelpark Arenberg 10 - box 2441, 3001 Leuven, Belgium}
}

\author{Florian Buettner}
\affiliation{%
  \institution{German Cancer Consortium (DKTK), partner site Frankfurt/Mainz, a partnership between DKFZ and UCT Frankfurt-Marburg}
  \country{Germany}
}
\affiliation{%
  \institution{German Cancer Research Center (DKFZ) Heidelberg}
  \country{Germany}
}
\affiliation{%
 \institution{Goethe University Frankfurt, Department of Medicine}
  \country{Germany}
}
\affiliation{%
 \institution{Goethe University Frankfurt, Department of Informatics}
  \country{Germany}
}
\affiliation{%
  \institution{Frankfurt Cancer Insititute}
  \country{Germany}
}
\author{M. Jorge Cardoso}
\affiliation{%
 \institution{School of Biomedical Engineering and Imaging Science, King's College London}
 \country{London, UK}}

 \author{Jianxu Chen}
\affiliation{%
  \institution{Leibniz-Institut für Analytische Wissenschaften – ISAS – e.V.}
  \country{Dortmund, Germany}}
  
\author{Veronika Cheplygina}
\affiliation{%
  \institution{Department of Computer Science, IT University of Copenhagen}
  \country{Copenhagen, Denmark}}
  
\author{Evangelia Christodoulou}
\affiliation{%
  \institution{German Cancer Research Center (DKFZ) Heidelberg, Division of Intelligent Medical Systems}
  \country{Germany}
}

\author{Beth A. Cimini}
\affiliation{%
  \institution{Imaging Platform, Broad Institute of MIT and Harvard}
  \country{Cambridge, MA, USA}}

\author{Gary S. Collins}
\affiliation{%
  \institution{Centre for Statistics in Medicine, University of Oxford}
  \country{Oxford, UK}}
  
\author{Sandy Engelhardt}
\affiliation{%
 \institution{Department of Internal Medicine III and Department of Cardiac Surgery, Heidelberg University Hospital}
 \country{Heidelberg, Germany}}

\author{Keyvan Farahani}
\affiliation{%
  \institution{Center for Biomedical Informatics and Information Technology, National Cancer Institute}
  \country{Bethesda, MD, USA}}
  
\author{Luciana Ferrer}
\affiliation{%
\institution{Instituto de Investigación en Ciencias de la Computación (ICC), CONICET-UBA}
  \country{Ciudad Universitaria, Ciudad Autónoma de Buenos Aires, Argentina}}
  
\author{Adrian Galdran}
\affiliation{%
  \institution{Universitat Pompeu Fabra}
  \country{Barcelona, Spain}}
\affiliation{%
  \institution{University of Adelaide}
  \country{Adelaide, Australia}}  

\author{Bram van Ginneken}
\affiliation{%
  \institution{Fraunhofer MEVIS}
  \country{Bremen, Germany}
}
\affiliation{%
  \institution{Radboud Institute for Health Sciences, Radboud University Medical Center}
  \country{Nijmegen, The Netherlands}
}

\author{Ben Glocker}
\affiliation{%
  \institution{Department of Computing, Imperial College London}
  \country{London, UK}}
  
\author{Patrick Godau}
\affiliation{%
  \institution{German Cancer Research Center (DKFZ) Heidelberg, Division of Intelligent Medical Systems}
  \country{Germany}
}
\affiliation{%
  \institution{Faculty of Mathematics and Computer Science, Heidelberg University}
  \country{Heidelberg, Germany}
}
\affiliation{%
  \institution{National Center for Tumor Diseases (NCT), NCT Heidelberg, a partnership between DKFZ and University Medical Center Heidelberg}
  \country{Germany}
}

\author{Robert Haase}
\affiliation{%
  \institution{Now with: Center for Scalable Data Analytics and Artificial Intelligence (ScaDS.AI), Leipzig University}
  \country{Leipzig, Germany}
}
\affiliation{%
  \institution{Technische Universität (TU) Dresden, DFG Cluster of Excellence "Physics of Life"}
  \country{Dresden, Germany}
}
\affiliation{%
  \institution{Center for Systems Biology}
  \country{Dresden, Germany}
}
  
\author{Fred Hamprecht}
\affiliation{%
  \institution{Heidelberg Collaboratory for Image Processing (HCI), Interdisciplinary Center for Scientific Computing (IWR)Heidelberg University}
  \country{Heidelberg, Germany}}

\author{Daniel A. Hashimoto}
\affiliation{%
  \institution{Department of Surgery, Perelman School of Medicine}
  \country{Philadelphia, PA, USA}}
\affiliation{%
  \institution{General Robotics Automation Sensing and Perception Laboratory, School of Engineering and Applied Science, University of Pennsylvania}
  \country{Philadelphia, PA, USA}}
  
  \author{Doreen Heckmann-Nötzel}
\affiliation{%
  \institution{German Cancer Research Center (DKFZ) Heidelberg, Division of Intelligent Medical Systems}
  \country{Germany}
}
\affiliation{%
  \institution{National Center for Tumor Diseases (NCT), NCT Heidelberg, a partnership between DKFZ and University Medical Center Heidelberg}
  \country{Germany}
}

\author{Peter Hirsch}
\affiliation{%
  \institution{Max-Delbrück Center for Molecular Medicine}
  \country{Regensburg, Germany}
}
  
\author{Michael M. Hoffman}
\affiliation{%
  \institution{Princess Margaret Cancer Centre, University Health Network}
  \country{Toronto, Canada}}
\affiliation{%
  \institution{Department of Medical Biophysics, University of Toronto}
  \country{Toronto, Canada}}
\affiliation{%
  \institution{Department of Computer Science, University of Toronto}
  \country{Toronto, Canada}}
\affiliation{%
  \institution{Vector Institute for Artificial Intelligence}
  \country{Toronto, Canada}}

\author{Merel Huisman}
\affiliation{%
  \institution{Department of Radiology and Nuclear Medicine, Radboud University Medical Center}
  \country{Nijmegen, The Netherlands}
}

\author{Fabian Isensee}
\affiliation{%
  \institution{German Cancer Research Center (DKFZ) Heidelberg, Division of Medical Image Computing and HI Applied Computer Vision Lab}
  \country{Germany}
}

\author{Pierre Jannin}
\affiliation{%
  \institution{Laboratoire Traitement du Signal et de l’Image – UMR\_S 1099, Université de Rennes 1}
  \country{Rennes, France}
}
\affiliation{%
  \institution{INSERM}
  \country{Paris Cedex, France}
}

\author{Charles E. Kahn}
\affiliation{%
  \institution{Department of Radiology and Institute for Biomedical Informatics, University of Pennsylvania}
  \country{Philadelphia, PA, USA}
}

\author{Dagmar Kainmueller}
\affiliation{%
  \institution{Max-Delbrück Center for Molecular Medicine in the Helmholtz Association (MDC), Biomedical Image Analysis and HI Helmholtz Imaging}
  \country{Berlin, Germany}
}
\affiliation{%
  \institution{University of Potsdam, Digital Engineering Faculty}
  \country{Potsdam, Germany}
}

\author{Bernhard Kainz}
\affiliation{%
  \institution{Department of Computing, Faculty of Engineering, Imperial College London}
  \country{London, UK}
}
\affiliation{%
  \institution{Department AIBE, Friedrich-Alexander-Universität (FAU)}
  \country{Erlangen-Nürnberg, Germany}
}

\author{Alexandros Karargyris}
\affiliation{%
  \institution{IHU Strasbourg}
  \country{Strasbourg, France}
}

\author{Alan Karthikesalingam}
\affiliation{%
  \institution{Google Health DeepMind}
  \country{London, UK}
}

\author{A. Emre Kavur}
\affiliation{%
  \institution{German Cancer Research Center (DKFZ) Heidelberg, Division of Intelligent Medical Systems, Division of Medical Image Computing, HI Applied Computer Vision Lab}
  \country{Germany}
}

\author{Hannes Kenngott}
\affiliation{%
  \institution{Department of General, Visceral and Transplantation Surgery, Heidelberg University Hospital}
  \country{Heidelberg, Germany}
}

\author{Jens Kleesiek}
\affiliation{%
  \institution{Institute for AI in Medicine, University Medicine Essen}
  \country{Essen, Germany}
}

\author{Andreas Kleppe}
\affiliation{%
  \institution{Institute for Cancer Genetics and Informatics, Oslo University Hospital}
  \country{Oslo, Norway}
}
\affiliation{%
  \institution{Department of Informatics, University of Oslo}
  \country{Oslo, Norway}
}

\author{Sven Koehler}
\affiliation{%
 \institution{Department of Internal Medicine III and Department of Cardiac Surgery, Heidelberg University Hospital}
 \country{Heidelberg, Germany}}
 
 \author{Florian Kofler}
\affiliation{%
  \institution{Helmholtz AI}
  \country{München, Germany}
}

\author{Annette Kopp-Schneider}
\affiliation{%
  \institution{German Cancer Research Center (DKFZ) Heidelberg, Division of Biostatistics}
  \country{Germany}
}
\author{Thijs Kooi}
\affiliation{%
  \institution{Lunit Inc}
  \country{Seoul, South Korea}
}

\author{Michal Kozubek}
\affiliation{%
  \institution{Centre for Biomedical Image Analysis and Faculty of Informatics, Masaryk University}
  \country{Brno, Czech Republic}
}

\author{Anna Kreshuk}
\affiliation{%
  \institution{Cell Biology and Biophysics Unit, European Molecular Biology Laboratory (EMBL)}
  \country{Heidelberg, Germany}
}

\author{Tahsin Kurc}
\affiliation{%
  \institution{Department of Biomedical Informatics, Stony Brook University}
  \country{Stony Brook, NY, USA}
}

\author{Bennett A. Landman}
\affiliation{%
  \institution{Electrical Engineering, Vanderbilt University}
  \country{Nashville, TN, USA}
}

\author{Geert Litjens}
\affiliation{%
  \institution{Department of Pathology, Radboud University Medical Center}
  \country{Nijmegen, The Netherlands}
}

\author{Amin Madani}
\affiliation{%
  \institution{Department of Surgery, University Health Network}
  \country{Philadelphia, PA, Canada}
}

\author{Klaus Maier-Hein}
\affiliation{%
  \institution{German Cancer Research Center (DKFZ) Heidelberg, Division of Medical Image Computing and HI Helmholtz Imaging,}
  \country{Germany}
}
\affiliation{%
  \institution{Pattern Analysis and Learning Group, Department of Radiation Oncology, Heidelberg University Hospital}
  \country{Heidelberg, Germany}
}

\author{Anne L. Martel}
\affiliation{%
  \institution{Physical Sciences, Sunnybrook Research Institute}
  \country{Toronto, Canada}
}
\affiliation{%
  \institution{Department of Medical Biophysics, University of Toronto}
  \country{Toronto, Canada}
}

\author{Peter Mattson}
\affiliation{%
  \institution{Google}
  \country{Mountain View, CA 94043, USA}
}

\author{Erik Meijering}
\affiliation{%
  \institution{School of Computer Science and Engineering, University of New South Wales}
  \country{Sydney, Kensington, Australia}
}

\author{Bjoern Menze}
\affiliation{%
  \institution{Department of Quantitative Biomedicine, University of Zurich}
  \country{Zurich, Switzerland}
}

\author{David Moher}
\affiliation{%
  \institution{Centre for Journalology, Clinical Epidemiology Program, Ottawa Hospital Research Institute}
  \country{Ottawa, Canada}
}
\affiliation{%
  \institution{School of Epidemiology and Public Health, Faculty of Medicine, University of Ottawa}
  \country{Ottawa, Canada}
}

\author{Karel G.M. Moons}
\affiliation{%
  \institution{Julius Center for Health Sciences and Primary Care, UMC Utrecht, Utrecht University}
  \country{Utrecht, The Netherlands}
}

\author{Henning Müller}
\affiliation{%
  \institution{Information Systems Institute, University of Applied Sciences Western Switzerland (HES-SO)}
  \country{Sierre, Switzerland}
}
\affiliation{%
  \institution{Medical Faculty, University of Geneva}
  \country{Geneva, Switzerland}
}

\author{Brennan Nichyporuk}
\affiliation{%
  \institution{MILA (Québec Artificial Intelligence Institute)}
  \country{Montréal, Canada}}

\author{Felix Nickel}
\affiliation{%
  \institution{Department of General, Visceral and Thoracic Surgery, University Medical Center Hamburg-Eppendorf}
  \country{Hamburg, Germany}
}

\author{M. Alican Noyan}
\affiliation{%
  \institution{Ipsumio}
  \country{Eindhoven, The Netherlands}
}

\author{Jens Petersen}
\affiliation{%
  \institution{German Cancer Research Center (DKFZ) Heidelberg, Division of Medical Image Computing}
  \country{Germany}
}

\author{Gorkem Polat}
\affiliation{%
  \institution{Graduate School of Informatics, Middle East Technical University}
  \country{Ankara, Turkey}
}

\author{Susanne M. Rafelski}
\affiliation{%
  \institution{Allen Institute for Cell Science}
  \country{Seattle, WA, USA}}

\author{Nasir Rajpoot}
\affiliation{%
  \institution{Tissue Image Analytics Laboratory, Department of Computer Science, University of Warwick}
  \department{Tissue Image Analytics Laboratory, Department of Computer Science}
  \country{Coventry, UK}
}

\author{Mauricio Reyes}
\affiliation{%
  \institution{ARTORG Center for Biomedical Engineering Research, University of Bern}
  \country{Bern, Switzerland}
}
\affiliation{%
  \institution{Department of Radiation Oncology, University Hospital Bern, University of Bern}
  \country{Bern, Switzerland}
}

\author{Nicola Rieke}
\affiliation{%
  \institution{NVIDIA GmbH}
  \country{München, Germany}
}

\author{Michael A. Riegler}
\affiliation{%
  \institution{Simula Metropolitan Center for Digital Engineering}
  \country{Oslo, Norway}
}
\affiliation{%
  \institution{UiT The Arctic University of Norway}
  \country{romsø, Norway}
}

\author{Hassan Rivaz}
\affiliation{%
  \institution{Department of Electrical and Computer Engineering, Concordia University}
  \country{Montreal, Canada}
}

\author{Julio Saez-Rodriguez}
\affiliation{%
  \institution{Institute for Computational Biomedicine, Heidelberg University}
  \country{Heidelberg, Germany}
}
\affiliation{%
  \institution{Faculty of Medicine, Heidelberg University Hospital}
  \country{Heidelberg, Germany}
}

\author{Clara I. Sánchez}
\affiliation{%
  \institution{Informatics Institute, Faculty of Science, University of Amsterdam}
  \country{Amsterdam, The Netherlands}
}

\author{Julien Schroeter}
\affiliation{%
  \institution{Centre for Intelligent Machines, McGill University}
  \country{Montreal, Canada}
}

\author{Anindo Saha}
\affiliation{%
  \institution{Diagnostic Image Analysis Group, Radboud University Medical Center}
  \country{Nijmegen, The Netherlands}
}

\author{M. Alper Selver}
\affiliation{%
  \institution{Electrical and Electronics Engineering Dept, Dokuz Eylül University}
  \country{Izmir, Turkey}
}

\author{Lalith Sharan}
\affiliation{%
 \institution{Department of Internal Medicine III and Department of Cardiac Surgery, Heidelberg University Hospital}
 \city{Heidelberg}
 \country{Germany}}

\author{Shravya Shetty}
\affiliation{%
  \institution{Google Health, Google}
  \country{Palo Alto, CA, USA}
}

\author{Maarten van Smeden}
\affiliation{%
  \institution{Julius Center for Health Sciences and Primary Care, University Medical Center Utrecht}
  \country{Utrecht, The Netherlands}
}

\author{Bram Stieltjes}
\affiliation{%
  \institution{Department of Radiology, University Hospital of Basel}
  \country{Basel, Switzerland}
}

\author{Ronald M. Summers}
\affiliation{%
  \institution{National Institutes of Health Clinical Center}
  \country{Bethesda, MD, USA}
}

\author{Abdel A. Taha}
\affiliation{%
\institution{Institute of Information Systems Engineering, TU Wien}
  \country{ Vienna, Austria}
}

\author{Aleksei Tiulpin}
\affiliation{%
\institution{Research Unit of Health Sciences and Technology, Faculty of Medicine, University of Oulu}
  \country{Oulu, Finland}
}
\affiliation{%
\institution{Neurocenter Oulu, Oulu University Hospital}
  \country{Oulu, Finland}
}

\author{Sotirios A. Tsaftaris}
\affiliation{%
  \institution{School of Engineering, The University of Edinburgh}
  \country{Edinburgh, Scotland}
}

\author{Ben Van Calster}
\affiliation{%
  \institution{Department of Development and Regeneration and EPI-centre, KU Leuven}
  \country{Leuven, Belgium}
}
\affiliation{%
  \institution{Department of Biomedical Data Sciences, Leiden University Medical Center}
  \country{Leiden, The Netherlands}
}

\author{Gaël Varoquaux}
\affiliation{%
  \institution{Parietal project team, INRIA Saclay-Île de France}
  \country{Palaiseau, France}
}

\author{Manuel Wiesenfarth}
\affiliation{%
  \institution{German Cancer Research Center (DKFZ) Heidelberg, Division of Biostatistics}
  \country{Germany}
}

\author{Ziv R. Yaniv}
\affiliation{%
  \department{National Institute of Allergy and Infectious Diseases, National Institutes of Health, Bioinformatics and Computational Bioscience Branch}
  \country{Bethesda, Maryland, USA}
}

\author{Paul Jäger}
\affiliation{%
  \institution{German Cancer Research Center (DKFZ) Heidelberg, Interactive Machine Learning Group and HI Helmholtz Imaging}
  \country{Germany}
}

\author{Lena Maier-Hein}
\affiliation{%
  \institution{German Cancer Research Center (DKFZ) Heidelberg, Division of Intelligent Medical Systems and HI Helmholtz Imaging}
  \country{Germany}
}
\affiliation{%
  \institution{Faculty of Mathematics and Computer Science and Medical Faculty, Heidelberg University}
  \country{Heidelberg, Germany}
}
\affiliation{%
  \institution{National Center for Tumor Diseases (NCT), NCT Heidelberg, a partnership between DKFZ and University Medical Center Heidelberg}
  \country{Germany}
}

\renewcommand{\shortauthors}{A. Reinke/M.D. Tizabi, et al.}

\addtocontents{toc}{\protect\setcounter{tocdepth}{-1}}

\begin{abstract}
\textbf{Abstract:} While the importance of automatic image analysis is continuously increasing, recent meta-research revealed major flaws with respect to algorithm validation. Performance metrics are particularly key for meaningful, objective, and transparent performance assessment and validation
of the used automatic algorithms, but relatively little attention has been given to the practical pitfalls when using specific metrics for a given image analysis task. These are typically related to (1) the disregard of inherent metric properties, such as the behaviour in the presence of class imbalance or small target structures, (2) the disregard of inherent data set properties, such as the non-independence of the test cases, and (3) the disregard of the actual biomedical domain interest that the metrics should reflect. This living dynamically document has the purpose to illustrate important limitations of performance metrics commonly applied in the field of image analysis. In this context, it focuses on biomedical image analysis problems that can be phrased as image-level classification, semantic segmentation, instance segmentation, or object detection task. The current version is based on a Delphi process on metrics conducted by an international consortium of image analysis experts from more than 60 institutions worldwide.
\end{abstract}


\maketitle
\tableofcontents
\addtocontents{toc}{\protect\setcounter{tocdepth}{2}}

\setlength{\parskip}{0.5em}
\newpage
\section{Purpose}
Validation of biological and medical image analysis algorithms is of the utmost importance for making scientific progress and for translating methodological research into practice. Validation metrics\footnote{not to be confused with distance metrics in the strict mathematical sense}, the measures according to which performance of algorithms is quantified, constitute a core component of validation design. While metrics can measure various quantities of interest, including speed, memory consumption or carbon footprint, most metrics applied today are \textit{reference-based metrics}, which have the purpose of measuring the agreement of an algorithm prediction with a given reference. The reference, in turn, serves as an approximation of the (typically unknown) ground truth. 

Knowing the properties of metrics in use and making educated choices is essential for meaningful and reliable validation in image analysis. Although several papers highlight specific strengths and weaknesses of common metrics \citep{kofler2021DICE, gooding2018comparative, vaassen2020evaluation, konukoglu2012discriminative, margolin2014evaluate}, an international survey~\cite{maier2018rankings} revealed the choice of inappropriate metrics as one of the core problems related to performance assessment in medical image analysis. Similar problems are present in other imaging domains~\cite{honauer2015hci, correia2006video}. 
Under the umbrella of Helmholtz Imaging (HI)\footnote{\ULurl{https://www.helmholtz-imaging.de/}}, three international initiatives have now joined forces to address these issues: the Biomedical Image Analysis Challenges (BIAS) initiative\footnote{\ULurl{https://www.dkfz.de/en/cami/research/topics/biasInitiative.html}}, the Medical Image Computing and Computer Assisted Interventions (MICCAI) Society's special interest group on challenges\footnote{\ULurl{https://miccai.org/index.php/special-interest-groups/challenges/}}, as well as the benchmarking working group of the MONAI framework\footnote{\ULurl{https://monai.io/}}.
A core mission is to provide researchers with guidelines and tools to choose the performance metrics in a problem- and context-aware manner. This dynamically updated document aims to illustrate important pitfalls and drawbacks of metrics commonly applied in the field of image analysis. The current version is based on a Delphi process on metrics conducted with an international consortium of medical image analysis experts. A Delphi process is a multi-stage survey process designed to pool the knowledge of several experts to arrive at a consensus decision~\cite{brown1968delphi}.

\begin{figure}[H]
\begin{tcolorbox}[title= Problem categories addressed by this paper, colback=white]
    \centering
    \includegraphics[width=1\linewidth]{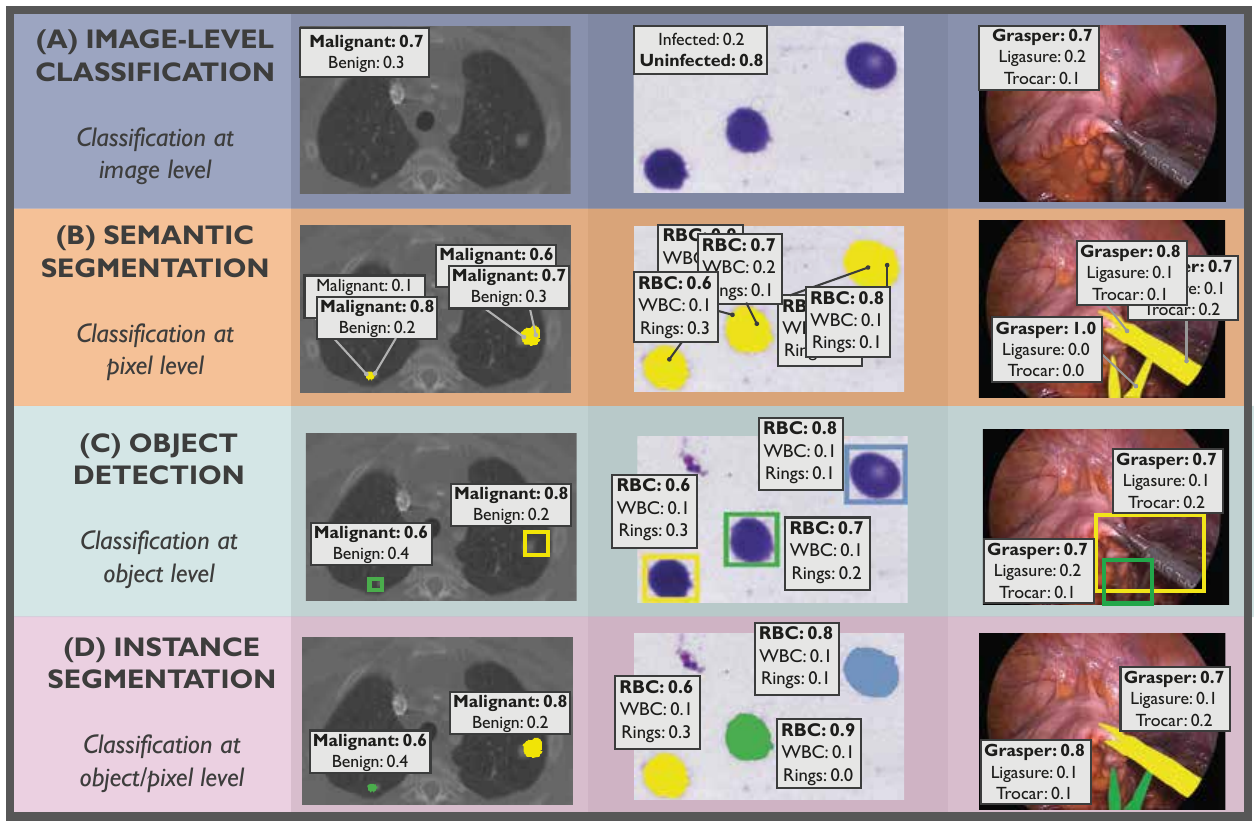}
    \caption{Problem categories covered in this paper and illustrated for three different application domains: radiology (left), cell biology (middle), surgery (right). The common denominator of the underlying research problems is the fact that they can be interpreted as classification tasks. The classification occurs at various scales, image level, object level and/or pixel level. Each of these tasks assigns a class label to the image or (multiple) components of it. (A)~The \textbf{image-level classification} task involves assigning a class label to the whole image. (B)~The \textbf{semantic segmentation} task involves assigning a class label to each individual pixel. (C)~The \textbf{object detection} task assigns a class label to identified objects. (D)~The \textbf{instance segmentation} task assigns a label to identified objects made up of multiple pixels. Gray boxes show the predicted class probabilities on image level, pixel level or object level. The class with the highest probability is shown in bold. Further abbreviations: Red Blood Cell (RBC), White Blood Cell (WBC).}
    \label{fig:problem-categories}
\end{tcolorbox}
\end{figure}

The Delphi consortium focused on problems reporting biomedical research that can be phrased as \textbf{image-level classification}, \textbf{semantic segmentation}, \textbf{instance segmentation} or \textbf{object detection} (Figure~\ref{fig:problem-categories}). Essentially, these can all be interpreted as a classification task at different scales and thus share many aspects in terms of validation (Figure~\ref{fig:metric-families}). For example, an object detection task can be interpreted as an object-/instance-level classification task, while a segmentation task can be interpreted as a pixel-level classification task. We will refer to these four different task types as \textit{problem categories}. Please note that we will use the term "pixel" even for three-dimensional (or n-dimensional) images for increased readability instead of referring to "pixels/voxels". Most of the examples are shown for two-dimensional images and can be translated to the n-dimensional case.

The manuscript is structured as follows. As a foundation, we first review the most commonly applied metrics for the problem categories addressed in this paper (Sec.~\ref{sec:fundamentals}). Since a common problem in the biomedical image analysis community is the selection of metrics from the wrong problem category, Sec.~\ref{sec:underlying-task} highlights pitfalls relevant in this context. The following sections then present pitfalls for image-level classification (Sec.~\ref{sec:classification}), image segmentation, including semantic and instance segmentation (Sec.~\ref{sec:segmentation}), and object detection, including instance segmentation (Sec.~\ref{sec:detection}). Finally, cross-topic pitfalls are highlighted (Sec.~\ref{sec:aggregation-combination}). An overview of all figures is presented in Table~\ref{tab:properties-pitfalls}.


\newpage
\begingroup
\small
\begin{longtable}{p{11cm} c}
    \caption{Overview of figures on pitfalls related to metrics classified into (1) pitfalls due to category-metric mismatch, (2) category-specific pitfalls and (3) cross-topic pitfalls. For each illustration, the corresponding figure and page number is given. Please note that the pitfalls are typically illustrated for only one or two problem categories but often also apply to other problem categories, as indicated in the table.}
    \label{tab:properties-pitfalls}\\
    \toprule
\textbf{Source of potential pitfall} & \textbf{Figure(s)} \\ \midrule

 \multicolumn{2}{c}{\textbf{\textit{Pitfalls due to category-metric mismatch}}}\\
\rowcolor{Gray2} Mismatch: Semantic segmentation $\leftrightarrow$ object detection & Fig.~\ref{fig:DSC-detection} (Page~\pageref{fig:DSC-detection}) \\ 
\rowcolor{Gray1} Mismatch: Semantic $\leftrightarrow$ instance segmentation & Fig.~\ref{fig:touching} (Page~\pageref{fig:touching})  \\
\rowcolor{Gray2} Mismatch: Image-level classification $\leftrightarrow$ object detection & Fig.~\ref{fig:roc} (Page~\pageref{fig:roc}) \\
\rowcolor{Gray1} No matching problem category & Fig.~\ref{fig:context-ratio} (Page~\pageref{fig:context-ratio})  \\
\midrule

\multicolumn{2}{c}{\textbf{\textit{Pitfalls in image-level classification}}}\\ \midrule
\rowcolor{Gray2} High class imbalance &  Fig.~\ref{fig:class-imbalance} (Page~\pageref{fig:class-imbalance}) \\ 
\rowcolor{Gray2} &  Fig.~\ref{fig:misleading-ba-j} (Page~\pageref{fig:misleading-ba-j}) \\ 
\rowcolor{Gray2} &  Fig.~\ref{fig:metric-landscapes} (Page~\pageref{fig:metric-landscapes}) \\ 
\rowcolor{Gray1} More than two classes available & Fig.~\ref{fig:multi-class} (Page~\pageref{fig:multi-class}) \\ 
\rowcolor{Gray2} Unequal importance of class confusions & Fig.~\ref{fig:unequal-class-confusions} (Page~\pageref{fig:unequal-class-confusions}) \\
\rowcolor{Gray1} Unequal importance of classes & Fig.~\ref{fig:unequal-class} (Page~\pageref{fig:unequal-class}) \\
\rowcolor{Gray2} Interdependencies between classes & Fig.~\ref{fig:multi-class-interdependencies} (Page~\pageref{fig:multi-class-interdependencies})\\ 
\rowcolor{Gray1} Stratification based on meta-information & Fig.~\ref{fig:stratification-gender} (Page~\pageref{fig:stratification-gender})  \\ 
\rowcolor{Gray2} Importance of cost-benefit analysis & Fig.~\ref{fig:cost-benefit} (Page~\pageref{fig:cost-benefit}) \\ 
\rowcolor{Gray1} Definition of class labels & Fig.~\ref{fig:positive-class} (Page~\pageref{fig:positive-class})  \\ 
\rowcolor{Gray2} Prevalence dependency & Fig.~\ref{fig:prevalence} (Page~\pageref{fig:prevalence}) \\ 
\rowcolor{Gray2} & Fig.~\ref{fig:prevalence-dependency} (Page~\pageref{fig:prevalence-dependency}) \\ 
\rowcolor{Gray2} & Fig.~\ref{fig:prevalence-dependency-rankings} (Page~\pageref{fig:prevalence-dependency-rankings}) \\ 
\rowcolor{Gray1} Importance of confidence awareness & Fig.~\ref{fig:calibration} (Page~\pageref{fig:calibration}) \\ 
\rowcolor{Gray1}  & Fig.~\ref{fig:brierscore-ordinal} (Page~\pageref{fig:brierscore-ordinal}) \\ 
\rowcolor{Gray1}  & Fig.~\ref{fig:ece-mce-bins} (Page~\pageref{fig:ece-mce-bins}) \\ 
\rowcolor{Gray1}  & Fig.~\ref{fig:ece-pitfalls} (Page~\pageref{fig:ece-pitfalls}) \\ 
\rowcolor{Gray2} Presence of ordinal classes & Fig.~\ref{fig:ordinal-grading} (Page~\pageref{fig:ordinal-grading}) \\ 
\rowcolor{Gray2} & Fig.~\ref{fig:ordinal-grading-label-shift} (Page~\pageref{fig:ordinal-grading-label-shift}) \\ 
\rowcolor{Gray1} Same metric scores for different confusion matrices & Fig.~\ref{fig:lr+ba} (Page~\pageref{fig:lr+ba}) \\ 
\rowcolor{Gray2} Upper and lower bounds not equally obtainable & Fig.~\ref{fig:kappa-max} (Page~\pageref{fig:kappa-max}) \\ 
\rowcolor{Gray2}  & Fig.~\ref{fig:lack-bounds} (Page~\pageref{fig:lack-bounds}) \\ 
\rowcolor{Gray1} Determination of a global threshold for all classes & Fig.~\ref{fig:auroc-classes-threshold} (Page~\pageref{fig:auroc-classes-threshold})\\
\rowcolor{Gray2} Model bias & Fig.~\ref{fig:auroc-bias} (Page~\pageref{fig:auroc-bias})\\
\rowcolor{Gray1} Small sample sizes & Fig.~\ref{fig:auroc-small-sample-sizes} (Page~\pageref{fig:auroc-small-sample-sizes})\\
\rowcolor{Gray2} Multi-threshold metric-related properties \scriptsize \textit{(here: pitfalls illustrated for object detection problems)}&  Fig.~\ref{fig:AP-small-conf-changes} (Page~\pageref{fig:AP-small-conf-changes}) \\
\rowcolor{Gray2} & Fig.~\ref{fig:AP-conf-not-important} (Page~\pageref{fig:AP-conf-not-important}) \\
\rowcolor{Gray2} & Fig.~\ref{fig:AP-FP-tail} (Page~\pageref{fig:AP-FP-tail})\\
\midrule

\multicolumn{2}{c}{\textbf{\textit{Pitfalls in semantic segmentation}}}\\ \midrule
\rowcolor{Gray2} Small size of structures relative to pixel size &  Fig.~\ref{fig:DSC-small} (Page~\pageref{fig:DSC-small})  \\
\rowcolor{Gray2}  &  Fig.~\ref{fig:clDice-small} (Page~\pageref{fig:clDice-small}) \\
\rowcolor{Gray2} \scriptsize \textit{(here: pitfall illustrated for object detection problems)}& Fig.~\ref{fig:boundary-mask-iou} (Page~\pageref{fig:boundary-mask-iou}) \\
\rowcolor{Gray1} High variability of structure sizes & Fig.~\ref{fig:high-variability} (Page~\pageref{fig:high-variability}) \\
\rowcolor{Gray1} & Fig.~\ref{fig:masd-assd} (Page~\pageref{fig:masd-assd}) \\
\rowcolor{Gray2} Complex shape of structures & Fig.~\ref{fig:DSC-shapes} (Page~\pageref{fig:DSC-shapes}) \\
\rowcolor{Gray2} & Fig.~\ref{fig:complex-shapes} (Page~\pageref{fig:complex-shapes}) \\
\rowcolor{Gray2} & Fig.~\ref{fig:length} (Page~\pageref{fig:length}) \\
\rowcolor{Gray1} Particular importance of structure volume & Fig.~\ref{fig:volume} (Page~\pageref{fig:volume}) \\
\rowcolor{Gray2} Particular importance of structure center & Fig.~\ref{fig:center} (Page~\pageref{fig:center})  \\
\rowcolor{Gray1} Particular importance of structure boundaries &  Fig.~\ref{fig:outline} (Page~\pageref{fig:outline}) \\
\rowcolor{Gray1} &  Fig.~\ref{fig:boundary-is-ss} (Page~\pageref{fig:boundary-is-ss}) \\
\rowcolor{Gray1} \scriptsize \textit{(here: pitfall illustrated for object detection problems)} & Fig.~\ref{fig:boundary-mask-iou} (Page~\pageref{fig:boundary-mask-iou}) \\
\rowcolor{Gray1}\scriptsize \textit{(here: pitfall illustrated for object detection problems)} & Fig.~\ref{fig:boundary-iou} (Page~\pageref{fig:boundary-iou}) \\
\rowcolor{Gray2} Possibility of multiple labels per unit & Fig.~\ref{fig:multi-labels} (Page~\pageref{fig:multi-labels}) \\
\rowcolor{Gray1} Noisy reference standard & Fig.~\ref{fig:low-quality} (Page~\pageref{fig:low-quality}) \\
\rowcolor{Gray2} Possibility of outliers in reference annotation & Fig.~\ref{fig:DSC-artifact} (Page~\pageref{fig:DSC-artifact}) \\
\rowcolor{Gray1} Possibility of reference or prediction without target structure(s) & Fig.~\ref{fig:empty} (Page~\pageref{fig:empty}) \\ 
\rowcolor{Gray2} Dependency on image resolution & Fig.~\ref{fig:DSC-grid-size} (Page~\pageref{fig:DSC-grid-size}) \\
\rowcolor{Gray1} Over- \textit{vs.} undersegmentation & Fig.~\ref{fig:DSC-overunder} (Page~\pageref{fig:DSC-overunder}) \\
\rowcolor{Gray2} Choice of global decision threshold & Fig.~\ref{fig:seg-threshold} (Page~\pageref{fig:seg-threshold}) \\
\rowcolor{Gray1} High class imbalance \scriptsize \textit{(here: pitfall illustrated for image-level classification problems)} &  Fig.~\ref{fig:class-imbalance} (Page~\pageref{fig:class-imbalance}) \\ 
\rowcolor{Gray1} &  Fig.~\ref{fig:misleading-ba-j} (Page~\pageref{fig:misleading-ba-j}) \\ 
\rowcolor{Gray1} &  Fig.~\ref{fig:metric-landscapes} (Page~\pageref{fig:metric-landscapes}) \\ 
\rowcolor{Gray2} More than two classes available \scriptsize \textit{(here: pitfall illustrated for image-level classification problems)} & Fig.~\ref{fig:multi-class} (Page~\pageref{fig:multi-class}) \\ 
\rowcolor{Gray1} Unequal importance of class confusions \scriptsize \textit{(here: pitfall illustrated for image-level classification problems)} & Fig.~\ref{fig:unequal-class-confusions} (Page~\pageref{fig:unequal-class-confusions}) \\
\rowcolor{Gray2} Unequal importance of classes \scriptsize \textit{(here: pitfall illustrated for image-level classification problems)} & Fig.~\ref{fig:unequal-class} (Page~\pageref{fig:unequal-class}) \\
\rowcolor{Gray1} Interdependencies between classes \scriptsize \textit{(here: pitfall illustrated for image-level classification problems)} & Fig.~\ref{fig:multi-class-interdependencies} (Page~\pageref{fig:multi-class-interdependencies})\\ 
\midrule
\multicolumn{2}{c}{\textbf{\textit{Pitfalls in object detection}}}\\ \midrule
\rowcolor{Gray1} Mathematical implications of center-based localization criteria &  Fig.~\ref{fig:hit-criteria} (Page~\pageref{fig:hit-criteria}) \\
\rowcolor{Gray1} &  Fig.~\ref{fig:center-vs-point} (Page~\pageref{fig:center-vs-point}) \\
\rowcolor{Gray2} Mathematical implications of \textit{IoU}-based localization criterion &  Fig.~\ref{fig:bb-2d-3d} (Page~\pageref{fig:bb-2d-3d}) \\
\rowcolor{Gray1} Mathematical implications of the choice of assignment strategies &  Fig.~\ref{fig:od-assignment} (Page~\pageref{fig:od-assignment}) \\
\rowcolor{Gray1}  &  Fig.~\ref{fig:od-ior} (Page~\pageref{fig:od-ior}) \\
\rowcolor{Gray2} Type of the provided annotations & Fig.~\ref{fig:disconnected} (Page~\pageref{fig:disconnected}) \\
\rowcolor{Gray1} Effect of small structures on localization criterion & Fig.~\ref{fig:boundary-mask-iou} (Page~\pageref{fig:boundary-mask-iou}) \\
\rowcolor{Gray1}\scriptsize \textit{(here: pitfall illustrated for semantic segmentation problems)}  & Fig.~\ref{fig:DSC-small} (Page~\pageref{fig:DSC-small}) \\
\rowcolor{Gray2} Perfect \textit{Boundary IoU} for imperfect prediction & Fig.~\ref{fig:boundary-iou} (Page~\pageref{fig:boundary-iou}) \\
\rowcolor{Gray1} Possibility of reference or prediction without target structure(s) & Fig.~\ref{fig:empty-ref-pred-od} (Page~\pageref{fig:empty-ref-pred-od}) \\ 
\rowcolor{Gray1} \scriptsize \textit{(here: pitfall illustrated for semantic segmentation problems)}& Fig.~\ref{fig:empty} (Page~\pageref{fig:empty}) \\ 
\rowcolor{Gray2} \textit{Average Precision} \textit{vs.} \textit{Free-response ROC} score & Fig.~\ref{fig:ap-froc} (Page~\pageref{fig:ap-froc}) \\
\rowcolor{Gray1} \textit{Free-response ROC score} is not standardized & Fig.~\ref{fig:froc-no-standard} (Page~\pageref{fig:froc-no-standard}) \\
\rowcolor{Gray2} Multi-threshold metric-related properties &  Fig.~\ref{fig:AP-small-conf-changes} (Page~\pageref{fig:AP-small-conf-changes}) \\
\rowcolor{Gray2} & Fig.~\ref{fig:AP-conf-not-important} (Page~\pageref{fig:AP-conf-not-important}) \\
\rowcolor{Gray2} & Fig.~\ref{fig:AP-FP-tail} (Page~\pageref{fig:AP-FP-tail})\\ 
\rowcolor{Gray1} High class imbalance \scriptsize \textit{(here: pitfall illustrated for image-level classification problems)} &  Fig.~\ref{fig:class-imbalance} (Page~\pageref{fig:class-imbalance}) \\ 
\rowcolor{Gray1} &  Fig.~\ref{fig:misleading-ba-j} (Page~\pageref{fig:misleading-ba-j}) \\ 
\rowcolor{Gray1} &  Fig.~\ref{fig:metric-landscapes} (Page~\pageref{fig:metric-landscapes}) \\ 
\rowcolor{Gray2} More than two classes available \scriptsize \textit{(here: pitfall illustrated for image-level classification problems)} & Fig.~\ref{fig:multi-class} (Page~\pageref{fig:multi-class}) \\ 
\rowcolor{Gray1} Unequal importance of class confusions \scriptsize \textit{(here: pitfall illustrated for image-level classification problems)} & Fig.~\ref{fig:unequal-class-confusions} (Page~\pageref{fig:unequal-class-confusions}) \\
\rowcolor{Gray2} Unequal importance of classes \scriptsize \textit{(here: pitfall illustrated for image-level classification problems)} & Fig.~\ref{fig:unequal-class} (Page~\pageref{fig:unequal-class}) \\
\rowcolor{Gray1} Interdependencies between classes \scriptsize \textit{(here: pitfall illustrated for image-level classification problems)} & Fig.~\ref{fig:multi-class-interdependencies} (Page~\pageref{fig:multi-class-interdependencies})\\ 
\rowcolor{Gray2} Particular importance of structure center \scriptsize \textit{(here: pitfall illustrated for semantic segmentation problems)} & Fig.~\ref{fig:center} (Page~\pageref{fig:center}) \\
\rowcolor{Gray1} High variability of structure sizes \scriptsize \textit{(here: pitfall illustrated for semantic segmentation problems)} & Fig.~\ref{fig:high-variability} (Page~\pageref{fig:high-variability}) \\
\rowcolor{Gray2} \scriptsize \textit{(here: pitfall illustrated for semantic segmentation problems)} & Fig.~\ref{fig:masd-assd} (Page~\pageref{fig:masd-assd}) \\
\rowcolor{Gray1} Complex shape of structures \scriptsize \textit{(here: pitfall illustrated for semantic segmentation problems)}& Fig.~\ref{fig:DSC-shapes} (Page~\pageref{fig:DSC-shapes}) \\
\rowcolor{Gray1} & Fig.~\ref{fig:complex-shapes} (Page~\pageref{fig:complex-shapes}) \\
\rowcolor{Gray1} & Fig.~\ref{fig:length} (Page~\pageref{fig:length}) \\
\rowcolor{Gray2} Possibility of multiple labels per unit \scriptsize \textit{(here: pitfall illustrated for semantic segmentation problems)} & Fig.~\ref{fig:multi-labels} (Page~\pageref{fig:multi-labels}) \\
\rowcolor{Gray1} Noisy reference standard \scriptsize \textit{(here: pitfall illustrated for semantic segmentation problems)}& Fig.~\ref{fig:low-quality} (Page~\pageref{fig:low-quality}) \\
\rowcolor{Gray2} Possibility of outliers in reference annotation \scriptsize \textit{(here: pitfall illustrated for semantic segmentation problems)}& Fig.~\ref{fig:DSC-artifact} (Page~\pageref{fig:DSC-artifact}) \\
\rowcolor{Gray1} Choice of global decision threshold \scriptsize \textit{(here: pitfall illustrated for semantic segmentation problems)} & Fig.~\ref{fig:seg-threshold} (Page~\pageref{fig:seg-threshold}) \\

\midrule
\multicolumn{2}{c}{\textbf{\textit{Pitfalls in instance segmentation}}}\\ \midrule
\rowcolor{Gray1} Segmentation quality \textit{vs.} detection quality &  Fig.~\ref{fig:pq-sq-vs-dq} (Page~\pageref{fig:pq-sq-vs-dq})  \\
\rowcolor{Gray2} Small size of structures relative to pixel size &  Fig.~\ref{fig:DSC-small} (Page~\pageref{fig:DSC-small})  \\
\rowcolor{Gray2}  &  Fig.~\ref{fig:clDice-small} (Page~\pageref{fig:clDice-small}) \\
\rowcolor{Gray2} & Fig.~\ref{fig:boundary-mask-iou} (Page~\pageref{fig:boundary-mask-iou}) \\
\rowcolor{Gray1} High variability of structure sizes & Fig.~\ref{fig:high-variability} (Page~\pageref{fig:high-variability})  \\
\rowcolor{Gray1} & Fig.~\ref{fig:masd-assd} (Page~\pageref{fig:masd-assd})  \\
\rowcolor{Gray2} Complex shape of structures & Fig.~\ref{fig:DSC-shapes} (Page~\pageref{fig:DSC-shapes}) \\
\rowcolor{Gray2} & Fig.~\ref{fig:complex-shapes} (Page~\pageref{fig:complex-shapes}) \\
\rowcolor{Gray2} & Fig.~\ref{fig:length} (Page~\pageref{fig:length}) \\
\rowcolor{Gray1} Particular importance of structure volume & Fig.~\ref{fig:volume} (Page~\pageref{fig:volume}) \\
\rowcolor{Gray2} Particular importance of structure center & Fig.~\ref{fig:center} (Page~\pageref{fig:center})  \\
\rowcolor{Gray1} Particular importance of structure boundaries &  Fig.~\ref{fig:boundary-mask-iou} (Page~\pageref{fig:boundary-mask-iou}) \\

\rowcolor{Gray1} & Fig.~\ref{fig:boundary-iou} (Page~\pageref{fig:boundary-iou}) \\
\rowcolor{Gray1} & Fig.~\ref{fig:outline} (Page~\pageref{fig:outline}) \\
\rowcolor{Gray1} &  Fig.~\ref{fig:boundary-is-ss} (Page~\pageref{fig:boundary-is-ss}) \\
\rowcolor{Gray2} Possibility of multiple labels per unit & Fig.~\ref{fig:multi-labels} (Page~\pageref{fig:multi-labels}) \\
\rowcolor{Gray1} Noisy reference standard & Fig.~\ref{fig:low-quality} (Page~\pageref{fig:low-quality}) \\
\rowcolor{Gray2} Possibility of outliers in reference annotation & Fig.~\ref{fig:DSC-artifact} (Page~\pageref{fig:DSC-artifact}) \\
\rowcolor{Gray1} Possibility of reference or prediction without target structure(s) & Fig.~\ref{fig:empty} (Page~\pageref{fig:empty}) \\ 
\rowcolor{Gray1} & Fig.~\ref{fig:empty-ref-pred-od} (Page~\pageref{fig:empty-ref-pred-od}) \\ 
\rowcolor{Gray2} Dependency on image resolution & Fig.~\ref{fig:DSC-grid-size} (Page~\pageref{fig:DSC-grid-size}) \\
\rowcolor{Gray1} Over- \textit{vs.} undersegmentation & Fig.~\ref{fig:DSC-overunder} (Page~\pageref{fig:DSC-overunder}) \\
\rowcolor{Gray2} Choice of global decision threshold & Fig.~\ref{fig:seg-threshold} (Page~\pageref{fig:seg-threshold}) \\
\rowcolor{Gray1} Mathematical implications of center-based localization criteria &  Fig.~\ref{fig:hit-criteria} (Page~\pageref{fig:hit-criteria}) \\
\rowcolor{Gray1} &  Fig.~\ref{fig:center-vs-point} (Page~\pageref{fig:center-vs-point}) \\
\rowcolor{Gray2} Mathematical implications of \textit{IoU}-based localization criterion &  Fig.~\ref{fig:bb-2d-3d} (Page~\pageref{fig:bb-2d-3d}) \\
\rowcolor{Gray1} Mathematical implications of the choice of assignment strategies &  Fig.~\ref{fig:od-assignment} (Page~\pageref{fig:od-assignment}) \\
\rowcolor{Gray1}  &  Fig.~\ref{fig:od-ior} (Page~\pageref{fig:od-ior}) \\
\rowcolor{Gray2} Effect of small structures on localization criterion & Fig.~\ref{fig:boundary-mask-iou} (Page~\pageref{fig:boundary-mask-iou}) \\
\rowcolor{Gray2} & Fig.~\ref{fig:DSC-small} (Page~\pageref{fig:DSC-small}) \\
\rowcolor{Gray1} Perfect \textit{Boundary IoU} for imperfect prediction & Fig.~\ref{fig:boundary-iou} (Page~\pageref{fig:boundary-iou}) \\
\rowcolor{Gray2} \textit{Average Precision} \textit{vs.} \textit{Free-response ROC} score & Fig.~\ref{fig:ap-froc} (Page~\pageref{fig:ap-froc}) \\
\rowcolor{Gray1} \textit{Free-response ROC score} is not standardized & Fig.~\ref{fig:froc-no-standard} (Page~\pageref{fig:froc-no-standard}) \\
\rowcolor{Gray2} Multi-threshold metric-related properties &  Fig.~\ref{fig:AP-small-conf-changes} (Page~\pageref{fig:AP-small-conf-changes}) \\
\rowcolor{Gray2} & Fig.~\ref{fig:AP-conf-not-important} (Page~\pageref{fig:AP-conf-not-important}) \\
\rowcolor{Gray2} & Fig.~\ref{fig:AP-FP-tail} (Page~\pageref{fig:AP-FP-tail})\\
\rowcolor{Gray1} High class imbalance \scriptsize \textit{(here: pitfall illustrated for image-level classification problems)} &  Fig.~\ref{fig:class-imbalance} (Page~\pageref{fig:class-imbalance}) \\ 
\rowcolor{Gray1} &  Fig.~\ref{fig:misleading-ba-j} (Page~\pageref{fig:misleading-ba-j}) \\ 
\rowcolor{Gray1} &  Fig.~\ref{fig:metric-landscapes} (Page~\pageref{fig:metric-landscapes}) \\ 
\rowcolor{Gray2} More than two classes available \scriptsize \textit{(here: pitfall illustrated for image-level classification problems)} & Fig.~\ref{fig:multi-class} (Page~\pageref{fig:multi-class}) \\ 
\rowcolor{Gray1} Unequal importance of class confusions \scriptsize \textit{(here: pitfall illustrated for image-level classification problems)} & Fig.~\ref{fig:unequal-class-confusions} (Page~\pageref{fig:unequal-class-confusions}) \\
\rowcolor{Gray2} Unequal importance of classes \scriptsize \textit{(here: pitfall illustrated for image-level classification problems)} & Fig.~\ref{fig:unequal-class} (Page~\pageref{fig:unequal-class}) \\
\rowcolor{Gray1} Interdependencies between classes \scriptsize \textit{(here: pitfall illustrated for image-level classification problems)} & Fig.~\ref{fig:multi-class-interdependencies} (Page~\pageref{fig:multi-class-interdependencies})\\ 
\midrule
\multicolumn{2}{c}{\textbf{\textit{Cross-topic pitfalls}}}\\ \midrule
\rowcolor{Gray1} Uninformative visualization & Fig.~\ref{fig:raw-metric-values-boxplot} (Page~\pageref{fig:raw-metric-values-boxplot}) \\ 
\rowcolor{Gray2} Metric aggregation for invalid algorithm output (e.g. \texttt{NaNs}) & Fig.~\ref{fig:missings} (Page~\pageref{fig:missings}) \\
\rowcolor{Gray2} & Fig.~\ref{fig:missings-hd} (Page~\pageref{fig:missings-hd}) \\
\rowcolor{Gray1} Hierarchical data aggregation & Fig.~\ref{fig:hier-aggr} (Page~\pageref{fig:hier-aggr}) \\
\rowcolor{Gray2} Aggregation per class & Fig.~\ref{fig:aggr-per-class} (Page~\pageref{fig:aggr-per-class}) \\
\rowcolor{Gray1} Metric combination & Fig.~\ref{fig:combination} (Page~\pageref{fig:combination}) \\ 
\rowcolor{Gray2} Ranking uncertainty & Fig.~\ref{fig:ranking-uncertainty} (Page~\pageref{fig:ranking-uncertainty}) \\ 
\rowcolor{Gray1} Insufficient biomedical relevance of metric score differences & Fig.~\ref{fig:ranking-relevance} (Page~\pageref{fig:ranking-relevance}) \\ 
\rowcolor{Gray2} Non-determinism of algorithms & Fig.~\ref{fig:non-determinism} (Page~\pageref{fig:non-determinism}) \\ 
 \bottomrule

\end{longtable}
\endgroup

\newpage
\section{Fundamentals}
\label{sec:fundamentals}
\begin{figure}[H]
\begin{tcolorbox}[title= Relationships between metric families, colback=white]
    \centering
    \includegraphics[width=1\linewidth]{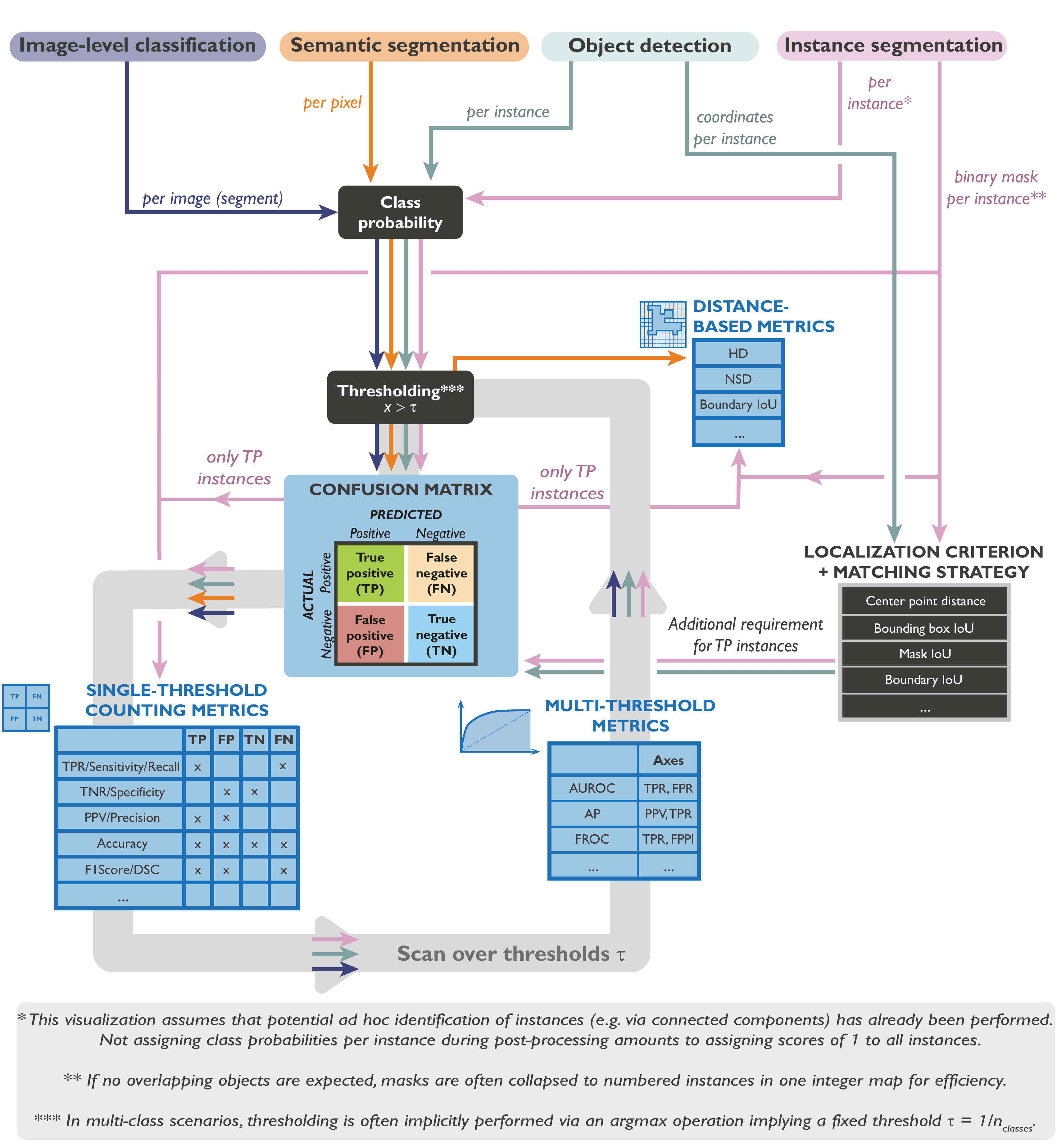}
    \caption{Most popular metric families and their relationships relevant for the problem categories addressed by this paper. The vast majority of metrics is directly or indirectly based on the cardinalities of the confusion matrix. The available metrics can be broadly classified into \textbf{counting metrics} that operate directly on the confusion matrix generated for a fixed threshold, \textbf{multi-threshold metrics} that operate on a dynamic confusion matrix (depending on threshold) and \textbf{distance-based metrics} that take into account the structure contour(s) or other spatial information, such as the structure center. Further abbreviations: \acf{AUROC}, \acf{AP}, \acf{DSC}, \acf{FPPI}, \acf{FPR}, \acf{HD}, \acf{IoU}, \acf{NSD}, \acf{PPV}, \acf{TNR}, \acf{TPR}.}
    \label{fig:metric-families}
\end{tcolorbox}
\end{figure}

The present work focuses on biomedical image analysis problems that can be interpreted as a classification task at image, object or pixel level. The vast majority of metrics for these problem categories is directly or indirectly based on epidemiological principles of \acf{TP}, \acf{FN}, \acf{FP}, \acf{TN}, i.e. the \textit{cardinalities} of the so-called confusion matrix, depicted in Figure~\ref{fig:metric-families}. The \ac{TP}/\ac{FN}/\ac{FP}/\ac{TN}, from now on referred to as cardinalities. In the case of more than two classes $C$ we also refer to the entries of the $C \times C$ confusion matrix as cardinalities. For simplicity and clarity in notation, we restrict ourselves to the binary case in most examples. Cardinalities can be computed for image (segment), object or pixel level. They are typically computed by comparing the prediction of the algorithm to a reference annotation. Modern neural network-based approaches typically require a threshold to be set in order to convert the algorithm output comprising predicted class scores (also referred to as continuous class scores) to a confusion matrix, as illustrated in Figure~\ref{fig:metric-families}. For the purpose of metric recommendation, the available metrics can be broadly classified as follows (see also \citep{cao2020mcc}): 

\begin{itemize}
    \item \textbf{Counting metrics} operate directly on the confusion matrix and express the metric value as a function of the cardinalities (see Figs.~\ref{fig:def-classification-1}, \ref{fig:def-classification-2}  and~\ref{fig:definition-overlap}). In the context of segmentation, they have typically been referred to as \textbf{overlap-based} metrics~\cite{taha2015metrics}. We distinguish \textbf{multi-class counting metrics}, which are defined for an arbitrary number of classes, from \textbf{per-class counting metrics}, which are computed by treating one class as foreground/positive class and all other classes as background. Popular examples for the former include \textit{\ac{MCC}}, \textit{Accuracy}, and \textit{Cohen's Kappa $\kappa$}, while examples for the latter are \textit{Sensitivity}, \textit{Specificity}, \textit{\ac{PPV}}, \textit{\ac{DSC}} and \textit{\ac{IoU}}.
    \item \textbf{Multi-threshold metrics} operate on a dynamic confusion matrix, reflecting the conflicting properties of interest, such as high \textit{Sensitivity} and high \textit{Specificity}. Popular examples include the \textit{\ac{AUROC}} (see Figure~\ref{fig:def-auc}) and \textit{\ac{AP}} (see Figure~\ref{fig:ap-example}).
    \item \textbf{Distance-based metrics} have been designed for semantic and instance segmentation tasks. They operate exclusively on the \ac{TP} and rely on the explicit definition of object boundaries (see Figs.~\ref{fig:definition-distance} and~\ref{fig:definition-nsd}). Popular examples are the \textit{\ac{HD}} and the \textit{\ac{NSD}} (see Figure~\ref{fig:definition-nsd}).
\end{itemize}

Depending on the context (e.g. image-level classification \textit{vs.} semantic segmentation task) and the community (e.g. medical imaging community \textit{vs.} computer vision community), identical metrics are referred to with different terminology. For example, \textit{Sensitivity}, \textit{\ac{TPR}} and \textit{Recall} refer to the same concept. The same holds true for the \textit{\ac{DSC}} and the \textit{F$_1$ Score}. The most relevant metrics for the problem categories in the scope of this paper are introduced in the following.

Most metrics are recommended to be applied per class (except the multi-class counting metrics), meaning that a potential multi-class problem is converted to multiple binary classification problems, such that each relevant class serves as the positive class once. This results in different confusion matrices depending on which class is used as the positive class. 

\newpage

\subsection{Image-level Classification}
\label{sec:fundamentals_ilc}

\textbf{Image-level classification} refers to the process of assigning one or multiple labels, or \textit{classes}, to an image. If there is only one class of interest (e.g. cancer \textit{vs.} no cancer), we speak of \textit{binary classification}, otherwise of \textit{categorical classification}. Modern algorithms usually output \textbf{predicted class probabilities} (or continuous class scores) between 0 and 1 for every image and class, indicating the probability of the image belonging to a specific class. By introducing a threshold (e.g. 0.5), predictions are considered as positive (e.g. cancer = true) if they are above the threshold or negative if they are below the threshold. Afterwards, predictions are assigned to the cardinalities (e.g. a cancer patient with prediction cancer = true is considered as \ac{TP}) \cite{davis2006relationship}. The most popular classification metrics are counting metrics, operating on a confusion matrix with fixed threshold on the class probabilities, and multi-threshold metrics, as detailed in the following. 

\begin{figure}[H]
\begin{tcolorbox}[title= Common binary classification metrics: Per-class counting-based, colback=white]
    \centering
    \includegraphics[width=1\linewidth]{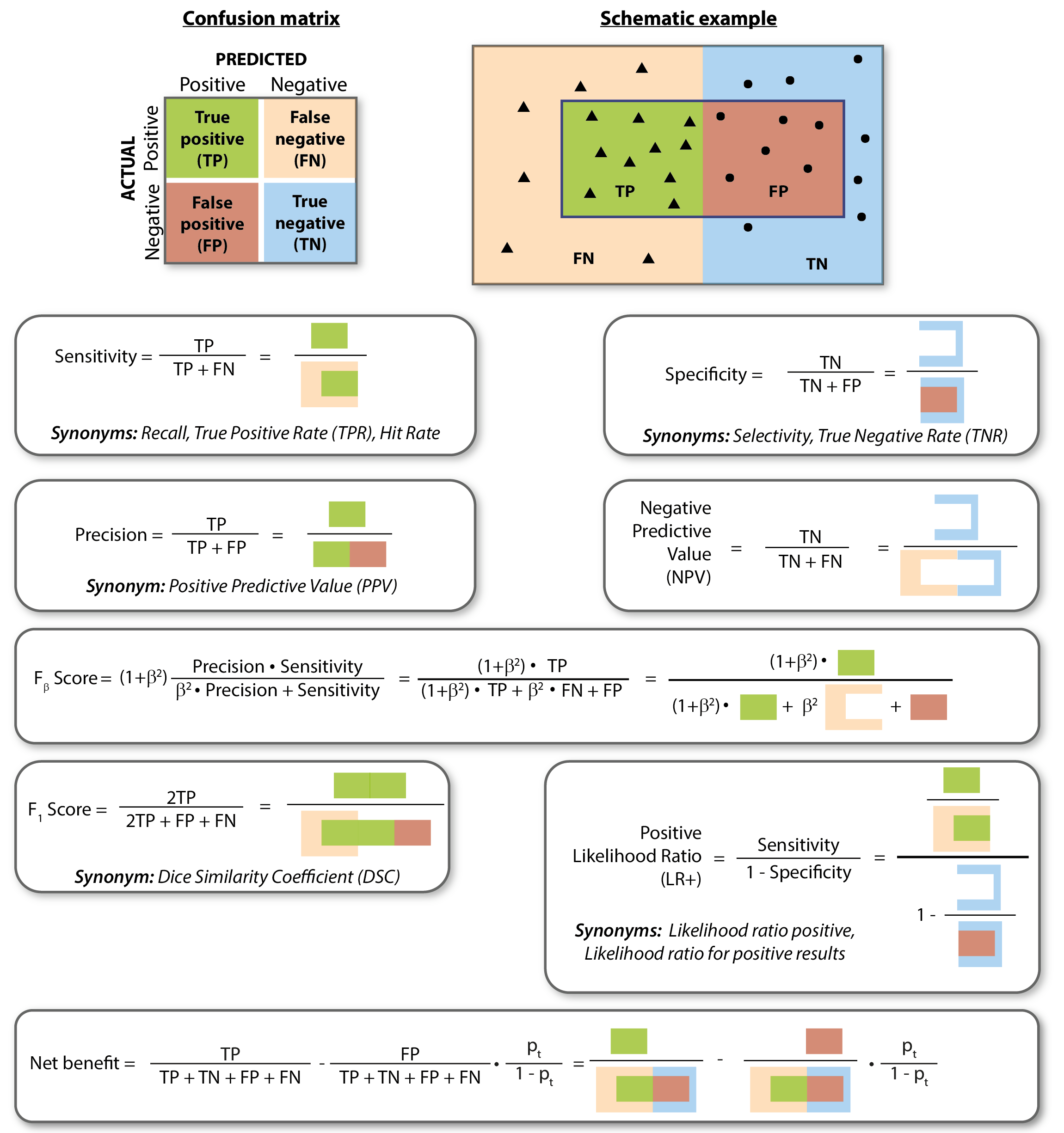}
    \caption{Overview of the most commonly used per-class counting classification measures that are based on the cardinalities of the confusion matrix, i.e. the true (T)/false (F) positives (P)/negatives (N) in the binary case.}
    \label{fig:def-classification-1}
\end{tcolorbox}
\end{figure}

\begin{figure}[H]
\begin{tcolorbox}[title= Common binary classification metrics: Multi-class counting-based, colback=white]
    \centering
    \includegraphics[width=0.9\linewidth]{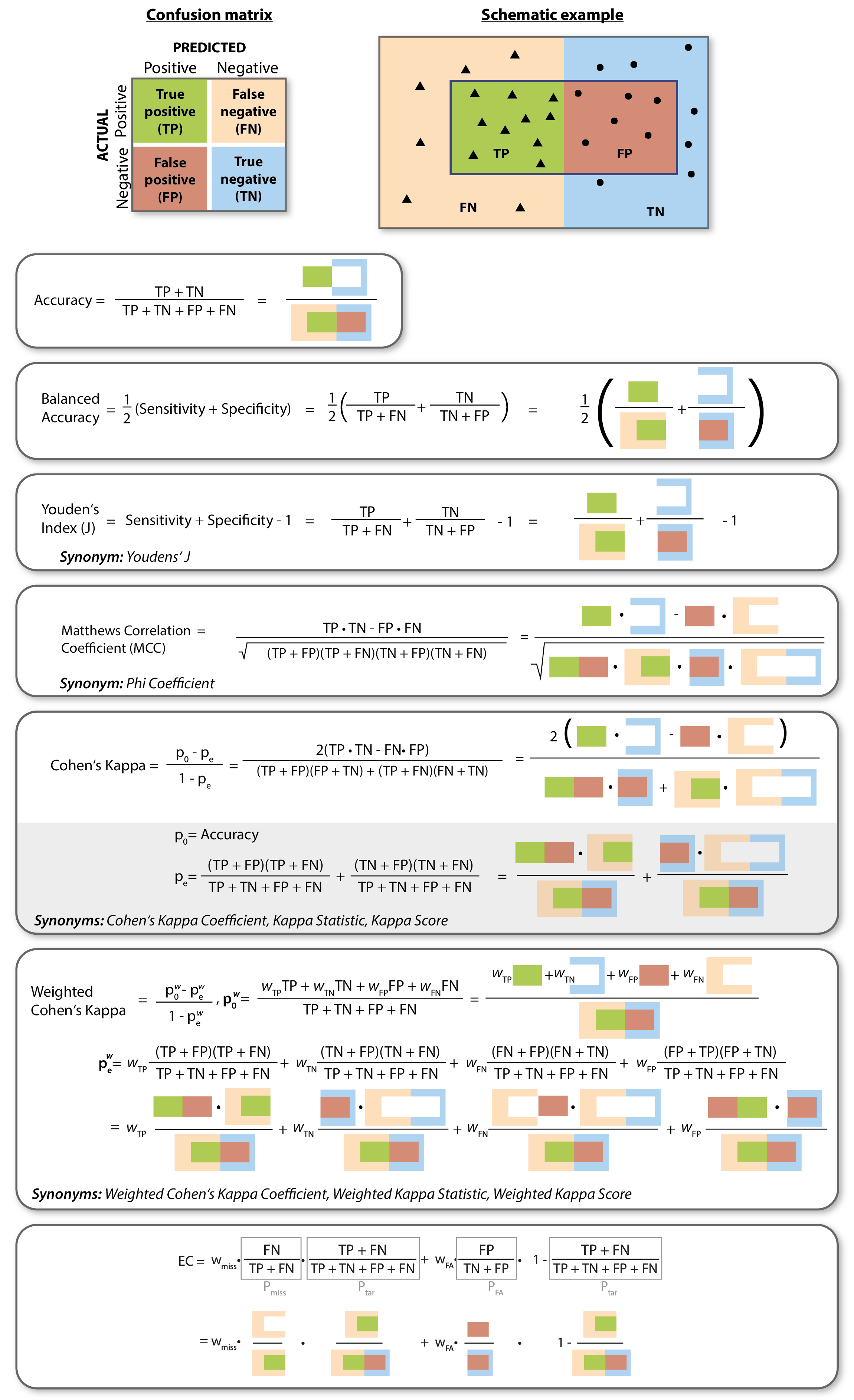}
    \caption{Overview of the most commonly used multi-class counting classification measures that are based on the cardinalities of the confusion matrix, i.e. the true (T)/false (F) positives (P)/negatives (N) in the binary case.}
    \label{fig:def-classification-2}
\end{tcolorbox}
\end{figure}

\paragraph{\textbf{Counting metrics}}
The most common binary counting metrics used for image-level classification are presented in Figs.~\ref{fig:def-classification-1} (per-class) and \ref{fig:def-classification-2} (multi-class). Please note that these metrics are also commonly used in segmentation and object detection tasks. For segmentation tasks, they are often referred to as overlap-based metrics. Each of the presented metrics covers specific properties. 

The \textit{Sensitivity} (also referred to as \textit{Recall}, \textit{\ac{TPR}} or \textit{Hit rate}) focuses on the actual positives (\ac{TP} and \ac{FN}) and represents the fraction of positives that were correctly detected as such. In contrast, \textit{\ac{PPV}} (or \textit{Precision}) divides the \ac{TP} by the total number of predicted positive cases, thus aiming to represent the probability of a positive prediction corresponding to an actual positive. A value of 1 would imply that all positive predicted cases are actually positives, but it might still be the case that positive cases were missed. Please note that the term \textit{Precision} has multiple meanings. In the context of computer assisted interventions, for example, it typically refers to the measured variance. Hence, the usage of its synonym \textit{\ac{PPV}} may be preferred.

In analogy to the \textit{Sensitivity} for positives, \textit{Specificity} (also referred to as \textit{Selectivity} or \textit{\ac{TNR}}) focuses on the negative cases by computing the fraction of negatives that were correctly detected as such. Similarly to the \textit{\ac{PPV}}, the \textit{\ac{NPV}} divides the \ac{TN} by the total number of predicted negative cases and measures how many of the predicted negative samples were actually negative. \textit{Specificity} and \textit{\ac{NPV}} require the definition of \ac{TN} cases, which is not always possible. In object detection tasks, for example (see Sec.~\ref{sec:fundamentals_od}), \ac{TN} are typically ill-defined and not provided. Therefore, these measures can not be computed in those cases. 

\textit{\ac{PPV}} and \textit{\ac{NPV}} provide quantities of direct interest to the medical practitioner, namely the probability of a certain event given the prediction of the classifier. However, unlike \textit{Sensitivity} and \textit{Specificity}, they are not only functions of the classifier but also of the study population. Hence, one cannot extrapolate from one population to another, e.g. from a case-control study to the general population without correcting for prevalence. In the case of the prevalence being unequal to 0.5, prevalence correction should be applied (see Figure~\ref{fig:prevalence}):

\begin{equation}
\text{\emph{PPV}} = \frac{\text{\emph{Sensitivity}} \cdot \text{\emph{Prevalence}}}{\text{\emph{Sensitivity}} \cdot \text{\emph{Prevalence}} + (1 - \text{\emph{Specificity}}) \cdot (1 - \text{\emph{Prevalence}})}
\end{equation}
\begin{equation}
\text{\emph{NPV}} = \frac{\text{\emph{Specificity}} \cdot (1 - \text{\emph{Prevalence}})}{\text{\emph{Specificity}} \cdot (1 - \text{\emph{Prevalence}}) + (1 - \text{\emph{Sensitivity}}) \cdot Prevalence}
\end{equation}

As illustrated in Figure~\ref{fig:class-imbalance}, reporting of a single metric such as \textit{Sensitivity}, \textit{\ac{PPV}} or \textit{Specificity} can be highly misleading because, for example, non-informative classifiers can achieve high values on imbalanced classes. The \textit{F$_1$ Score} (also known as \textit{\ac{DSC}} in the context of segmentation), overcomes this issue by representing the harmonic mean of \textit{\ac{PPV}} and \textit{Sensitivity} and therefore penalizing extreme values of either metric~\cite{hicks2021evaluation}, while being relatively robust against imbalanced data sets~\cite{Sun2009imbalanced}. The \textit{F$_1$ Score} is a specification of the \textit{F$\beta$ score}, which adds a weighting between \textit{\ac{PPV}} and \textit{Sensitivity}. For higher values of $\beta$, \textit{Sensitivity} is given a higher weight over \textit{\ac{PPV}}, which might be desired depending on the application. More specifically, the \textit{F$\beta$ Score} weights between \ac{FP} and \ac{FN} samples. Another metric which combines the metrics above is the \ac{LR+}, which is the ratio of the \textit{Sensitivity} and 1-\textit{Specificity} (\textit{\ac{FPR}}). This metric is described as the "probability that a positive test would be expected in a patient divided by the probability that a positive test would be expected in a patient without a disease" \citep{shreffler2020diagnostic}. It is defined as odds ratio and therefore invariant to the prevalence \citep{vsimundic2009measures}.

All of the metrics presented so far are bounded between 0 and 1 with 1 representing a perfect value and 0 the worst possible prediction of this metric. However, all of them rely on the definition of the positive class, which may be straightforward in some cases but can be based on a rather arbitrary choice in others. Notably, metric values may be completely different depending on the choice of positive class \cite{powers2020evaluation}.

To overcome the need for selecting one class as the positive class, other metrics have been suggested that can be based on all entries of a \textit{multi-class} confusion matrix, in which each class is assigned a row and a column of the matrix. The \textit{Accuracy} is one of the most commonly used metrics and measures the ratio between all correct predictions (\ac{TP} and \ac{TN}) and the total number of samples. \textit{Accuracy} is not robust against imbalanced data sets (see Figure~\ref{fig:class-imbalance}), and is therefore often replaced by the more robust \textit{\ac{BA}} that averages the \textit{Sensitivity} over all classes \cite{grandini2020metrics}. For two classes, the metric averages \textit{Sensitivity} and \textit{Specificity}. An alternative is \textit{Youden's Index J} or \textit{\ac{BM}}, similarly summing up \textit{Sensitivity} per class (or for two classes, summing up \textit{Sensitivity} and \textit{Specificity}). Note that \textit{Youden's Index J} and \textit{\ac{BA}} are closely related and can directly be calculated from the other as $J = 2BA - 1$ and $BA = (J+1)/2$. Both \ac{BA} and \textit{Youden's Index J} are invariant to the prevalence.

The \textit{\ac{MCC}}, also known as \textit{Phi Coefficient}, measures the correlation between the actual and predicted class. The metric is bounded between -1 and 1, with high positive values referring to a good prediction which can only be achieved when all cardinalities are good, i.e. with a low number of \ac{FP}/\ac{FN} and high \ac{TP}/\ac{TN}. Another popular metric is \textit{Cohen's Kappa $\kappa$}, which calculates the agreement between the reference and prediction while incorporating information on the agreement by chance. It is therefore a form of chance-corrected \textit{Accuracy}. Similarly to \textit{\ac{MCC}}, it incorporates all values of the confusion matrix and is bounded between -1 and 1. In contrast to \textit{\ac{MCC}}, negative values do not indicate anti-correlation, but less agreement than expected by chance. \textit{Cohen's Kappa $\kappa$} can be generalized by introducing a weighting scheme for the cardinalities in the \textit{Weighted Cohen's Kappa $\kappa$} metric. For those three metrics, a value of 0 refers to a prediction which is not better than random guessing. 

All of the presented binary counting metrics can be transferred to the multi-class case~\cite{hossin2015review, grandini2020metrics}, where \textit{\ac{MCC}} and \textit{(Weighted) Cohen's Kappa $\kappa$} have explicit definitions, whereas the others become the implicit result of an aggregation across a rotating one-versus-the-rest binary perspective for each of the classes.

The \textit{\acf{EC}} is a generalization of the probability of error (which, in turn, is \textit{1 - Accuracy}) for cases in which errors cannot all be considered to have equally severe consequences. It is defined as the expectation of the cost, where the cost incurred on a certain sample depends on the sample's class and the decision made for that sample. In practice, the expectation can be estimated as a simple average of the costs over the validation samples. \textit{\ac{EC}} describes the weighted sum of error rates. It can also be used to measure discrimination and calibration in one score. A variant of \textit{\ac{EC}} normalizes \textit{\ac{EC}} by the \textit{\ac{EC}} of a naive system.

\begin{figure}[H]
\begin{tcolorbox}[title= Prevalence dependency, colback=white]
    \centering
    \includegraphics[width=1\linewidth]{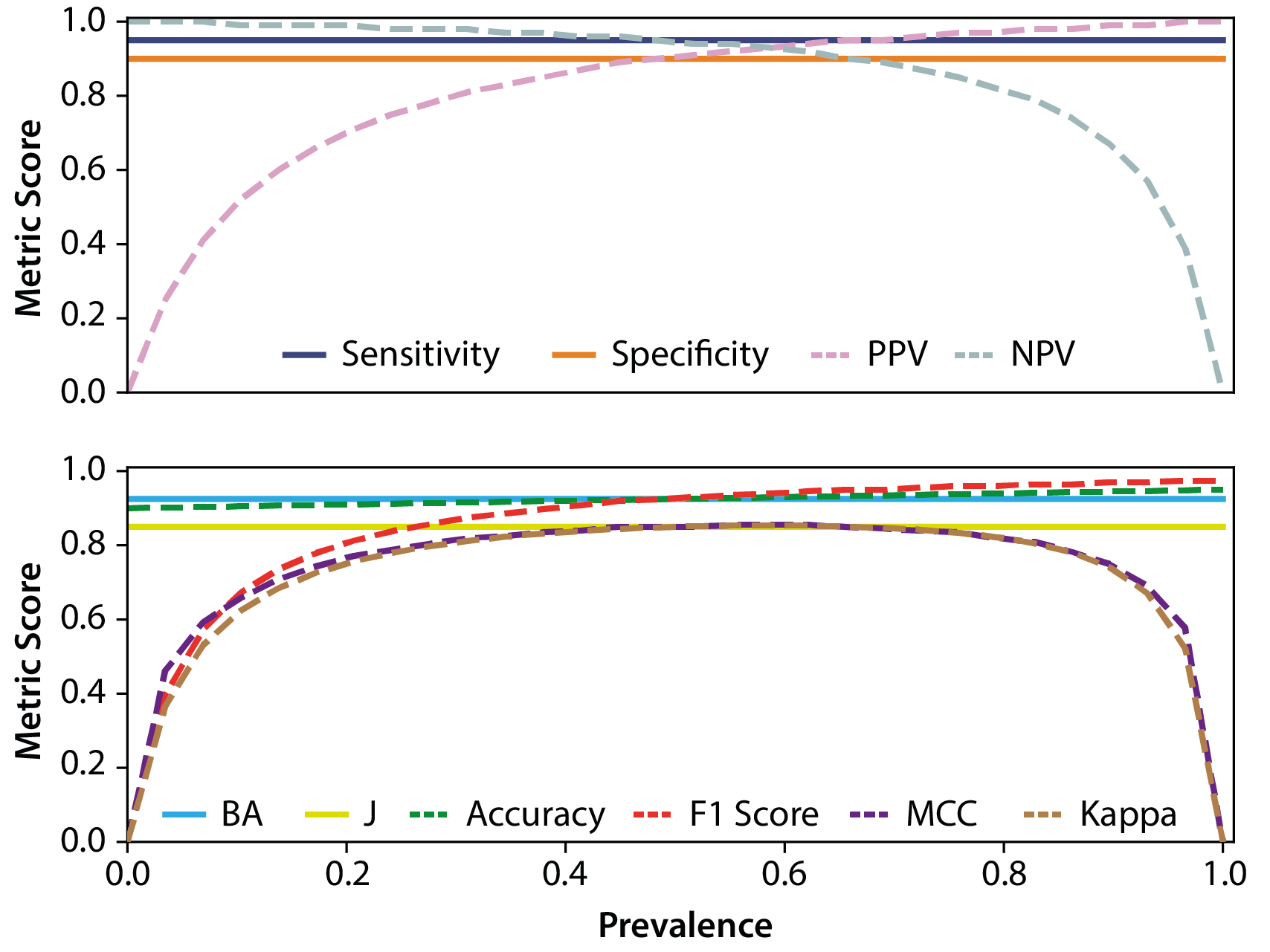}
    \caption{Overview of metrics plotted as a function of the prevalence for fixed \textit{Sensitivity} of 0.95 and \textit{Specificity} of 0.90. Prevalence-dependent metrics are shown as dotted lines. Used abbreviations: \textit{\acf{PPV}}, \textit{\acf{NPV}}, \textit{\acf{BA}}, \textit{Youden's Index J (J)}, \textit{\acf{MCC}}, \textit{Cohen's Kappa (Kappa)}.}
    \label{fig:metrics-vs-prevalence}
\end{tcolorbox}
\end{figure}

Many of the presented metrics depend on the prevalence. Figure~\ref{fig:metrics-vs-prevalence} illustrates how their values change based on the prevalence.

\begin{figure}[H]
\begin{tcolorbox}[title= Common classification metrics: Multi-threshold-based, colback=white]
    \centering
    \includegraphics[width=0.9\linewidth]{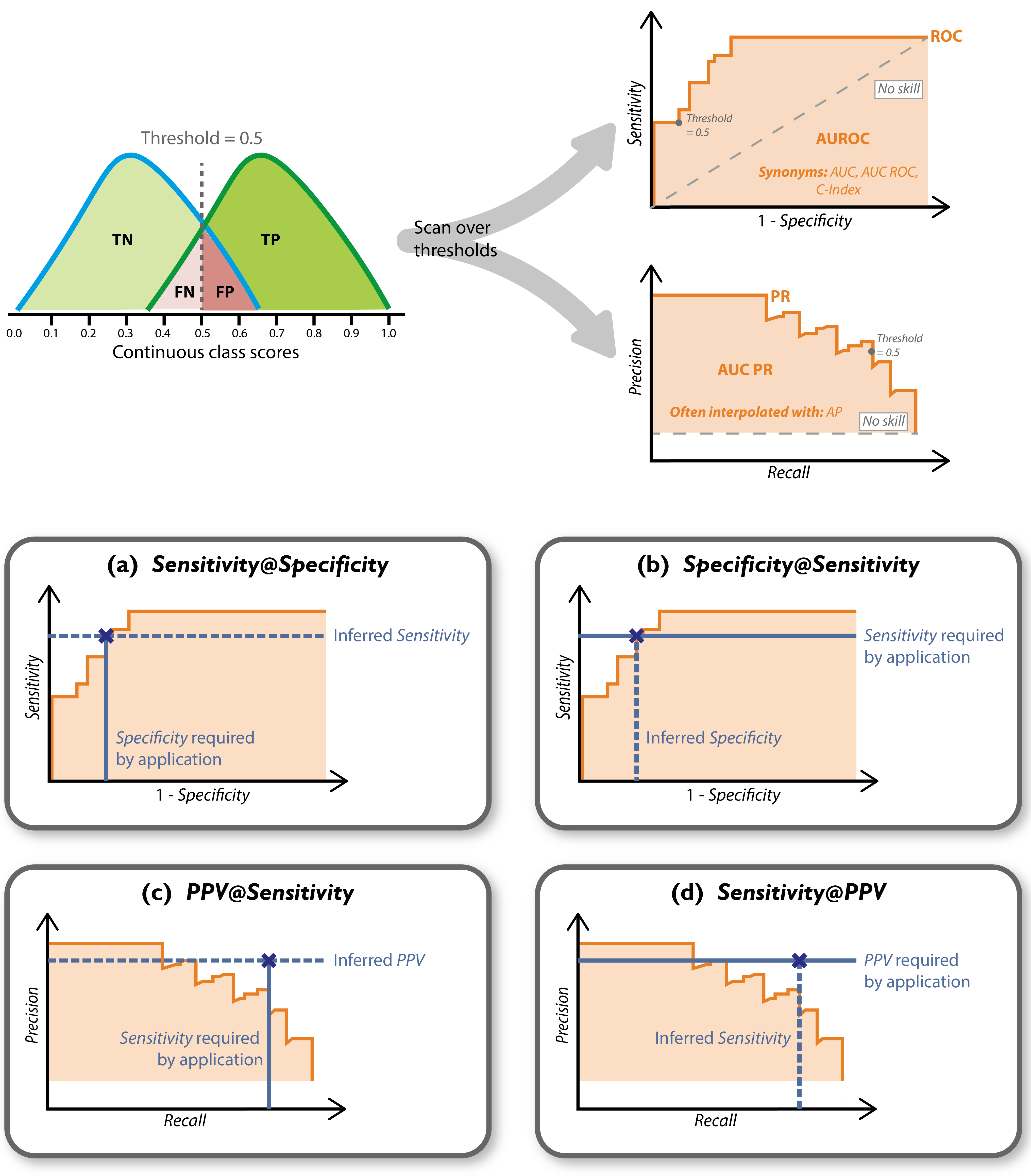}
    \caption{Principle of multi-threshold metrics (top) and per-class counting metrics with application-driven thresholds (bottom). Rather than being based on a static threshold (e.g. for generating the confusion matrix), multi-threshold-based metrics integrate over a range of thresholds. Prominent examples are the \textit{\acf{AUROC}} (also known as \textit{\acf{AUC}} or \textit{AUC \acf{ROC}}) and the Area under the \textit{\acf{PR}} curve (\textit{AUC PR}). Cardinalities, i.e. the true (T)/false (F) positives (P)/negatives (N), are computed based on a threshold (e.g. 0.5) of predicted class probabilities (left). Based on those values, \textit{Sensitivity} (also known as \textit{Recall}) and \textit{1 - Specificity}/\textit{\acf{PPV}} are calculated and plotted against each other (right). The procedure is repeated for several thresholds, resulting in the \textit{\ac{ROC}}/\textit{\ac{PR}} curve. The area under the \textit{\ac{ROC}}/\textit{\ac{PR}} curve is referred to as \textit{\ac{AUROC}}/\textit{AUC PR}. The latter is often interpolated by the \textit{\acf{AP}} metric as detailed in Figure~\ref{fig:ap-example}. The dashed gray lines refer to a classifier with no skill level (random guessing). In the case of an application-driven threshold (e.g. required \textit{Sensitivity} of 0.9), the metrics \textit{Sensitivity@Specificity}, \textit{Specificity@Sensitivity}, \textit{\ac{PPV}@Sensitivity} and \textit{Sensitivity@\ac{PPV}} can be calculated on the basis of the \textit{\ac{ROC}}/\textit{\ac{PR}} curves. Please note that we use the synonyms \textit{Precision} instead of \textit{\ac{PPV}} and \textit{Recall} instead of \textit{Sensitivity} for the \textit{\ac{PR}} curve, given the common use of these terms.}
    \label{fig:def-auc}
\end{tcolorbox}
\end{figure}

\paragraph{\textbf{Multi-threshold metrics}}
The classical counting metrics presented above rely on fixed thresholds to be set on the predicted class probabilities (if available), resulting in them being based on the cardinalities of the confusion matrix.\textbf{ Multi-threshold metrics} overcome this limitation by calculating metric scores based on multiple thresholds. For instance, to emphasize how well a prediction distinguishes between the positive and negative class, the \textit{\ac{AUROC}} can be utilized. The \textit{\ac{ROC}} curve plots the \textit{\ac{FPR}}, which is equal to $1-\text{\emph{Specificity}}$, against the \textit{Sensitivity} for multiple thresholds of the predicted class probabilities, contrarily to just choosing one fixed threshold. For computation of the ROC curve, the class scores can be ordered in descending order and each score regarded as a potential threshold. For each threshold, the resulting \textit{Sensitivity} and \textit{Specificity} are computed, and the resulting tuple is added to the \textit{\ac{ROC}} curve as one point (cf. Figure~\ref{fig:def-auc}); note that the lower the threshold, the higher the \textit{Sensitivity} but the lower (potentially) the \textit{Specificity}. This leads to a monotonic increase of the curve. To interpolate between all points, i.e. to approximate the values between the calculated \textit{Sensitivity} and \textit{Specificity} tuples, a simple linear interpolation can be employed by drawing a line between each pair of points~\cite{davis2006relationship}. An optimal classifier would lead to \textit{Sensitivity} and \textit{Specificity} of 1 (1-\textit{Specificity} of 0), therefore corresponding to a single point $(0,1)$ on the \textit{\ac{ROC}} curve. In contrast, a classifier with no skill level (random guessing) would result in a diagonal line from $(0,0)$ to $(1,1)$ (dashed line in Figure~\ref{fig:def-auc}). The area under the \textit{\ac{ROC}} curve is referred to as \textit{\ac{AUROC}}, also called \textit{AUC ROC} or simply \textit{AUC}.

\textit{\ac{AUROC}} comes with two advantages: threshold and scale invariance. \textit{\ac{AUROC}} measures the quality of the predictions regardless of the threshold, as it is calculated over a number of thresholds. Furthermore, \textit{\ac{AUROC}} does not focus on the absolute values of predictions, but rather on how well they are ranked. However, those properties are not always desired. If a specific penalization of \ac{FP} or \ac{FN} is desired (cf. Figure~\ref{fig:unequal-class}), \textit{\ac{AUROC}} is not the best metric choice as it is invariant to the threshold. If the predicted class probabilities are intended to be well calibrated, the scale invariance feature will prevent from doing so.

Per definition, \textit{\ac{AUROC}} measures the complete area under the \textit{\ac{ROC}} curve. If only a specific range is of interest, a partial or ranged \textit{\ac{AUROC}} can also be computed \cite{ma2013use}. Similarly, metrics can be assessed at a certain point of the \textit{\ac{ROC}} curve, for example the \textit{Sensitivity} value at a specific score of the \textit{Specificity} (e.g. 0.9), also referred to as \textit{Sensitivity@Specificity}. This approach can similarly be used for other curve measures, e.g. the \textit{\ac{PR}} curve, introduced in Sec.~\ref{sec:fundamentals_od}, and is illustrated in Figure~\ref{fig:def-auc}. Please note that we will use the synonyms \textit{Precision} instead of \textit{\ac{PPV}} and \textit{Recall} instead of \textit{Sensitivity} in the case of the \textit{\ac{PR}} curve, given the common use of these terms.

\newpage

\paragraph{\textbf{Calibration metrics}} While most research in biomedical image analysis focuses on the discrimination capabilities of classifiers, a complementary property of relevance is the \textit{calibration} of predicted class scores (also known as \textit{confidence scores}). Intuitively speaking, a system is well-calibrated if the predicted class scores (i.e., the output of the model) reflect the true probabilities of the outcome. In practice, this means that calibrated scores match the empirical success rate of associated predictions. For a binary classification task, calibration implies that of all the data samples assigned a predicted score of, for example, $0.8$ for the positive class, empirically, $80\%$ belong to this class \citep{maier2022metrics}. 

Calibration is typically assessed by either calculating calibration errors (for example as done for the \textit{\acf{ECE}} or \textit{\acf{MCE}} \cite{naeini2015obtaining}) or by proper scoring rules (also referred to as \textit{overall performance measures}~\cite{steyerberg2010assessing}), which measure discrimination and calibration in a single score (for example the \textit{\acf{BS}}). A detailed overview of calibration metrics is given in \citep{maier2022metrics}.

\newpage

\newpage
\subsection{Semantic Segmentation}
\label{sec:fundamentals_ss}
\textbf{Semantic segmentation} is commonly defined as the process of partitioning an image into multiple segments/regions. To this end, one or multiple labels are assigned to every pixel such that pixels with the same label share certain characteristics. Semantic segmentation can therefore also be regarded as pixel-level classification. As in image-classification problems, predicted class probabilities are typically calculated for each pixel deciding on the class affiliation based on a threshold over the class scores \cite{asgari2021deep}. In semantic segmentation problems, the pixel-level classification is typically followed by a post-processing step, in which connected components are defined as objects, and object boundaries are created accordingly. Semantic segmentation metrics can roughly be classified into three classes: (1) counting metrics or overlap-based metrics, for measuring the overlap between the reference annotation and the prediction of the algorithm, (2) distance-based metrics, for measuring the distance between object boundaries, and (3) problem-specific metrics, measuring, for example, the volume of objects.

\begin{figure}[H]
\begin{tcolorbox}[title= Common segmentation metrics: Overlap-based, colback=white]
    \centering
    \hspace{-0.5cm}
    \includegraphics[width=0.8\linewidth]{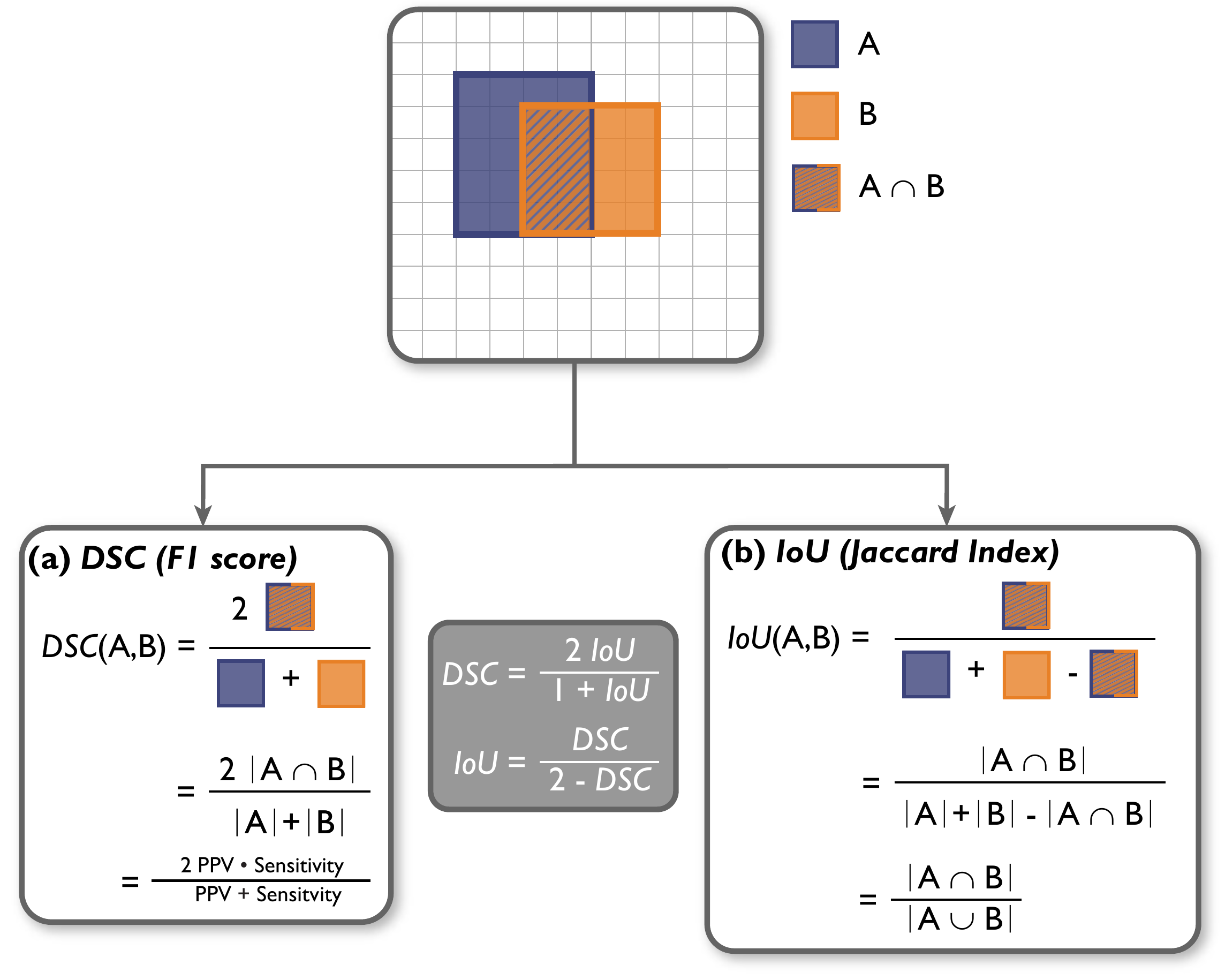}
    \caption{Most commonly used overlap-based segmentation metrics: \textbf{(a)} the \textit{\acf{DSC}} and \textbf{(b)} the \textit{\acf{IoU}}, with $\vert A\vert$ denoting the cardinality of set $A$, $A \cap B$ the intersection between sets $A$ and $B$, and $A \cup B$ the union of sets $A$ and $B$. The \textit{\ac{DSC}} can furthermore be computed as the harmonic mean of \textit{\acf{PPV}} and \textit{Sensitivity}. Note that the \textit{\ac{DSC}} is equivalent to the \textit{F$_1$ Score} and the \textit{\ac{IoU}} is equivalent to the \textit{Jaccard index}.}
    \label{fig:definition-overlap}
\end{tcolorbox}
\end{figure}

\begin{figure}[H]
\begin{tcolorbox}[title= Common segmentation metrics: Boundary-based (Part 1), colback=white]
    \centering
    \hspace{-0.5cm}
    \includegraphics[width=1\linewidth]{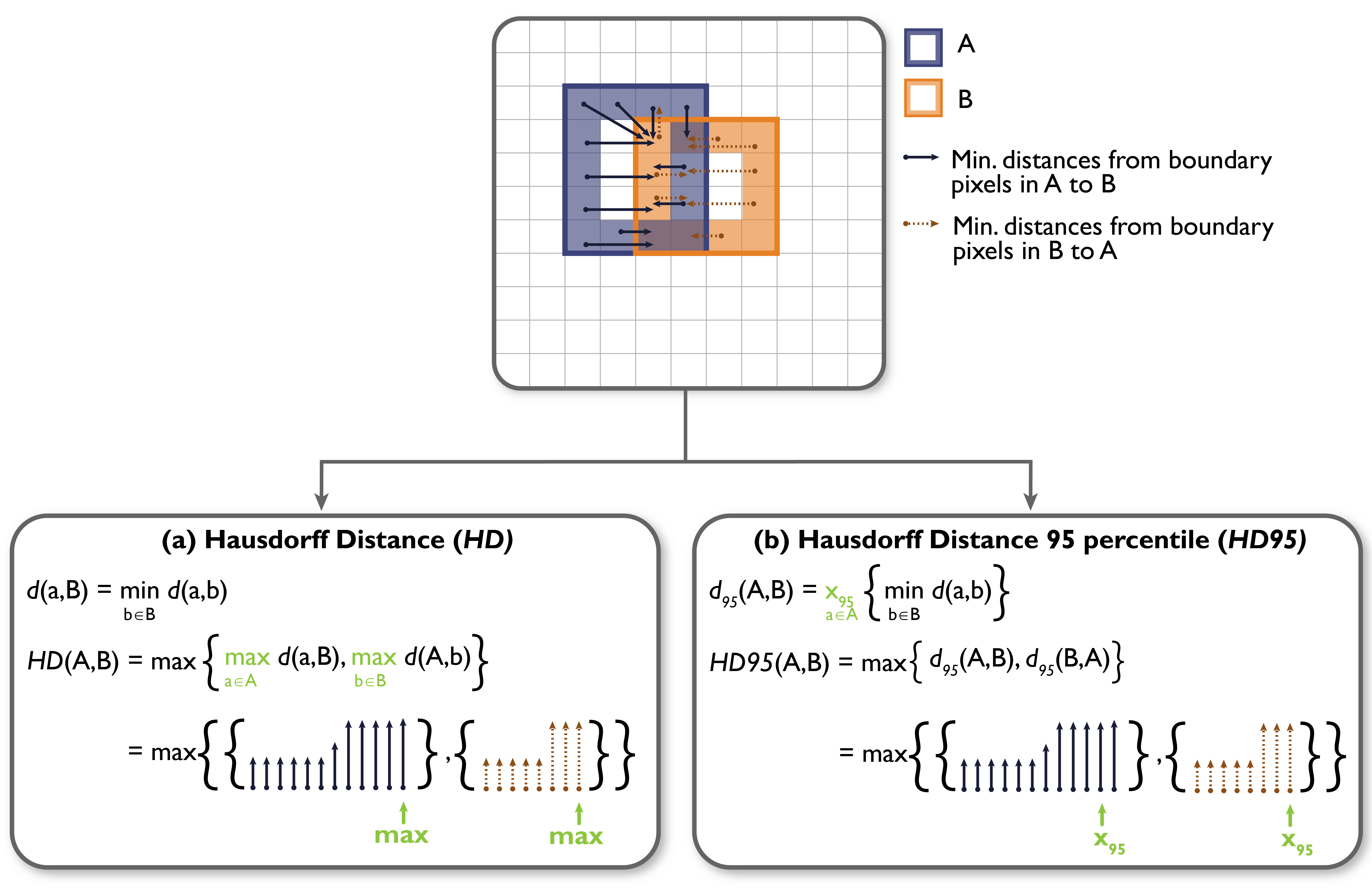}
    \caption{Most commonly used distance-based segmentation metrics: \textbf{(a)} the \textit{\acf{HD}} and \textbf{(b)} the 95\% percentile (denoted as $x_{95}$) \textit{\acf{HD95}}. The Euclidean distance between boundary pixels $a$ and $b$ is defined as $d(a,b)$. Only True Positives (TP) are considered.}
    \label{fig:definition-distance}
\end{tcolorbox}
\end{figure}

\begin{figure}[H]
\begin{tcolorbox}[title= Common segmentation metrics: Boundary-based (Part 2), colback=white]
    \centering
    \hspace{-0.5cm}
    \includegraphics[width=1\linewidth]{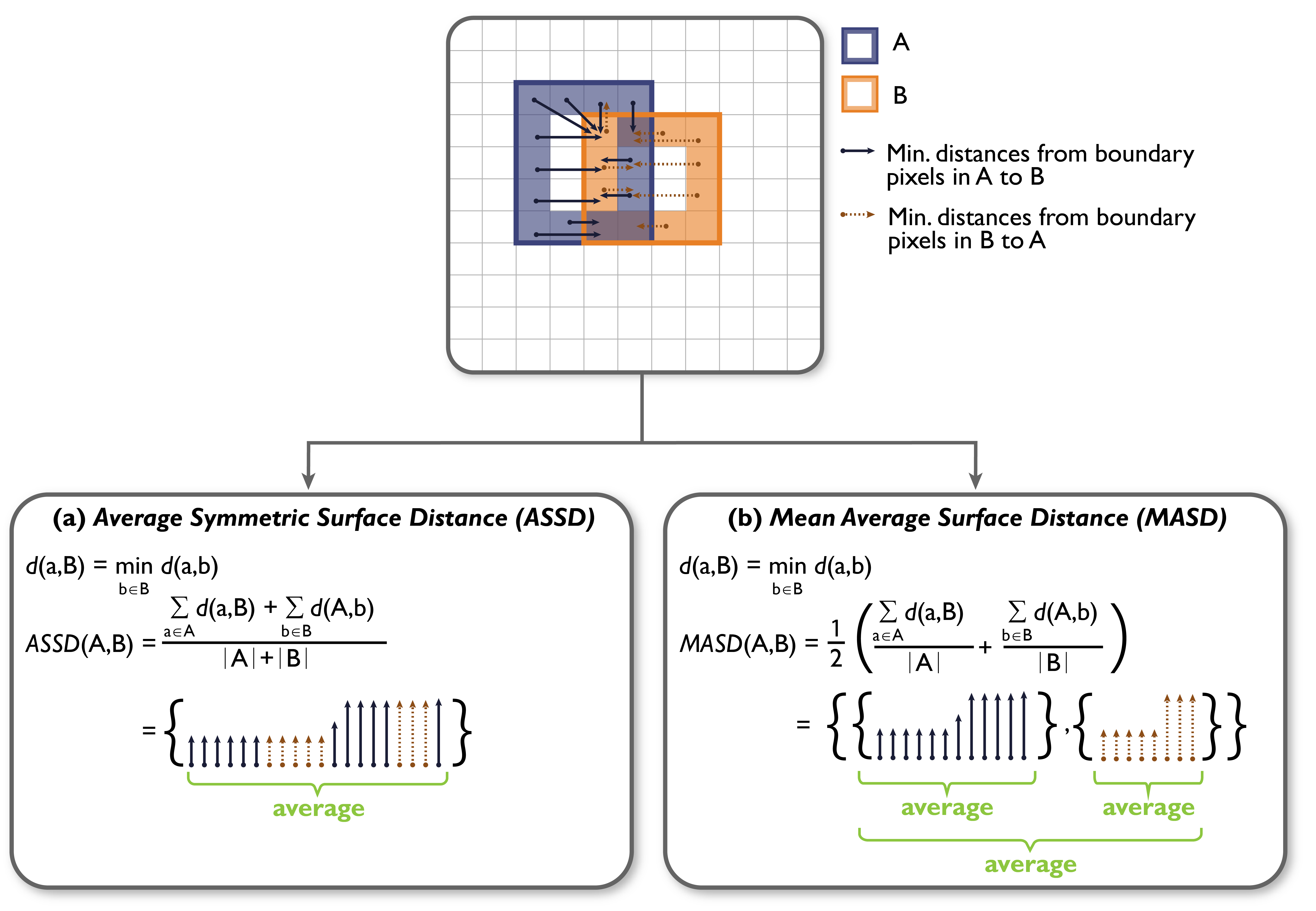}
    \caption{Most commonly used distance-based segmentation metrics: \textbf{(a)} the \textit{\acf{ASSD}} and \textbf{(b)} the \textit{\acf{MASD}}. The Euclidean distance between boundary pixels $a$ and $b$ is defined as $d(a,b)$. Only \acf{TP} are considered.}
    \label{fig:definition-distance-assd-masd}
\end{tcolorbox}
\end{figure}

\paragraph{\textbf{Counting metrics}} The most frequently used segmentation metrics are \textbf{counting metrics}. In the context of segmentation they are also referred to as \textbf{overlap metrics}, as they essentially measure the overlap between a reference mask and the algorithm prediction. According to a comprehensive analysis of biomedical image analysis challenges~\cite{maier2018rankings}, the \textit{\ac{DSC}} \citep{dice1945measures} is the by far most widely used metric in the field of medical image analysis. As illustrated in Figure~\ref{fig:definition-overlap}, it yields a value between 0 (no overlap) and 1 (full overlap). The \textit{\ac{DSC}} is identical to the \textit{F$_1$ Score} and closely related to the \textit{\ac{IoU}}, which is identical to the \textit{Jaccard Index}:

\noindent\hspace{-1cm}\begin{minipage}{.5\linewidth}
\begin{equation}
\label{eq:iou-dsc}
\text{\emph{IoU}} = \frac{\text{\emph{DSC}}}{2-\text{\emph{DSC}}}
\end{equation}
\end{minipage}
\begin{minipage}{.5\linewidth}
\begin{equation}
\label{eq:dsc-iou}
\text{\emph{DSC}} = \frac{2 \text{\emph{IoU}}}{1+\text{\emph{IoU}}}
\end{equation}
\end{minipage}
\paragraph{\textbf{Distance-based metrics}} Overlap-based metrics are often complemented by \textbf{distance-based metrics} that operate exclusively on the \ac{TP} and compute one or several distances between the reference and the prediction. Apart from a few exceptions, distance-based metrics are often \textbf{boundary-based metrics} which focus on assessing the accuracy of object boundaries. According to~\cite{maier2018rankings}, the \textit{\ac{HD}} and its 95\% percentile variant (\textit{\ac{HD95}})~\citep{huttenlocher1993comparing} are the most commonly used boundary-based metrics. The \textit{\ac{HD}} calculates the maximum of all shortest distances for all points from one object boundary to the other, which is why it is also known as the \textit{Maximum Symmetric Surface Distance} \cite{yeghiazaryan2018family}. The \textit{\ac{HD95}} calculates the 95\% percentile instead of the maximum, therefore disregarding outliers  (see Figure~\ref{fig:definition-distance}). Another popular metric is the \textit{\ac{ASSD}}, measuring the average of all distances for every point from one object to the other and vice versa \citep{van20073d, yeghiazaryan2018family}. However, if one boundary is much larger than the other, it will impact the score much more. This is avoided by the \textit{\ac{MASD}} \cite{benevs2015performance}, which treats both structures equally by computing the average distance from structure $A$ to structure $B$ and the average distance from structure $B$ to structure $A$ and averaging both (see Figure~\ref{fig:definition-distance-assd-masd}). For the \textit{\ac{HD}(95)}, \textit{\ac{MASD}} and \textit{\ac{ASSD}} metrics, a value of 0 refers to a perfect prediction (distance of 0 to the reference boundary), while there exists no fixed upper bound. The \textit{Boundary \ac{IoU}} (cf. Figure~\ref{fig:definition-boundary-iou}) is another option for measuring the boundary quality of a prediction. It measures the overlap between prediction and reference up to a certain width (which is controlled by the width parameter $d$) (see Sec.~\ref{sec:fundamentals_od} for details) and is bounded between 0 and 1.

\begin{figure}[H]
\begin{tcolorbox}[title= Common segmentation metrics: \acf{NSD}, colback=white]
    \centering
    \hspace{-0.5cm}
    \includegraphics[width=0.9\linewidth]{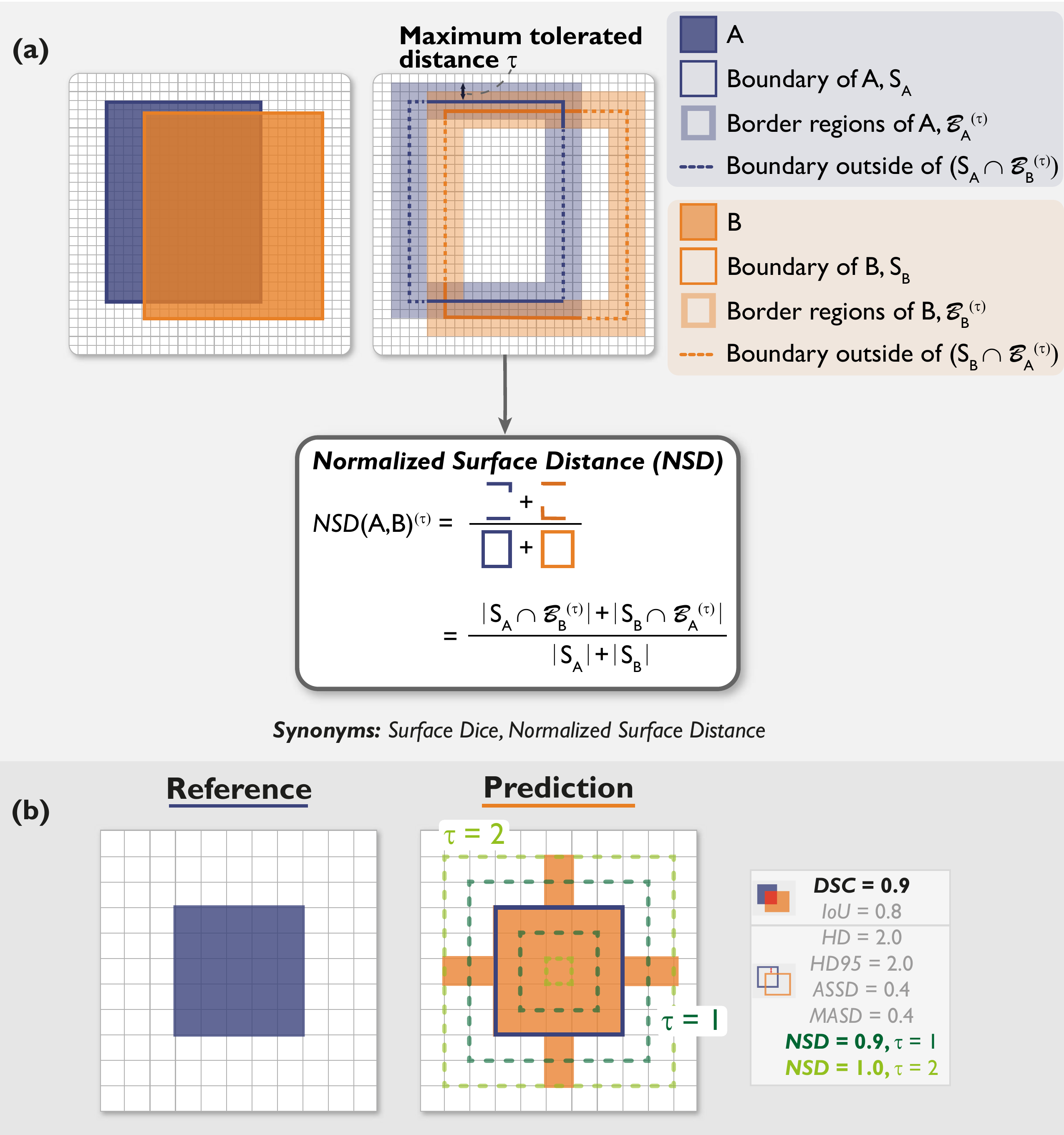}
    \caption{\textbf{(a)} The \textit{\acf{NSD}} is an \textbf{uncertainty-aware} segmentation metric that measures the overlap between two boundaries. The parameter $\tau$ represents the tolerated difference between the prediction and the reference boundary $S$ and defines the border regions $\mathcal{B^{(\tau)}}$ for each structure, i.e. the pixels within the range of $\tau$ from the boundary. They are defined as all pixels within distance $\tau$ from the boundary $S$. The threshold can be based on the domain-related requirements and/or the inter-rater variability, for example. \textbf{(b)} Example showing how the \textit{\ac{NSD}} can handle outliers in the prediction by adjusting the tolerance value to $\tau=2$ pixels. The \textit{\acf{DSC}} and other metrics (\textit{\acf{IoU}}, \textit{\acf{HD}(95)}, \textit{\acf{ASSD}}), in contrast, penalize these tolerated errors.}
    \label{fig:definition-nsd}
\end{tcolorbox}
\end{figure}

A major problem related to boundary-based metrics are the error-prone reference annotations (see Figs.~\ref{fig:low-quality} and \ref{fig:DSC-artifact}). In fact, domain experts often disagree on the definition and annotation of objects and their boundaries \cite{joskowicz2019inter}. While the  \textit{\ac{HD}(95)} and \textit{\ac{ASSD}} are not robust with respect to uncertain reference annotations, the \textit{\ac{NSD}} was explicitly designed for this purpose as a hybrid metric between boundary-based and counting-based approaches. Known uncertainties in the reference as well as acceptable deviations of the predicted boundary from the reference are captured by a threshold $\tau$ \cite{nikolov2021clinically}, as shown in Figure~\ref{fig:definition-nsd}. Only boundary parts within the border regions defined by $\tau$ are counted as \ac{TP}. The metric is bounded between 0 (no boundary overlap) and 1 (full boundary overlap), so that it can be interpreted similarly to the classical \textit{\ac{DSC}} (though restricted to the boundary). Please note that $\tau$ is another important hyperparameter which should be chosen wisely, based on inter-rater agreement, for example.

\begin{figure}[H]
\begin{tcolorbox}[title= Common segmentation connectivity metrics: \acf{clDice}, colback=white]
    \centering
    \hspace{-0.5cm}
    \includegraphics[width=\linewidth]{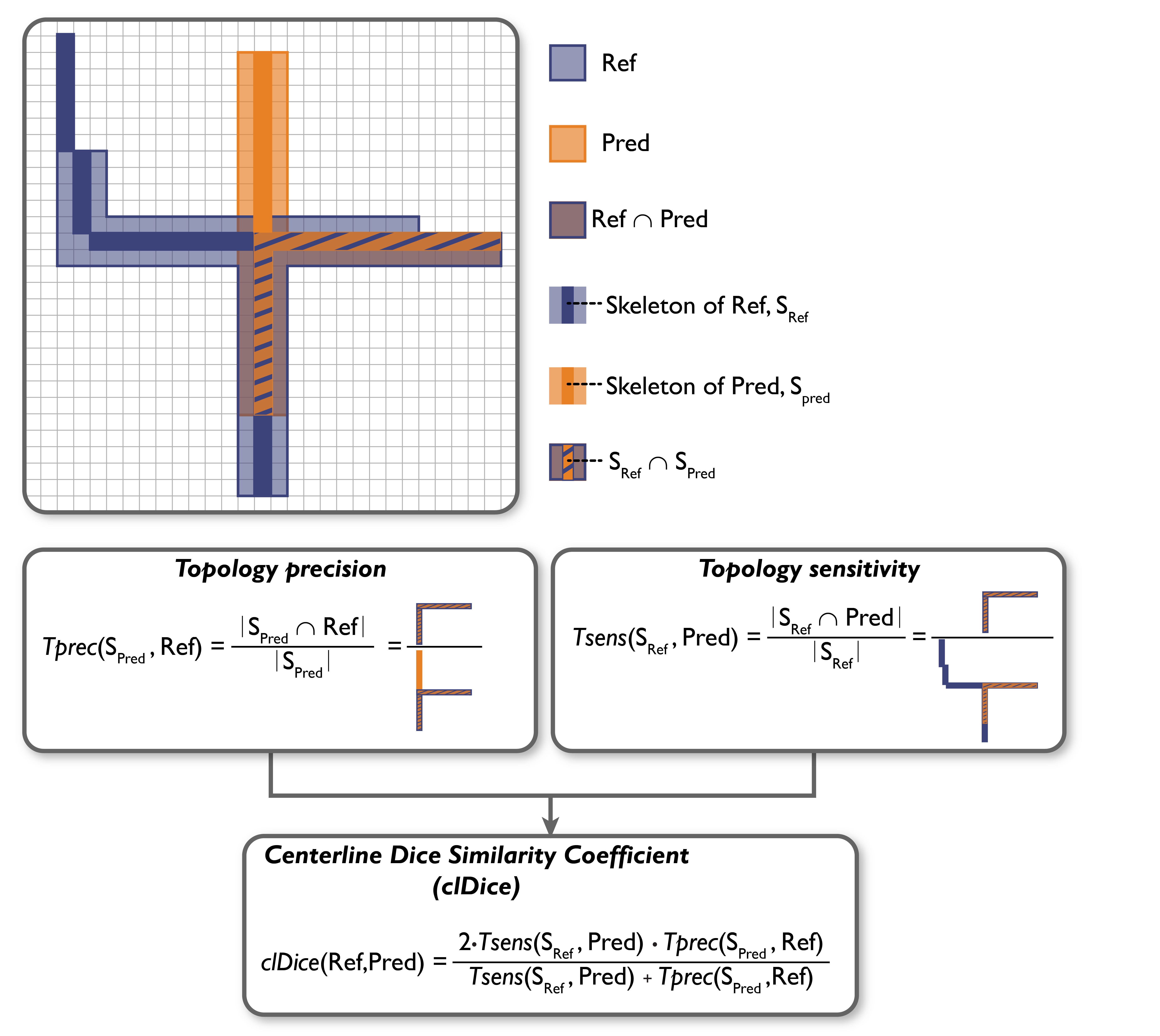}
    \caption{The \textit{\acf{clDice}} measures the \textbf{connectivity} of structures. As designed for tubular structures, the metric relies on the skeletons of the structures, given by $S_{Ref}$ and $S_{Pred}$ for structures $Ref$ and $Pred$ in this case. The \textit{\ac{clDice}} is defined as the harmonic mean of the \textit{Topology precision} and \textit{Topology sensitivity}. $\vert S_{Ref}\vert$ denotes the cardinality of the skeleton $S_{Ref}$ and $S_{Pred} \cap Ref$ refers to the intersection between the skeleton $S_{Pred}$ and mask $Ref$ (similar for $S_{Ref} \cap Pred$).}
    \label{fig:definition-cldice}
\end{tcolorbox}
\end{figure}

\paragraph{\textbf{Problem-specific segmentation metrics}} 
While overlap-based metrics and distance-based metrics are the standard metrics used by the general computer vision community, biomedical applications often have special domain-specific requirements. In medical imaging, for example,  the actual volume of an object may be of particular interest (for example tumor volume). In this case, \textbf{volume metrics} such as the \textit{Absolute} or \textit{Relative Volume Error} and the \textit{Symmetric Relative Volume Difference} can be computed \cite{nai2021comparison}. However, they are less common than overlap metrics, as the location of objects is not considered at all (see Figure~\ref{fig:center}). If the structure center or center line is of particular interest (e.g. in cells or vessels), \textbf{connectivity metrics} come into play, which measure the agreement of the center line between two objects. This is of special interest if linear or tube-like objects are present in a data set, for example in brain vascular analysis. In these cases, over- or undersegmentation are typically not of special interest, with the focus rather being on connectivity or network topology. For this purpose, the \textit{\ac{clDice}} \cite{shit2021cldice} has been designed. For computation, the reference and prediction skeletons need to be extracted from the binary masks. Multiple approaches have been proposed, as described in \cite{shit2021cldice}. As shown in Figure~\ref{fig:definition-cldice}, the \textit{\ac{clDice}} is then defined as the harmonic mean of \textit{Topology Precision} and \textit{Topology Sensitivity}. The \textit{Topology Precision} measures the ratio of the overlap of the skeleton of the prediction skeleton and the reference mask and the skeleton of the prediction. Similarly to \textit{Precision}, it measures the \ac{FP} samples. On the other hand, \textit{Topology Sensitivity} incorporates \ac{FN} samples by measuring the ratio of the overlap of the skeleton of the reference and the prediction mask and the reference skeleton. 

\newpage
\newpage

\newpage

\newpage

\newpage
\subsection{Object Detection}
\label{sec:fundamentals_od}
\textbf{Object detection} refers to the detection of one or multiple objects (or: instances) of a particular class (e.g. lesion) in an image \cite{lin2014microsoft}. The following description assumes single-class problems, but translation to multi-class problems is straightforward, as validation for multiple classes on object level is performed individually per class. Notably, as multiple predictions and reference instances may be present in one image, the predictions need to include localization information, such that a matching between reference and predicted objects can be performed. Important design choices with respect to the validation of object detection methods include:
\begin{enumerate}
    \item \textit{How to represent an object?} Representation is typically composed of location information and a class affiliation. The former may take the form of a bounding box (i.e. a list of coordinates), a pixel mask, or the object's center point. Additionally, modern algorithms typically assign a confidence value to each object, representing the probability of a prediction corresponding to an actual object of the respective class. Note that a confusion matrix is later computed for a fixed threshold on the predicted class probabilities.\footnote{Please note that we will use the term confidence scores analogously to predicted class probabilities in the context of object detection and instance segmentation.}
    \item \textit{How to decide whether a reference instance was correctly detected?} This step is achieved by applying the \textit{localization criterion}. This may, for example, be based on comparing the object centers of the reference and prediction or computing their overlap (Figs.~\ref{fig:hit-criteria} and~\ref{fig:bb-2d-3d}).
    \item \textit{How to resolve assignment ambiguities?} The above step might lead to ambiguous matchings, such as two predictions being assigned to the same reference object. Several strategies exist for resolving such cases.
\end{enumerate}
\noindent
The following sections provide details on (1) applying the localization criterion, (2) applying the assignment strategy and (3) computing the actual performance metrics.

\paragraph{\textbf{Localization criterion}}
As one image may contain multiple objects or no object at all, the \textbf{localization criterion} or \textbf{hit criterion} measures the (spatial) similarity between a prediction (represented by a bounding box, pixel mask, center point or similar) and a reference object. It defines whether the prediction \textit{hit/detected} (\ac{TP}) or \textit{missed} (\ac{FP}) the reference. Any reference object not detected by the algorithm is defined as \ac{FN}. Please note that \ac{TN} are not defined for object detection tasks, which has several implications on the applicable metrics, as detailed below. 

\begin{figure}[H]
\begin{tcolorbox}[title= Common localization criterion: Boundary Intersection over Union (Boundary IoU), colback=white]
    \centering
    \includegraphics[width=1\linewidth]{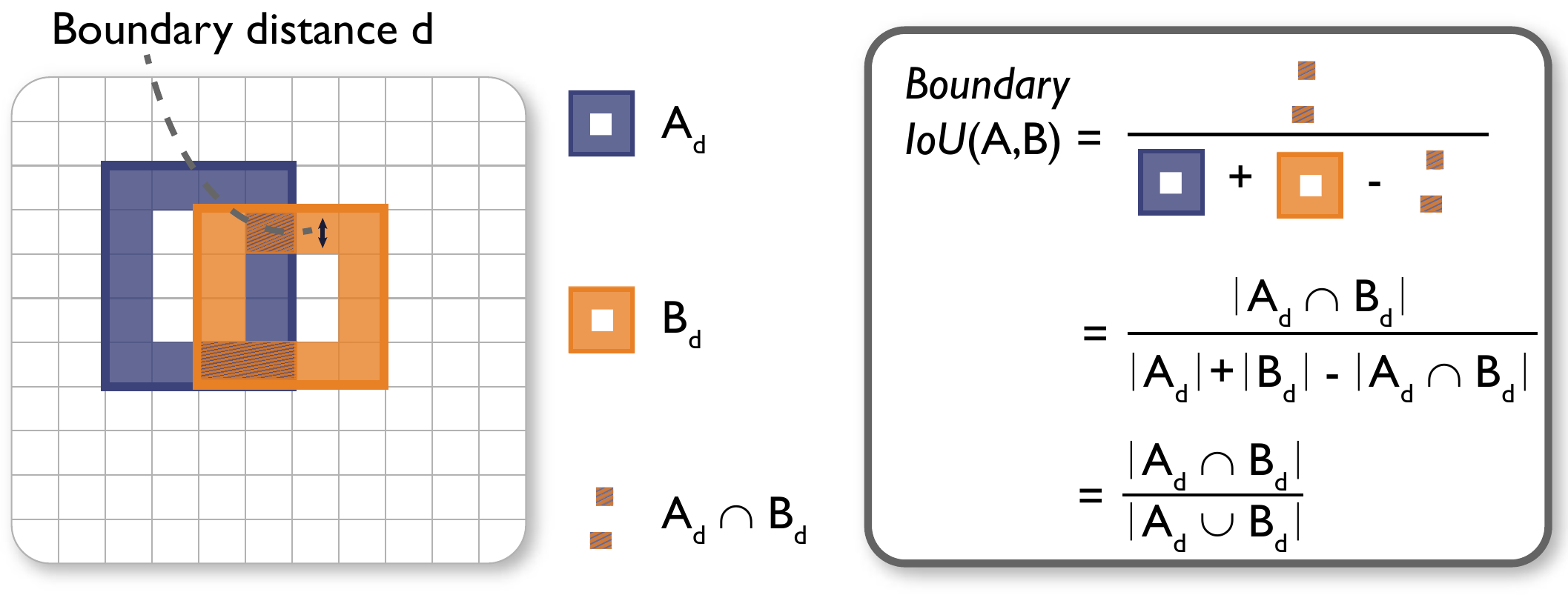}
    \caption{The \textit{Boundary Intersection over Union (Boundary IoU)} is an \textbf{uncertainty-aware} metric that measures the overlap between two boundaries. The overlap is computed for the structures $A$ and $B$ for mask pixels within a certain distance $d$ from the structure boundaries ($A_d$ and $B_d$), by computing the \textit{\ac{IoU}} between $A_d$ and $B_d$.}
    \label{fig:definition-boundary-iou}
\end{tcolorbox}
\end{figure}

There are multiple ways to define the localization or hit criterion (see Figs.~\ref{fig:definition-overlap}, \ref{fig:definition-boundary-iou} and \ref{fig:hit-criteria}). Popular center-based localization criteria are (a) \textit{the center-cover criterion}, for which the reference object is considered hit if the center of the reference object is inside the predicted detection, (b) the \textit{distance-based hit criterion}, which considers a \ac{TP} if the distance $d$ between the center of the reference and the detected object is smaller than a certain threshold $\tau$ and (c) the \textit{center-hit criterion}, which holds true if the center of the predicted object is inside the reference bounding box or mask.

The most commonly used overlap-based hit criterion is determined by computing the \textit{\ac{IoU}} \citep{jaccard1912distribution} (cf. Figure~\ref{fig:definition-overlap}b). The prediction is considered as \ac{TP} if the overlap is larger than a certain threshold (e.g. 0.3 or 0.5) and as \ac{FP} otherwise. If bounding boxes are considered, the \textit{\ac{IoU}} is computed between the reference and predicted bounding boxes (\textit{Box IoU}). For more fine-grained annotations in form of a pixel mask, the \textit{\ac{IoU}} may be computed for the complete mask (\textit{Mask IoU}). The \textit{Mask IoU} is less sensitive to structure boundary quality in larger objects (cf. Sec.~\ref{sec:segmentation}). This is due to the fact that boundary pixels will increase linearly while pixels inside the structure will increase quadratically with an increase in structure size. The \textit{Boundary IoU} measures the \textit{\ac{IoU}} of two structures for mask pixels within a certain distance $d$ from the structure boundaries \cite{cheng2021boundary}, as illustrated in Figure~\ref{fig:definition-boundary-iou}.

\begin{figure}[H]
\begin{tcolorbox}[title= Discarding information of provided annotations, colback=white]
    \centering
    \includegraphics[width=0.7\linewidth]{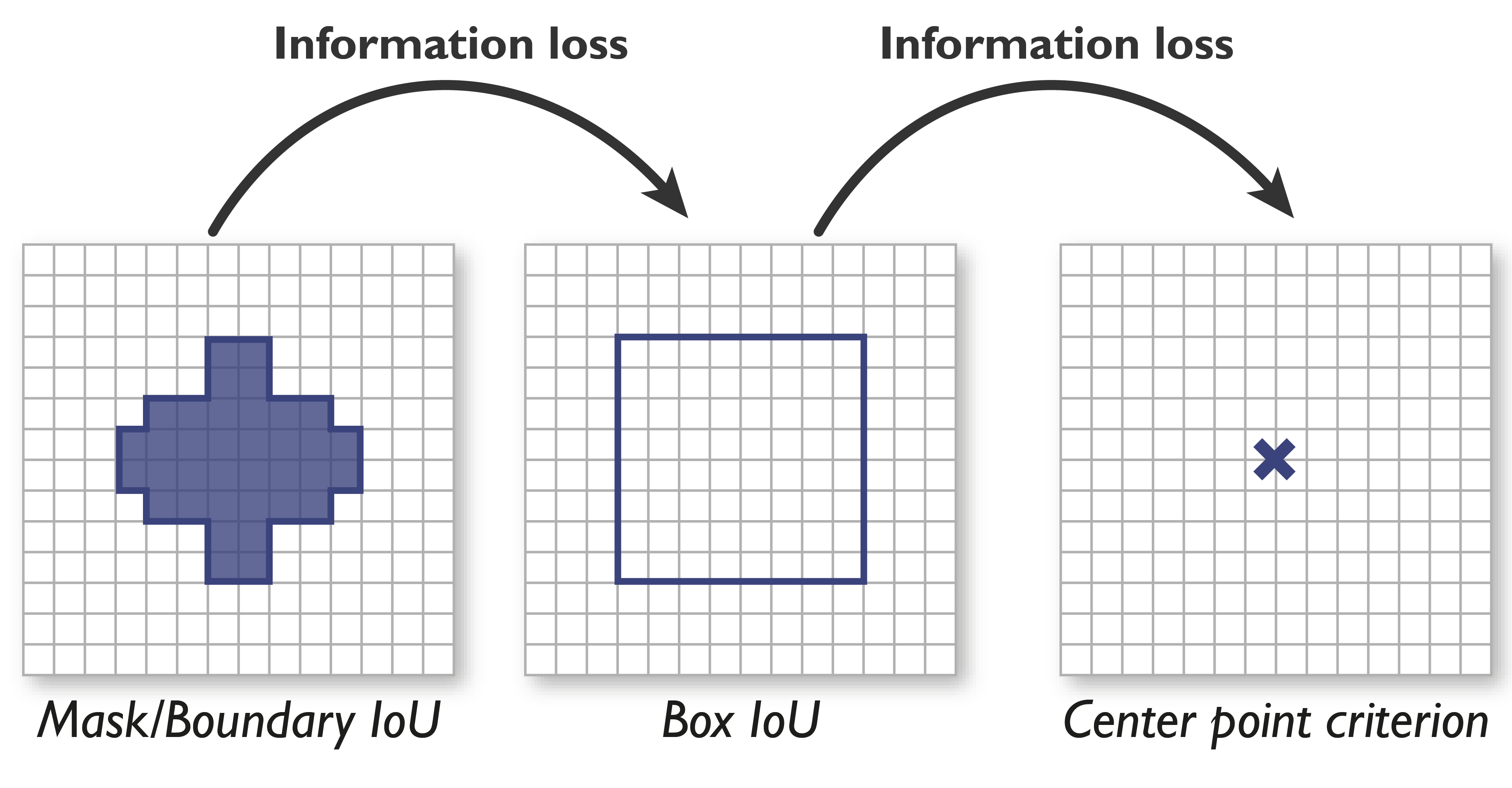}
    \caption{Selection of a localization criterion that discards spatial information should be well motivated by the given task.}
    \label{fig:od-information-loss}
\end{tcolorbox}
\end{figure}

The localization criterion should be carefully chosen according to the underlying motivation and research question and depending on the available coarseness of annotations. However, it should be noted that annotations of a lower resolution will result in an information loss, as illustrated in Figure~\ref{fig:od-information-loss}. For example, the \textit{Box IoU} is sometimes used although pixel-mask annotations are available because algorithms are expected to output rough localization in the shape of boxes. Such a simplification might cause problems if structures are not well-approximated by a box shape, or if structures can overlap causing multi-component masks (cf. Sec.~\ref{sec:detection}, Figure~\ref{fig:disconnected}). Lastly, it should be noted that the decision for a cutoff value on the localization criterion leads to instabilities in the validation (e.g. see Figure~\ref{fig:bb-2d-3d}). For this reason, it is common practice in the computer vision community to average metrics over multiple cutoff values (default for \textit{\ac{IoU}} criteria: 0.50:0.05:0.95 \cite{lin2014microsoft}). Generally speaking, the cutoff values should be chosen according to the driving biomedical question. For example, if particular interest lies on the exact outlines, higher thresholds should be chosen. On the other hand, for noisy reference standards, a low cutoff value is preferable.

\newpage

\begin{figure}[H]
\begin{tcolorbox}[title= Example process of resolving a matching ambiguity, colback=white]
    \centering
    \includegraphics[width=0.7\linewidth]{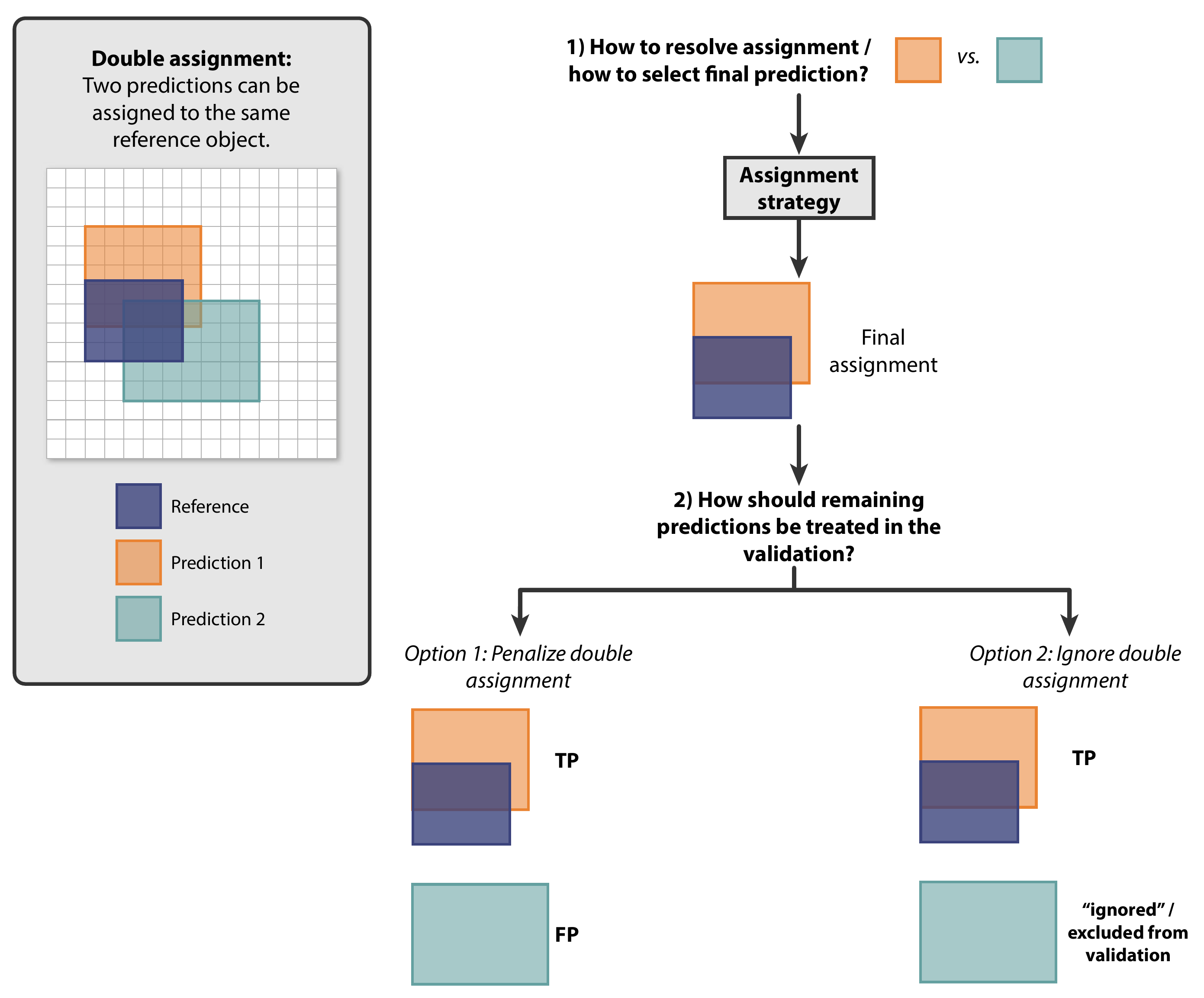}
    \caption{In case of multiple predictions that may be assigned to the same reference object, an assignment strategy needs to be chosen. This includes a decision on how to treat the remaining predictions. For example, they can be penalized with a \acf{FP} assignment or may be ignored. Used abbreviations: \textit{\acf{IoU}}, \textit{\acf{IoR}} and \acf{TP}.}
    \label{fig:od-assignment-strategy}
\end{tcolorbox}
\end{figure}

\paragraph{\textbf{Assignment strategy}}
The localization criterion alone is not sufficient to extract the final confusion matrix based on a fixed threshold for the predicted class probabilities (confidence scores), as ambiguities can occur. For example, two predictions may have been assigned to the same reference object in the localization step, or vice versa. These ambiguities need to be resolved in a further \textbf{assignment step}, as exemplarily shown in Figure~\ref{fig:od-assignment-strategy}. 

This assignment and thus the resolving of potential assignment ambiguities can be done via different strategies. The most common strategy in the computer vision community is the \textit{Greedy by Score} strategy \cite{everingham2015pascal}. All predictions in an image are ranked by their predicted class probability and iteratively (starting with the highest probability) assigned to the reference object with the highest localization criterion for this prediction. The selected reference object is subsequently removed from the process since it can not be matched to any other prediction (unless double assignments are allowed). The \textit{Hungarian Matching} \cite{kuhn1955hungarian} is associated with a cost function, usually depending on the localization criterion, which is minimized to find the optimal assignment of predictions and reference. In the biomedical domain, more sophisticated matching strategies are often avoided by setting the localization criterion threshold to \textit{\ac{IoU}} > 0.5 and only allowing non-overlapping object predictions (which inherently avoids matching conflicts). In the case of a high ratio of touching reference objects and common non-split errors, meaning that one prediction overlaps with multiple reference objects, the Intersection over Reference (\textit{IoR}) \cite{matula2015cell} might be considered as an alternative to \textit{\ac{IoU}} \cite{matula2015cell} (see Figure~\ref{fig:od-ior}).

\newpage

\begin{figure}[H]
\begin{tcolorbox}[title= Common object detection metrics: Average Precision (AP), colback=white]
    \centering
    \includegraphics[width=1\linewidth]{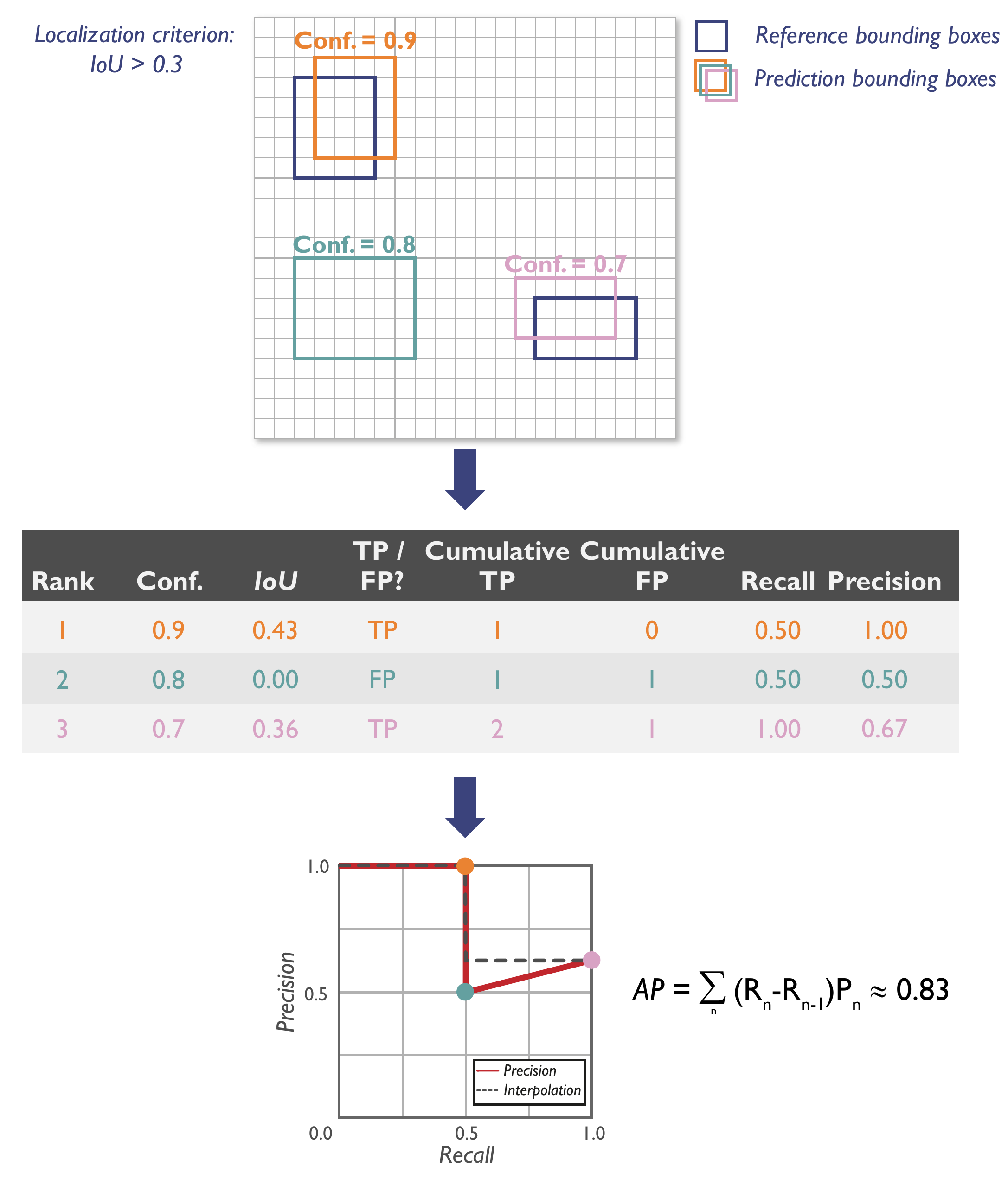}
    \caption{Exemplary computation of the \textit{\acf{AP}} in object detection tasks. Predictions are ranked according to their predicted class probabilities, represented by the confidence (conf.). Based on the \textit{\acf{IoU}} or a similar localization/hit criterion, it is determined whether the prediction is a \acf{TP} or \acf{FP} (here: \textit{\ac{IoU}} > 0.3). For the creation of the \textit{\acf{PR}} curve, \textit{Precision} and \textit{Recall} are computed for the accumulated \ac{TP} and \ac{FP} for every confidence score (conf.). The \textit{\ac{AP}} interpolates the points of the \textit{\ac{PR}} curve as shown by the dashed gray line. Calculating the \textit{\ac{AP}} for multiple classes and averaging the values per class yields the \textit{\acf{mAP}} score.}
    \label{fig:ap-example} 
\end{tcolorbox}
\end{figure}

\newpage
\paragraph{\textbf{Metric computation}}
Similar to image-level classification and semantic segmentation algorithms, object detection algorithms are commonly assessed with counting metrics, assuming a fixed confusion matrix, (cf. Figs.~\ref{fig:def-classification-1} and \ref{fig:def-classification-2}). However, one of the most popular object detection metrics is the multi-threshold metric \textit{\acf{AP}} \citep{lin2014microsoft}, which is the area under the \textit{\acf{PR}} curve for a certain interpolation scheme. The \textit{\ac{PR}} curve is computed similarly to the \textit{\ac{ROC}} curve by scanning over confidence thresholds and computing the \textit{Precision (\ac{PPV})} and \textit{Recall (Sensitivity)} for every threshold (cf. Figure~\ref{fig:def-auc}). Note in this context that the popular \textit{\ac{ROC}} curve is not applicable in object detection tasks because \ac{TN} are not available. Also, while the \textit{\ac{ROC}} curve is monotonically rising, this behavior may not be expected from the \textit{\ac{PR}} curve, which typically features a zigzag shape, as illustrated in Figure~\ref{fig:ap-example}. Specifically "as the level of \textit{Recall} varies, the \textit{Precision} does not necessarily change linearly due to the fact that \ac{FP} replaces \ac{FN} in the denominator of the \textit{Precision} metric." \cite{davis2006relationship}. A linear interpolation would therefore be overly optimistic, which is why more complex interpolation is needed, as detailed in \cite{davis2006relationship}. 

The area under the \textit{\ac{PR}} curve is typically calculated as the \textit{\ac{AP}} implying a conservative simplification of curve interpolation,

\begin{equation}
    \text{\emph{AP}} = \sum_i (R_i - R_{i-1})P_i,
\end{equation}
with $R_i$ and $P_i$ denoting the \textit{Recall} and \textit{Precision} at the $i$th threshold\footnote{\url{https://scikit-learn.org/stable/modules/generated/sklearn.metrics.average_precision_score.html}} (cf. dashed gray line in Figure~\ref{fig:ap-example}). For the \textit{\ac{PR}} curve, an optimal model would lead to \textit{Recall (Sensitivity)} and \textit{Precision (\ac{PPV})} of 1, therefore being the point $(1,1)$ on the \textit{\ac{PR}} curve. Conversely, a model with no skill level (random guessing) would result in a horizontal line with a precision proportional to the portion of positive samples, i.e. the prevalence (dashed line in Figure~\ref{fig:def-auc}). For computation of the metric (for a given class), the predictions are sorted in descending order of the confidence for each prediction (for that class). For each possible confidence threshold, the cardinalities are computed and the resulting tuple of \textit{Recall (Sensitivity)} and \textit{Precision} is added to the curve (cf. Figure~\ref{fig:ap-example}). In the case of multiple classes, the \textit{\ac{AP}} is calculated per class and averaged to obtain the \textit{\ac{mAP}} score.

\begin{figure}[H]
\begin{tcolorbox}[title= Common classification metrics: \acf{FROC}, colback=white]
    \centering
    \includegraphics[width=1\linewidth]{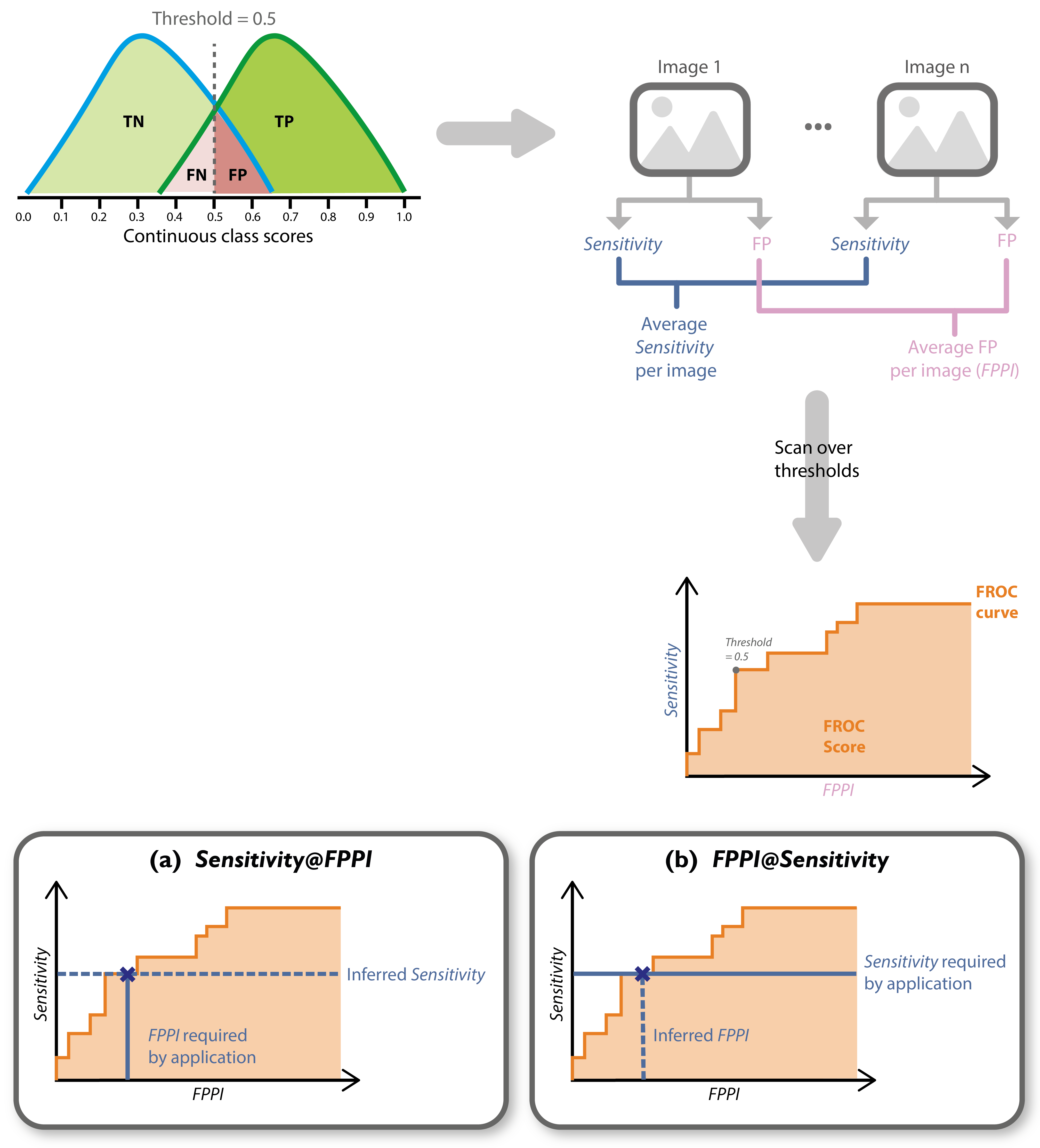}
    \caption{Principle of the \textit{\acf{FROC}} curve. Rather than being based on a static threshold (e.g. for generating the confusion matrix), the \textit{\ac{FROC}} curve is generated by applying a range of thresholds. The \textit{\ac{FROC} Score} measures the area under the \textit{\ac{FROC}} curve. In contrast to \textit{\acf{AUROC}}, \textit{\ac{FROC}} operates on object level by plotting the \textit{Sensitivity} against the \textit{\acf{FPPI}} per threshold (e.g. 0.5). In the case of an application-driven threshold (e.g. required \textit{Sensitivity} of 0.9), the metrics \textit{Sensitivity@\ac{FPPI}} and \textit{\ac{FPPI}@Sensitivity} can be calculated on the basis of the  the \textit{\ac{FROC}} curve.}
    \label{fig:def-froc}
\end{tcolorbox}
\end{figure}

In contrast to drawing the \textit{\ac{PR}} curve and computing the \textit{\ac{AP}}, \textit{\ac{FROC}} curve is often favoured in the clinical context due to its easier interpretability. It operates at object level and plots the average number of \textit{\ac{FPPI}} (in contrast to the \textit{\ac{FPR}}) against the \textit{Sensitivity} \cite{chakraborty2019analysis, van2010comparing} (cf. Figure~\ref{fig:def-froc}). The \textit{\ac{FROC} Score}, measured as the area under the \textit{\ac{FROC}} curve, however, is not bounded between 0 and 1 and the employed \textit{\ac{FPPI}} scores vary across studies, such that there exists no standardized definition of an area under the respective curve. Overall, the decision between the two metrics often boils down to a decision between a standardized and technical validation versus an interpretable and application-focused validation. 


\newpage

\newpage

\begin{figure}[H]
\begin{tcolorbox}[title= Validating objects per data set \textit{vs.} validating objects per image, colback=white]
    \centering
    \includegraphics[width=1\linewidth]{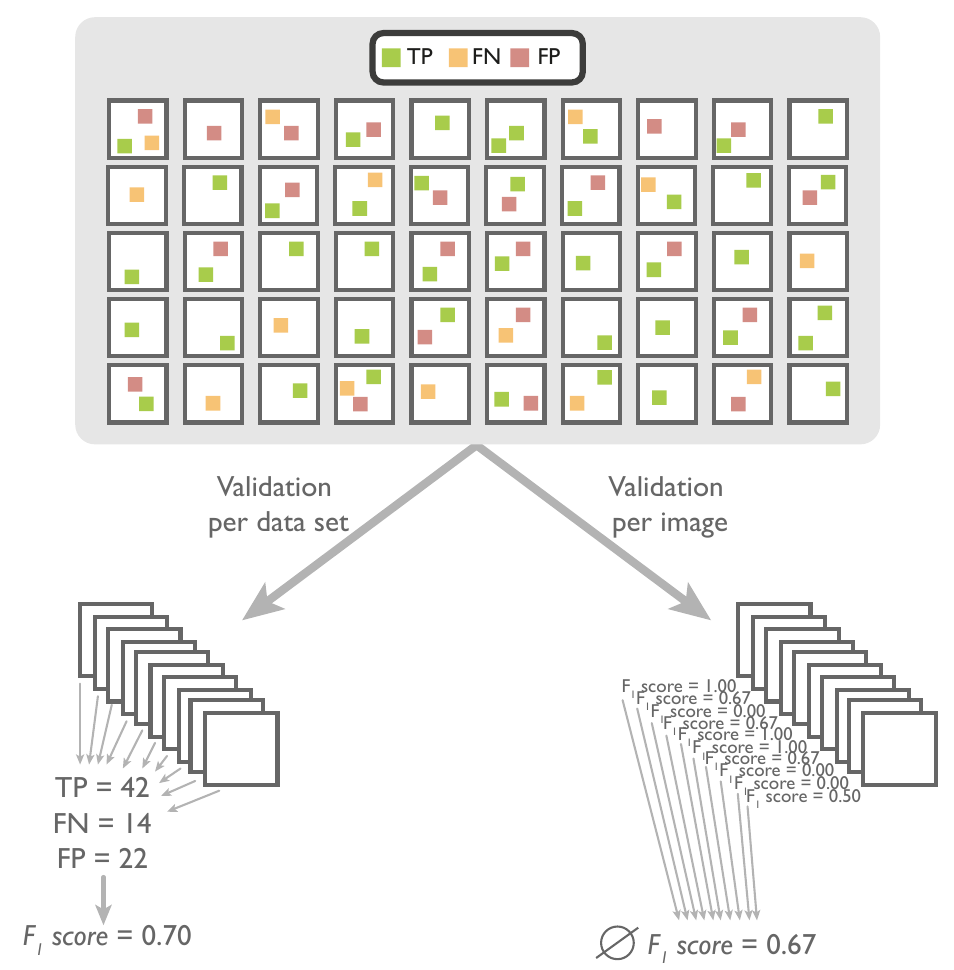}
    \caption{Validation on object level can be performed per data set (left) or per image (right). For the per-data set validation of objects, the cardinalities are calculated over the whole data set. For the per-image validation of objects, metric scores are computed per image and aggregated afterwards. $\varnothing$ refers to the average \textit{F$_1$ Score}.}
    \label{fig:perimage-perdataset} 
\end{tcolorbox}
\end{figure}

\newpage
\paragraph{Counting cardinalities at different scales}
In image-level classification problems, validation is naturally performed on the entire data set, while segmentation typically relies on computing metrics for each image and then aggregating metric values. This latter approach is not applicable in object detection in a straightforward manner because of the relatively small number of samples per image (typically a few objects rather than thousands of pixels). Figure~\ref{fig:perimage-perdataset} illustrates the per-image and the per-data set validation of objects. In the per-image aggregation approach, special care needs to be taken in the case of an empty reference or prediction, as detailed in Sec.~\ref{sec:detection} (Figure~\ref{fig:empty-ref-pred-od}).

A few comments on how to compute evaluation metrics while aggregating matched objects per image: 

\begin{itemize}
    \item \textbf{Counting metrics:} Computing per-class metrics such as Sensitivity, \ac{PPV}, or F$_1$ Score (see Figure~\ref{fig:def-classification-1}) works intuitively (as depicted in Figure~\ref{fig:perimage-perdataset}) by computing a score per image and averaging scores over the data set while considering the \texttt{\ac{NaN}} conventions shown in Figure~\ref{fig:empty-ref-pred-od}.
    \item \textbf{Multi-threshold metrics:} While scanning over all thresholds in the data set (analogously to per data set evaluation), for every threshold the Precision (\ac{PPV}) and Recall (Sensitivity) scores are computed per image/patient (while considering the \texttt{\ac{NaN}} conventions in Figure~\ref{fig:empty-ref-pred-od}) and averaged over the data set. The averaged Precision and Recall score pairs per threshold form the \ac{PR} curve. The \ac{AP} (see Figure~\ref{fig:ap-example}) can be computed for this curve analogously to the per-data set evaluation. The \ac{FROC} curve computation works analogously while replacing Precision per image with \ac{FPPI}. 
    \item \textbf{Single working point evaluation:} A single working point (e.g. \ac{PPV}@Sensitivity) can be retrieved based on the curves resulting from multi-threshold evaluation described above.
\end{itemize}

\newpage

\subsection{Instance Segmentation}
\label{sec:fundamentals_is}
In contrast to semantic segmentation, \textbf{instance segmentation} problems distinguish different instances of the same class (e.g. different lesions). Similarly to object detection problems, the task is to detect individual instances of the same class, but detection performance is measured by pixel-level correspondences (as in semantic segmentation problems). Optionally, instances can be applied to one of multiple classes. Validation metrics in instance segmentation problems often combine common detection metrics with segmentation metrics applied per instance.  
 
\begin{figure}[H]
\begin{tcolorbox}[title= Common instance segmentation metrics: \acf{PQ}, colback=white]
    \centering
    \hspace{-0.5cm}
    \includegraphics[width=\linewidth]{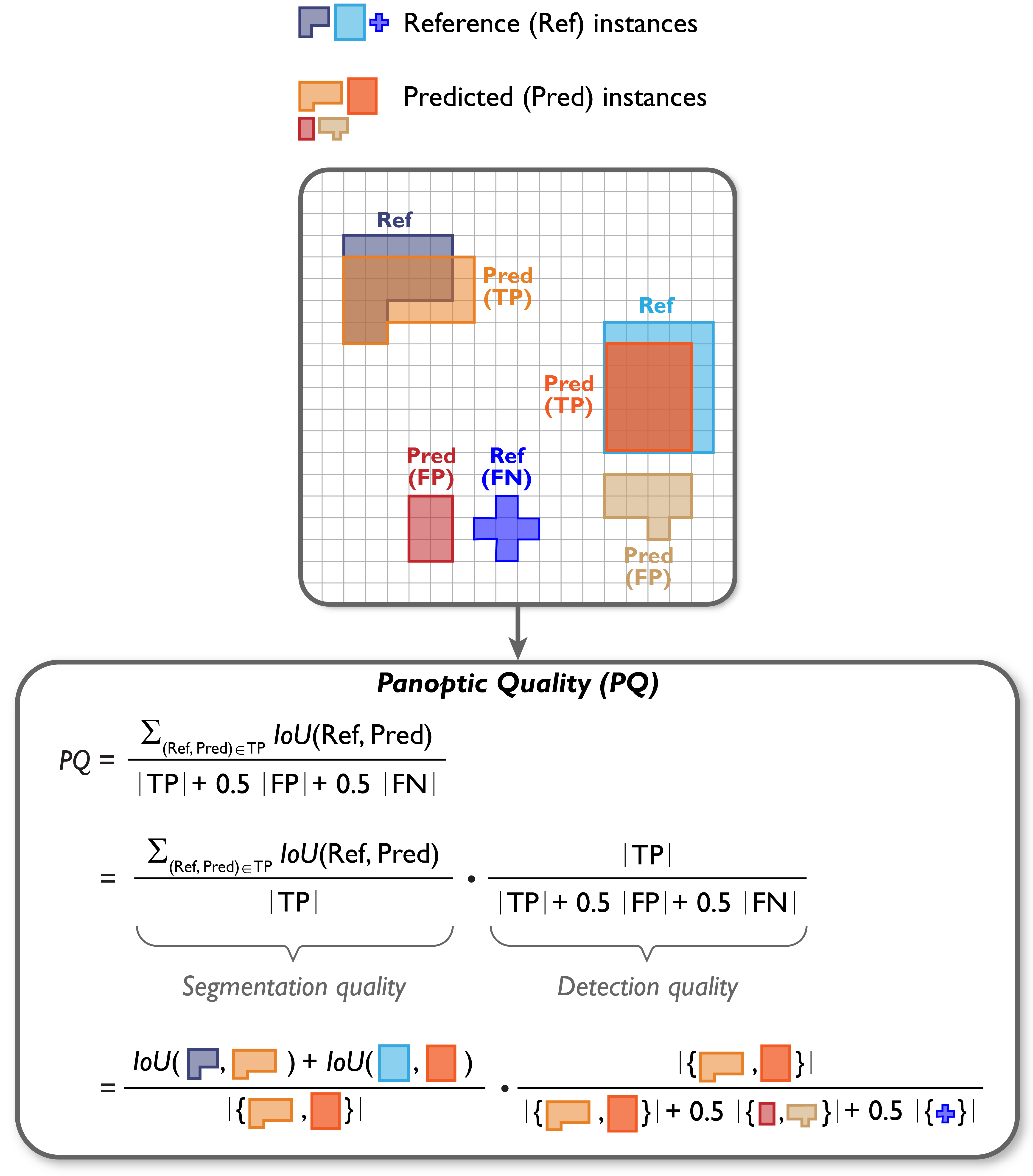}
    \caption{The \textit{\acf{PQ}} measures the \textbf{segmentation and detection quality} of a prediction in one score. The segmentation quality is assessed via the \textit{\acf{IoU}} for every predicted \acf{TP} instance. The detection quality is measured via the \textit{F$_1$ score} by taking into account the \acf{FP} and \acf{FN} predictions.}
    \label{fig:definition-pq}
\end{tcolorbox}
\end{figure}

If detection and segmentation performance should be assessed simultaneously in a single score, the \textit{\ac{PQ}} metric can be utilized \cite{kirillov2019panoptic}. As shown in Figure~\ref{fig:definition-pq}, the segmentation quality is assessed by averaging the \textit{\ac{IoU}} scores for all \ac{TP} instances. While this part alone would neglect \ac{FP} and \ac{FN} predictions, it is multiplied with the detection quality, which is equal to the \textit{F$_1$ score}. For perfect segmentation results, i.e. an average \textit{\ac{IoU}} of 1, the \textit{\ac{PQ}} would simply equal the \textit{F$_1$ Score}. In other words, while segmentation quality in the \textit{F$_1$ Score} is merely integrated via a hard cutoff during localization (e.g. \textit{\ac{IoU}} > 0.5) affecting the counts of \ac{TP} versus \ac{FP} and \ac{FN}, \textit{\ac{PQ}} allows for a continuous measurement of segmentation quality by 1) employing a lose prior localization criterion (e.g. \textit{\ac{IoU}} > 0, i.e. any prediction overlapping with a reference instance is treated as \ac{TP}) and 2) replacing the simple counting of \ac{TP} by adding up the continuous \textit{\ac{IoU}} scores per \ac{TP}. \textit{\ac{PQ}} was introduced for panoptic segmentation problems that are a combination of semantic and instance segmentation \cite{kirillov2019panoptic}. It is computed per class (including background classes) and can therefore be directly used for instance segmentation problems by only assessing the foreground instances. If needed, the background quality can also be assessed.

\begin{figure}[H]
\begin{tcolorbox}[title= Pitfall: Assessing segmentation and detection quality simultaneously, colback=white]
    \centering
    \includegraphics[width=1\linewidth]{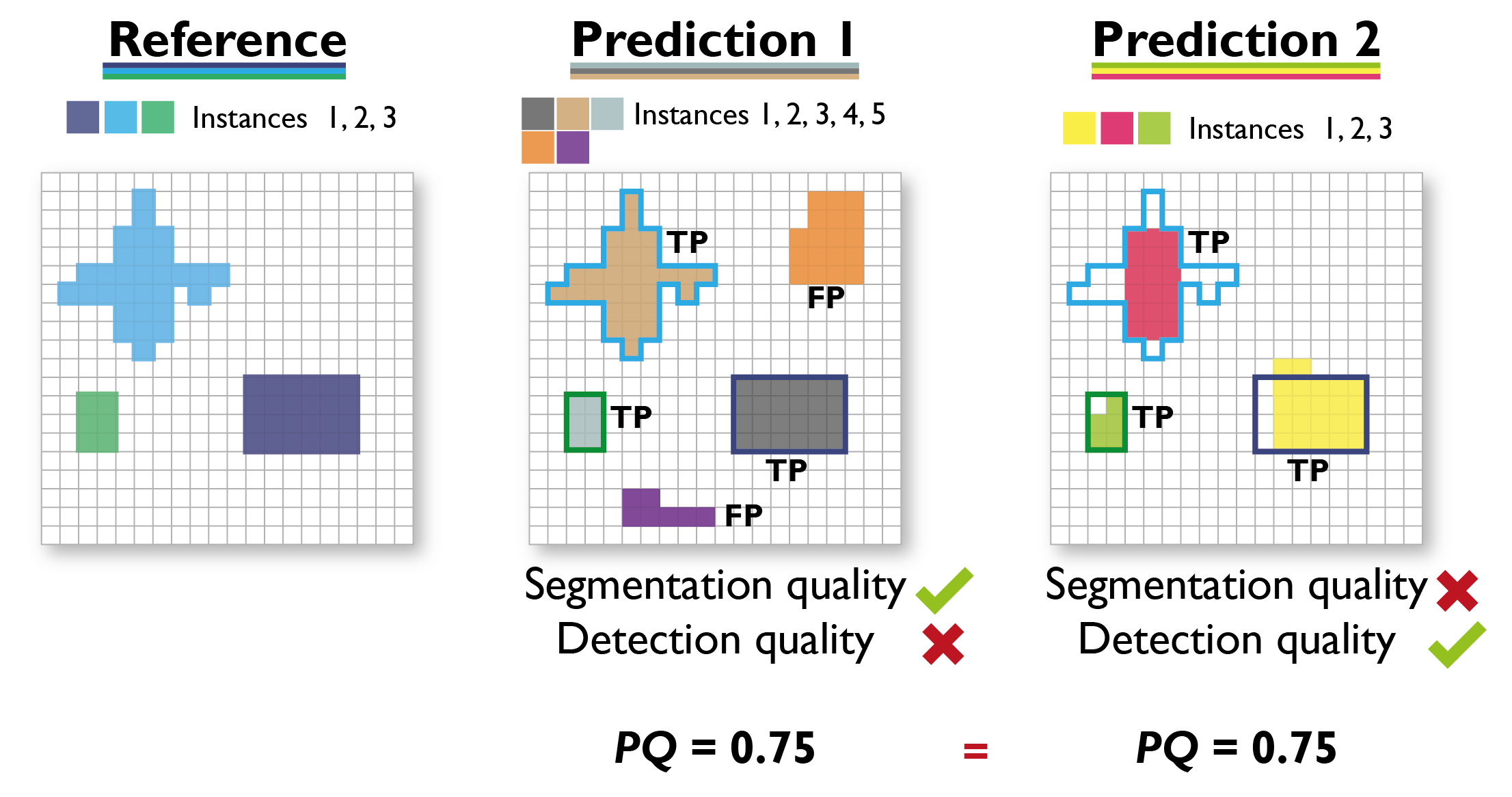}
    \caption{Effect of assessing segmentation and detection quality in a single score. \textit{Prediction 1} achieves a high segmentation but low detection quality (with several \acf{FP} predictions); conversely, \textit{Prediction 2} only predicts \acf{TP} and no \ac{FP} instances but achieves low segmentation quality). However, both yield the same \acf{PQ} score.}
     \label{fig:pq-sq-vs-dq}
\end{tcolorbox}
\end{figure}

\newpage
\textit{\ac{PQ}} covers segmentation and detection quality in a single score. However, this can be misleading, as shown in Fig.~\ref{fig:pq-sq-vs-dq}. \textit{Prediction 1} with a perfect segmentation and poor detection achieves a \textit{\ac{PQ}} score similar to that of \textit{Prediction 2}, which detects all objects correctly (without any \ac{FP} or \ac{FN}), but only provides moderate segmentation results. If segmentation and detection quality should be assessed individually, two separate metrics - one for segmentation quality and one for detection quality - should be preferred. 

\newpage
 It should be noted that instance segmentation problems are often phrased as semantic segmentation problems with an additional post-processing step, such as connected component analysis \cite{rosenfeld1966sequential}. In practice, predicted class probabilities, yielded by modern segmentation algorithms, are often discarded in the post-processing step and are thus not available for subsequent validation. Figure~\ref{fig:class-scores} illustrates how to overcome this potential problem.

\begin{figure}[H]
\begin{tcolorbox}[title= Retaining class scores when extracting instances from semantic segmentation output, colback=white]
    \centering
    \includegraphics[width=1\linewidth]{images/Detection/class_scores.png}
    \caption{Retaining class scores when extracting instances from semantic segmentation output. Often, semantic segmentation methods are used for instance segmentation problems by post-processing the segmentation output (e.g. via connected component analysis). Predicted class probabilities are often lost during this process and therefore not available for further validation processes. However, instance-wise class scores can be retained by taking the maximum or average over all pixel-wise class scores of every instance.}
    \label{fig:class-scores} 
\end{tcolorbox}
\end{figure}

\newpage
\section{Pitfalls due to category-metric mismatch}
\label{sec:underlying-task}
Performance metrics are typically expected to reflect a domain-specific validation goal (e.g. clinical goal). Previous research, however, suggests, that this is often not the case \cite{saha2021anatomical}. Before choosing validation metrics, the correct problem category needs to be defined. In the following, we will describe pitfalls related to metrics not being applied to the appropriate problem category.

\begin{figure}[H]
\begin{tcolorbox}[title= Pitfall: Mismatch semantic segmentation $\leftrightarrow$ object detection, colback=white]
    \centering
    \includegraphics[width=1\linewidth]{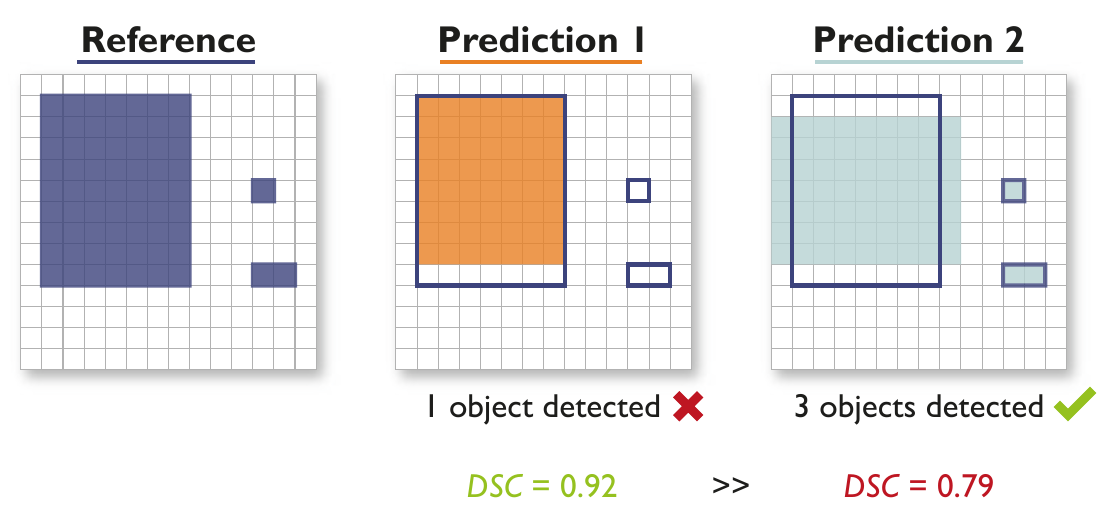}
    \caption{Effect of using a segmentation metric for object detection. In this example, the prediction of an algorithm only detecting one of three structures (\textit{Prediction 1}) leads to a higher \textit{\acf{DSC}} compared to that of another algorithm (\textit{Prediction 2}) detecting all structures.}
    \label{fig:DSC-detection}
\end{tcolorbox}
\end{figure}

\paragraph*{\textbf{Mismatch semantic segmentation $\leftrightarrow$ object detection}}
A common problem is that segmentation metrics, such as the \textit{\ac{DSC}}, are applied to \textit{object detection} tasks \citep{carass2020evaluating,jager2020challenges}, as illustrated in Figure~\ref{fig:DSC-detection}. From a clinical perspective, for example, the algorithm producing \textit{Prediction 2} and covering all three structures of interest (e.g. tumors) would be clinically much more valuable compared to the one producing a highly accurate segmentation for one structure but missing the other two in \textit{Prediction 1}. This is not reflected in the metric values, which are substantially higher for \textit{Prediction 1}. In general, the \textit{\ac{DSC}} is strongly biased against single objects, therefore not appropriate for the detection of multiple structures \citep{yeghiazaryan2018family, kirillov2019panoptic}. 

\begin{figure}[H]
\begin{tcolorbox}[title= Pitfall: Mismatch semantic $\leftrightarrow$ instance segmentation, colback=white]
    \centering
    \includegraphics[width=1\linewidth]{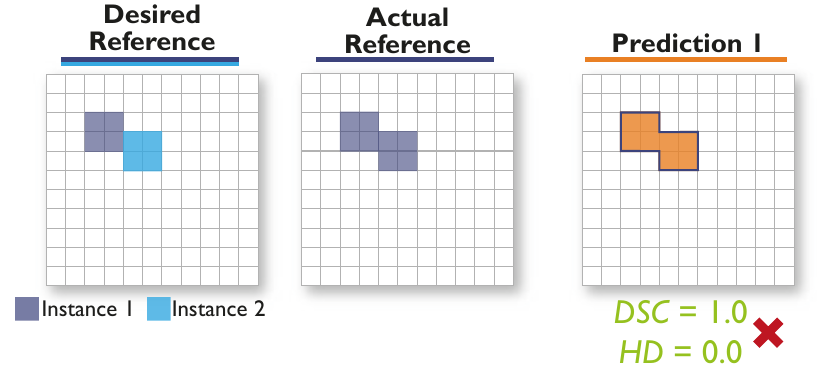}
    \caption{Effect of using segmentation metrics for object detection problems. The pixel-level \acf{DSC} of a prediction recognizing every structure (\textit{Prediction 2}) is lower than that of a prediction that only recognizes one of the three structures (\textit{Prediction 1}).}
    \label{fig:touching}
\end{tcolorbox}
\end{figure}

\begin{figure}[H]
\begin{tcolorbox}[title= Pitfall: Mismatch image-level classification $\leftrightarrow$ object detection, colback=white]
    \centering
    \includegraphics[width=1\linewidth]{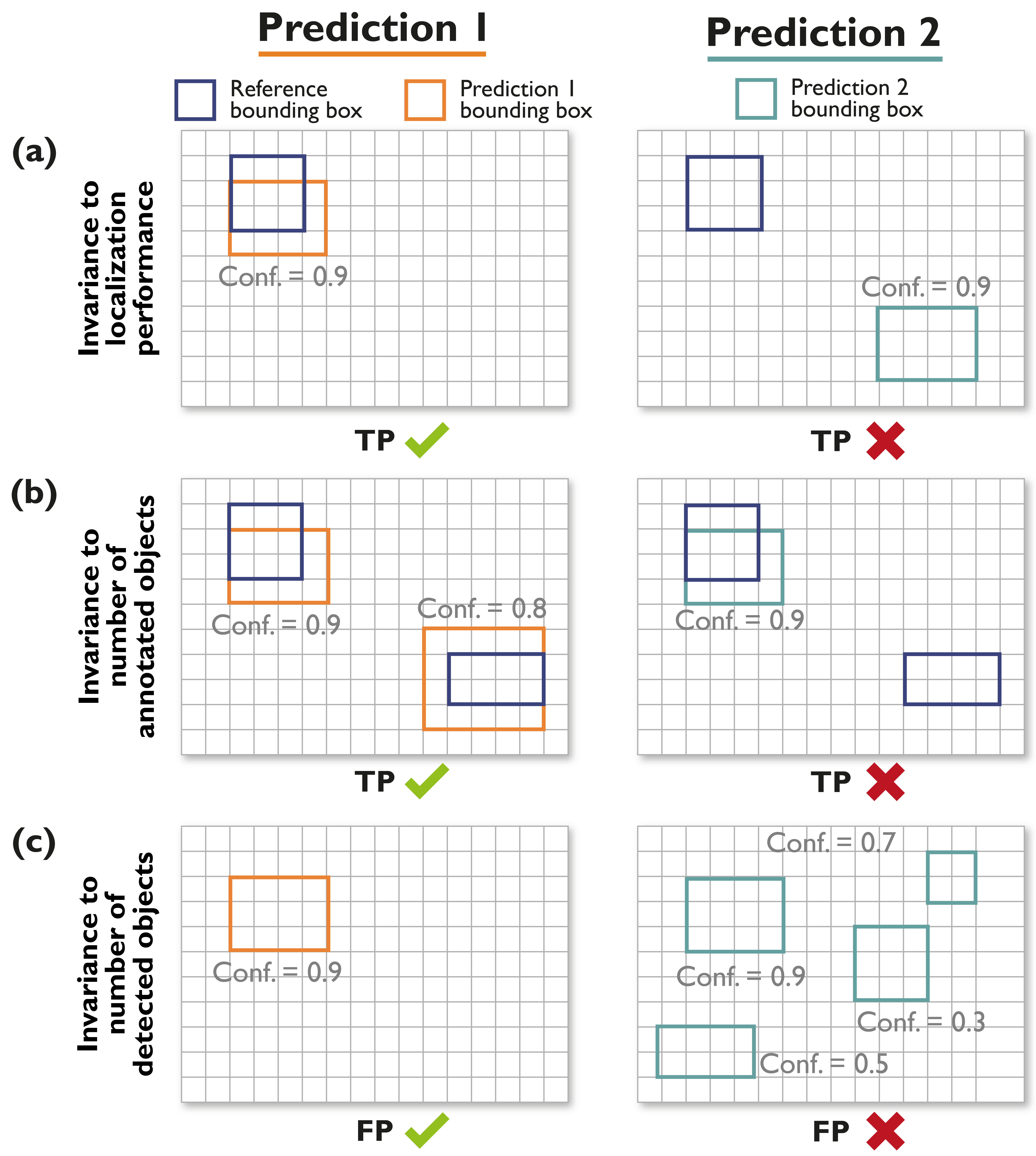}
    \caption{Image-level classification metrics such as the \acf{AUROC} curve can be used to validate object detection models by first aggregating predictions to one image-level score (per class). This validation scheme discards the information on the object matching (localization, number of objects etc.), which leads to several problems:
    \textbf{(a)} The image-level \textit{\ac{ROC}} curve does not measure the localization performance. Both \textit{Prediction 1} and \textit{2} are considered as \acf{TP} due to their score being very high, although \textit{Prediction 2} is not hitting the annotated object. \textbf{(b)} The image-level \textit{\ac{ROC}} is invariant to the number of annotated objects in an image. The curve does not discriminate between a model detecting all positives (\textit{Prediction 1}) and a model detecting only one of the positives (\textit{Prediction 2}), as long as the maximum score is the same. \textbf{(c)} The image-level \textit{\ac{ROC}} is invariant to the number of detections in an image. The curve does not discriminate between a model with many False Positives (FP), (\textit{Prediction 2}) and a model with just one \ac{FP} (\textit{Prediction 1}), as long as the maximum score is the same. The class probabilities are represented by confidence scores (conf.).}
    \label{fig:roc}
\end{tcolorbox}
\end{figure}

\newpage
\paragraph*{\textbf{Mismatch semantic $\leftrightarrow$ instance segmentation}}
In segmentation problems, the driving research question should decide whether semantic or instance segmentation should be chosen for validation. This is particularly relevant when multiple objects within one image overlap or touch, as often occurring in cell images. For semantic segmentation problems, overlapping or touching objects may end up merged into a single object without clear boundaries or distinction between the single objects. Instance segmentation problems, on the other hand, ensure that the borders of touching or overlapping structures can be accurately assigned and that objects can be differentiated. If instance segmentation is preferred, the labels need to be chosen accordingly. An example is shown in Figure~\ref{fig:touching}: The desired annotation consists of two different instances, but only semantic labels are available (middle). A prediction will only be as accurate as the reference, hence detecting only one instance but yielding a perfect metric score although the desired task is not solved.
\newpage
\paragraph{\textbf{Mismatch image-level classification $\leftrightarrow$ object detection}}
Tasks that should be validated at image level are sometimes erroneously approached with object detection models instead of image-level classification models \cite{jaeger2020retina}. Object detection models are designed to handle different objects in an image rather than the complete image and will naturally introduce problems in a validation setting on image level. Object detection tasks are dependent on choosing a proper localization criterion, which is not needed for an image-level classification problem. For example, a \textit{\ac{ROC}} curve, typically used for assessing the performance of image-level classification algorithms, does not consider the localization step needed in object detection tasks, as it was designed to validate at image rather than object level. It therefore does not take into account whether a detected object is at the correct location in the image. Moreover, when validation on image level is conducted by using an object detection model, the detection with the largest class probability (confidence score) of all detections in one image is usually taken, neglecting all other predictions. This does not capture the performance of the model accurately. Figure~\ref{fig:roc} illustrates some of the resulting problems: 

\begin{enumerate}
    \item The image-level \textit{\ac{ROC}} curve \textbf{does not measure the localization performance}. As can be seen from Figure~\ref{fig:roc}a, the validation is performed per image, not per object, therefore not considering whether an object is actually hit (see \textit{Prediction 2}).
    \item The image-level \textit{\ac{ROC}} curve is \textbf{invariant to the number of annotated objects}. As can be seen from Figure~\ref{fig:roc}b, the curve can not discriminate between a model detecting all objects in an image (\textit{Prediction 1}) or just detecting one object (\textit{Prediction 2}), as long as the largest score is the same across predictions.
    \item The image-level \textit{\ac{ROC}} curve is \textbf{invariant to the number of detected objects}. As can be seen from Figure~\ref{fig:roc}c, the curve can not discriminate between a model detecting many \ac{FP} objects in an image (\textit{Prediction 2}) or only detecting one \ac{FP} (\textit{Prediction 1}), as long as the largest score is the same.
\end{enumerate}

\newpage

\begin{figure}[H]
\begin{tcolorbox}[title= Pitfall: Metrics may be poor proxies for computing properties of interest, colback=white]
    \centering
    \includegraphics[width=\linewidth]{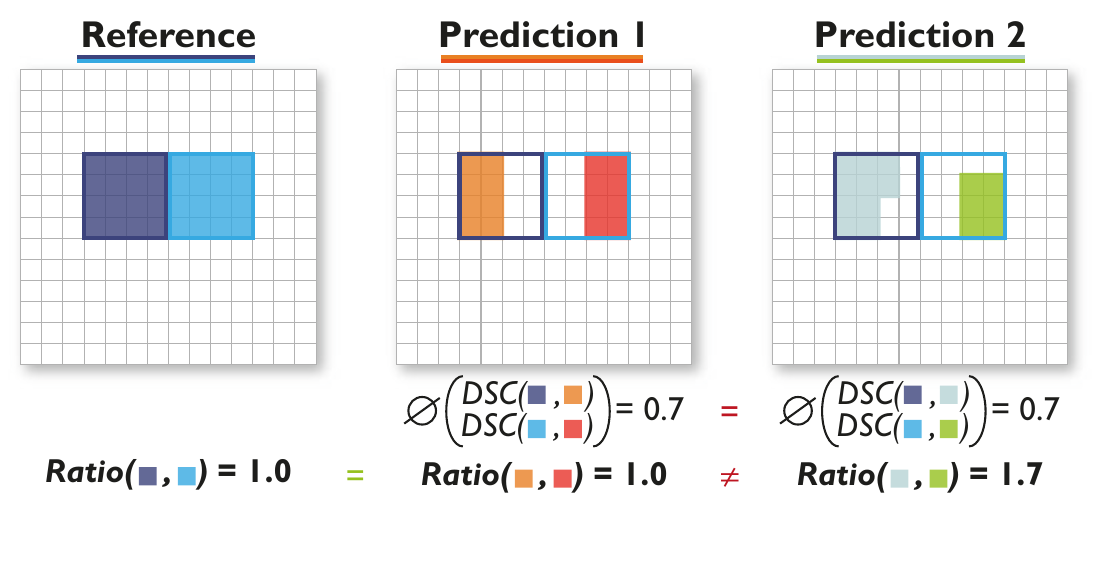}
    \caption{In the absence of a matching problem category for the problem at hand, it may not be possible to find a common metric that ideally captures the domain interest. In this example, accuracy of the ratio between two volumes is the property of interest (e.g., the percentage of blood volume ejected in each cardiac cycle \cite{bamira2018imaging}). Using overlap-based segmentation metrics (here: \textit{\acf{DSC}}) to measure the volumetric ratio may be misleading. \textit{Predictions 1} and \textit{2} result in similar averaged \textit{\ac{DSC}} metric values although they result in a different ratio between structure volumes, which is the parameter of interest. $\varnothing$ refers to the average \textit{\ac{DSC}} values.}
     \label{fig:context-ratio}
\end{tcolorbox}
\end{figure}

\newpage
\paragraph{\textbf{No matching problem category}} Metrics should reflect a domain-specific validation goal. This goal may not align with commonly used technical measures like the \textit{\ac{DSC}}. Figure~\ref{fig:context-ratio} shows an example with the property of interest being the accuracy of the ratio between two structure volumes, indicating, for example, the percentage of blood volume ejected in each cardiac cycle \cite{bamira2018imaging}. Both predictions will result in similar averaged \textit{\ac{DSC}} scores, although the ratio of the volumes vastly differs. A common segmentation metric thus does not reflect the actual research question in this case.

\newpage
\section{Pitfalls related to image-level classification}
\label{sec:classification}
\begin{figure}[H]
\begin{tcolorbox}[title= Pitfall: Class imbalance, colback=white]
    \centering
    \includegraphics[width=0.9\linewidth]{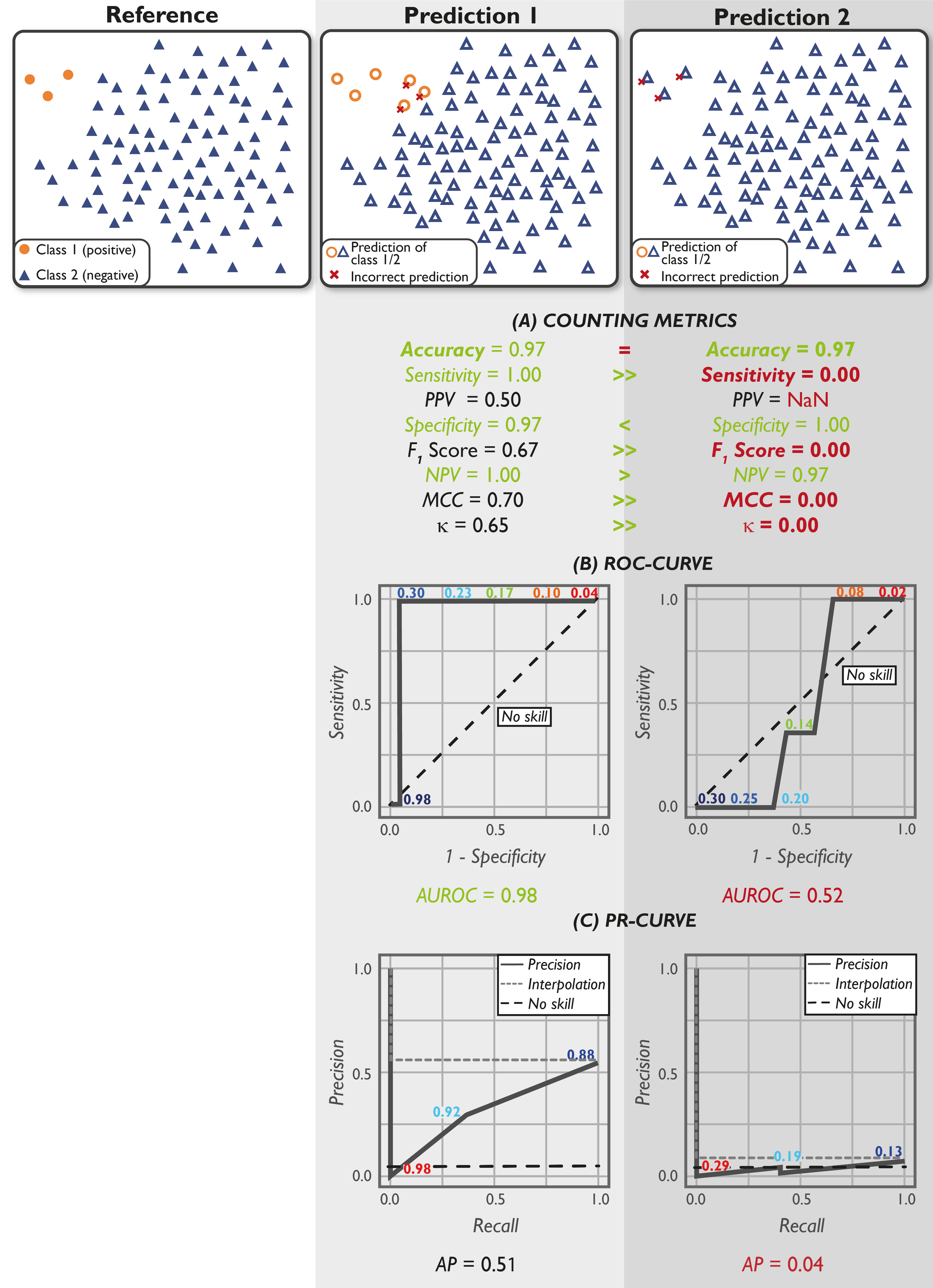}
    \caption{Effect of class imbalance. Not every metric is designed to reflect class imbalance (e.g. \textit{Accuracy}). In the case of underrepresented classes, such a  metric yields a high value even if the classifier performs very poorly for one of the classes (here: \textit{Prediction 2}). Multi-threshold metrics, such as the \textit{\acf{AUROC}} and the \textit{\acf{AP}}, reveal the weakness, indicating that \textit{Prediction 2} is not better than random guessing. For comparison, a no skill classifier (random guessing) is shown as a black dashed line. For the \textit{\acf{PR} curves}, the interpolation applied to compute the \textit{\ac{AP}} metric is shown by a dashed grey line. Thresholds used for curve generation are provided as small numbers in the curve. Further abbreviations: \textit{\acf{PPV}}, \textit{\acf{NPV}}, \textit{\acf{MCC}}, \textit{Cohen's Kappa} $\kappa$.}
    \label{fig:class-imbalance}
\end{tcolorbox}
\end{figure}

\begin{figure}[H]
\begin{tcolorbox}[title= Pitfall: Misleading \textit{\ac{BA}} for underrepresented classes, colback=white]
    \centering
    \includegraphics[width=0.7\linewidth]{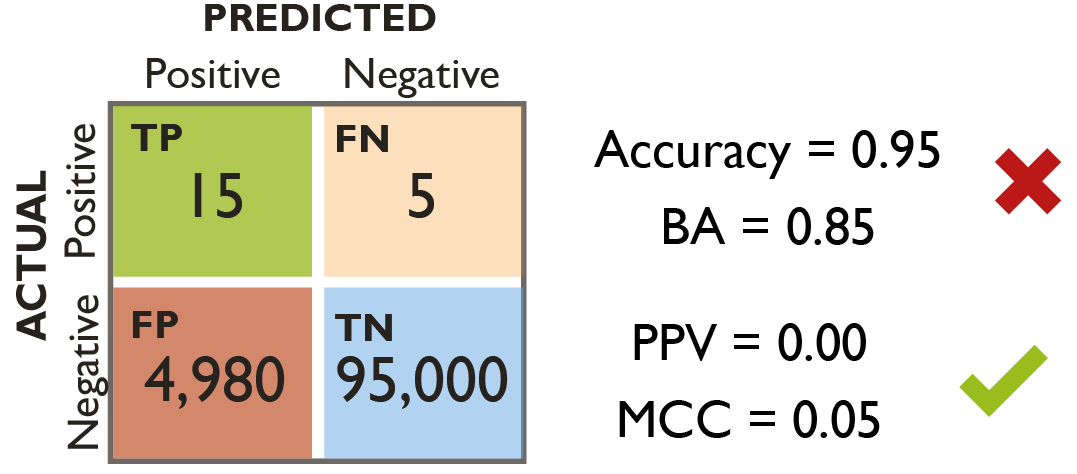}
    \caption{In the case of underrepresented classes, common metrics may yield misleading values. In the given example, Accuracy and \textit{\acf{BA}} have a high score despite the high amount of \acf{FP} samples. The class imbalance is only uncovered by metrics considering predictive values (here: \textit{\acf{PPV}} and \textit{\acf{MCC})}.}
    \label{fig:misleading-ba-j}
\end{tcolorbox}
\end{figure}

\begin{figure}[H]
\begin{tcolorbox}[title= Pitfall: Metric landscapes in the case of class imbalance, colback=white]
    \centering
    \includegraphics[width=0.9\linewidth]{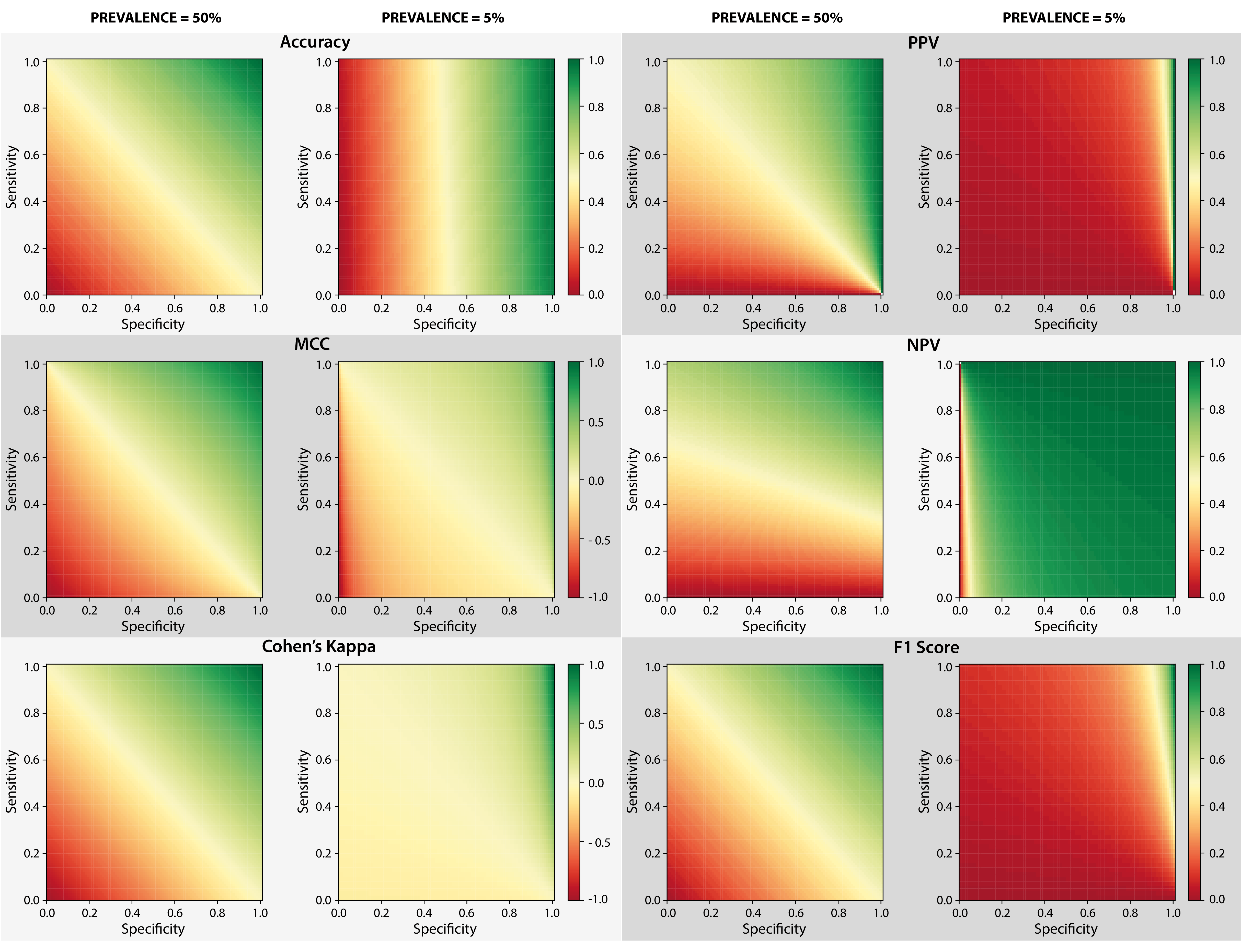}
    \caption{Effect of class imbalance on prevalence-dependent metrics. The metric landscape is shown as a function of the prevalence (data balance; here: prevalence = 50\% and prevalence = 5\%), \textit{Sensitivity} and \textit{Specificity}. Used abbreviations: \textit{\acf{MCC}}, \textit{\acf{PPV}}, \textit{\acf{NPV}}. Plots generated based on \cite{brown2018classifiers}.}
    \label{fig:metric-landscapes}
\end{tcolorbox}
\end{figure}

\begin{figure}[H]
\begin{tcolorbox}[title= Pitfall: Hierarchical structure of classes, colback=white]
    \centering
    \includegraphics[width=1\linewidth]{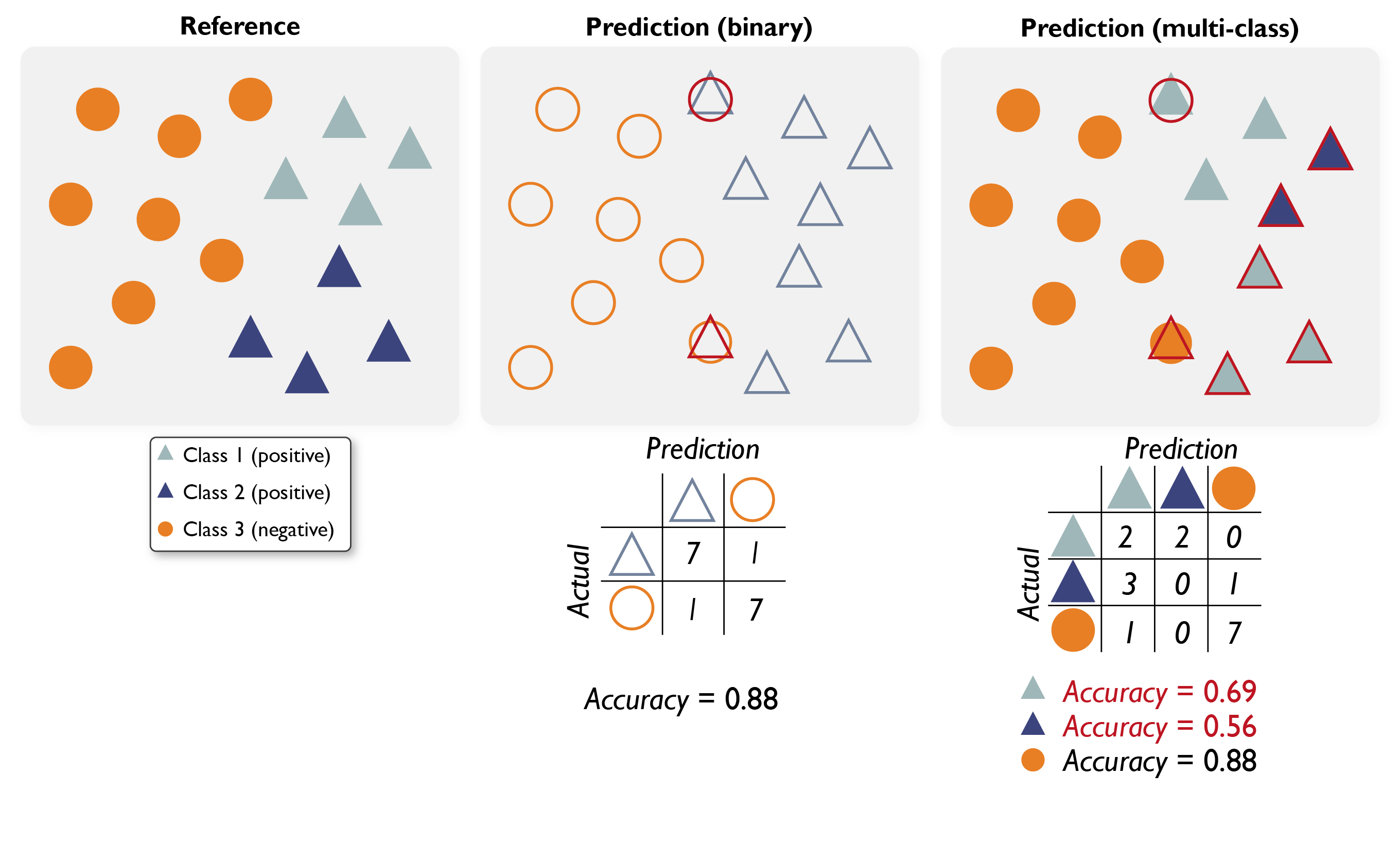}
    \caption{Classes in categorical classification may be hierarchically structured, for example in the shape of multiple positive classes and one negative class. The phrasing of the problem as binary \textit{vs.} multi-class hugely affects the validation result. Binary classification (middle), differentiating triangles from circles, yields a good \textit{Accuracy}, while per-class validation yields a poor score because the two circle classes cannot be distinguished well. Incorrect predictions are indicated by a red cross.}
    \label{fig:multi-class}
\end{tcolorbox}
\end{figure}

\begin{figure}[H]
\begin{tcolorbox}[title= Pitfall: Different metrics measure complementary properties, colback=white]
    \centering
    \includegraphics[width=1\linewidth]{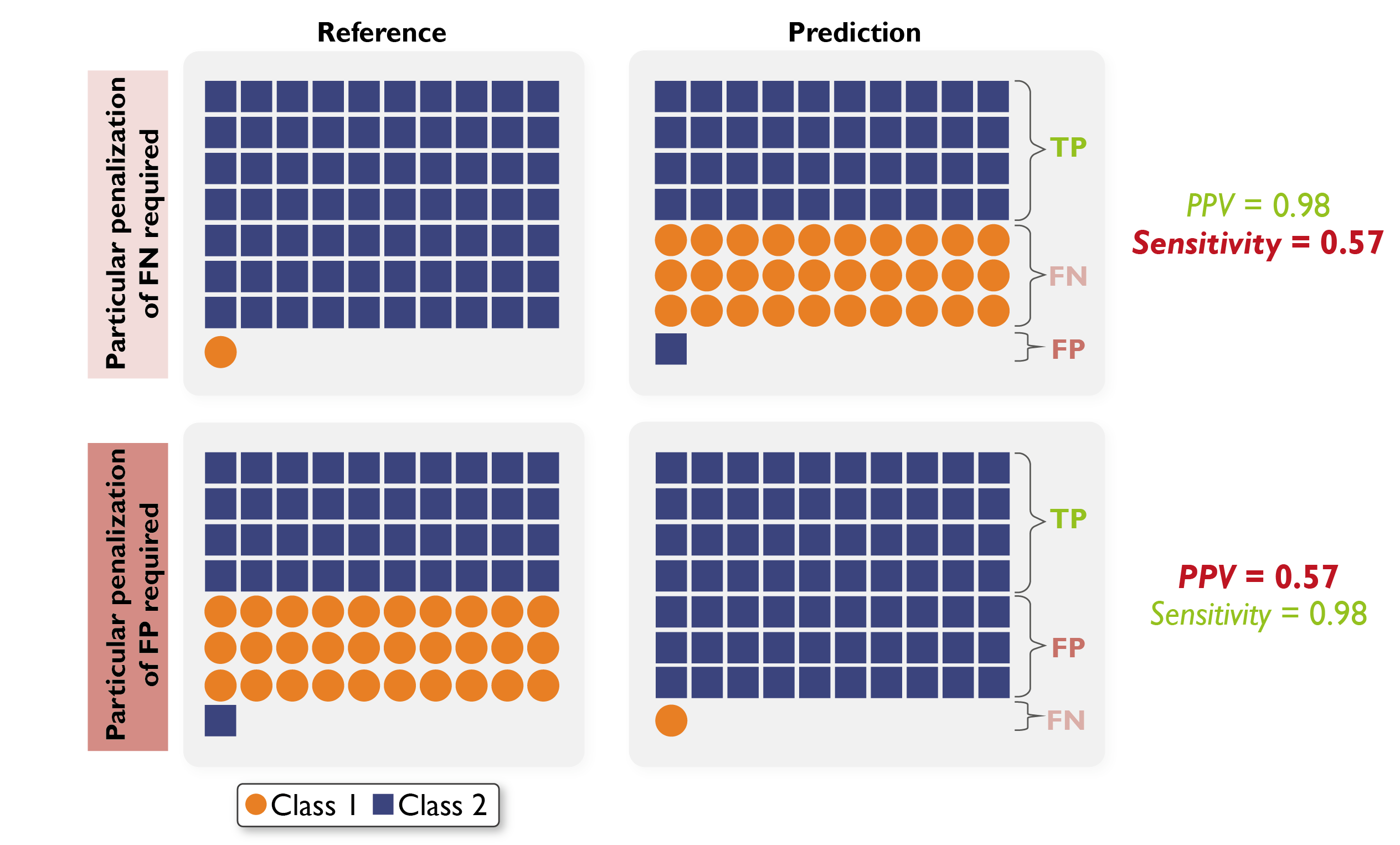}
    \caption{Effect of using metrics that are not suitable for penalizing \acf{FN}/\acf{FP}. The definition of the \textit{\acf{PPV}} metric does not incorporate \ac{FN} and is therefore not well-suited for penalizing \ac{FN}, as required in cancer screening tasks, for example. Analogously, the \textit{Sensitivity} is not well-suited for penalizing \ac{FP}, as required in many retrieval tasks, for example.}
    \label{fig:unequal-class-confusions}
\end{tcolorbox}
\end{figure}

\begin{figure}[H]
\begin{tcolorbox}[title= Pitfall: Unequal handling of classes, colback=white]
    \centering
    \includegraphics[width=1\linewidth]{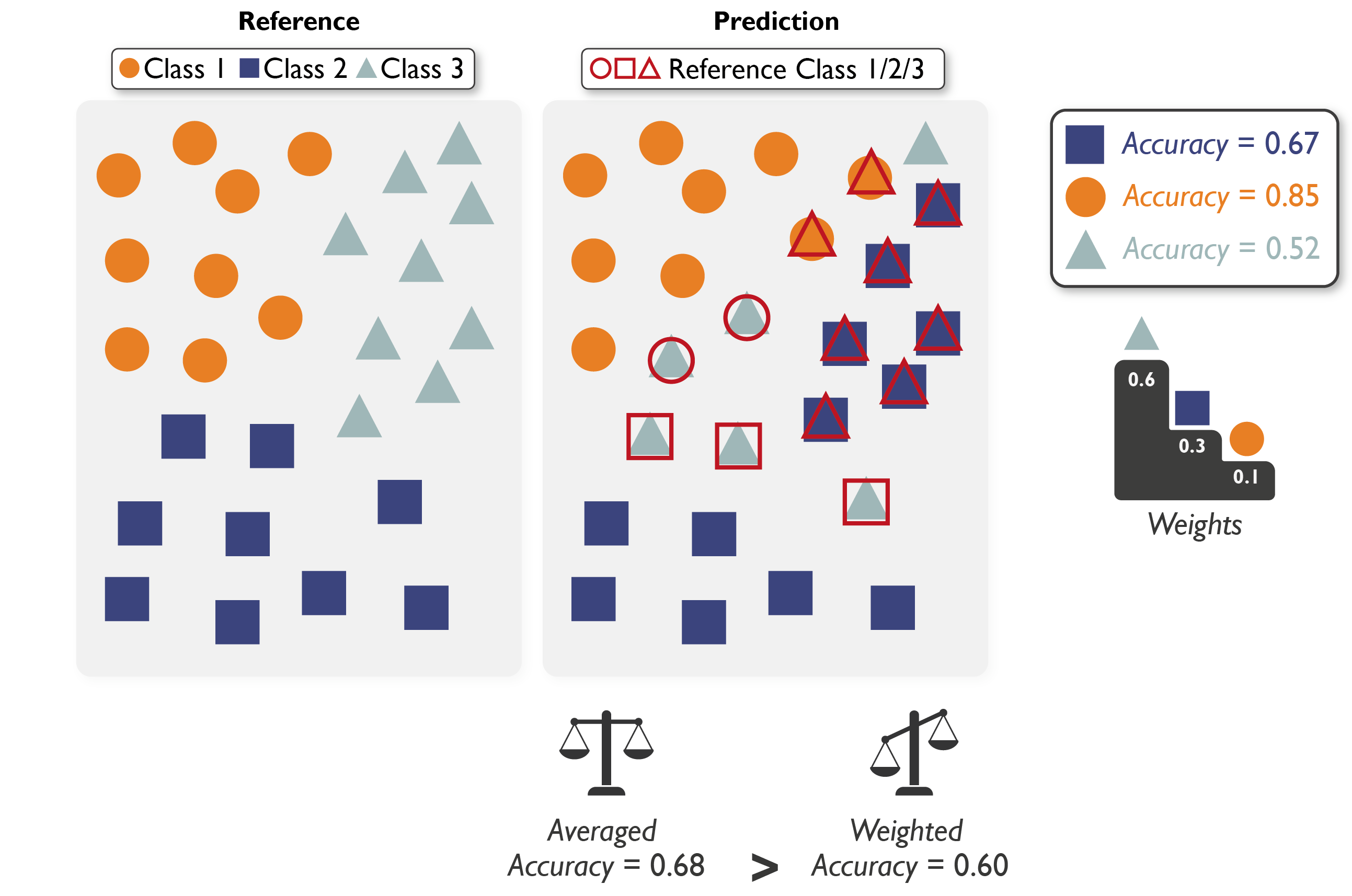}
    \caption{Effect of unequal handling of classes. Simple averaging (macro-averaging) of the \textit{Accuracy} ignores the unequal importance of classes, given by pre-defined weights of classes. Incorrect predictions are indicated by a red square.}
    \label{fig:unequal-class}
\end{tcolorbox}
\end{figure}

\begin{figure}[H]
\begin{tcolorbox}[title= Pitfall: Interdependencies between classes, colback=white]
    \centering
    \includegraphics[width=0.7\linewidth]{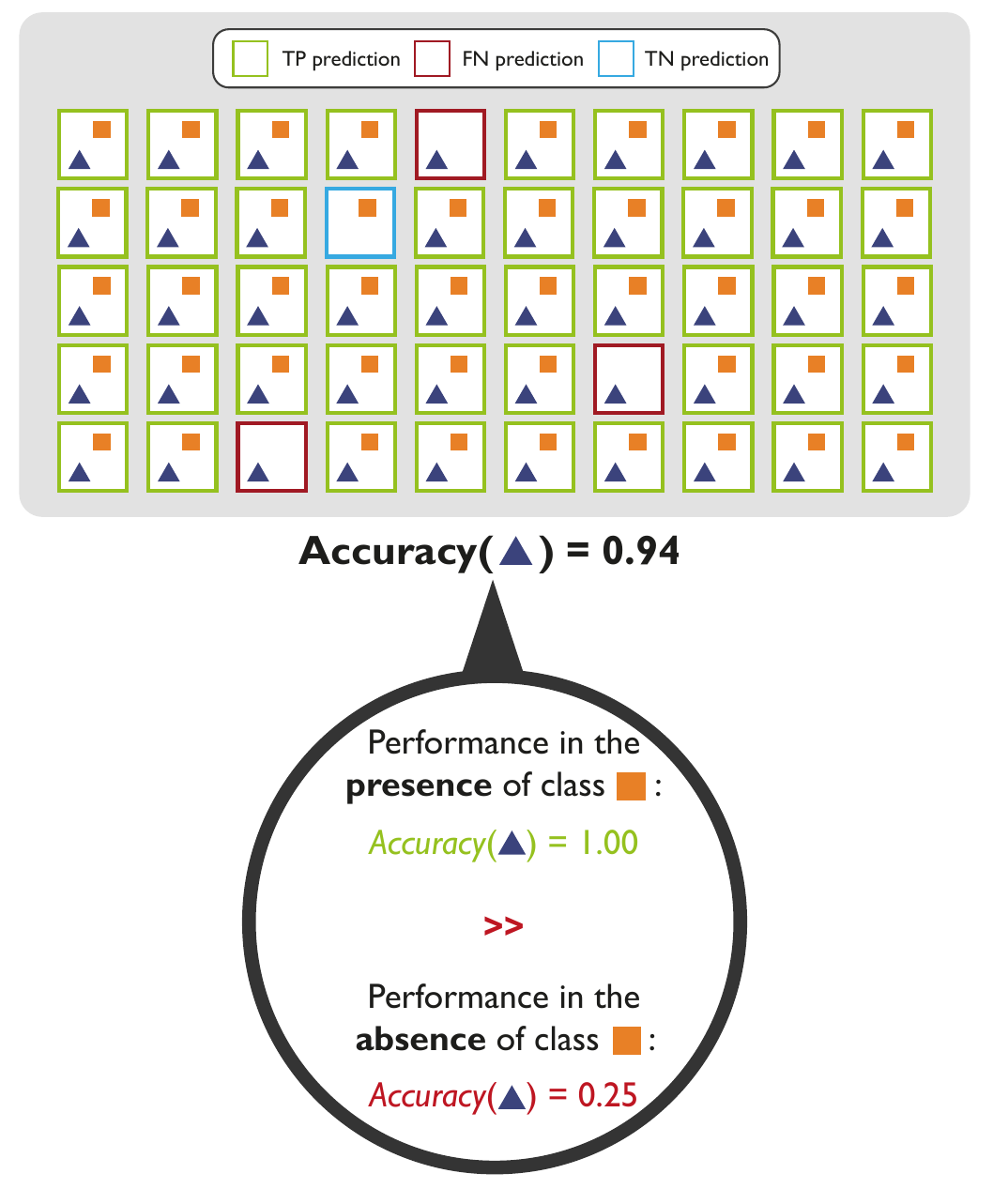}
    \caption{Effect of interdependencies between classes. A prediction may show a near-perfect \textit{Accuracy} score of 0.94 for the dark blue triangle as it frequently appears in conjunction with the orange square. By calculating the \textit{Accuracy} in the \textit{presence} and \textit{absence} of the square class, it can be seen that the algorithm only works well in the presence of the orange class. Incorrect predictions are indicated by a red cross.}
    \label{fig:multi-class-interdependencies}
\end{tcolorbox}
\end{figure}

\begin{figure}[H]
\begin{tcolorbox}[title= Pitfall: Meta-information ignored, colback=white]
    \centering
    \includegraphics[width=0.6\linewidth]{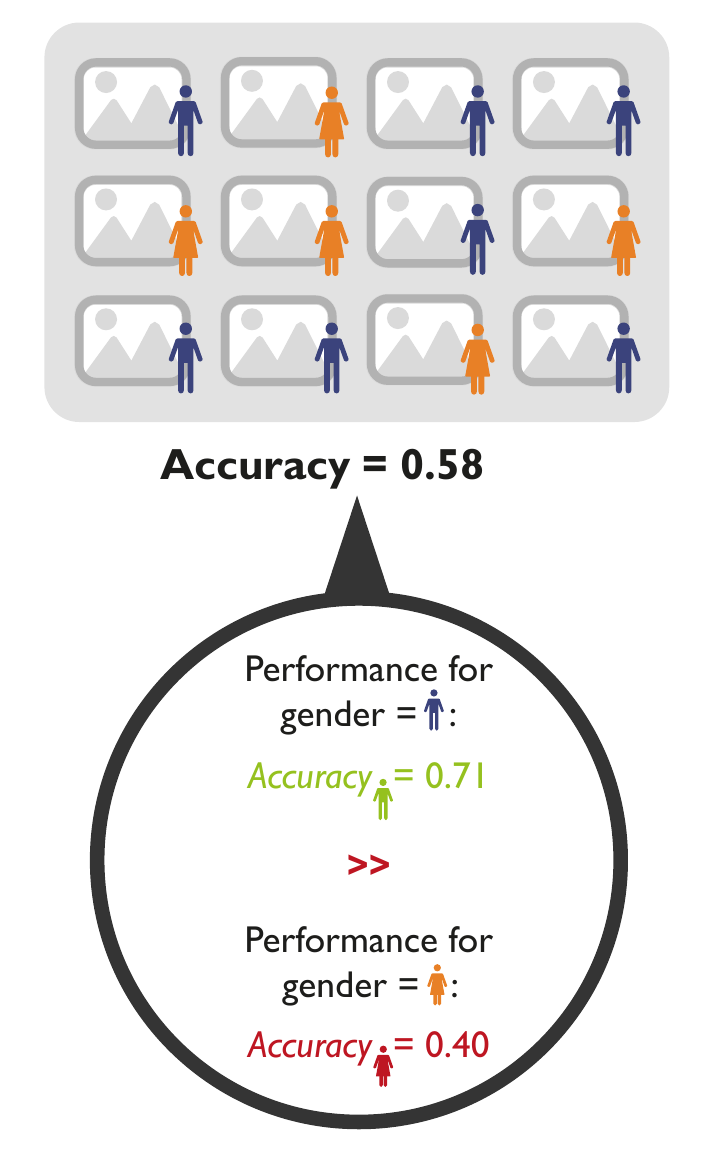}
    \caption{Effect of disregarding relevant meta-information (here: gender). Ignoring the available meta-information of the patient's gender per image, the \textit{Accuracy} does not reveal that the algorithm performs much better for men compared to women.}
    \label{fig:stratification-gender}
\end{tcolorbox}
\end{figure}

\begin{figure}[H]
\begin{tcolorbox}[title= Common metrics disregard cost-benefit analysis, colback=white]
    \centering
    \includegraphics[width=1\linewidth]{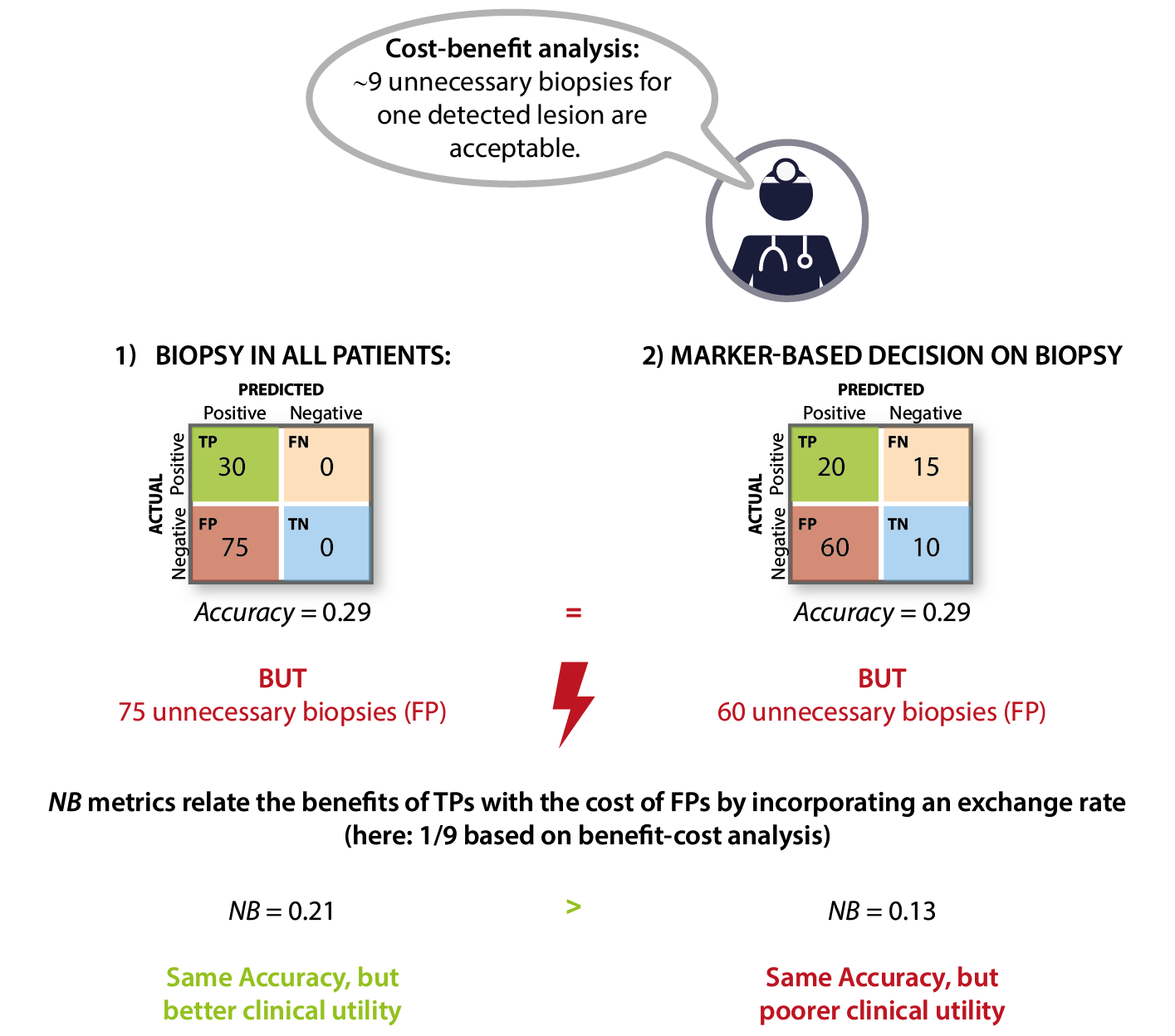}
    \caption{Effect of neglecting a cost-benefit analysis. In a cost-benefit analysis, clinicians are able to define a risk-specific exchange rate to be used in the computation of the \textit{\acf{NB}} metric. Common metrics such as the Accuracy do not include this analysis and would rate both decision types similarly, while \textit{\ac{NB}} indicates that having all patients undergo biopsy actually yields a better clinical outcome \citep{vickers2016net}.}
     \label{fig:cost-benefit}
\end{tcolorbox}
\end{figure}

\begin{figure}[H]
\begin{tcolorbox}[title= Pitfall: Per-class counting metrics depend on the definition of the positive class, colback=white]
    \centering
    \includegraphics[width=1\linewidth]{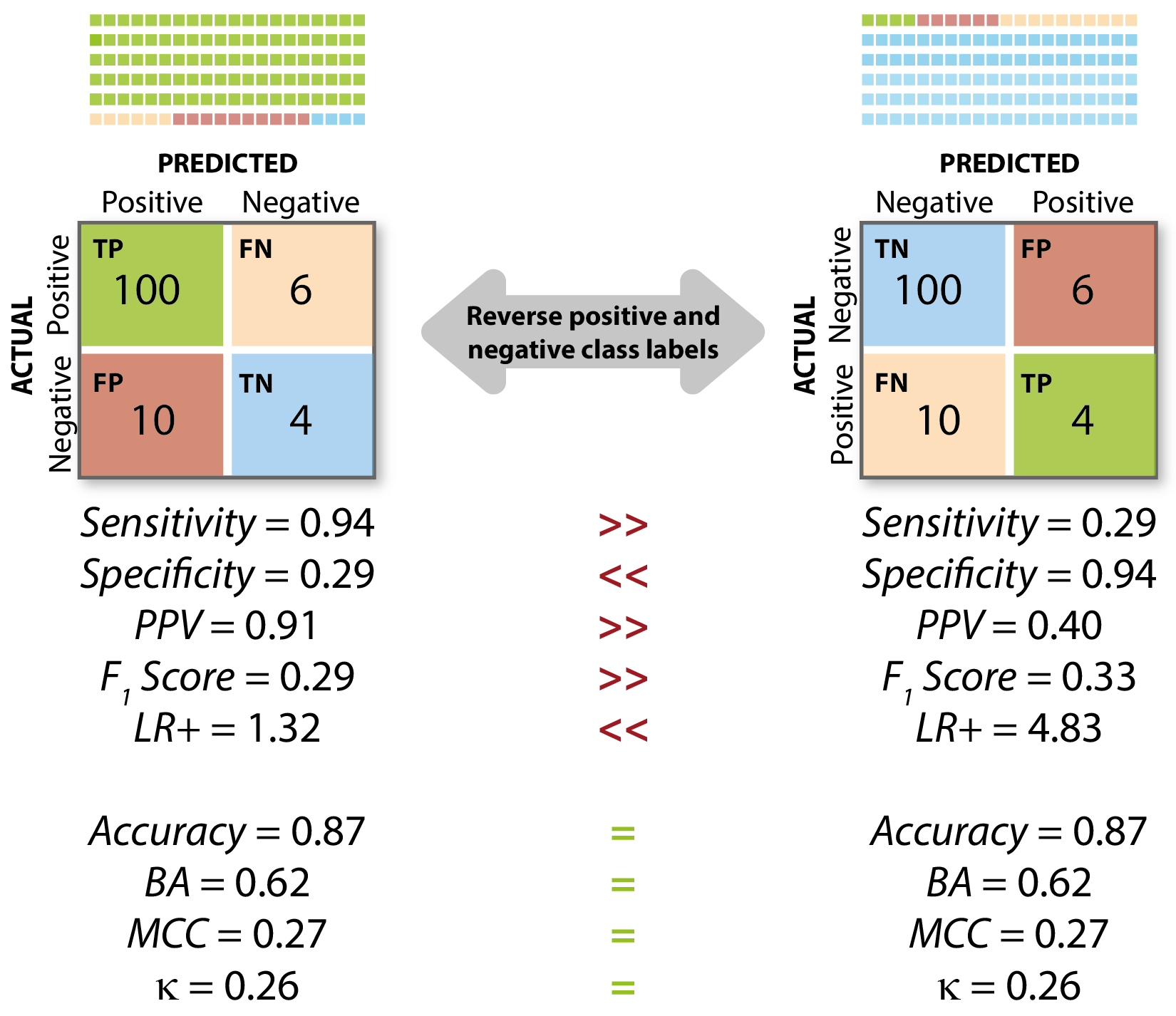}
    \caption{Effect of defining the positive class. Per-class counting metrics (here: \textit{Sensitivity}, \textit{Specificity}, \textit{\acf{PPV}}, \textit{F$_1$ Score}, and \textit{\acf{LR+}}) heavily depend on the definition of the positive class in binary classification problems. When reversing the class labels, the scores of per-class counting metrics change significantly. Multi-class metrics (here: \textit{Accuracy}, \textit{\acf{BA}}, \textit{\acf{MCC}}, and \textit{Cohen's Kappa $\kappa$}) do not rely on this definition and yield the exact same values in both cases. Used abbreviations: \acf{TP}, \acf{FP}, \acf{FN}, \acf{TN}.}
    \label{fig:positive-class}
\end{tcolorbox}
\end{figure}

\begin{figure}[H]
\begin{tcolorbox}[title= Pitfall: Missing prevalence correction, colback=white]
    \centering
    \includegraphics[width=1\linewidth]{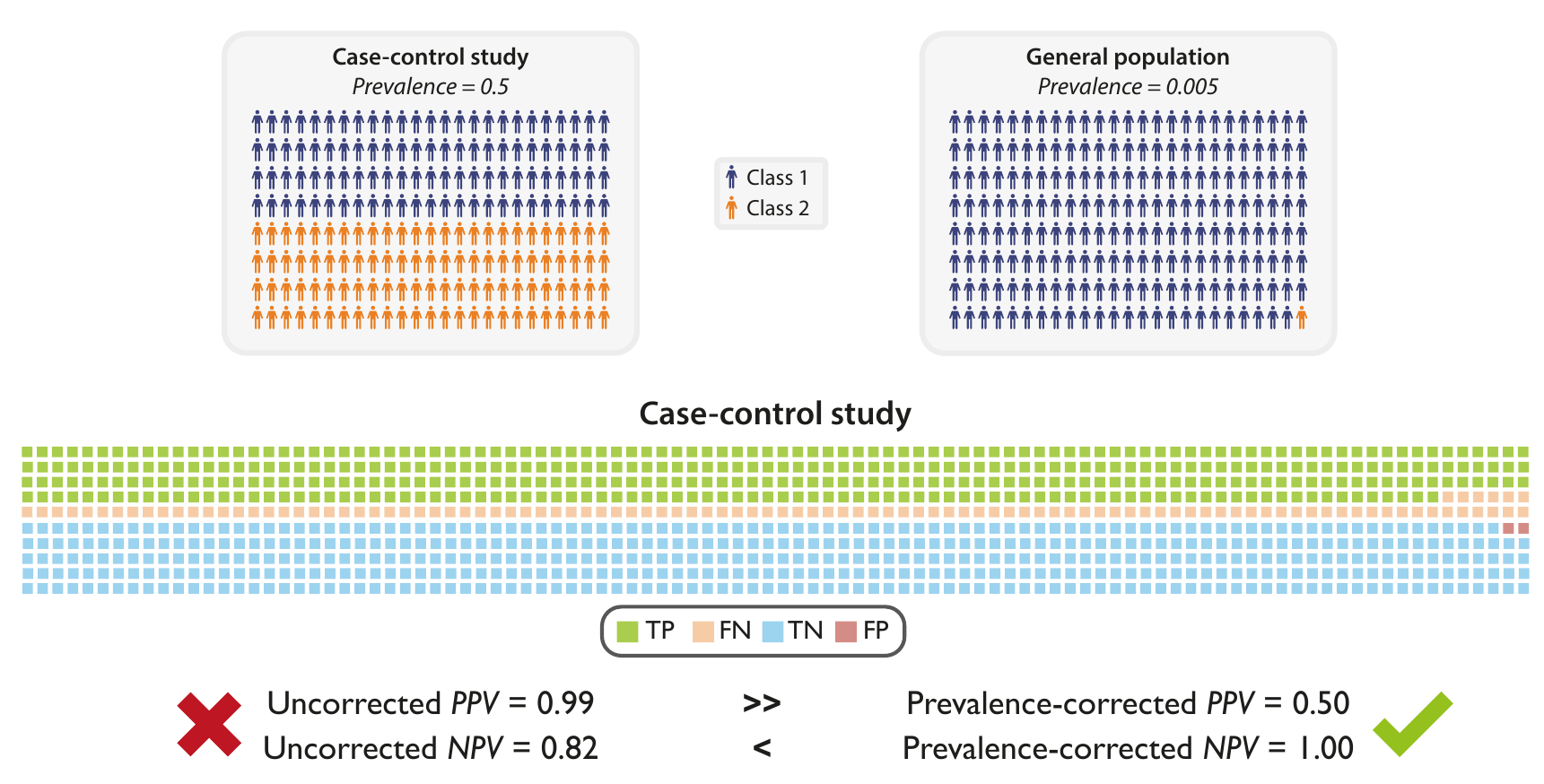}
    \caption{Effect of missing prevalence correction. For case-control studies, the \textit{\acf{PPV}} and \textit{\acf{NPV}} are required to apply prevalence correction based on the general population. Falsely used case-control prevalence leads to incorrect metric scores compared to using the general population prevalence (which is often significantly lower).}
    \label{fig:prevalence}
\end{tcolorbox}
\end{figure}

\begin{figure}[H]
\begin{tcolorbox}[title= Prevalence-dependent metrics cannot be compared across data sets with different prevalences, colback=white]
    \centering
    \includegraphics[width=1\linewidth]{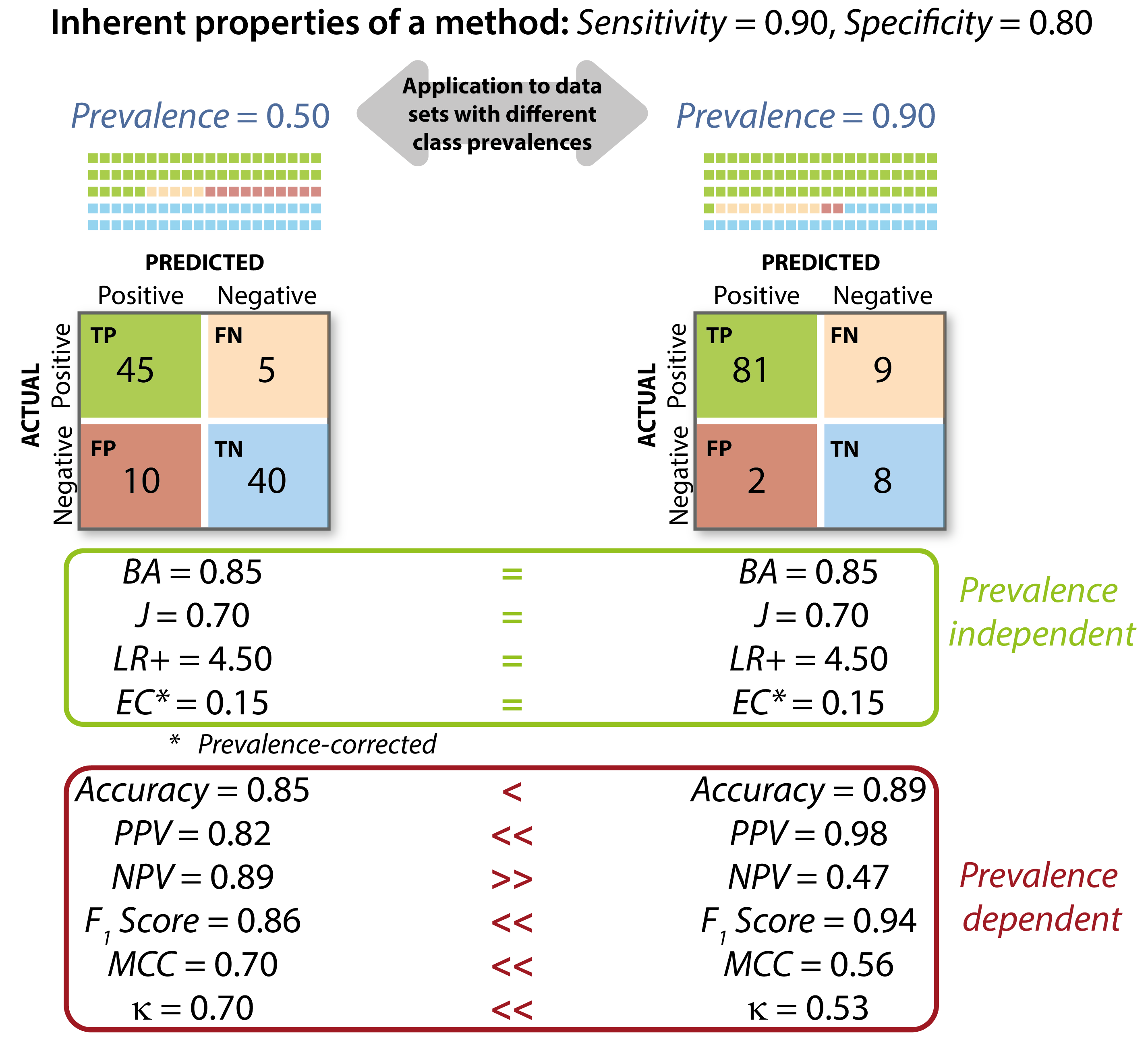}
    \caption{Effect of prevalence dependency. An algorithm with specific inherent properties (here: Sensitivity of 0.9 and Specificity of 0.8) may perform completely differently on different data sets if the prevalences differ (here: 50\% (left) and 90\% (right)) and prevalence-dependent metrics are used for validation (here: Accuracy, \acf{PPV}, \textit{\acf{NPV}}, \textit{F$_1$ Score}, \textit{\acf{MCC}}, \textit{Cohen's Kappa} $\kappa$). In contrast, prevalence-independent metrics (here: \textit{\acf{BA}, Youden's Index J, \acf{LR+}}, and \textit{\acf{EC}}) can be used to compare validation results across different data sets. Used abbreviations: \acf{TP}, \acf{FN}, \acf{FP} and \acf{TN}.}
    \label{fig:prevalence-dependency}
\end{tcolorbox}
\end{figure}

\begin{figure}[H]
\begin{tcolorbox}[title= Pitfall: Rankings in the case of prevalence dependency, colback=white]
    \centering
    \includegraphics[width=1\linewidth]{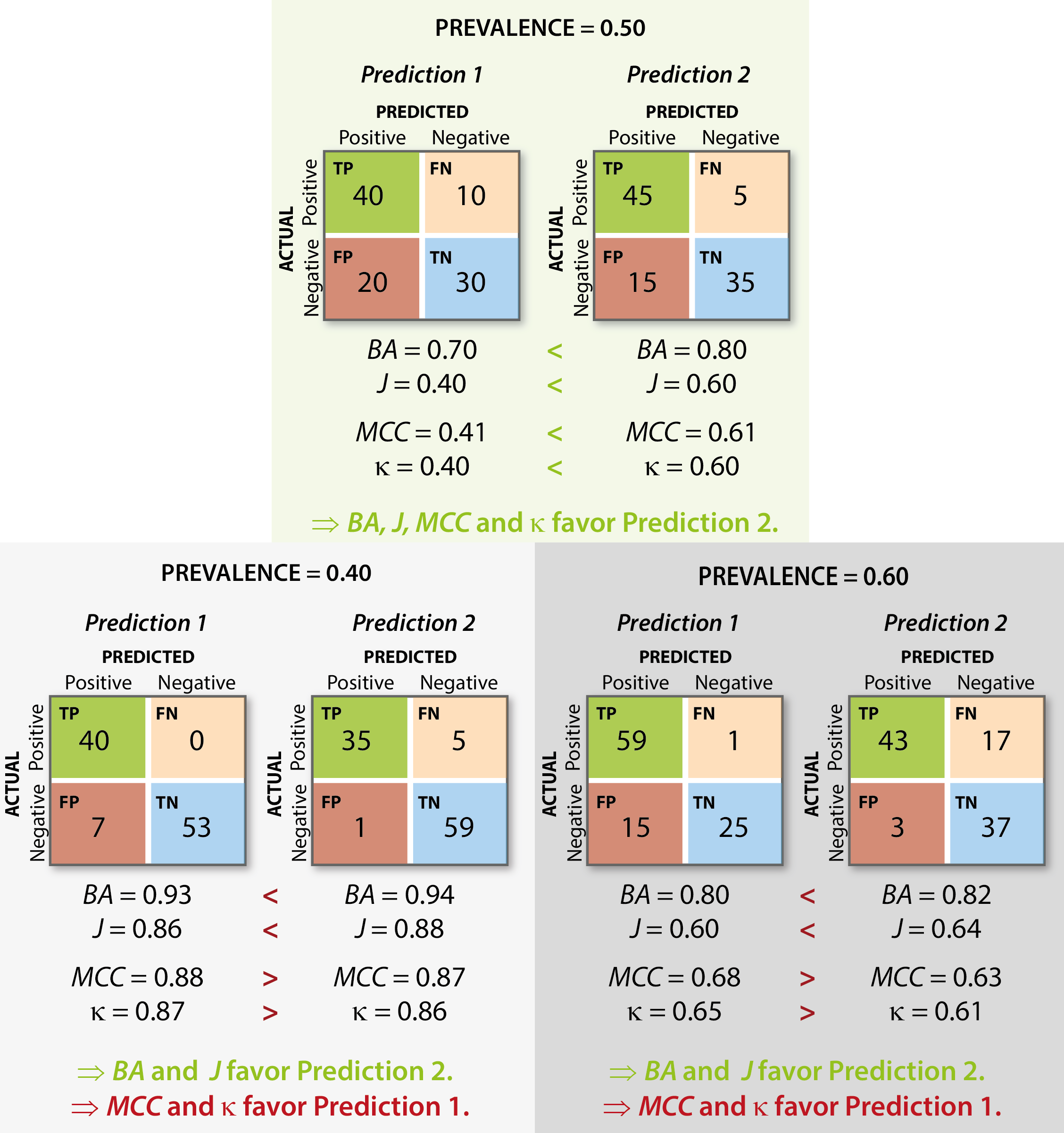}
    \caption{Effect of prevalence dependency on rankings. For a prevalence unequal to 0.5, the rankings generated by \textit{\acf{BA}} and \textit{Youden's Index J (J)} (here preferring \textit{Prediction 2} over \textit{Prediction 1}) differ from those generated by \textit{\acf{MCC}} and \textit{Cohen's Kappa $\kappa$} (preferring \textit{Prediction 1} over \textit{Prediction 1}). Rankings will only be the same for a prevalence of 0.5 (favoring \textit{Prediction 2}). Used abbreviations: \acf{TP}, \acf{FN}, \acf{FP} and \acf{TN}.}
    \label{fig:prevalence-dependency-rankings}
\end{tcolorbox}
\end{figure}

\begin{figure}[H]
\begin{tcolorbox}[title= Pitfall: Perfect discrimination but poor calibration, colback=white]
    \centering
    \includegraphics[width=1\linewidth]{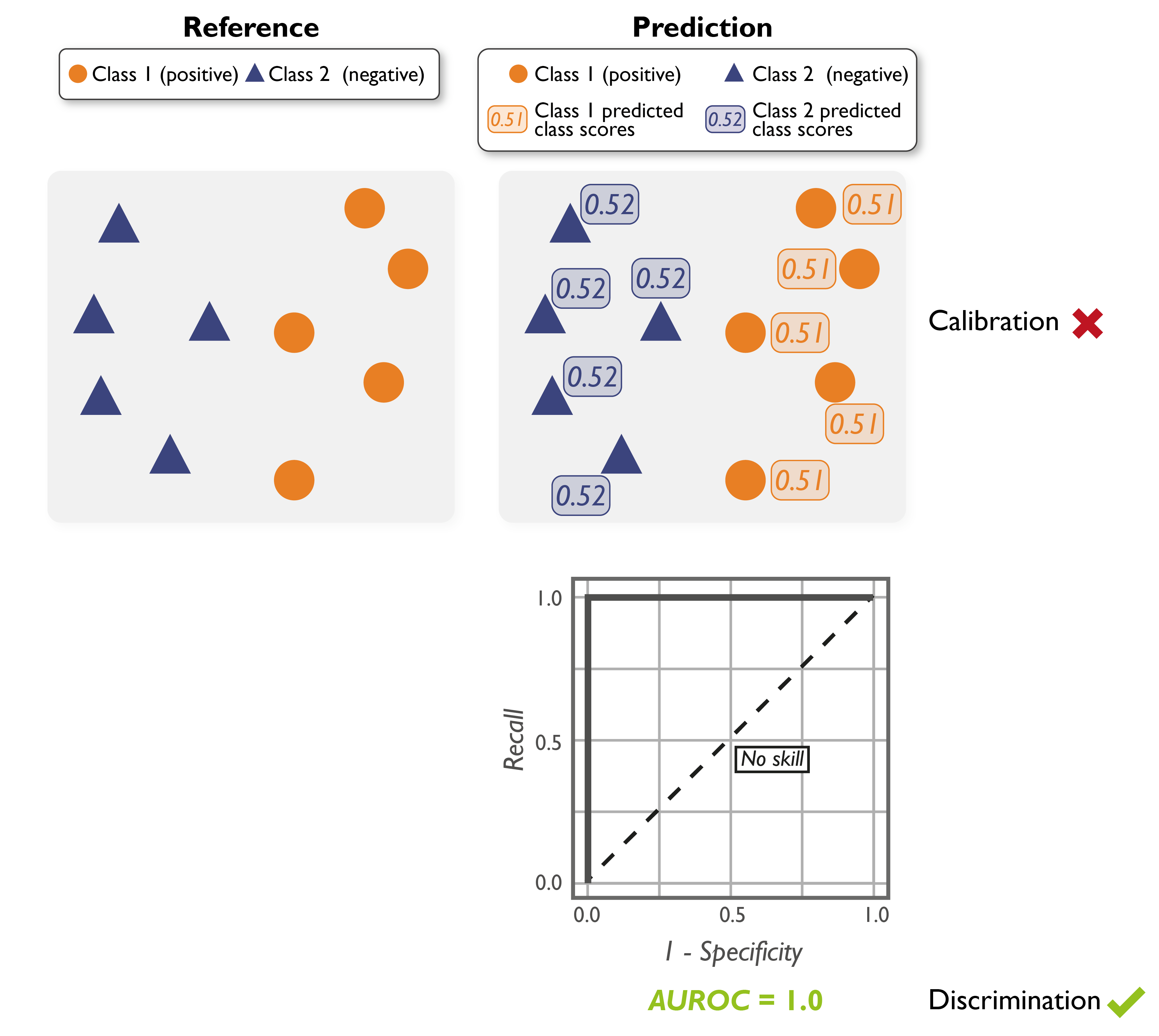}
    \caption{Effect of a perfect discrimination but a poor calibration. The predicted classes are perfectly discriminated and yield a perfect \acf{AUROC} score of 1. However, the predicted class scores (0.52 for the positive and 0.51 for the negative class) are not well calibrated as there is very little correlation between the predicted class scores and actual probability.}
    \label{fig:calibration}
\end{tcolorbox}
\end{figure}

\begin{figure}[H]
\begin{tcolorbox}[title= Pitfall: \textit{\acf{BS}} for ordinal classes, colback=white]
    \centering
    \includegraphics[width=1\linewidth]{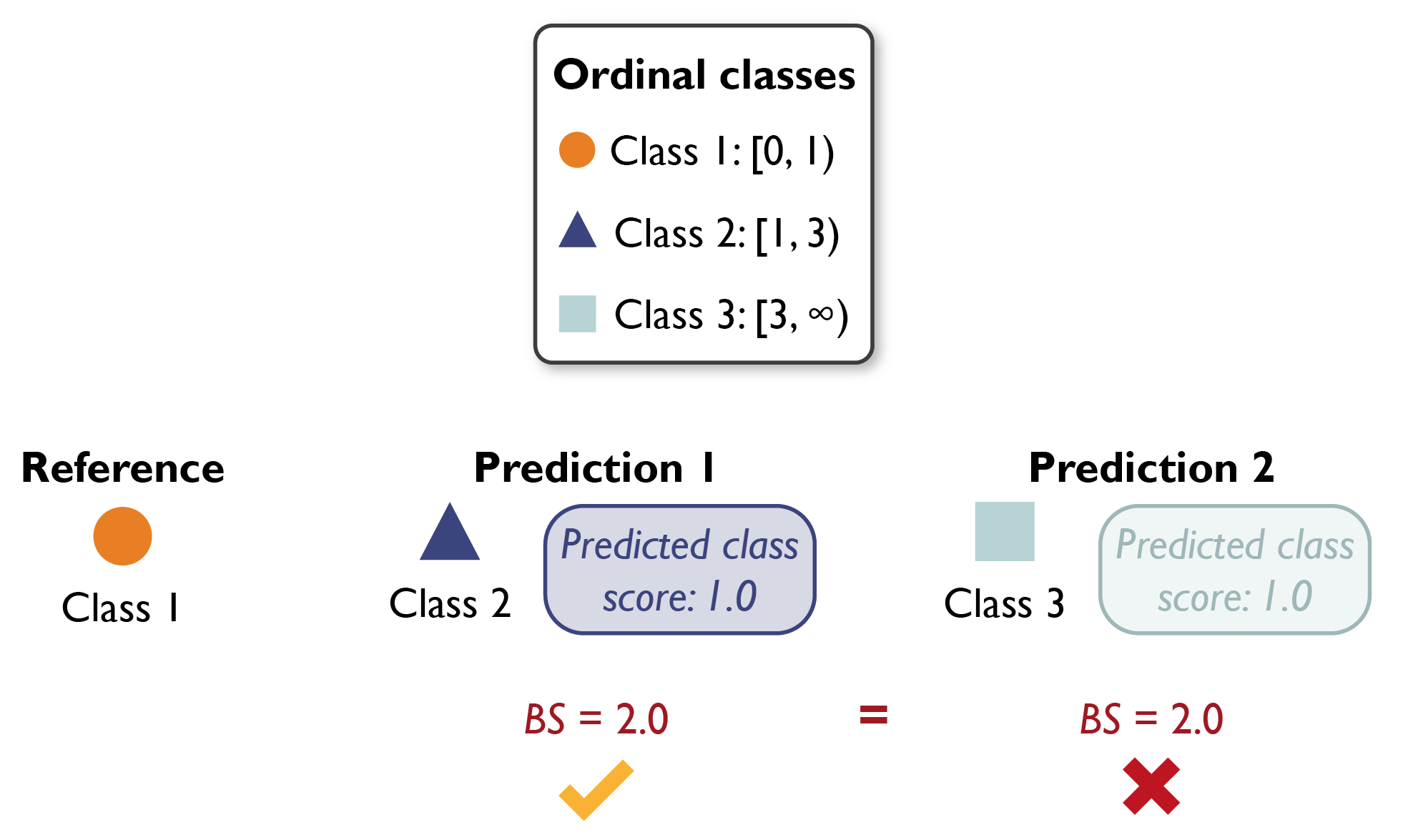}
    \caption{Effect of ordinal classes on the \textit{\acf{BS}}. While \textit{Predictions 1} and \textit{2} 
    both predict the wrong class, yielding the same \textit{\ac{BS}} of 2.0, \textit{Prediction 1} should be favored due to the ordinal scale: Class 2 is closer to the actual outcome Class 1. This fact is obscured by the \textit{\ac{BS}} values.}
    \label{fig:brierscore-ordinal}
\end{tcolorbox}
\end{figure}

\begin{figure}[H]
\begin{tcolorbox}[title= Pitfall: Bins for calibration errors not standardized, colback=white]
    \centering
    \includegraphics[width=1\linewidth]{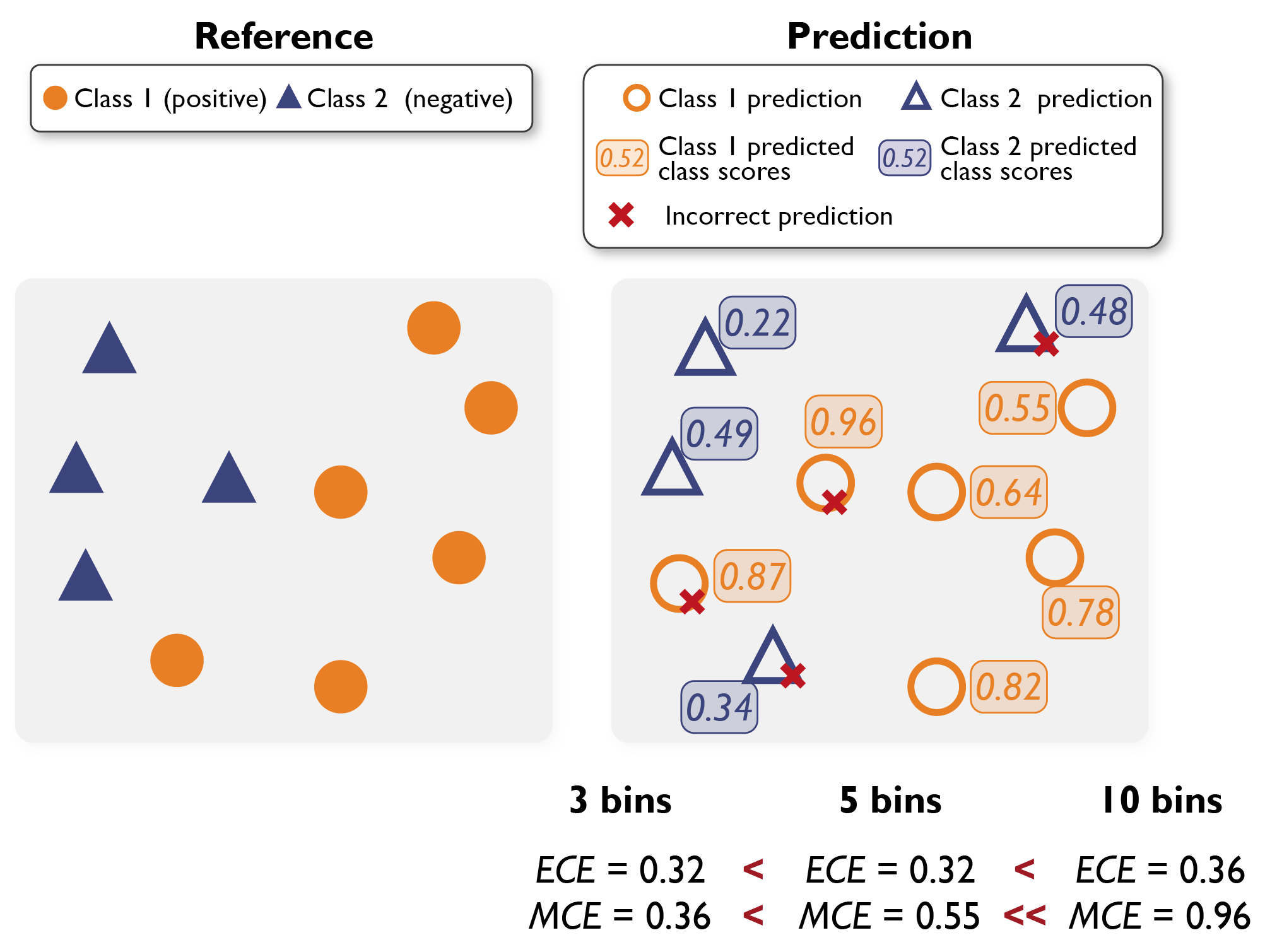}
    \caption{Effect of choosing different bins for calculating the \textit{\acf{ECE}} and \textit{\acf{MCE}}. Three different strategies are chosen for the binning of the interval [0, 1] of the predicted class scores of the \textit{Prediction}. The resulting metric scores are substantially affected by the number of bins \citep{guoCalibrationModernNeural2017}.}
    \label{fig:ece-mce-bins}
\end{tcolorbox}
\end{figure}

\begin{figure}[H]
\begin{tcolorbox}[title= Pitfall: Definition of calibration and sample size dependency, colback=white]
    \centering
    \includegraphics[width=1\linewidth]{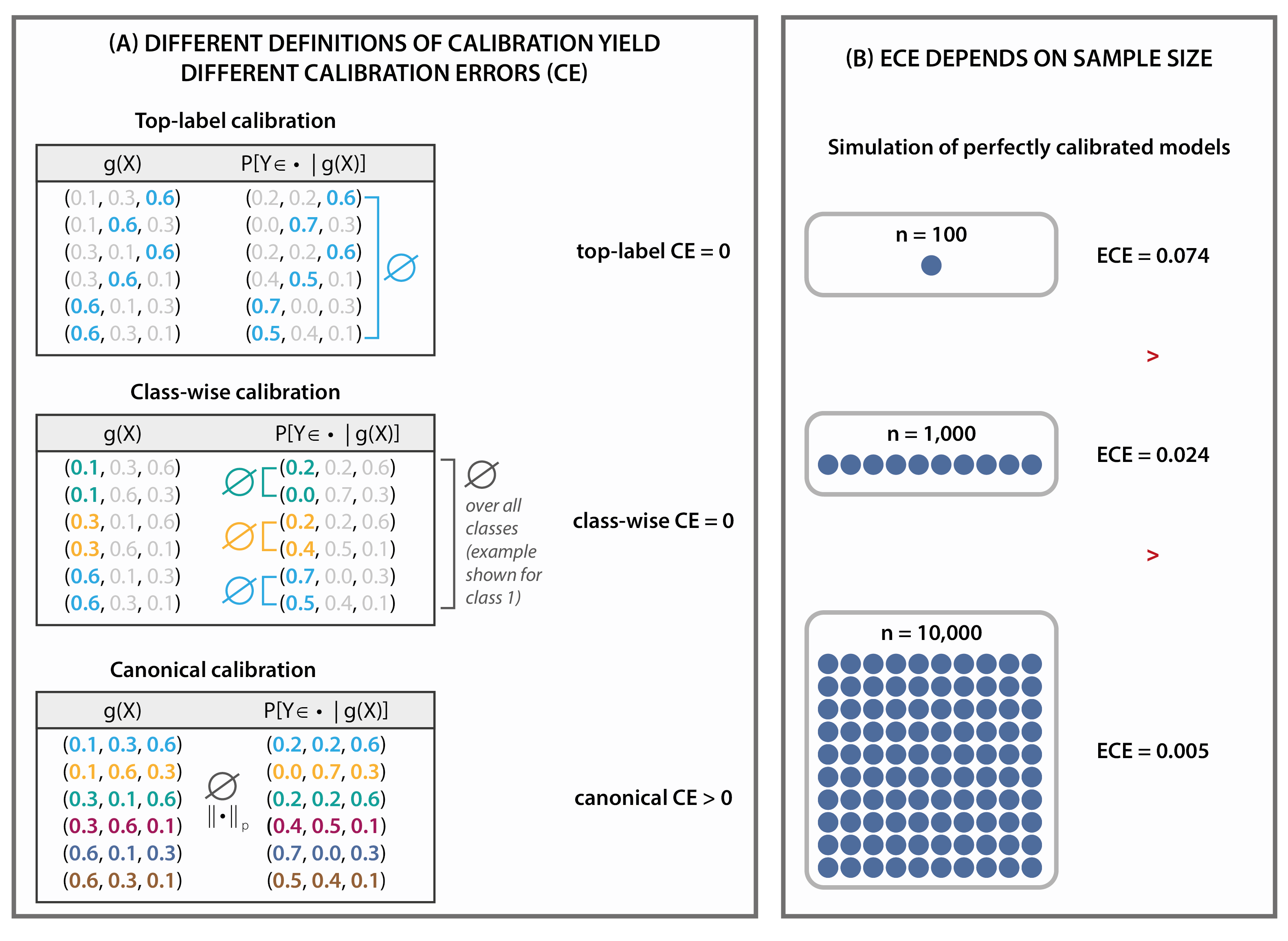}
    \caption{\textbf{(a)} Effect of different definitions of calibration on the \textit{\acf{CE}}. For top-label calibration, only the maximum values of the predicted class scores $f \left( X \right)$ are considered, while all other values are neglected. In the computation of the \textit{\ac{ECE}}, for each distinct output value of $f \left( X \right)$ (only 0.6 in this case), $\mathbb{P}_{Y \mid f \left( X \right)}$ is determined as the average over the empirical rates of this output (0.6, 0.7, 0.6, 0.5, 0.7, 0.5 in this case). The top-label calibration condition (i.e., matching the two scores) results in a perfect \textit{\ac{CE}} $ = 0$ in this scenario. Similarly, in class-wise calibration, the predicted class scores are compared per class, a requirement that is also fulfilled by the depicted system. Only the canonical calibration, which comes with the strict requirement that the model output must match the full probability distribution (implying the comparison of entire vectors rather than single values) indicates a miscalibrated system (\textit{\ac{CE}} > 0). This figure is inspired by ~\cite{vaicenavicius2019evaluating}. \textbf{(b)} Effect of the sample size on the \textit{\ac{ECE}}. The values of the \textit{\ac{ECE}} depend on the sample size. Even for a simulated perfectly calibrated model, the \textit{\ac{ECE}} score will be higher than zero for small sample sizes.} 
    \label{fig:ece-pitfalls}
\end{tcolorbox}
\end{figure}

\begin{figure}[H]
\begin{tcolorbox}[title= Pitfall: Multi-class counting metrics for ordinal grading, colback=white]
    \centering
    \includegraphics[width=0.8\linewidth]{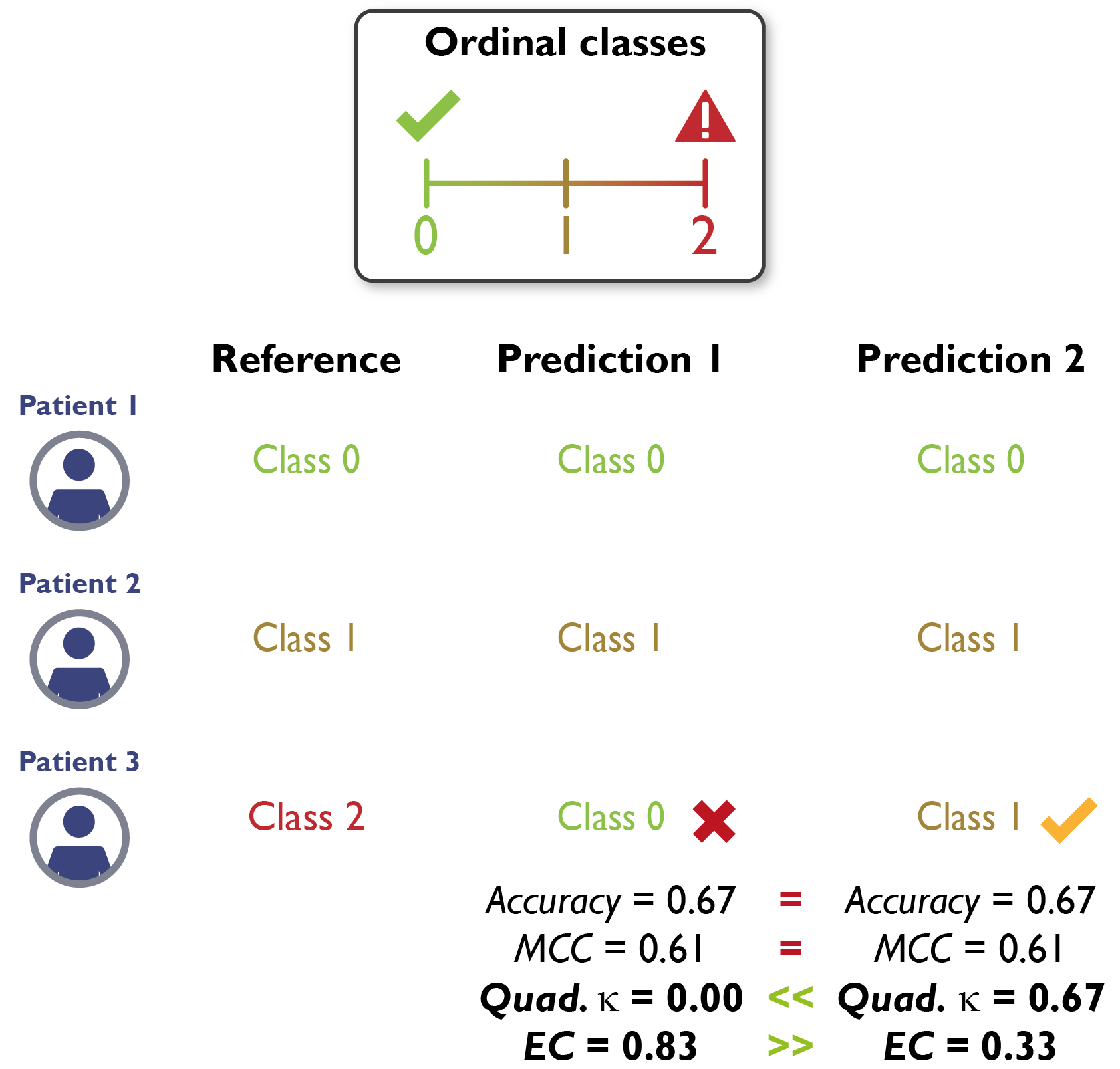}
    \caption{When predicting the severity of a disease for three patients in an ordinal classification problem, \textit{Prediction 1} assumes a much lower severity for \textit{Patient 3} than actually observed. This critical issue is overlooked by common metrics (here: \textit{Accuracy} and \textit{\acf{MCC}}), which measure no difference to \textit{Prediction 2}, which assesses the severity much better. Metrics with pre-defined weights (here: \textit{quadratic (quad.) weighted Cohen's Kappa ($\kappa$)} and \textit{\acf{EC}}) correctly penalize \textit{Prediction 1} much more than \textit{Prediction 2}.}
    \label{fig:ordinal-grading}
\end{tcolorbox}
\end{figure}

\begin{figure}[H]
\begin{tcolorbox}[title= Pitfall: Label shifts in ordinal grading tasks, colback=white]
    \centering
    \includegraphics[width=1\linewidth]{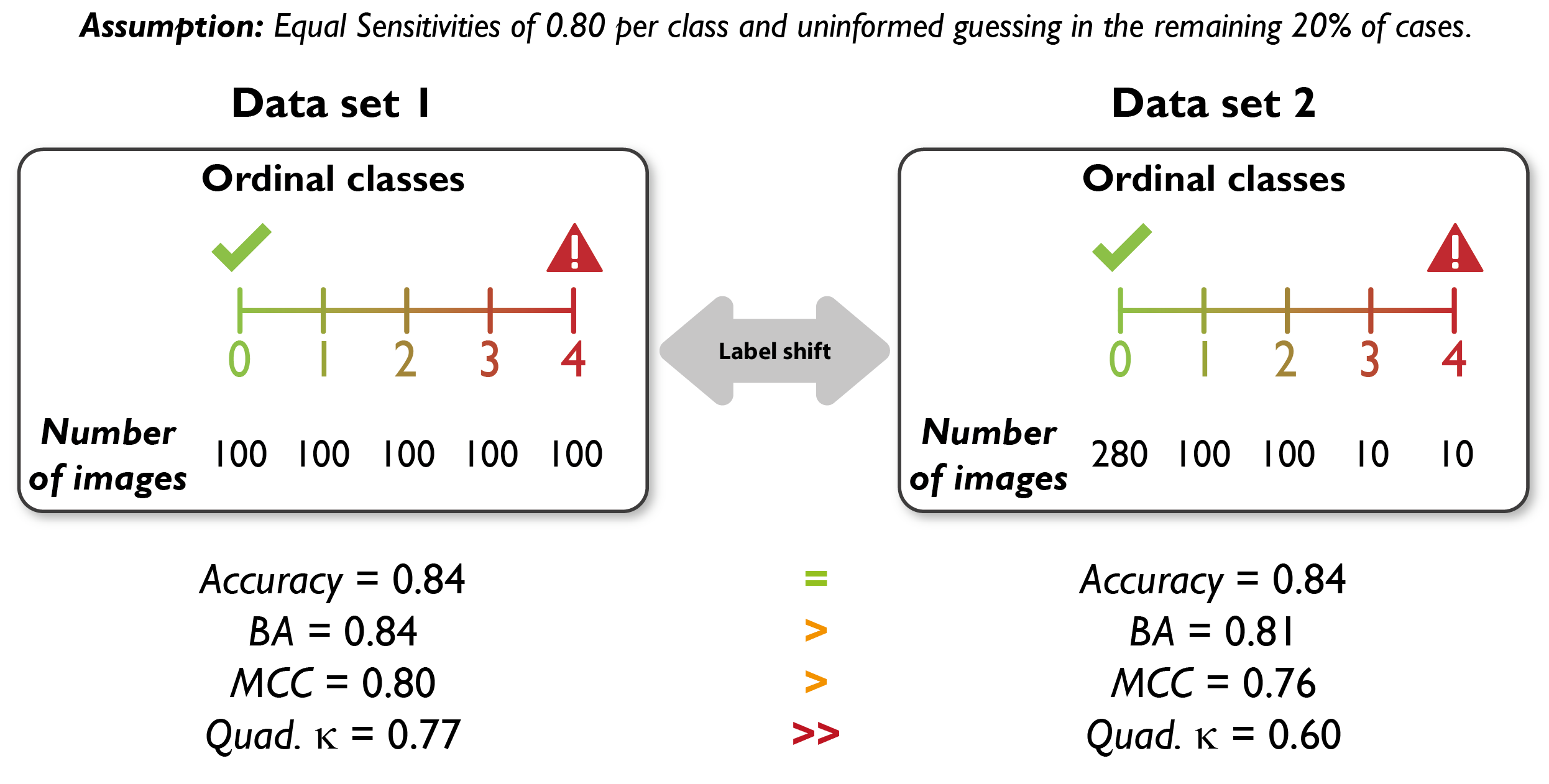}
    \caption{Effect of label shifts in ordinal grading tasks. In a setting with two data sets with the same total number of data, but a varying number of images per label, the \textit{Accuracy} is not affected by the label shift. However, the \textit{\acf{BA}} and \textit{\acf{MCC}} are slightly affected, while the \textit{quadratic-weighted Cohen's Kappa (Quad. $\kappa$)} changes substantially. Equal Sensitivities of 0.8 per class and uninformed guessing in the remaining 20\% of cases were assumed.}
    \label{fig:ordinal-grading-label-shift}
\end{tcolorbox}
\end{figure}

Most issues related to classification metrics are related to one of the following properties of the underlying biomedical problem: 

\begin{itemize}
    \item High class imbalance (Figures~\ref{fig:class-imbalance}  - \ref{fig:metric-landscapes})
    \item Presence of more than two classes (Figure~\ref{fig:multi-class}) 
    \item Unequal severity of class confusions (Figure~\ref{fig:unequal-class-confusions})
    \item Unequal handling of classes (Figure~\ref{fig:unequal-class})
    \item Interdependencies between classes (Figure~\ref{fig:multi-class-interdependencies})
    \item Lack of stratification (Figure~\ref{fig:stratification-gender})
    \item Importance of cost-benefit analysis (Figure~\ref{fig:cost-benefit})
    \item Definition of class labels (Figure~\ref{fig:positive-class})
    \item Prevalence dependency (Figures~\ref{fig:prevalence} - \ref{fig:prevalence-dependency-rankings})
    \item Importance of confidence awareness (Figures~\ref{fig:calibration} - \ref{fig:ece-pitfalls})
    \item Presence of ordinal classes (Figures~\ref{fig:ordinal-grading} - \ref{fig:ordinal-grading-label-shift})
\end{itemize}

\begin{figure}[H]
\begin{tcolorbox}[title= Pitfall: Same metric scores for different confusion matrices, colback=white]
    \centering
    \includegraphics[width=0.9\linewidth]{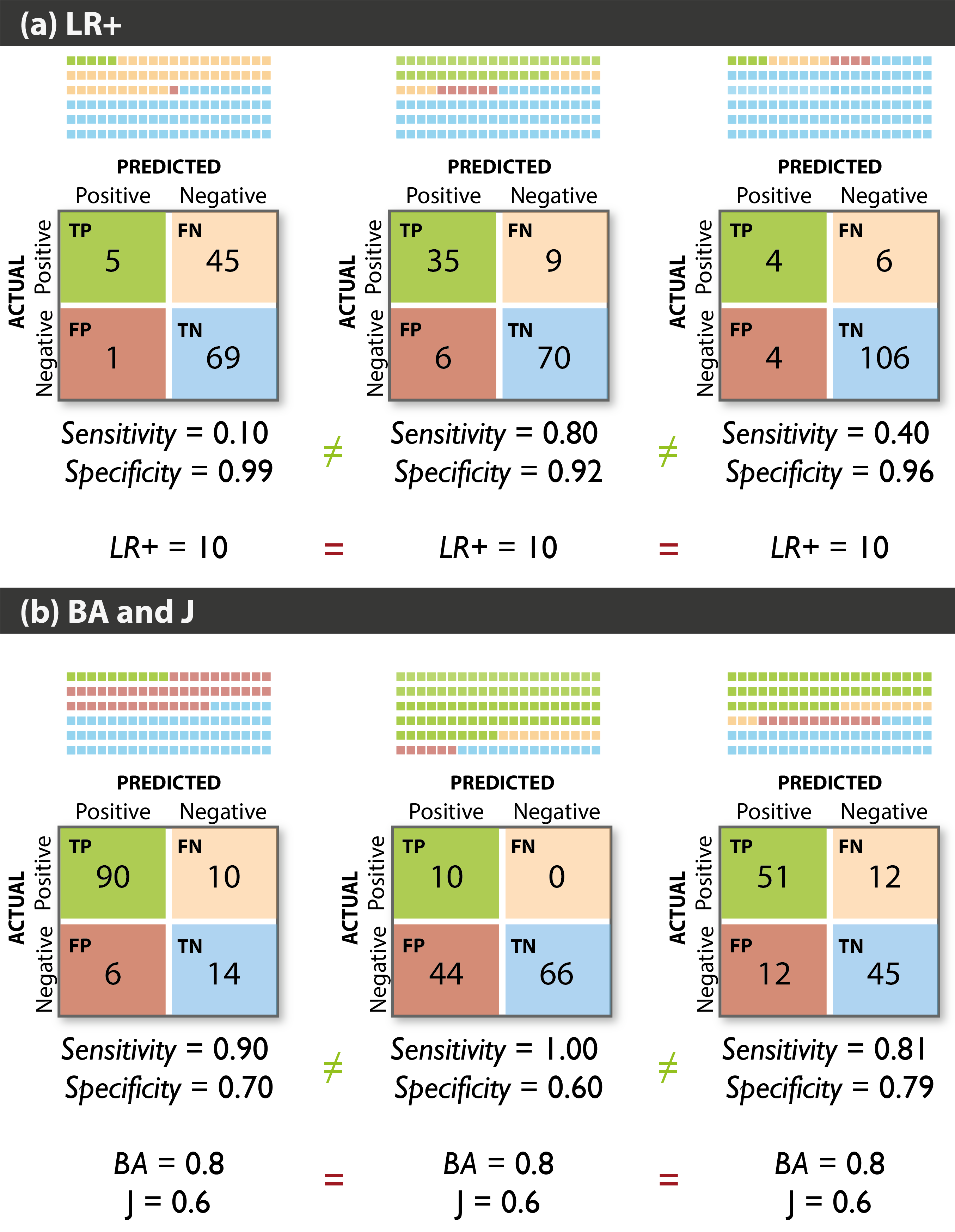}
    \caption{Effect of the same (a) \textit{\acf{LR+}}, (b) \textit{\acf{BA}}, and \textit{Youden Index J} values for different combinations of cardinalities (\acf{TP}, \acf{FN}, \acf{FP} and \acf{TN}). A data set of 120 samples with different confusion matrices results in different \textit{Sensitivity} and \textit{Specificity} values. The \textit{\ac{LR+}}, \textit{\ac{BA}}, and \textit{Youden Index J}, however, can still yield the exact same scores.}
    \label{fig:lr+ba}
\end{tcolorbox}
\end{figure}

\begin{figure}[H]
\begin{tcolorbox}[title= Pitfall: Upper bound not equally obtainable in Cohen's $\kappa$, colback=white]
    \centering
    \includegraphics[width=0.95\linewidth]{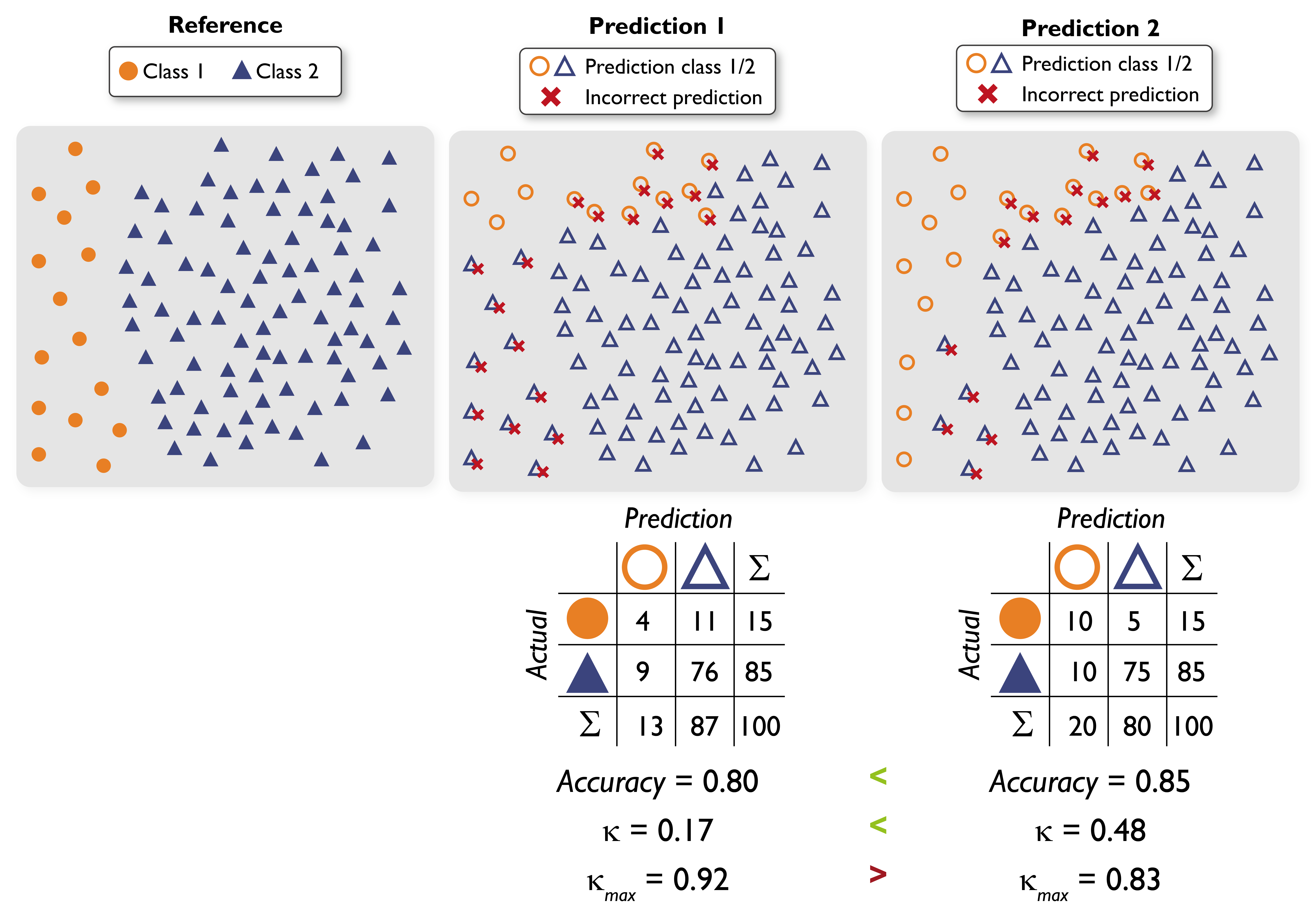}
    \caption{Effect of different numbers of predicted positive and negative samples in comparison to the actual number of positive and negative samples. A prediction with a similar count of positive and negative samples to the actual distribution (\textit{Prediction 1}) reaches higher maximum \textit{Cohen's $\kappa$} values compared to a prediction with a dissimilar distribution (\textit{Prediction 2}), although the overall \textit{Accuracy} and \textit{Cohen's $\kappa$} is lower. Incorrect predictions are indicated by a red cross.}
    \label{fig:kappa-max}
\end{tcolorbox}
\end{figure}

\begin{figure}[H]
\begin{tcolorbox}[title= Theoretical bounds of metrics may not be achievable in practice, colback=white]
    \centering
    \includegraphics[width=0.7\linewidth]{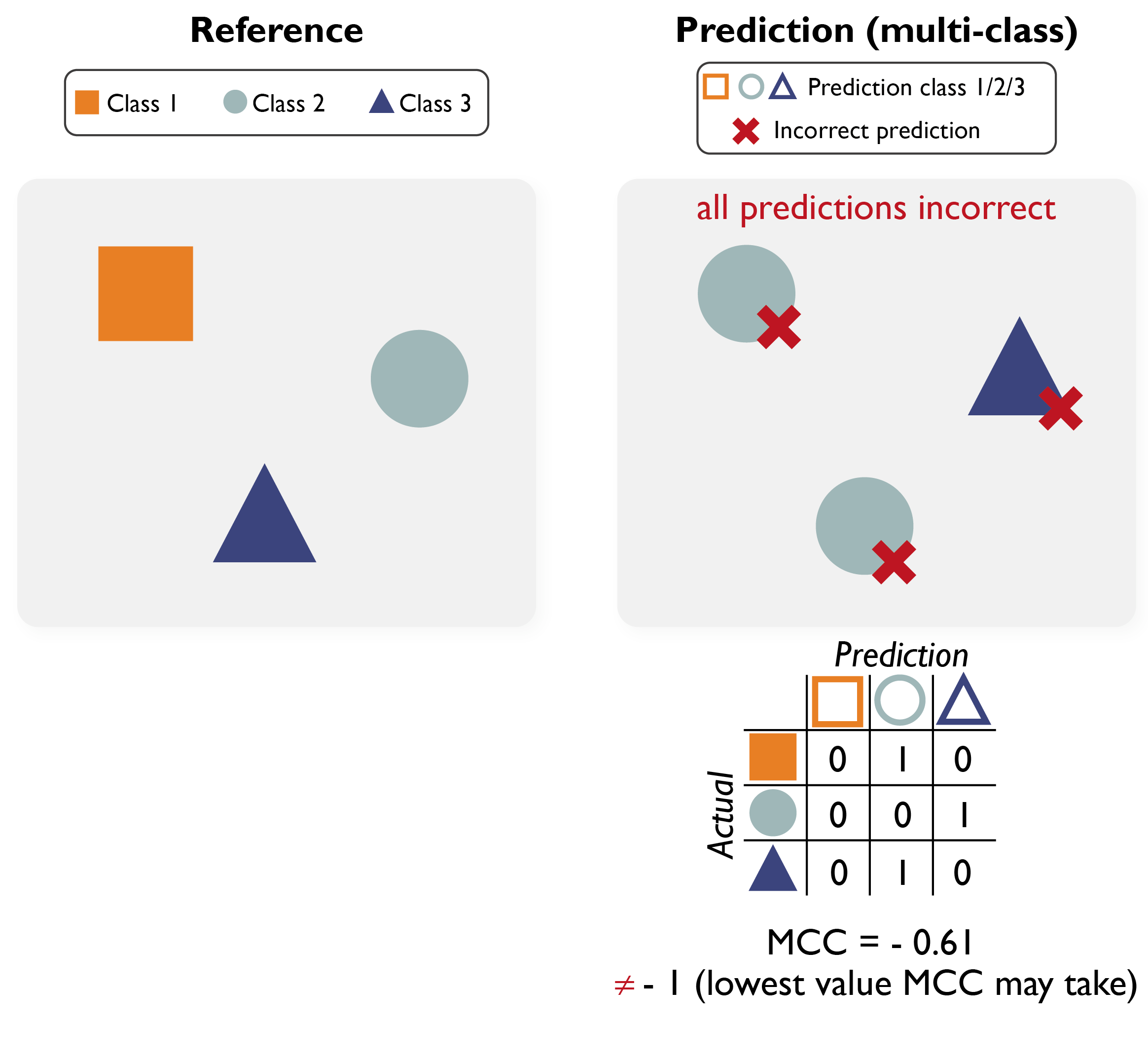}
    \caption{Effect of theoretical bounds that may not be achievable in practice. In this multi-class example, all samples were predicted incorrectly. However, the theoretical lowest value for \textit{\acf{MCC}} (-1) cannot be achieved in this situation, rendering interpretation difficult.}
    \label{fig:lack-bounds}
\end{tcolorbox}
\end{figure}

\begin{figure}[H]
\begin{tcolorbox}[title= Pitfall: Per-class tuning of the decision threshold yields misleading results, colback=white]
    \centering
    \includegraphics[width=0.85\linewidth]{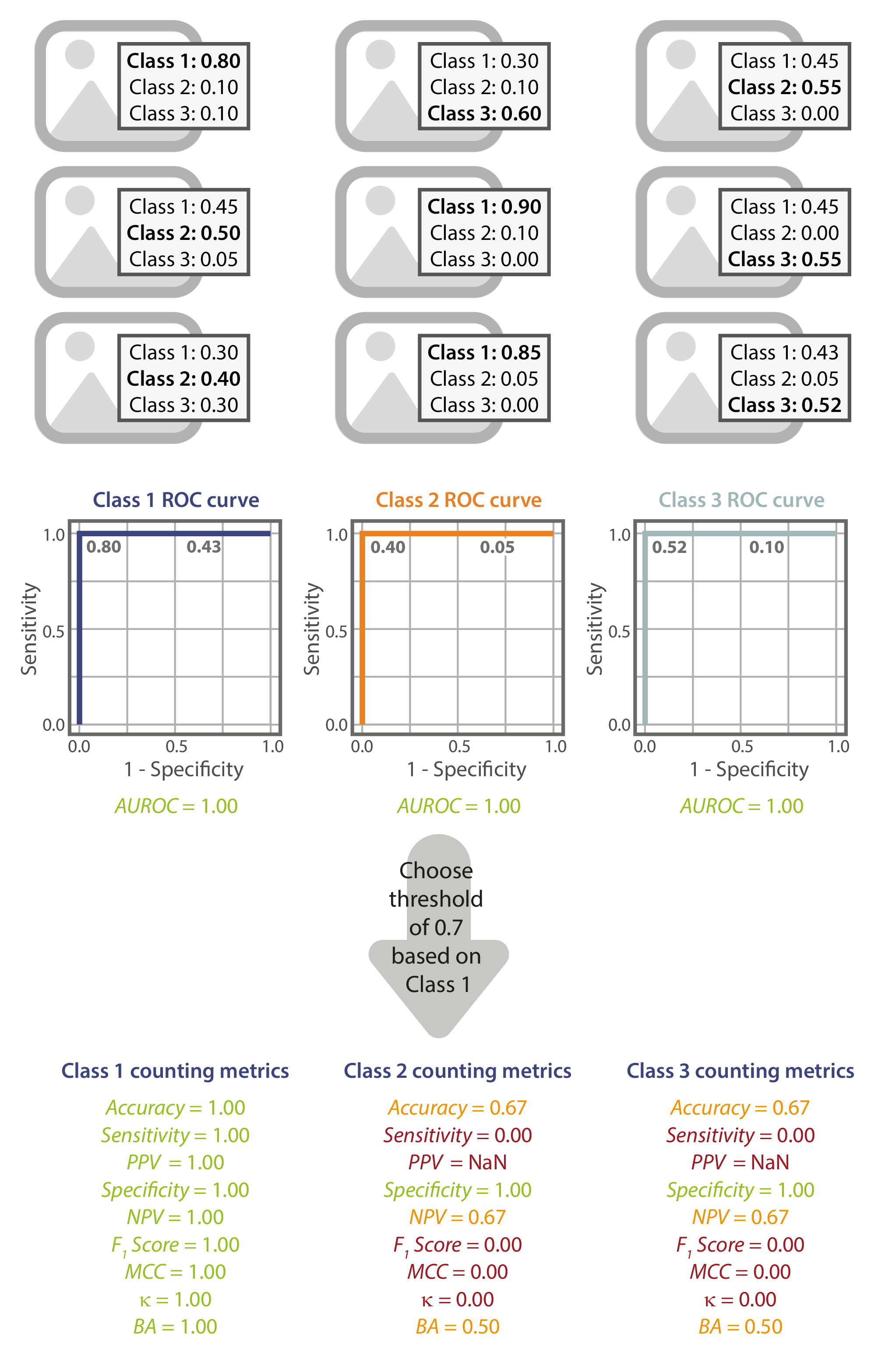}
    \caption{Effect of the determination of a global threshold for all classes based on a single class. In a data set of three classes and nine images, the \textit{\acf{AUROC}} score is 1.0 for every class. In practice, however, a global decision threshold needs to be set in multi-class problems, which typically renders substantially worse results. Here, the optimal threshold for \textit{Class 1} yields poor results for \textit{Classes 2} and \textit{3} (see e.g., \citep{krause2018grader, de2018clinically}). Used abbreviations: \textit{\acf{PPV}}, \textit{\acf{NPV}}, \textit{\acf{MCC}}, \textit{Cohen's Kappa} $\kappa$, and \textit{\acf{BA}}.}
    \label{fig:auroc-classes-threshold}
\end{tcolorbox}
\end{figure}

\begin{figure}[H]
\begin{tcolorbox}[title= Pitfall: \textit{\acf{AUROC}} may obscure model bias, colback=white]
    \centering
    \includegraphics[width=0.9\linewidth]{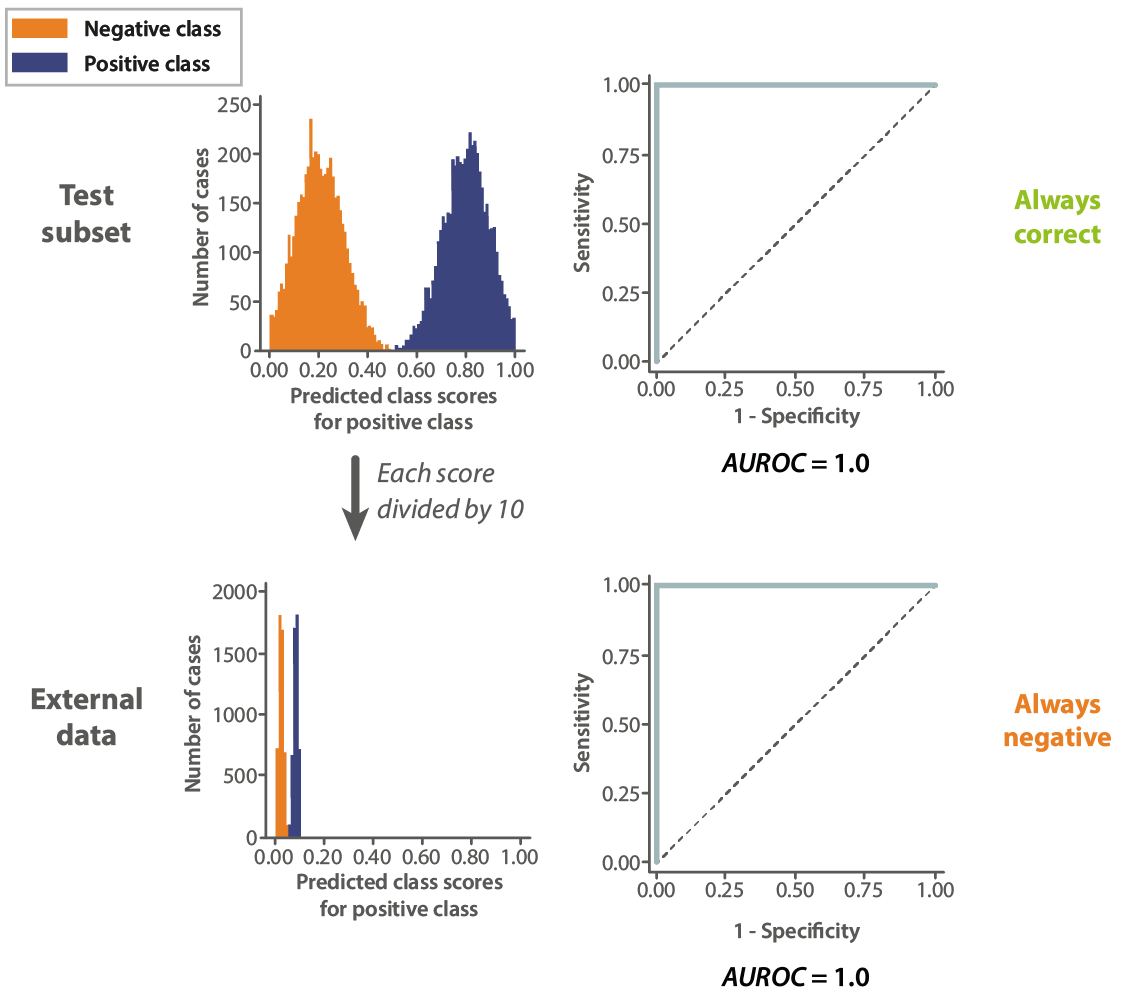}
    \caption{Calculating the \textit{\acf{AUROC}} may obscure model bias. In this example, the model works well, i.e., yields very distinct predicted class scores, on a test subset ('always correct'). When generalizing to a different dataset, assuming similar conditions for the given model's predicted class scores (in this example obtained by linear rescaling), \textit{\ac{AUROC}} still indicates a perfect performance, although the predicted class scores for the external dataset are all very low, indicating that all predictions would be of the negative class with high confidence ('always negative'). Figure courtesy of \cite{kleppe2022area}.}
    \label{fig:auroc-bias}
\end{tcolorbox}
\end{figure}

\begin{figure}[H]
\begin{tcolorbox}[title= Pitfall: \textit{\acf{AUROC}} for small sample sizes, colback=white]
    \centering
    \includegraphics[width=0.85\linewidth]{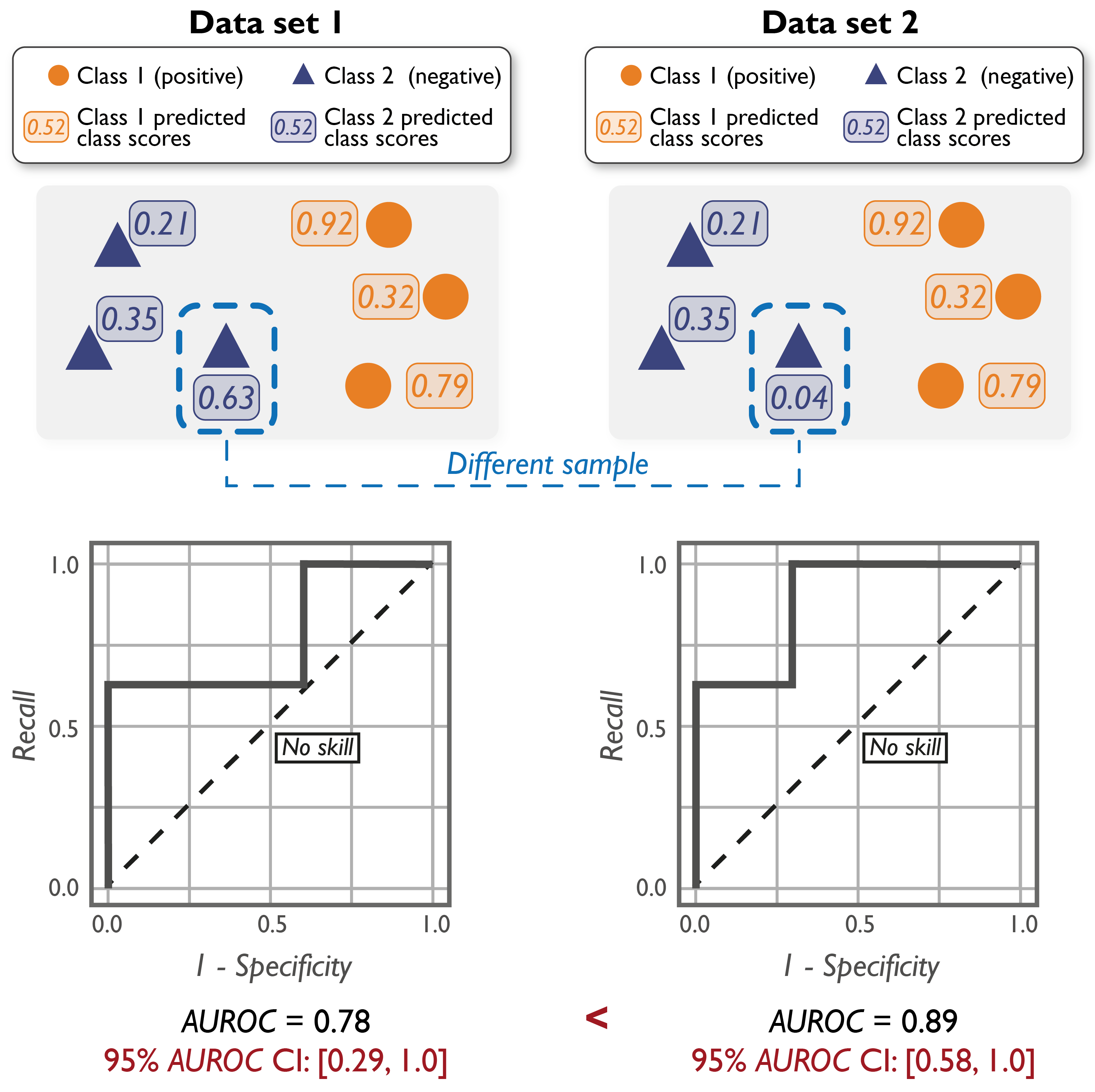}
    \caption{Effect of calculating the \textit{\acf{AUROC}} for very small sample sizes. The \textit{\ac{AUROC}} is very unstable for small sample sizes. \textit{Data sets 1} and \textit{2} only contain six samples each, for which only one predicted score differs between sets. Drawing the \textit{\acf{ROC}} curve and calculating the \textit{\ac{AUROC}} leads to a large difference in scores between both data sets. The 95\% \acf{CI} reveals that there is a large range of possible \textit{\ac{AUROC}} values. \ac{CI}s were calculated based on \cite{delong1988comparing}.}
    \label{fig:auroc-small-sample-sizes}
\end{tcolorbox}
\end{figure}

Furthermore, metric-specific limitations may arise (Figures~\ref{fig:lr+ba} - \ref{fig:auroc-small-sample-sizes}).
Please note that all of these also apply to semantic/instance segmentation or object detection problems. The discourse focuses on the most commonly used image-level classification metrics, as presented in Figs.~\ref{fig:def-classification-1}, \ref{fig:def-classification-2} and~\ref{fig:def-auc}. For most of the problems, it focuses on the \textit{Accuracy}, \textit{Sensitivity} or \textit{\ac{PPV}}, because those are the most common metrics for image-level classification in biomedical image analysis. Please note that we do not recommend their indiscriminate use, as they come with limitations (discussed in the following paragraphs), but rather wish to spotlight the problems and pitfalls of those most commonly used metrics. 

To preserve the clarity of the illustrations, the most important of the presented metric values may be highlighted with color. Green metric values correspond to a "good" metric value (e.g. a high \textit{Sensitivity} score), whereas red values correspond to a "bad" value (e.g. a low \textit{Sensitivity}). Green check marks indicate desirable behavior of metrics, red crosses indicate undesirable behavior. Please note that a low metric value is not automatically a "bad" score. A metric value should always be put into perspective and compared to inter-rater variability. For simplicity, we still use the terms "good" and "bad/poor" throughout the section. Finally, our illustrations do not provide the concrete class probabilities of the presented classifiers.

\paragraph*{\textbf{High class imbalance}} \textit{Accuracy} is one of the commonly applied metrics in classification problems, presumably because it is particularly straightforward to interpret. However, the metric is not designed to handle imbalanced data sets, which often occur across all domains. Figure~\ref{fig:class-imbalance} provides an example in which the positive class (orange circle) is heavily underrepresented. While \textit{Prediction 1} gives a reasonable separation of the classes, \textit{Prediction 2} results in the same \textit{Accuracy} value (0.97) although the algorithm only provides the majority vote as a result. In this specific example, \textit{Sensitivity}, \textit{\ac{PPV}} and \textit{F1 score} reveal the issue, as does \textit{Matthews correlation coefficient (MCC)}, a metric designed to handle class imbalance which reflects that \textit{Prediction 2} is not better than a random guess (0.00) \citep{chicco2020advantages}. As many classification measures are easily computable using the number of \ac{TP}, \ac{TN}, \ac{FP} and \ac{FN} samples, it is highly recommended to report these \ac{TP}, \ac{TN}, \ac{FP} and \ac{FN} values explicitly and then compute multiple metrics \citep{hicks2021evaluation}. 

Plotting the \textit{\ac{ROC}} and \textit{\ac{PR}} curves (Figs.~\ref{fig:class-imbalance}b and c) also reveals the limitations of \textit{Prediction 2}, which yields an \textit{\ac{AUROC}} of 0.52 and \textit{\ac{AP}} of 0.04, indicating that the prediction is not better than random guessing.

Some metrics inherently address the problem of performance overestimation or distortion caused by class imbalance. \textit{\acf{MCC}} and \textit{Cohen's Kappa} $\kappa$, for example, are designed in such a way that they assign a (prevalence-aware) random guessing algorithm the value $0$. Similarly, \textit{\acf{AUROC}} assigns the value $0.5$ and \textit{Balanced Accuracy} yields a value of $\frac{1}{C}$, where $C$ is the number of classes. For all other metrics (see Figure~\ref{fig:class-imbalance}), it is generally recommended to put the resulting values into perspective of with respect to a (prevalence-aware) random guessing algorithm baseline.

\newpage 

\newpage
Given their prevalence invariance (see Figure~\ref{fig:prevalence-dependency}), metrics such as \textit{\ac{BA}} are a good choice in many scenarios. However, they may be misleading in the case of underrepresented classes (see also use case 2 from \cite{chicco2021matthews}). In such a scenario, \textit{\ac{BA}} yields high scores although the prediction is far from perfect. Here, the \textit{\ac{MCC}} and \textit{\ac{PPV}} reveal that the prediction is quite inefficient by assessing predictive values (see Figure~\ref{fig:misleading-ba-j}).

\newpage
Prevalence-dependent metrics yield different metric scores for imbalanced data sets. Figure~\ref{fig:metric-landscapes} shows the metric landscapes for a balanced data set, meaning a prevalence of 50\%, in comparison to those for imbalanced data (prevalence of 5\%) \cite{brown2018classifiers}.

\newpage
\paragraph{\textbf{More than two classes available}} Many binary metrics can directly be translated to the multi-class case by expanding the confusion matrix to all classes. These classes are often hierarchically structured, for example in the shape of one negative class (e.g. no pathology) and multiple positive classes (e.g. different types of pathologies). Figure~\ref{fig:multi-class} shows an example of a classification into triangles and circles, for which the circle class is further separated into two distinct classes (green and orange). The binary performance into triangle \textit{vs.} circle, shown in the middle, is good (\textit{Accuracy} of 0.88). But when considering the three classes separately, the prediction struggles to identify the color of the circles, causing their per-class accuracy scores to drop significantly.

\newpage 
\paragraph{\textbf{Unequal severity of class confusions}} In biomedical applications, classes are often not equally important. Consider the task of colon polyp detection in the gastrointestinal tract, for example. To provide the patient with the best care, it is crucial to detect all of these precancerous lesions. This requires a particular penalization of those samples containing a polyp which have been marked as 'no polyp' (\ac{FN}), and the metrics need to be chosen accordingly. The \textit{\ac{PPV}}, for example, does not include the \ac{FN} in its definition, hence would not be appropriate for this research question. \textit{Sensitivity}, on the other hand, would show the desired poor performance in the presence of many \ac{FN} predictions, as seen in the top row of Figure~\ref{fig:unequal-class-confusions}. 

For image retrieval, the task of finding images for a specific content, it is not important to find every single existing image, but the images found should be correct. In this setting, the \ac{FP} (assigning an incorrect image as correct) need to be penalized. Since it includes the computation of \ac{FP}, in this case, the \textit{\ac{PPV}} would be a good metric. In contrast, \textit{Sensitivity} does not consider \ac{FP}, therefore being inappropriate in this context (see bottom row of Figure~\ref{fig:unequal-class-confusions}). Penalization in both cases is especially important in cases of imbalanced data sets (see Figure~\ref{fig:class-imbalance}).

\newpage
\paragraph{\textbf{Unequal handling of classes}}
When it comes to multi-class problems, different approaches may be chosen to compute the metric values. One possibility is to first compute the metric values per class and aggregate them subsequently. Special care has to be taken in the case of unequal importance of the different classes. For example, identifying whether a patient harbors a pathology in general might be more important than identifying the specific type of pathology. In this case, one should not just average over all class metric scores, but instead apply a sufficient weighting scheme. In the example of Figure~\ref{fig:unequal-class}, the triangle class is the most important class but also the one with the lowest per-class \textit{Accuracy}. Simple averaging, so-called macro-averaging, would ignore that property and thus result in a higher aggregated \textit{Accuracy} than merited. This effect can be compensated with the \textit{Weighted Accuracy}.

\newpage 
\paragraph{\textbf{Interdependencies between classes}} If multiple classes are visible in the data set, one should carefully account for interdependencies between the classes. Interdependencies can happen in cases of multi-colinearities in which two classes are correlated, either inherently, such as for the body mass index (BMI) and the body fat percentage, or in the case of dependent data settings, for example multiple images per patient or the presence of confounders. An algorithm aiming to classify the dark blue triangle class in Figure~\ref{fig:multi-class-interdependencies} may result in a nearly perfect \textit{Accuracy} of 0.94, but only because the dark blue triangle almost always appears in conjunction with the orange square. Computing the \textit{Accuracy} for those images individually without the square class would lead to a much lower performance.

\newpage 
\paragraph{\textbf{Stratification based on meta-information}} Different kinds of meta-information may be available for a data set, including the presence and relevance of artifacts or artificial structures (e.g. metal artifacts in CT images or text overlay in endoscopic data) as well as specifics of acquisition protocols (e.g. acquisition angle or viewpoint) or grid size (cf. Figure~\ref{fig:DSC-grid-size}). Another typical example is the gender of a patient, as shown in Figure~\ref{fig:stratification-gender}. In this case, the \textit{Accuracy} is computed over twelve cases, disregarding the available meta-information (gender). Stratification based on gender will reveal that the prediction performs much worse for women compared to men. These aspects are currently investigated in the fairness literature \cite{barocas2017fairness}.

\newpage
\paragraph{\textbf{Importance of cost-benefit analysis}} Most common performance metrics fall short when it comes to determining whether an algorithm can lead to improved clinical decisions. Knowing that an algorithm performs well in terms of calibration, discrimination, and accuracy, for instance, does not necessarily imply that decisions guided by the algorithm will lead to improved clinical outcomes. Motivated by the idea of taking into account the tradeoff between the benefit resulting from the detection of a \ac{TP} case and the cost of a \ac{FP}, a risk threshold should be defined for a specific prediction problem. \ac{NB} is a metric that incorporates the tradeoff between costs and benefits resulting from the selection of a specific risk threshold in the validation of an algorithm, and thus facilitates informed medical decision-making \citep{vickers2016net}. In Figure~\ref{fig:cost-benefit}, we present a similar example as the one presented by \citep{vickers2016net}. In this example, nine unnecessary biopsies are deemed acceptable to detect one lesion, resulting in an exchange rate of 1:9. Two scenarios are to be considered here. On the one hand, applying the biopsy to every patient would ensure that all patients with the disease receive a biopsy (benefit) but also result in 75 patients with no disease undergoing biopsy (harm). On the other hand, a marker-based biopsy decision would result in only 20 patients with the disease undergoing biopsy, thus missing 15 of them, while reducing the number of unnecessarily biopsied healthy patients to 60. Only considering the Accuracy would rate both scenarios similarly. However, incorporating the cost-benefit analysis and the resulting exchange rate, \ac{NB} would indicate that performing biopsy on all patients yields better clinical outcome. Please note that defining risk cutoffs and exchange rates is highly subjective. This is why \ac{NB} is often considered over various thresholds.

\newpage 
\paragraph{\textbf{Definition of class labels}} Binary classification problems typically involve the step of defining one class as positive and the other as negative. The final scores of per-class counting metrics rely on the definition of the positive class. Figure~\ref{fig:positive-class} shows an example in which the class labels are reversed in the same data set. The scores of per-class counting metrics dramatically change when the class labels are switched. On the other hand, multi-class counting metrics remain stable as they rely on the entire confusion matrix and are invariant to changes in the class labels.

\newpage
\paragraph{\textbf{Prevalence dependency}}
The \textit{\ac{PPV}} and the \textit{\acf{NPV}} are common measures to validate classification performances. In many cases, binary classification is considered, for example presence or absence of a disease. 
In contrast to \textit{Sensitivity} and \textit{Specificity}, in case-control studies, \textit{\ac{PPV}} and \textit{\ac{NPV}} should be seen as the conditional probability of a disease being present based on a test result and the prevalence in a general population \citep{molinaro2015diagnostic}. 

However, \textit{\ac{PPV}} and \textit{\ac{NPV}} are frequently used incorrectly. This is due to the fact that many practitioners assume the same prevalence in an analyzed case-control study group as in the general population. However, a study group is often heavily biased, either due to the study design or due to the observation of patient groups from specialized clinics. Thus, the assumed disease prevalence in scientific literature is often higher than that found in the general population (cf. Figure~\ref{fig:prevalence}). This problem is amplified by default implementations (e.g. in \texttt{scipy}~\cite{virtanen2020scipy}) which disregard wider population prevalence and calculate prevalence from the study group. Without prevalence correction, this can lead to misleading results, confusion among patients and ill-informed policy-making.  

\newpage
Metrics depending on the prevalence cannot easily be compared across data sets with different prevalences, as shown in Figure~\ref{fig:prevalence-dependency}. Here, two situations with the same \textit{Sensitivity} and \textit{Specificity} values but with different prevalences (50\% \textit{vs.} 90\%) are presented. Only metrics that do not rely on the prevalence (such as \textit{\ac{BA}} and \textit{\ac{EC}}) can be used for a comparison across data sets. 

\newpage
For a prevalence unequal to 50\%, \textit{\ac{BA}} and \textit{Youden's Index J} may lead to different rankings compared to \textit{\ac{MCC}} and \textit{Cohen's Kappa $\kappa$} \cite{chicco2021matthews}, as illustrated in Figure~\ref{fig:prevalence-dependency-rankings}. In this example, predictions for three different prevalences (50\%, 40\% and 60\%) are shown. Only in the case of a 50\% prevalence will the rankings generated by all metrics be the same, with all preferring \textit{Prediction 2}. For different prevalences, rankings may differ: While \textit{\ac{BA}} and \textit{Youden's Index J} prefer \textit{Prediction 2} over \textit{Prediction 1}, \textit{\ac{MCC}} and \textit{Cohen's Kappa $\kappa$} favor the \textit{Prediction 1}.

\newpage
\paragraph{\textbf{Discrimination \textit{vs.} calibration}} A model's \textit{discrimination} capability refers to how well it separates samples with and without the class of interest. If all predicted class scores for the samples with the class of interest are higher than for the others, the model discriminates perfectly, reflected in an \ac{AUROC} score of 1. A \textit{calibrated model} outputs predicted class scores that match the empirical success rate (e.g. outputs with score 0.8 for a specific class, empirically belong to this class in $80\%$ of the cases) \cite{cook2007use}. However, it may happen a perfectly discriminating model, such as the one presented in Figure~\ref{fig:calibration}, is not well calibrated. The \textit{Prediction} gives a confidence score of 0.52 for all samples of the positive class (orange circle) and a score of 0.51 for all samples of the negative class (blue triangle). Although this leads to a perfect discrimination, the calibration is very poor and the predicted probabilities would not be helpful in practice at all.

Similarly to standard validation metric measuring the discrimination power of algorithms, calibration metrics come with limitations. For example, the \textit{\ac{BS}} \citep{brier1950verification, gneiting2007strictly} is defined as the squared difference between a predicted class score $f_t$ and the actual outcome $o_t$ for one of $n$ events $t$, typically defined as 1 if the event occurred and 0 otherwise. \\

However, the \textit{\ac{BS}} should be used with caution in the case of ordinal classes, in which the classes underlie an order, such as in the example in Figure~\ref{fig:brierscore-ordinal}. Here, three classes are present which denote patient survival times in years. The first class refers to zero to one year, the second class to two to three years, and the third class denotes a survival time of three or more years. While the actual survival is zero to one year (class 1; orange circle), \textit{Predictions 1} and \textit{2} are predicting classes 2 (blue triangle) and 3 (green square), both with a probability of 100\%. Although both predictions are incorrect, \textit{Prediction 1} would be preferable given the ordinal scale of variables: The predicted time frame of one to two years of survival time is closer to the actual survival time of zero to one year than that of \textit{Prediction 2}. However, both predictions would yield the same \textit{\ac{BS}}.

\newpage
Calibration errors \citep{guo2017calibration}, on the other hand, compute the difference between the accuracy of a prediction and the average predicted class scores. Classifier outputs are generally continuous, which often reduces the number of available samples per prediction to one. Strategies for alleviating the sparse sampling problem include binning the continuous scale \citep{maier2022metrics}, as done by the \textit{\ac{ECE}} and \textit{\ac{MCE}} \citep{guo2017calibration, naeini2015obtaining}. However, the binning itself is not strictly defined, neither in regard to whether the bins should be equidistant nor in the number of bins. Depending on the binning strategy, the scores of the \textit{\ac{ECE}} and \textit{\ac{MCE}} will differ, as shown for the example in Figure~\ref{fig:ece-mce-bins}.

\newpage
Calibration can be defined on different levels. While top-label and class-wise calibration only consider certain parts of the predicted class score vectors, solely the canonical definition of calibration computes the calibration errors based on the full probability vectors. As shown in Figure~\ref{fig:ece-pitfalls}(a), the \ac{CE} may imply a perfect calibration although the prediction is not perfectly calibrated, as indicated by the canonical \ac{CE} \citep{gruber2022better,vaicenavicius2019evaluating}. Moreover, the \ac{CE} depends on the sample size \citep{gruber2022better}. Although a model may be perfectly calibrated, the \ac{ECE} only converges towards a perfect score with an increasing number of data samples, as shown in Figure~\ref{fig:ece-pitfalls}(b).

\newpage
\paragraph{\textbf{Classification metrics in the case of ordinal classes}} In the case of ordinal classes, similarly to calibration metrics, the scores of classification metrics should be interpreted with caution, as shown in Figure~\ref{fig:brierscore-ordinal}. The classification or grading of patients according to a scale of disease severity would be a typical use case. In Figure~\ref{fig:ordinal-grading}, the severity of the disease is given by three classes, with class 0 denoting the lowest severity and class 2 the highest. \textit{Predictions 1} and \textit{2} agree in the grading of the first four patients. However, \textit{Prediction 1} classifies \textit{Patient 3} into the lowest severity, and thus performs much more poorly than \textit{Prediction 2}, which is only one class of severity below the reference. However, both \textit{Accuracy} and \textit{\ac{MCC}} do not spot this difference. Underestimating the severity of disease in a patient, however, would lead to serious consequences in clinical practice and thus needs to be heavily penalized. Only metrics with pre-defined weights would highly penalize \textit{Prediction 1} (here: \textit{\ac{EC}} and \textit{quadratic-weighted Cohen's Kappa}).

\newpage
\paragraph{\textbf{Label shift in the case of ordinal classes}} In ordinal class scenarios, one should also be aware of potential label shifts, in which the distribution or number of labels changes over time or data sets \citep{zhang2021dive}. Figure~\ref{fig:ordinal-grading-label-shift} shows the metric scores for two data sets with the same labels and same number of data, but a differing number of images per label. While \textit{Data set 1} is balanced, \textit{Data set 2} underwent a label shift with different per-class prevalences. While the \textit{Accuracy} values do not change, the \textit{\ac{BA}} and \textit{\ac{MCC}} are slightly affected by the label shift. The most dramatic change is observed for the \textit{quadratic-weighted Cohen's Kappa}, with a drop of 0.17 after label shift. Thus, when comparing the performance of a model on data sets with varied regional or temporal origins, one should take precautions when utilizing the \textit{quadratic-weighted Cohen's Kappa}\footnote{Equal Sensitivities of 0.8 per class and uninformed guessing in the remaining 20\% of cases were assumed. See \url{https://github.com/agaldran/kappa_pitfall/blob/main/kappa_pitfall.ipynb} for details. }.  

\paragraph{\textbf{Same metric values for different confusion matrices}} Per definition, the \textit{\ac{LR+}}, \textit{Youden Index J} and \textit{\ac{BA}} are calculated from \textit{Sensitivity} and \textit{Specificity}. However, the exact same \ac{LR+} or \ac{BA} values can occur for very different specifications of the confusion matrix. Figure~\ref{fig:lr+ba}a shows three examples of predictions, yielding quite different confusion matrices. For example, the leftmost prediction shows a very high number of \acp{TN} and \acp{FN}, while the prediction in the middle only has very few false predictions. The prediction on the right correctly classifies most of the negative samples. Despite the different \textit{Sensitivity} and \textit{Specificity} values, \ac{LR+} will be the same for all three examples \citep{dujardin1994likelihood}. The same can occur for the \ac{BA}, as shown in Figure~\ref{fig:lr+ba}b. In both cases, the substantial differences between the predictions remain hidden unless \textit{Sensitivity} and \textit{Specificity} are reported in addition. Similar issues hold true for other metrics that rely on multiple measures. An \ac{AUROC} of 0.9, for example, may correspond to various appearances of the \ac{ROC} curve, and various underlying distributions of positive and negative samples and predictions \citep{wald2014area}.

\newpage 
\paragraph{\textbf{Upper bound in \textit{Cohen's $\kappa$} calculation}} \textit{Cohen's $\kappa$} measures the agreement between ratings while incorporating information on the \textit{Accuracy} by chance. It therefore investigates how well a prediction follows the distribution of the actual class. The maximum \textit{Cohen's $\kappa$} helps interpreting the calculated \textit{$\kappa$} score by symbolizing the corner case in which either the \ac{FP} or \ac{FN} are equal to 0 \cite{umesh1989interjudge}:
\begin{equation}
    \begin{split}
        \kappa_{max} &= \frac{p_{max}-p_e}{1-p_e}, \\
        p_{max} &= \min \left( \frac{TP+FN}{TP+TN+FP+FN}, \frac{TP+FP}{TP+TN+FP+FN}\right) \\
        &+ \min \left( \frac{TN+FN}{TP+TN+FP+FN}, \frac{TN+FP}{TP+TN+FP+FN} \right)
    \end{split}
\end{equation}
The maximum \textit{Cohen's $\kappa$} score will be lower as the number of the predicted positive and negative samples diverges more from the actual number of positive and negative samples. This is shown in Figure~\ref{fig:kappa-max} with two predictions. \textit{Prediction 1} achieves lower \textit{Accuracy} and \textit{Cohen's $\kappa$} scores compared to \textit{Prediction 2}, as it only predicts a very low number of \ac{TP}. However, the predicted distribution of samples in \textit{Prediction 1} is closer to the actual distribution of samples (13 circle predictions \textit{vs.} 15 actual circles and 87 triangle predictions \textit{vs.} 85 actual triangles). The distribution of samples of \textit{Prediction 2} differs more from the actual distribution, yielding a lower \textit{Cohen's $\kappa_{max}$} value\footnote{\url{https://www.knime.com/blog/cohens-kappa-an-overview}}.
\newpage
\newpage
Similarly, the theoretical lower bound for the \ac{MCC} is not always achievable, such as in the example in Figure~\ref{fig:lack-bounds}.

\paragraph{\textbf{Determination of a global threshold for all classes}} In a scenario with multiple classes, a single cutoff value for a threshold needs to be chosen for all classes. Multi-threshold metrics, such as \textit{\ac{AUROC}}, may be overly optimistic as the optimal threshold range for one class may differ from the optimal threshold range for another. Figure~\ref{fig:auroc-classes-threshold} illustrates the case of three classes that yield three perfect \textit{\ac{AUROC}} scores. However, when choosing a single threshold of 0.8 based on class 1 for all classes, the respective counting metrics will yield very poor results for classes 2 and 3.

\newpage

\paragraph{\textbf{Model bias}} Image analysis models might be affected by image features that human experts naturally ignore and that are not truly relevant for the prediction task. Such model biases might be revealed by analyzing data external to those used for development \cite{kleppe2021designing}. However, multi-threshold metrics such as \textit{\ac{AUROC}} will inherently obscure model biases that affect the predicted class scores but not their ranking, if model calibration is not considered. Such a pitfall is illustrated in Figure~\ref{fig:auroc-bias} (similar to \cite{kleppe2022area}). The example assumes a model that works perfectly on a subset of a larger dataset used for training. Generalizing to a different dataset when using a model with a severe bias is simulated by linear rescaling, i.e., dividing all model scores by 10, meaning that all predicted class scores will be 0.1 or below, while the ranking of the scores remains the same. Since the ranking remains the same, the \textit{\ac{AUROC}} value will not change. However, given the low predicted class scores, the model would be very confident that all cases actually belong to the negative class. Here, the substantial model bias is thus not captured by multi-threshold metrics such as \textit{\ac{AUROC}}, which would yield perfect metric values on the external dataset. In general, any metric calculation that adapts to the score distribution of the test dataset will obscure such model biases, including metrics such as \textit{Specificity} at a particular level of \textit{Sensitivity} (see \cite{kleppe2022area} for details).


\newpage
\paragraph{\textbf{Small sample sizes}} Small sample sizes are a common issue in the biomedical image analysis domain. Caution should be applied when calculating, for example, the \textit{\ac{AUROC}} in the presence of only very few images. Figure~\ref{fig:auroc-small-sample-sizes} provides an example of six images (three positive, three negative samples) for two data sets and the respective predicted class scores of an algorithm. The data sets only differ in a single image. However, this apparently minor difference will result in a difference in the \textit{\ac{AUROC}} scores of 0.11. By calculating the 95\% \ac{CI}, we see that the \acp{CI} are very large and do not allow for proper interpretation of \textit{\ac{AUROC}} scores in the case of very small sample sizes.

\newpage
\section{Pitfalls related to segmentation} 
\label{sec:segmentation}

\begin{figure}[H]
\begin{tcolorbox}[title= Pitfall: Small structures, colback=white]
    \centering
    \includegraphics[width=0.85\linewidth]{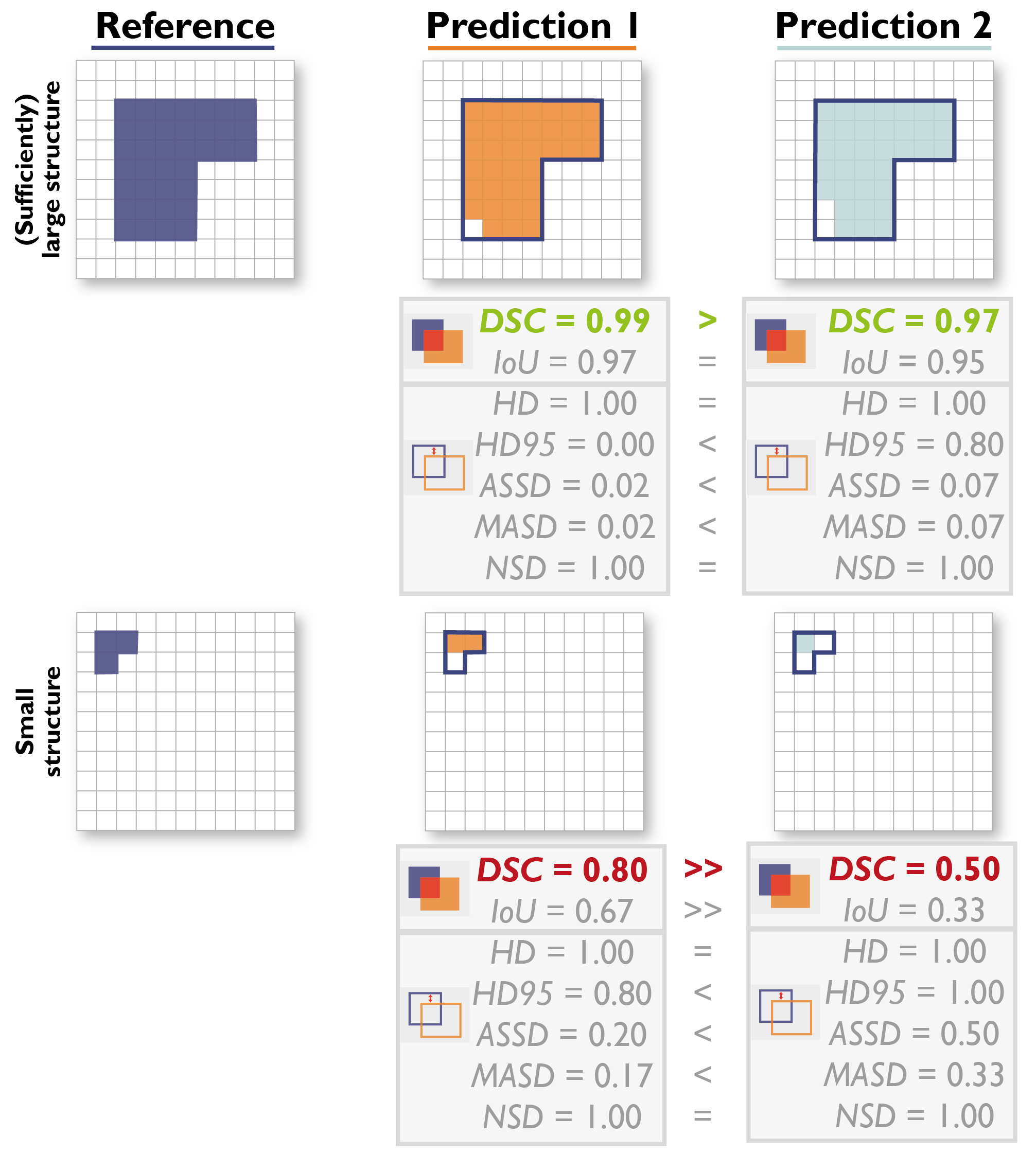}
    \caption{Effect of structure size on the \textit{\acf{DSC}}. The predictions of two algorithms (\textit{Prediction 1/2}) differ in only a single pixel. In the case of the small structure (bottom row), this has a substantial effect on the corresponding metric value (similar for the \textit{\acf{IoU}}). The effects are considerably lower for the boundary-based metrics (\textit{\acf{HD}}, \textit{\acf{HD95}}, \textit{\acf{ASSD}} and \textit{\acf{NSD}}.}
    \label{fig:DSC-small}
\end{tcolorbox}
\end{figure}

\begin{figure}[H]
\begin{tcolorbox}[title= Pitfall: Small tubular structures, colback=white]
    \centering
    \includegraphics[width=0.85\linewidth]{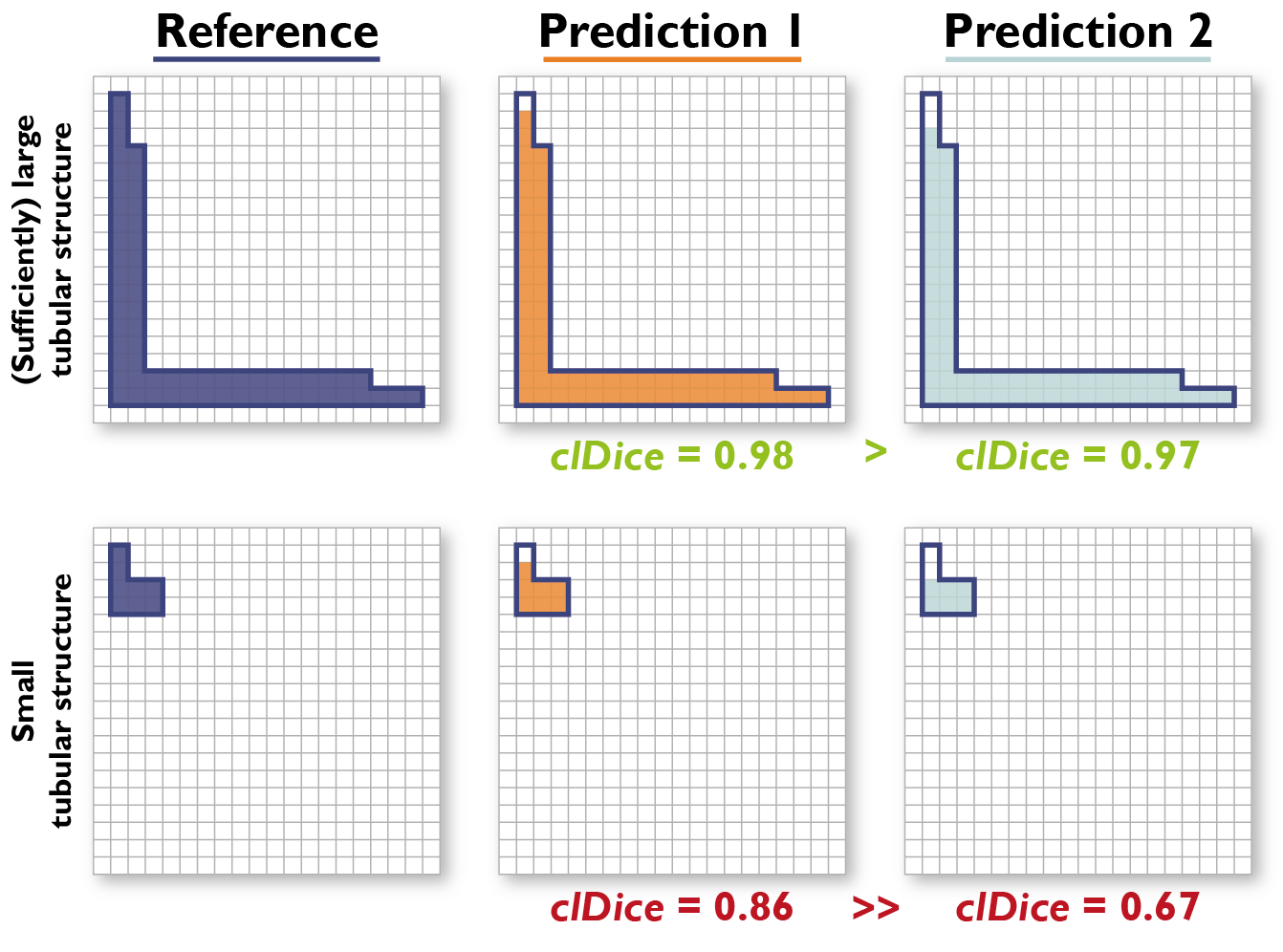}
    \caption{Effect of structure size on the \textit{\acf{clDice}}. The predictions of two algorithms (\textit{Prediction 1/2}) differ in only a single pixel. In the case of the small tubular structure (bottom row), this has a substantial effect on the corresponding \textit{\ac{clDice}} value.}
    \label{fig:clDice-small}
\end{tcolorbox}
\end{figure}

\begin{figure}[H]
\begin{tcolorbox}[title= Pitfall: High variability of structure sizes, colback=white]
    \centering
    \includegraphics[width=\linewidth]{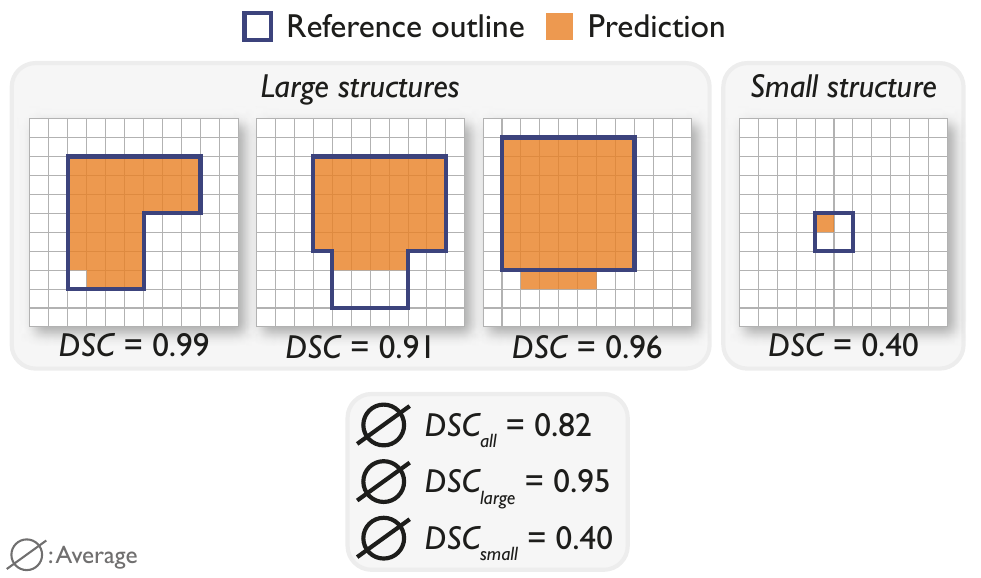}
    \caption{Effect of high variability of structure sizes across images. As shown in Figure~\ref{fig:DSC-small}, \textit{\acf{DSC}} scores penalize errors in small structures much more, leading to large and small structures influencing the overall averaged performance to different extents. Only by computing the average for small and large structures separately can it be seen that the prediction in the present example is much poorer for the small structure. $\varnothing$ refers to the average \textit{\ac{DSC}} values.}
    \label{fig:high-variability}
\end{tcolorbox}
\end{figure}

\begin{figure}[H]
\begin{tcolorbox}[title= Pitfall: High variability of sizes for the \acf{MASD} and \acf{ASSD}, colback=white]
    \centering
    \includegraphics[width=0.7\linewidth]{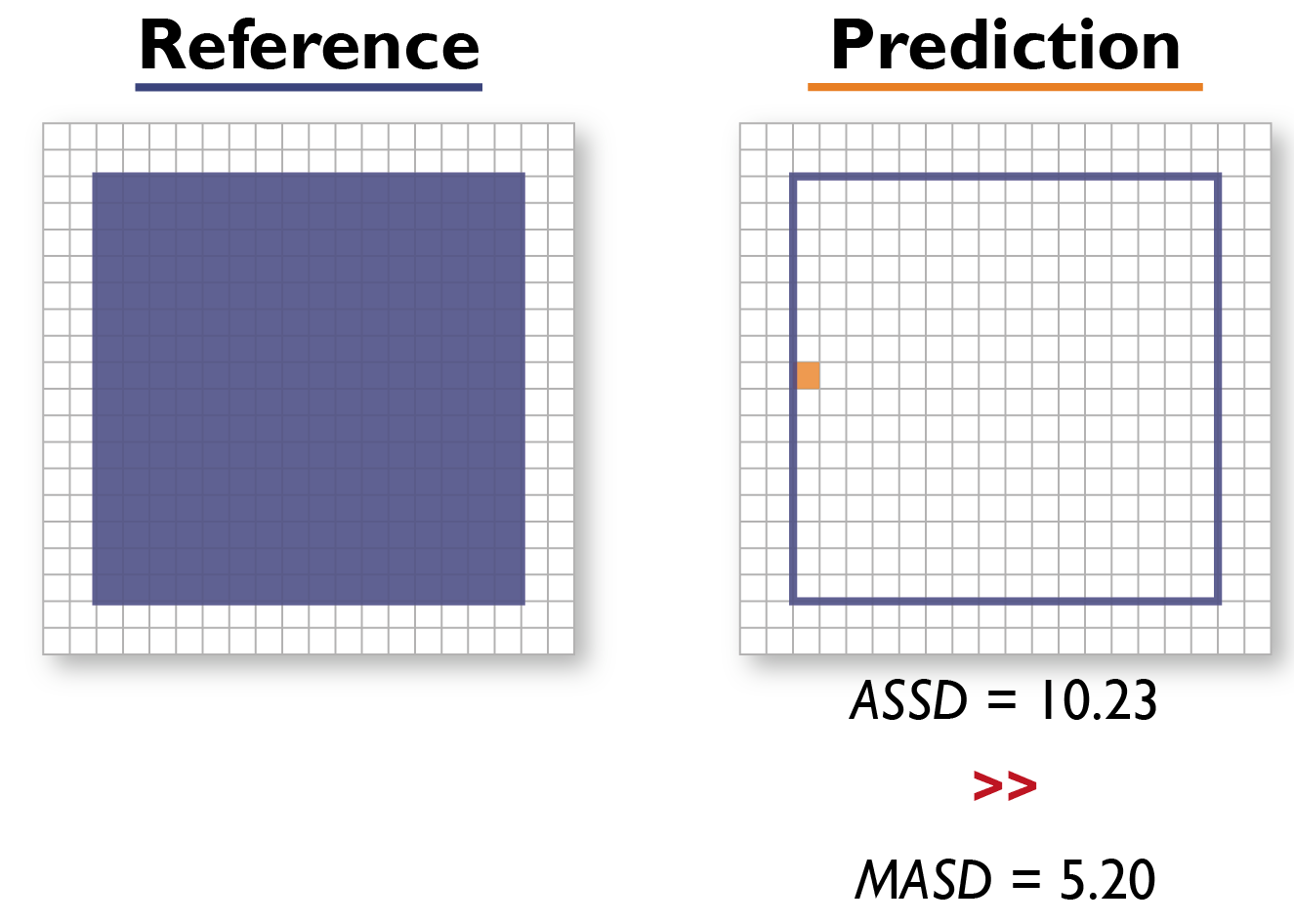}
    \caption{Corner case in which \acf{MASD} yields an undesired result. If the \textit{Prediction} is very small (here: one pixel) and located close to the reference boundary, the \acf{MASD} will be much lower compared to the \acf{ASSD}.}
     \label{fig:masd-assd}
\end{tcolorbox}
\end{figure}

\begin{figure}[H]
\begin{tcolorbox}[title= Pitfall: Shape unawareness, colback=white]
    \centering
    \includegraphics[width=1\linewidth]{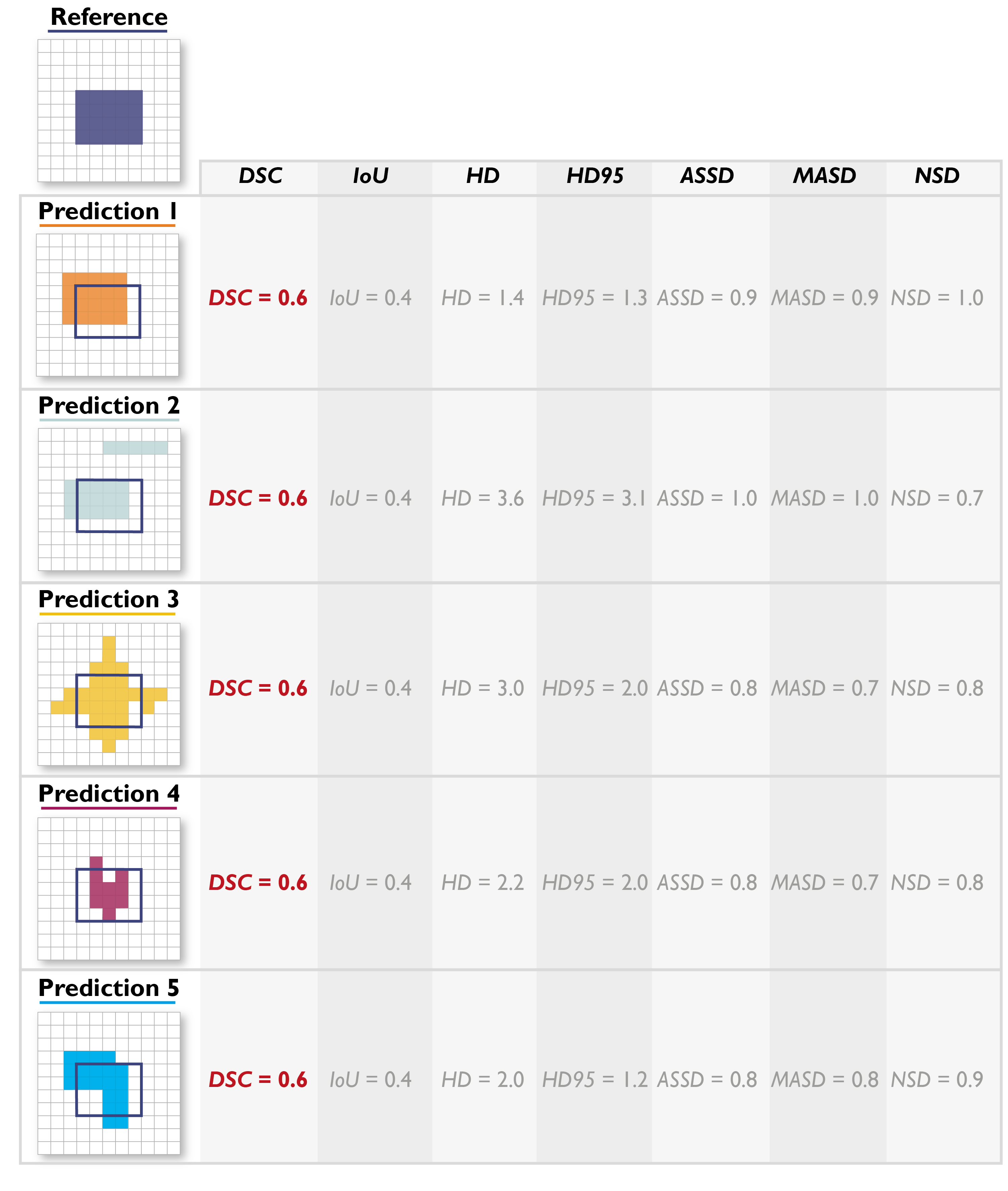}
    \caption{Effect of different shapes. The shapes of the predictions of five algorithms (\textit{Predictions 1-5}) differ substantially, but lead to the exact same \textit{\acf{DSC}} and \textit{\acf{IoU}}, while boundary-based metrics (\textit{\acf{HD}}, \textit{\acf{HD95}}, \textit{\acf{ASSD}} and \textit{\acf{NSD}}) consider the shape differences.}
     \label{fig:DSC-shapes}
\end{tcolorbox}
\end{figure}

\begin{figure}[H]
\begin{tcolorbox}[title=Pitfall: Common metrics disregard complex shapes, colback=white]
    \centering
    \includegraphics[width=1\linewidth]{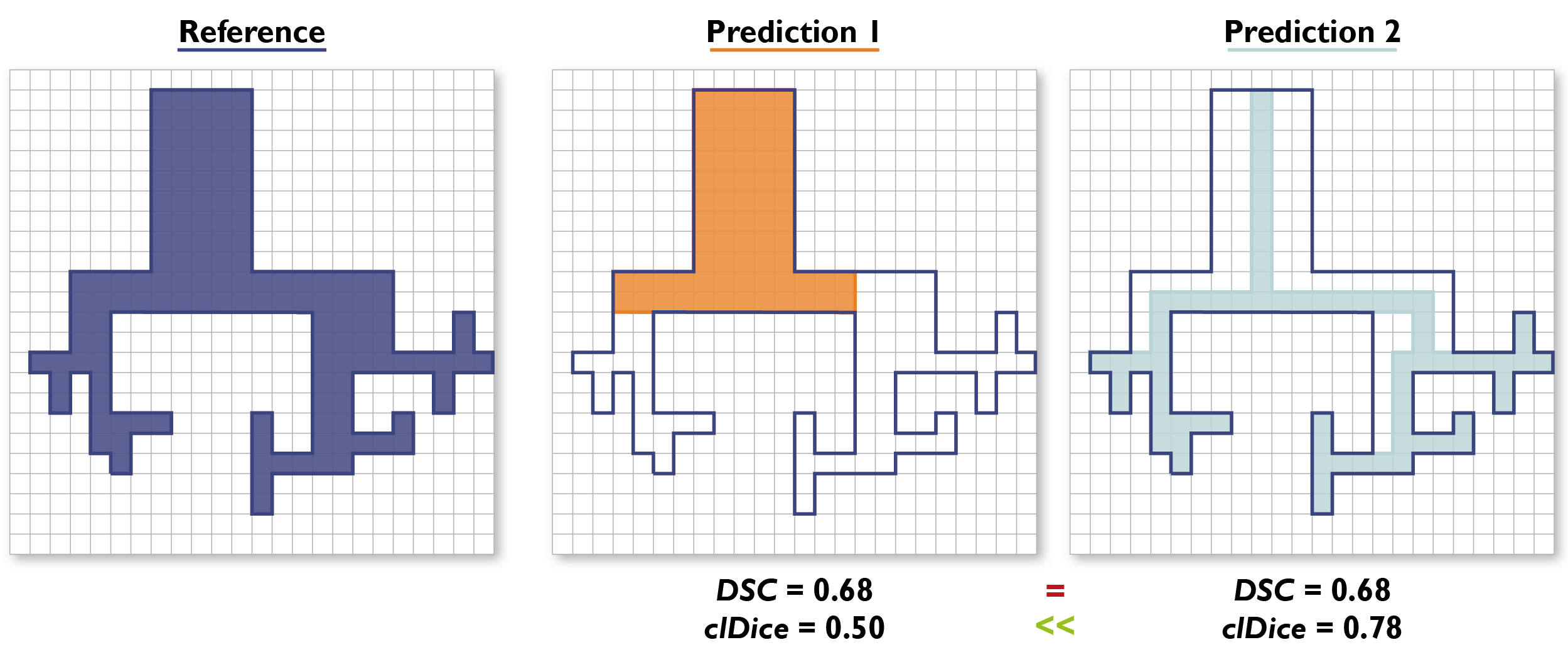}
    \caption{Effect of complex shapes. Common overlap-based metrics such as the \textit{\acf{DSC}} are unaware of complex structure shapes and treats \textit{Predictions 1} and \textit{2} equally. The \textit{\acf{clDice}} uncovers that \textit{Prediction 1} misses the fine-granular branches of the reference and favors \textit{Prediction 2}, which focuses on the object's center line and better captures its fine branches.}
     \label{fig:complex-shapes}
\end{tcolorbox}
\end{figure}

\begin{figure}[H]
\begin{tcolorbox}[title=Pitfall: \textit{clDice} does not capture spatial shift in vessel width, colback=white]
    \centering
    \includegraphics[width=1\linewidth]{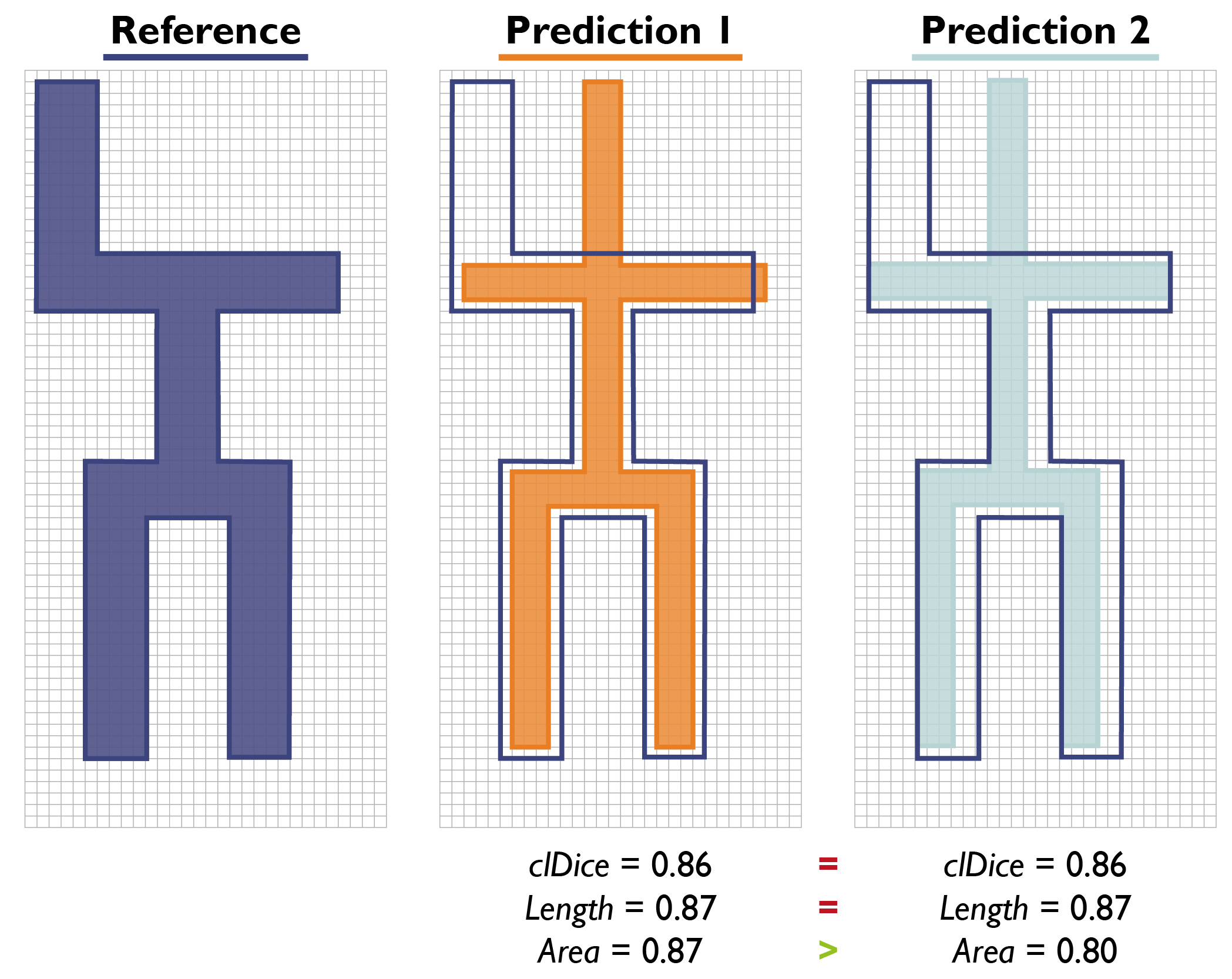}
    \caption{Effect of shifted prediction in vessel-like structures. While \textit{\acf{clDice}} and \textit{Length} properly capture the general center line overlap of \textit{Predictions 1} and \textit{2}, they do not recognize that \textit{Prediction 2} is shifted to the left. This is only revealed by calculating the vessel application-specific metric \textit{Area}.}
     \label{fig:length}
\end{tcolorbox}
\end{figure}

\begin{figure}[H]
\begin{tcolorbox}[title= Pitfall: Ignoring structure location when focusing on volume, colback=white]
    \centering
    \includegraphics[width=1\linewidth]{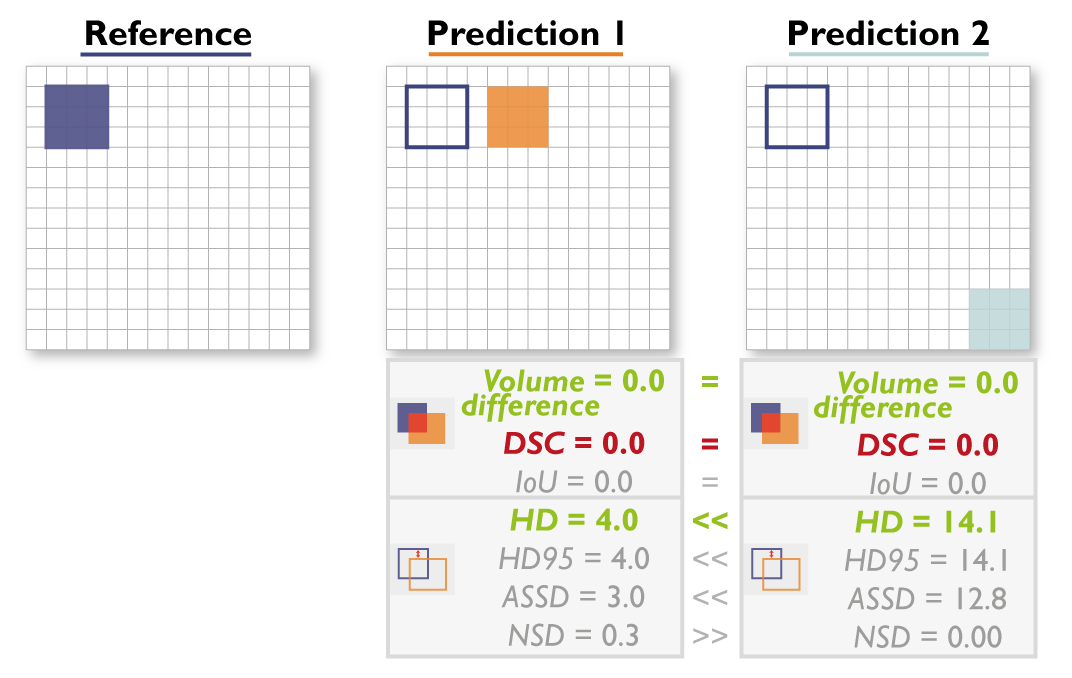}
    \caption{Effect of only focusing on the volume of an object. Both \textit{Predictions 1} and \textit{2} result in the correct volume difference of 0, but do not overlap with the reference (\textit{\acf{DSC}} and \textit{\acf{IoU}} of 0). Only the boundary-based measures (\textit{\acf{HD}}, \textit{\acf{HD95}}, \textit{\acf{ASSD}} and \textit{\acf{NSD}}) recognize the mislocalization.}
     \label{fig:volume}
\end{tcolorbox}
\end{figure}

\begin{figure}[H]
\begin{tcolorbox}[title= Pitfall: Center unawareness of overlap metrics, colback=white]
    \centering
    \includegraphics[width=1\linewidth]{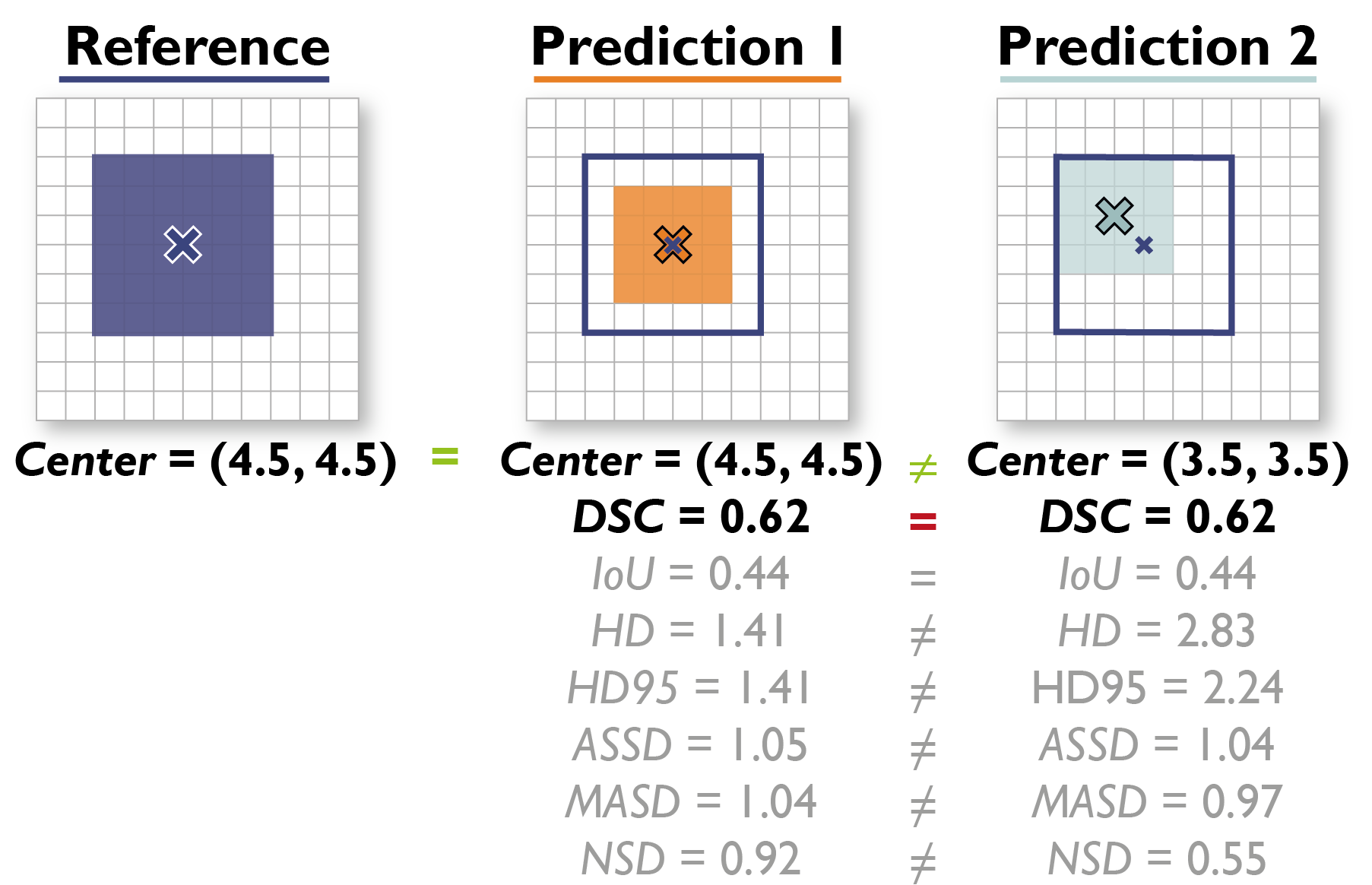}
    \caption{The most common counting-based metrics are poor proxies for the center point alignment. Here, \textit{Predictions 1} and \textit{2} yield the same \textit{\acf{DSC}} value although \textit{Prediction 1} approximates the location of the object much better.}
     \label{fig:center}
\end{tcolorbox}
\end{figure}

\begin{figure}[H]
\begin{tcolorbox}[title= Pitfall: Holes in the segmentation ignored by boundary-based metrics, colback=white]
    \centering
    \includegraphics[width=1\linewidth]{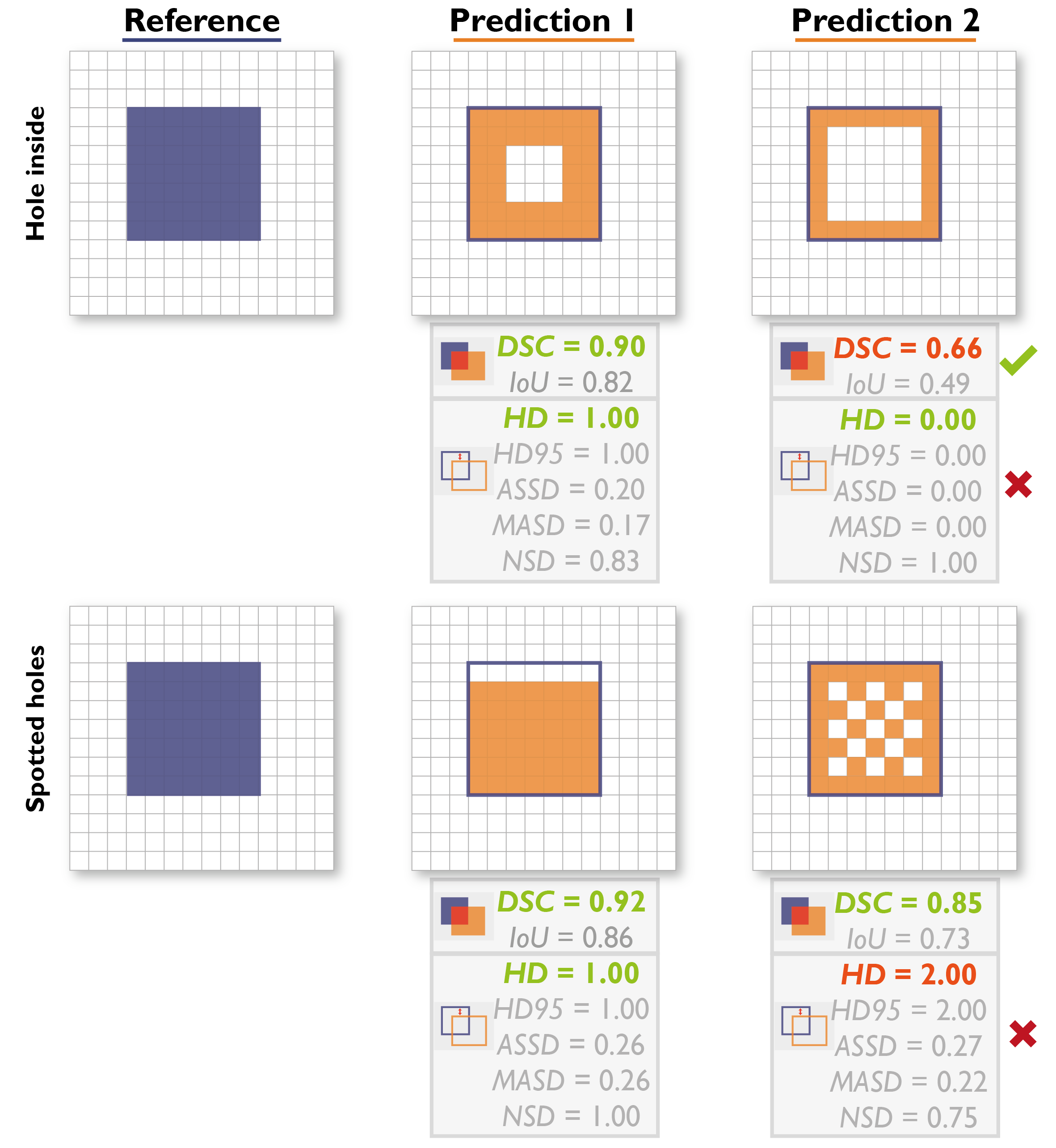}
    \caption{Boundary-based metrics commonly ignore the overlap between structures and are thus insensitive to holes in structures. \textbf{Upper part:} Here, \textit{Predictions 1} and \textit{2} feature holes within the object. The boundary-based metrics (\textit{\acf{HD}}, \textit{\acf{HD95}}, \textit{\acf{ASSD}}, \textit{\acf{NSD}}) do not recognize this problem, yielding very good or even perfect metric scores of 0.00 for the \textit{\ac{HD}(95)/\ac{ASSD}} and 1.00 for the \textit{\ac{NSD}} (\textit{Prediction 2}), whereas the overlap-based metrics (\textit{\acf{DSC}}, \textit{\acf{IoU}}) reflect the fact that the inner area is missed by the predictions. \textbf{Lower part:} Here, \textit{Predictions 1} and \textit{2} feature a spotted pattern within the object. Although the boundary of \textit{Prediction 2} is perfect, the holes are penalized by the boundary-based metrics compared to \textit{Prediction 1}. \textit{Prediction 1} shows an imperfect boundary. Depending on the surface-based metric used, slight deviations in the boundary (here in \textit{Prediction 1}) may be tolerated, reflected by calculating the \textit{\ac{NSD}} for $\tau=1$.}
     \label{fig:outline}
\end{tcolorbox}
\end{figure}

\begin{figure}[H]
\begin{tcolorbox}[title= Pitfall: Comparing the wrong boundaries, colback=white]
    \centering
    \includegraphics[width=1\linewidth]{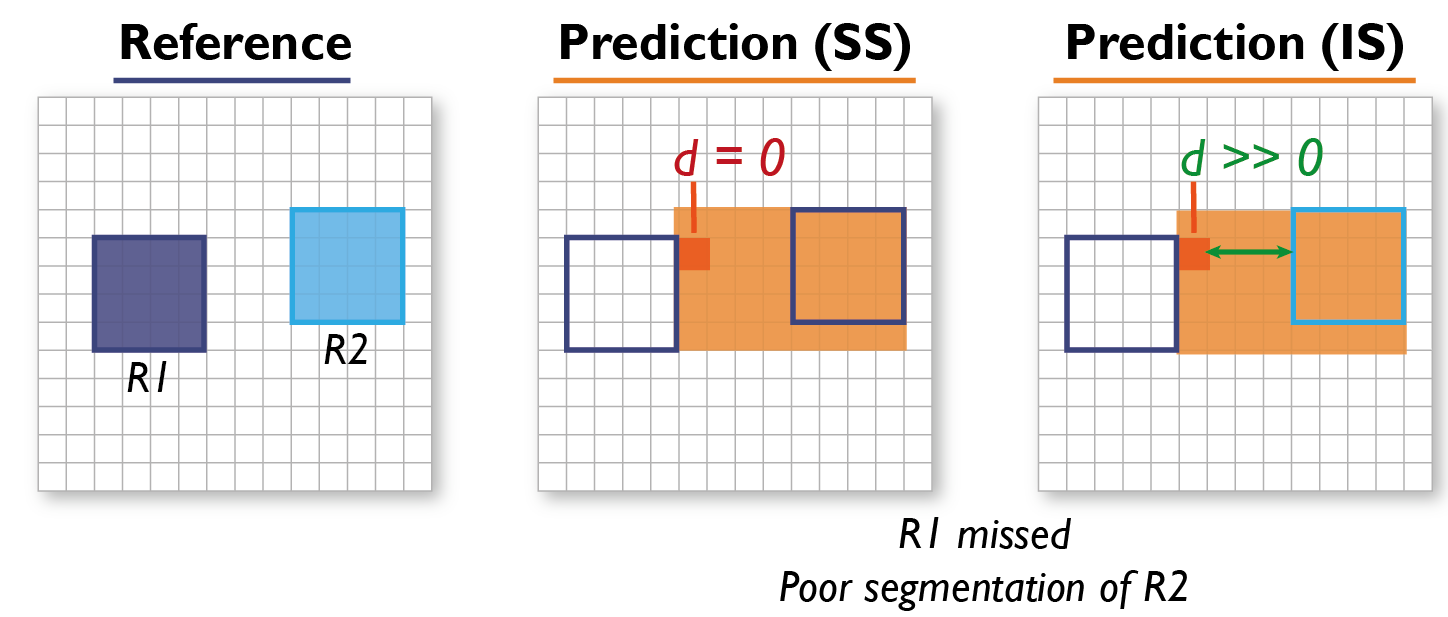}
    \caption{\textbf{Boundary-based metrics in semantic/instance segmentation problems.} If multiple structures of the same type can be seen within the same image (here: reference objects \textit{R1} and \textit{R2}), it is generally advisable to phrase the problem as instance segmentation (IS; right) rather than semantic segmentation (SS; left). This way, issues with boundary-based metrics resulting from comparing a given structure boundary to the boundary of the wrong instance in the reference can be avoided. In the provided example, the distance of the red boundary pixel to the reference, as measured by a boundary-based metric in SS problems, would be zero, because different instances of the same structure cannot be distinguished. This problem is overcome by phrasing the problem as IS. In this case, (only) the boundary of the matched instance (here: R2) is considered for distance computation.}
     \label{fig:boundary-is-ss}
\end{tcolorbox}
\end{figure}

\begin{figure}[H]
\begin{tcolorbox}[title= Pitfall: Multiple labels per pixel, colback=white]
    \centering
    \includegraphics[width=1\linewidth]{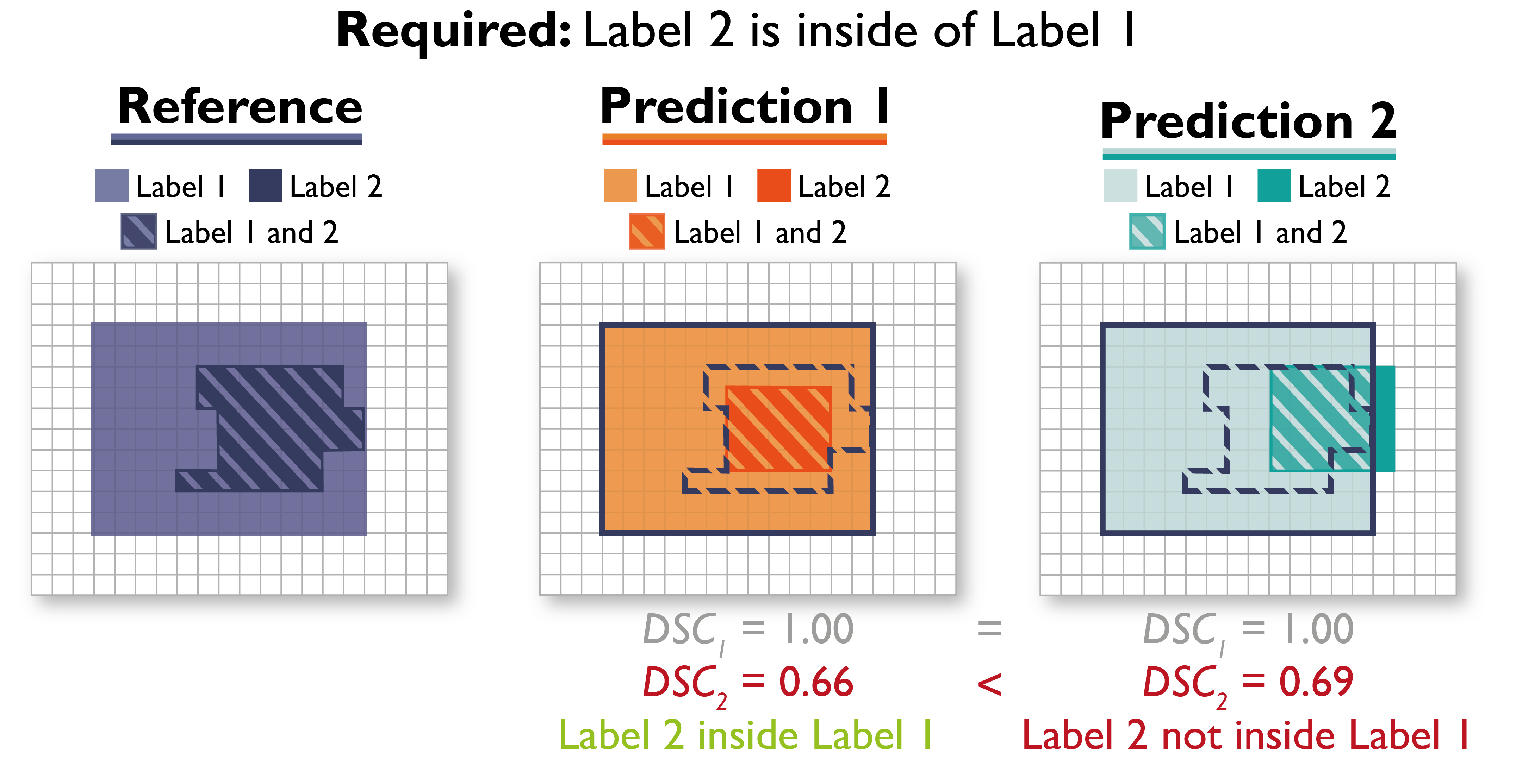}
    \caption{Effect of multiple labels per pixel. The requirement of \textit{Label 2} being inside of \textit{Label 1} is violated by \textit{Prediction 2}. Nevertheless, \textit{Prediction 2} shows a higher \textit{\acf{DSC}} score compared to \textit{Prediction 1}, which adheres to the requirement.}
    \label{fig:multi-labels}
\end{tcolorbox}
\end{figure}

\begin{figure}[H]
\begin{tcolorbox}[title= Pitfall: Inter-rater variability , colback=white]
    \centering
    \includegraphics[width=1\linewidth]{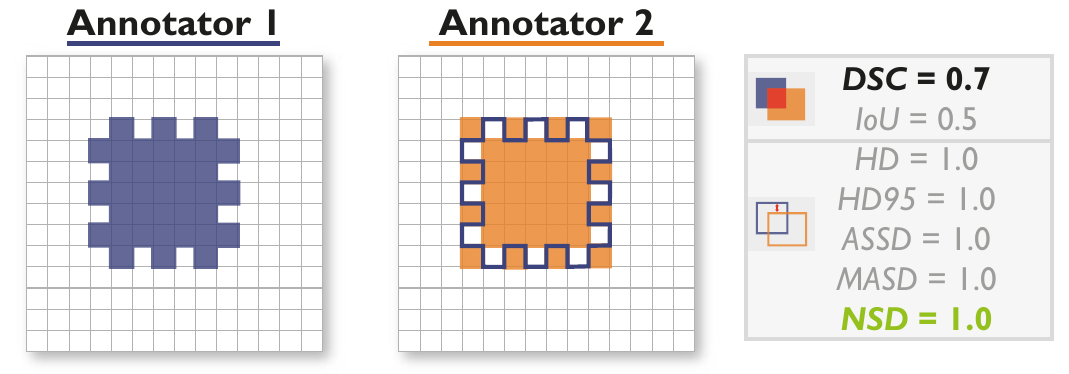}
    \caption{Effect of inter-rater variability between two annotators. Assessing the performance of \textit{Annotator 2} while using \textit{Annotator 1} for creating the reference annotation leads to a low \textit{\acf{DSC}} score because inter-rater variability is not taken into account by common overlap-based metrics. In contrast, the \textit{\acf{NSD}}, applied with a threshold of $\tau = 1$, captures this variability. It should be noted, however, that this effect occurs primarily in small structures as overlap-based metrics tend to be robust to variations in the object boundaries in large structures. Further abbreviations: \textit{\acf{IoU}}, \textit{\acf{HD}}, \textit{\acf{HD95}}, \textit{\acf{ASSD}}.}
    \label{fig:low-quality}
\end{tcolorbox}
\end{figure}

\begin{figure}[H]
\begin{tcolorbox}[title= Pitfall: Noise and artifacts, colback=white]
    \centering
    \includegraphics[width=0.9\linewidth]{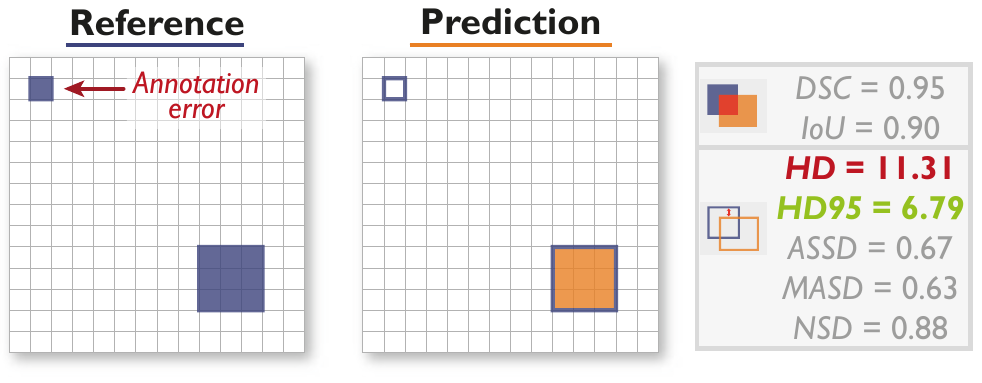}
    \caption{Effect of annotation errors/noise. A single erroneously annotated pixel may lead to a large decrease in performance, especially in the case of the \textit{\acf{HD}} when applied to small structures. The \textit{\acf{HD95}}, on the other hand, was designed to deal with spatial outliers. Further abbreviations: \textit{\acf{DSC}}, \textit{\acf{IoU}}, \textit{\acf{ASSD}}, \textit{\acf{NSD}}.}
    \label{fig:DSC-artifact}
\end{tcolorbox}
\end{figure}

\begin{figure}[H]
\begin{tcolorbox}[title= Pitfall: Reference or prediction without target structure(s), colback=white]
    \centering
    \includegraphics[width=1\linewidth]{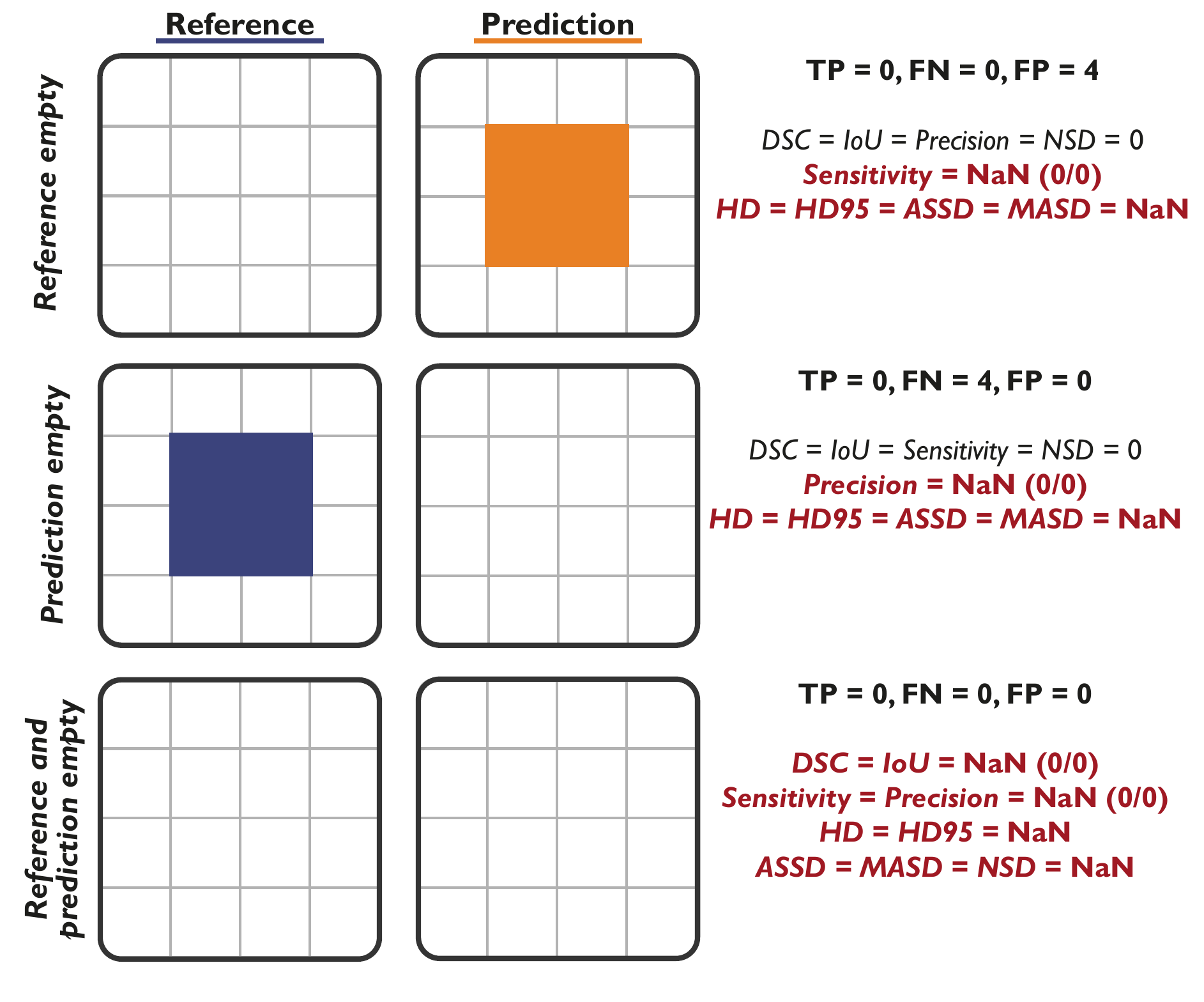}
    \caption{Effect of empty references or predictions when applying common metrics per image (here for semantic segmentation). Empty images lead to division by zero for many common metrics as the numbers of the true (T)/false (F) positives (P)/negatives (N) turn zero.}
    \label{fig:empty}
\end{tcolorbox}
\end{figure}

All pitfalls compiled for this work and relevant for semantic or instance segmentation are summarized in Table~\ref{tab:properties-pitfalls}. This section focuses on limitations for semantic segmentation, but some of them are also transferable to other problem categories, as indicated in the table. Limitations of metrics are typically related to the following properties:
\begin{itemize}
    \item Small size of structures relative to pixel size (Figures~\ref{fig:DSC-small} -~\ref{fig:clDice-small})
    \item High variability of structure sizes (Figures~\ref{fig:high-variability} -~\ref{fig:masd-assd})
    \item Complex shapes of structures (Figures~\ref{fig:DSC-shapes} -~\ref{fig:length})
    \item Particular importance of structure volume (Figure~\ref{fig:volume})
    \item Particular importance of structure center (Figure~\ref{fig:center})
    \item Particular importance of structure boundaries (Figure~\ref{fig:outline} -~\ref{fig:boundary-is-ss})
    \item Possibility of multiple labels per unit (Figure~\ref{fig:multi-labels})
    \item High inter-rater variability (Figure~\ref{fig:low-quality})
    \item Possibility of outliers in reference annotation (Figure~\ref{fig:DSC-artifact})
    \item Possibility of reference or prediction without the target structure (Figure~\ref{fig:empty})
\end{itemize}

\begin{figure}[H]
\begin{tcolorbox}[title= Pitfall: Grid size resolution, colback=white]
    \centering
    \includegraphics[width=1\linewidth]{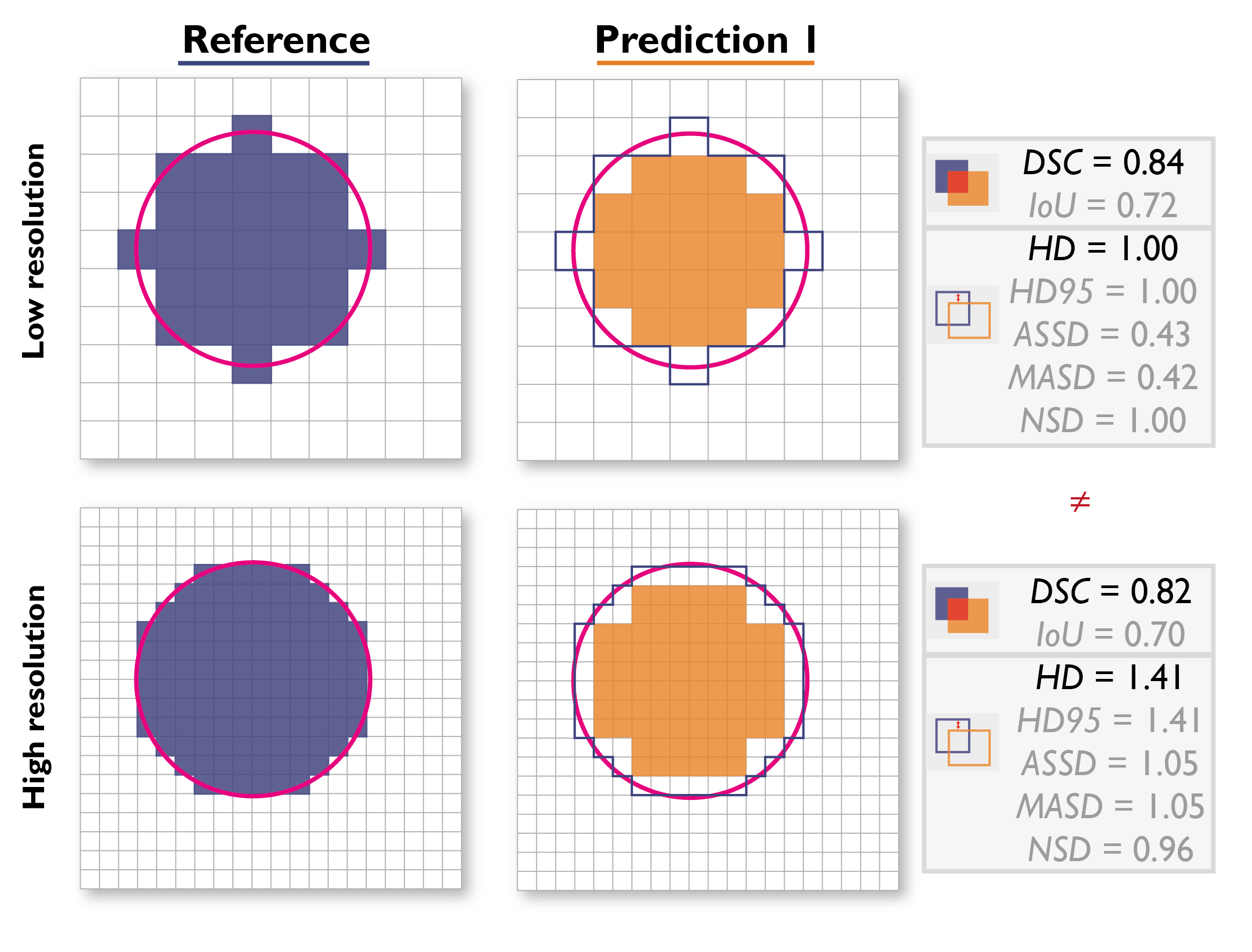}
    \caption{Effect of different grid sizes. Differences in the grid size (resolution) of an image highly influence the image and the reference annotation (dark blue shape (reference) \textit{vs.} pink outline (desired circle shape)). A prediction of the exact same shape (\textit{Prediction 1}) leads to different metric scores due to the different resolution. Abbreviations: \textit{\acf{DSC}}, \textit{\acf{IoU}}, \textit{\acf{HD}}, \textit{\acf{HD95}}, \textit{\acf{ASSD}}, \textit{\acf{NSD}}.}
     \label{fig:DSC-grid-size}
\end{tcolorbox}
\end{figure}

\begin{figure}[H]
\begin{tcolorbox}[title= Pitfall: Oversegmentation \textit{vs.} undersegmentation, colback=white]
    \centering
    \includegraphics[width=1\linewidth]{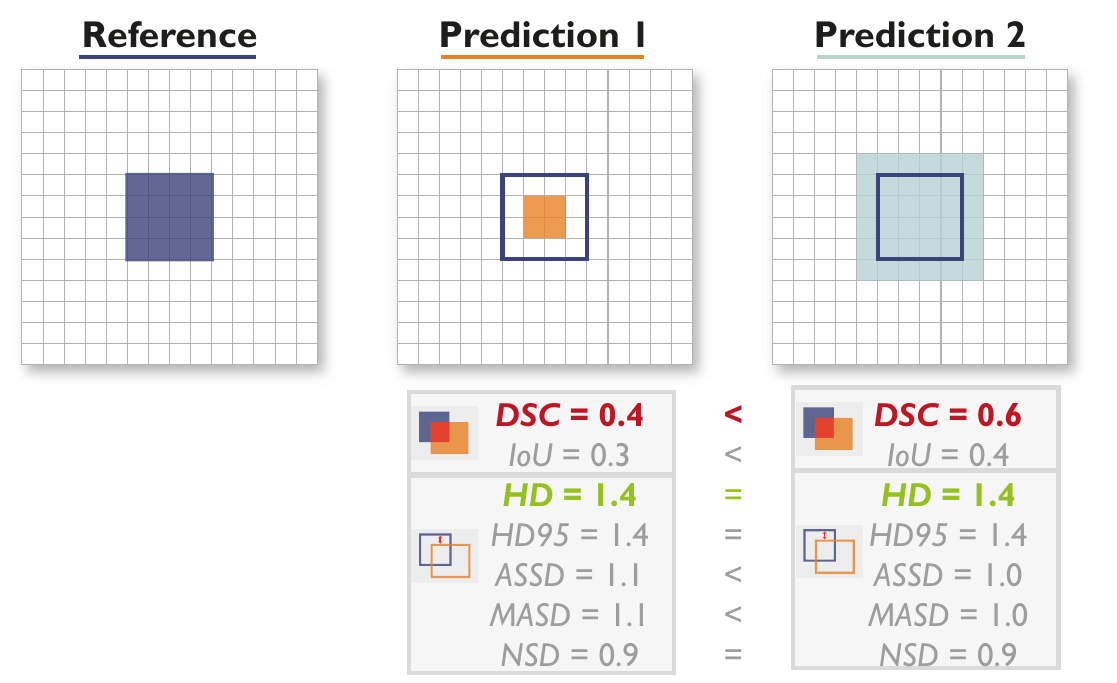}
    \caption{Effect of undersegmentation \textit{vs.} oversegmentation. The outlines of the predictions of two algorithms (\textit{Prediction 1/2}) differ in only a single layer of pixels (\textit{Prediction 1}: undersegmentation, \textit{Prediction 2}: oversegmentation). This has no (or only a minor) effect on the \textit{\acf{HD}/(95\%)}, the \textit{\acf{NSD}} and the \textit{\acf{ASSD}}, but yields a substantially different \textit{\acf{DSC}} or \textit{\acf{IoU}} score.}
     \label{fig:DSC-overunder}
\end{tcolorbox}
\end{figure}

\begin{figure}[H]
\begin{tcolorbox}[title= Pitfall: Choice of global decision threshold, colback=white]
    \centering
    \includegraphics[width=1\linewidth]{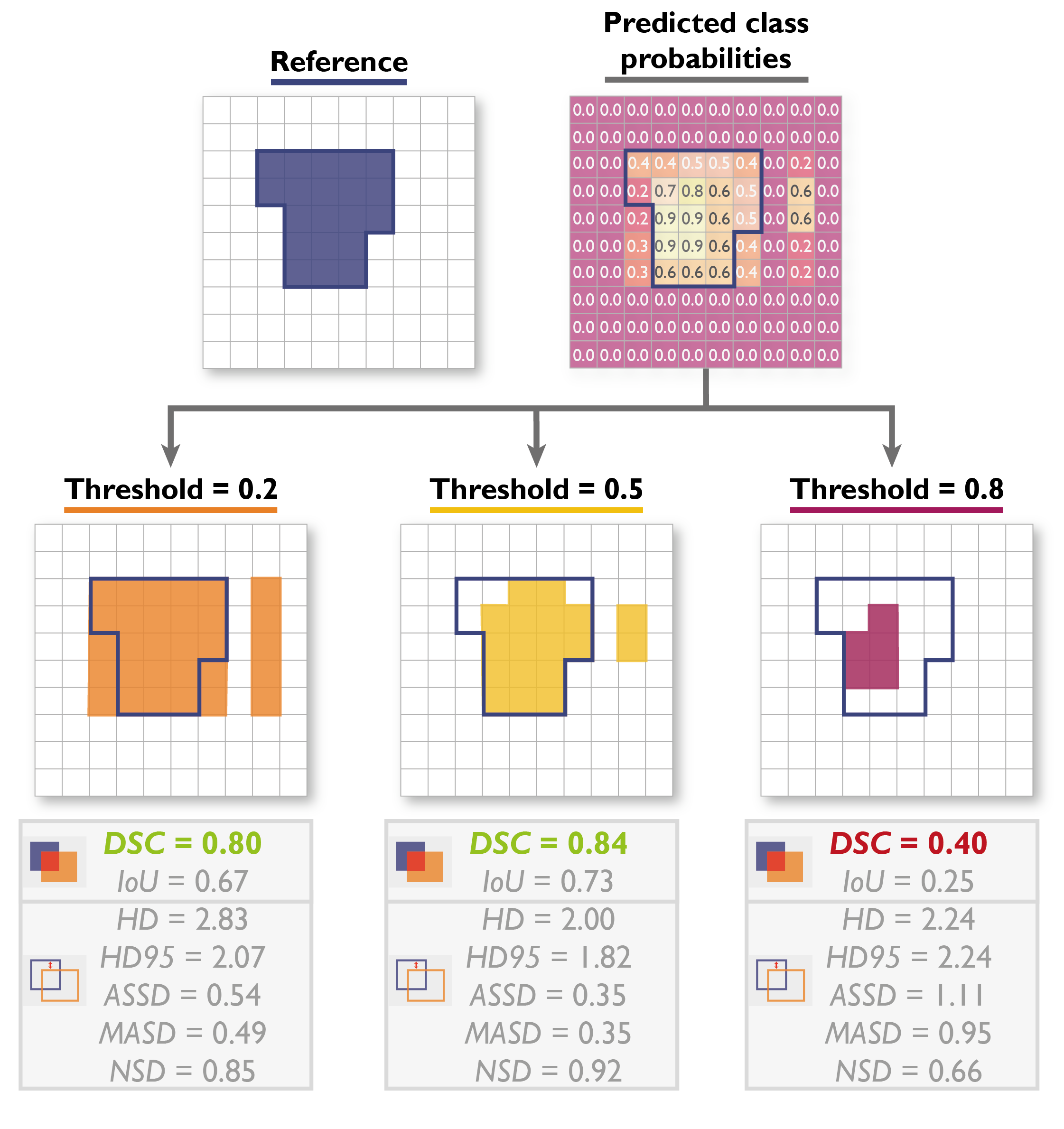}
    \caption{Effect of the choice of a global decision threshold for binarization in segmentation problems. The results and metric values for three decision thresholds (0.2, 0.5 and 0.8), from which predicted class probabilities per pixel are translated into a segmentation mask, are provided. Based on the threshold chosen, the predicted objects and resulting metric values change dramatically. }
     \label{fig:seg-threshold}
\end{tcolorbox}
\end{figure}

Further pitfalls are related to technical peculiarities, such as the image resolution (Figure~\ref{fig:DSC-grid-size}), preference for over- \textit{vs.} undersegmentation (Figure~\ref{fig:DSC-overunder}) and the choice of global decision threshold for creating the confusion matrix (Figure~\ref{fig:seg-threshold}).

The limitations are presented for the most commonly used overlap segmentation metrics, namely \textit{\ac{DSC}}, \textit{\ac{IoU}}, and the most common boundary-based metrics, namely \textit{\ac{HD}}, \textit{\ac{HD95}}, \textit{\ac{ASSD}}, \textit{\ac{MASD}} and \textit{\ac{NSD}}. The \textit{\ac{NSD}} calculation is based on a user-defined threshold (cf. Figure~\ref{fig:definition-nsd}). Results differ for different thresholds. Unless stated otherwise, we set the threshold to $\tau = 1$.\\

To preserve clarity of the illustrations, specific values may only be highlighted for one metric from each metric family, if the other metrics share similar properties (e.g. \textit{\ac{DSC}} and \textit{\ac{IoU}} share the same properties). Green metric values correspond to a "good" value (e.g. a high \textit{\ac{DSC}} or a low \textit{\ac{HD}} score), whereas red values correspond to a "bad" value (e.g. a low \textit{\ac{DSC}} or a high \textit{\ac{HD}} score). Green check marks indicate metric scores reflecting the research question, red crosses show those that do not. Please note that a low \textit{\ac{DSC}} value (or similar) is not automatically a "bad" score. A metric value should always be put into perspective and compared to inter-rater variability. We only use the terms "good" and "bad/poor" for simplicity.

\newpage
\paragraph*{\textbf{Small size of structures relative to pixel size}} Segmentation of small structures, such as brain lesions or cells imaged at low magnification, is essential for many image processing applications. In these cases, the \textit{\ac{DSC}} or \textit{\ac{IoU}} may not be appropriate metrics, as illustrated in Figure~\ref{fig:DSC-small} (cf. \cite{cheng2021boundary}). In fact, a single-pixel difference between two predictions can have a large impact on the metric values. Given that the correct outlines (e.g. of pathologies) are often unknown and taking into account the potentially high inter-observer variability related to generating reference annotations~\citep{joskowicz2019inter}, it is typically not desirable for few pixels to influence the metrics as much. This problem is particularly amplified in cases of large variability of structure sizes (cf. Figure~\ref{fig:high-variability}). The same problem arises for other versions of the \textit{\ac{DSC}} or \textit{\ac{IoU}}. For example, the \textit{\ac{clDice}} is often used in the case of tubular structures. Similarly to the original \textit{\ac{DSC}} metric, a single-pixel difference will have a larger influence on the \textit{\ac{clDice}} for small tube structures compared to larger ones. This pitfall also applies to object detection tasks. It should be noted that once a data set exclusively contains only very tiny structures, one may consider this problem to be an object detection rather than a segmentation problem.

\newpage

\newpage
\paragraph*{\textbf{High variability of structure sizes}} The size of target structures may vary substantially, both within an image and across images. For example, in medical instrument segmentation in laparoscopic video data, an image frame may contain full-sized instruments as well as only the tip of an instrument just entering the scene \cite{ross2021comparative}. In these cases, metrics need to be chosen carefully. As shown in the example above (Figure~\ref{fig:DSC-small}), metrics such as the \textit{\ac{DSC}} or \textit{\ac{IoU}} are typically not well-suited for very small structures. Furthermore, size stratification -- the aggregation of metric values for objects of similar sizes to uncover differences between them -- should be employed. Figure~\ref{fig:high-variability} shows an exemplary data set of four images, containing three large structures and one small structure. When aggregating over all \textit{\ac{DSC}} values, the average \textit{\ac{DSC}} is 0.82. Computing the average for large and small structures separately, however, shows that the performance is much lower for the small structures compared to the large ones, demonstrating the large influence of the low metric values of small objects. This pitfall also applies to object detection tasks and other metrics.

\newpage
The \ac{ASSD} and \ac{MASD} often result in very similar metric scores, as both compute the average over distances from the boundaries. While \ac{ASSD} calculates the general average over all distances, the \ac{MASD} computes the average distance for each of the structures and calculates the mean over those averages. \ac{MASD}, however, substantially rewards situations in which the predicted structure is significantly smaller than the reference object and is located near the surface of the reference, as shown in the example of Figure~\ref{fig:masd-assd}. In this case, the average distance from the prediction to the reference is approximately zero. Thus, only the average distance from the reference to the prediction decides the score. For \ac{MASD}, this distance is halved, which leads to a significantly better score compared to that obtained with \ac{ASSD}. 

\paragraph*{\textbf{Complex shapes of structures}} Metrics measuring the overlap between objects are not designed to uncover differences in shapes. This is an important problem in many applications such as radiotherapy, for which identifying and treating all parts of the tumor is essential to avoid recurrence \cite{burnet2004defining}. Figure~\ref{fig:DSC-shapes} illustrates that completely different object shapes may lead to the exact same \textit{\ac{DSC}} and \textit{\ac{IoU}} values. Boundary-based measures are able to detect the changes in shapes \cite{taha2015metrics}. Note that this pitfall also applies to object detection tasks.

\newpage

\newpage
Shape unawareness is especially harmful in the case of very complex shapes, for example in bronchi. These structures typically feature many small branches, as depicted in a simplified manner in Figure~\ref{fig:complex-shapes}. In this example, \textit{Prediction 1} only shows the root of the structure, while \textit{Prediction 2} focuses on its center line, which would be the preferred option. However, common overlap-based metrics such as the \textit{\ac{DSC}} yield the same value for both predictions. The \textit{\ac{clDice}} on the other hand, designed to capture the center line focus, recognizes the difference and penalizes \textit{Prediction 1} with a substantially lower value.

\newpage
While the \textit{\ac{clDice}} can typically validate tree-like structures such as vessels better than other overlap-based measures, it still faces limitations. Clinicians are sometimes particularly interested in predicting the length of a vessel, whereas its width may be of lesser importance. As shown in Figure~\ref{fig:length}, \textit{\ac{clDice}} captures the length of the vessel-like structure in a similar manner as another proposed metric called \textit{Length} \cite{gegundez2011function}. However, it does not recognize that \textit{Prediction 2} is spatially shifted to the left. This shift could be captured by domain-specific metrics such as the \textit{Area} \cite{gegundez2011function}, introducing a tolerance factor to vessel width.

\newpage 
\paragraph*{\textbf{Particular importance of structure volume}} Depending on the domain focus, a surgeon, radiologist or similar may be especially interested in the volume of a segmented structure. The most commonly used metrics may, however, result in predictions at entirely wrong locations if boundary or overlap are not considered. Figure~\ref{fig:volume} shows two predictions of a 3x3 square structure, both of them being at the wrong position. While the volume difference is correct for both predictions, the overlap is zero. Only boundary-based metrics will indicate the magnitude of mislocalization of the predicted objects.

\newpage
\paragraph*{\textbf{Particular importance of structure center}} The structure center point or center line may be more important than an accurate boundary or overlap of the structure, as for example in nerve segmentation \cite{mlynarski2020anatomically}. In these cases, the accuracy of the center point or line should be examined via an additional metric to make sure the center is correct for the prediction. Figure~\ref{fig:center} shows two predictions yielding the same \textit{\ac{DSC}} values, as they have the same overlap to the reference annotation. However, only \textit{Prediction 1} is centered around the same point as the reference, while \textit{Prediction 2} is shifted slightly towards the upper left corner and thus centered incorrectly. This pitfall also applies to object detection tasks. It should be noted that once the center location is of particular importance to the task, one may consider it an object detection rather than a segmentation problem.

\newpage
\paragraph*{\textbf{Particular importance of structure boundaries}} While boundary-based metrics such as the \textit{\ac{HD}(95)}, \textit{\ac{ASSD}} and others can help to detect shape differences between the reference and the predicted object, they do not focus on the object itself. As shown in Figure~\ref{fig:outline}(top), the boundary-based metrics do not recognize a prediction with a large hole inside as poor (\textit{Prediction 2}). Furthermore, in Figure~\ref{fig:outline}(bottom) \cite{taha2015metrics}, those metrics do not punish the spotted pattern within the object.  It should be noted that this behavior may also be desirable. For example, it may be highly difficult to decide whether a necrotic core (hole) is present in a tumor or not. A boundary-based metric would not punish errors resulting from such annotation uncertainties.

\newpage
Please note that boundary-based metrics are not appropriate under several circumstances. In scenarios in which multiple structures of the same type are present within the same image (e.g., in multiple sclerosis lesion segmentation), for example, a potential pitfall is related to comparing a predicted structure boundary to the boundary of the wrong instance in the reference, as shown in Figure~\ref{fig:boundary-is-ss}. In this example, the prediction misses the reference object \textit{R1}, while poorly segmenting \textit{R2}. In the case of a semantic segmentation, the distance between the red pixel and the object boundary would be zero, which may not be desired. If phrased as an instance segmentation problem, the distance would be higher, and thus penalized.

\newpage
\paragraph*{\textbf{Possibility of multiple labels per unit}} In several biomedical imaging scenarios, multiple labels per pixel may be possible. A prominent example would be the tumor core inside the tumor \citep{menze2014multimodal}. Often, however, prior knowledge related to such scenarios (e.g. a tumor core cannot lie outside the tumor) is not reflected by common metrics, which simply calculate the agreement of the reference and prediction per class. Figure~\ref{fig:multi-labels} shows two predictions for a multi-label example. The \textit{\ac{DSC}} value of \textit{Label 2}, which is required to be inside of \textit{Label 1}, is higher for \textit{Prediction 2} although \textit{Label 2} is also found outside the\textit{ Label 1} area. For simplicity, we only show the results for the \textit{\ac{DSC}} metric. This pitfall also applies to object detection tasks.

\newpage

\paragraph*{\textbf{Imperfect reference standard}} A high quality reference annotation is crucial to determine the performance of a supervised learning algorithm. A prediction can be almost perfect, but low quality reference images will still result in a bad metric score. Especially in the medical domain, the inter-rater variability is often very high as domain knowledge is required and experts themselves often disagree \cite{joskowicz2019inter}. Figure~\ref{fig:low-quality} shows two masks from different annotators approximating the same structure. Although the annotations differ only slightly at the boundary, the \textit{\ac{DSC}} score is 0.7. With such inter-rater variability, a \textit{\ac{DSC}} score of 1 would not be achievable in practice. To address this issue, the  \textit{\ac{NSD}} metric can be applied as an alternative or additional metric, as it is designed to allow a certain tolerance of outline pixels based on the threshold $\tau$. This pitfall can also be translated to object detection and image-level classification tasks.

\newpage
\paragraph*{\textbf{Possibility of outliers in reference annotation}} The presence of spatial outliers, such as noise or reference annotation artifacts, may severely impact performance metric values. Figure~\ref{fig:DSC-artifact} demonstrates how a single erroneous pixel in the reference annotation (or the prediction) leads to a substantial decrease in the measured performance, especially in the case of the \textit{\ac{HD}}. Using the 95\% percentile instead of the maximum (\textit{\ac{HD95}}) to compute the distance substantially improves the metric score as it can handle outliers. Please note that the presented example may also be seen vice versa, with a prediction including single pixel errors. It should further be noted that whether or not outliers should be considered depends on the respective research question.  

\newpage
\paragraph*{\textbf{Possibility of reference/prediction without target structure(s)}} A given data set may contain reference annotations without the target structure(s). For example, the data set may consist of healthy and sick patients. A healthy patient will not have a tumor in the image, yielding an empty reference if the tumor is the targeted structure. An algorithm should be careful not to classify a healthy patient as tumourous as this may lead to unnecessary medical interventions. Similarly, a patient with a tumor should not be classified as healthy (empty prediction). These cases require special care to be taken in the validation, because some metrics may be undefined due to division by zero errors or similar. It is necessary to either choose appropriate metrics that consider empty references (or predictions) or account for it in the metric implementation. For example, boundary-based metrics such as the \textit{\ac{HD}(95)} and \textit{\ac{ASSD}} will be \texttt{\ac{NaN}} if one of the structures is empty. Figure~\ref{fig:empty} shows three examples, for which several counting- and boundary-based metrics were computed. The top row depicts the case of an empty reference and a prediction of an object. Given the number of \ac{TP} and \ac{FN} being 0, this will result in a division by zero in the \textit{Sensitivity} calculation, yielding a \texttt{\ac{NaN}} score. A similar case is given in the second row, showing an empty prediction for a given target structure in the reference annotation, yielding an undefined \textit{Precision}. When both reference and prediction are empty (bottom row), all scores will be undefined. 
Please note that this example is shown for a validation per image, as done for segmentation tasks. For classification and object detection tasks, the validation is typically performed over the whole data set, which would possibly preclude this problem. The presented pitfall also applies to object detection tasks.

\newpage
\paragraph*{\textbf{Technical peculiarities}} Several technical peculiarities also have an impact on metric behavior. For example, the image resolution and pixel sizes highly influence the reference annotation and the predicted shapes in image processing tasks. Figure~\ref{fig:DSC-grid-size} illustrates how the reference annotation differs between a low resolution image (top) and a high resolution image (bottom) compared to a circle. The latter is more exact. A prediction of the same size will therefore lead to different corresponding metric values, independent of the type of the metric. This pitfall also applies to object detection tasks.

\newpage
In some applications such as  radiotherapy, it may be highly relevant whether an algorithm tends to over- or undersegment the target structure. The \textit{\ac{DSC}} metric, however, does not represent over- and undersegmentation equally \citep{yeghiazaryan2018family}. As depicted in Figure~\ref{fig:DSC-overunder}, a difference of a single layer of pixels in the outline yields different \textit{\ac{DSC}} scores (oversegmentation preferred) \cite{taha2015metrics}. Other boundary-based performance values such as the \textit{\ac{HD}} are invariant to these properties. 

Another technical peculiarity is the choice of global decision threshold. Most methods in modern image analysis output continuous class scores. While it is quite common to provide those scores in image-level classification and object detection tasks, segmentation architectures often do not output class probabilities per pixel. However, fuzzy segmentation masks are getting more and more common (for instance, see \cite{nida2019melanoma, alzu2020parallel, kaftan2008fuzzy}) and the choice of a global decision threshold $\tau$ is very important for the algorithm's result. Figure~\ref{fig:seg-threshold} (cf. \cite{nair2018thesis}) shows the predicted class probabilities for a reference annotation. For a binarization typically required for segmentation outputs, a threshold needs to be defined based on which a pixel is assigned to a class (here: a pixel with class probability < $\tau$ corresponds to the background class, otherwise to the foreground class). The resulting segmentation masks are shown for the thresholds 0.2, 0.5 and 0.8. It can be seen that the respective masks completely differ across the thresholds. Consequently, metric values will also vastly change.

\newpage

\newpage
\section{Pitfalls related to  object detection} 
\label{sec:detection}
\begin{figure}[H]
\begin{tcolorbox}[title= Pitfall: Center-based \textit{vs.} Point inside Mask localization criteria, colback=white]
    \centering
    \includegraphics[width=0.9\linewidth]{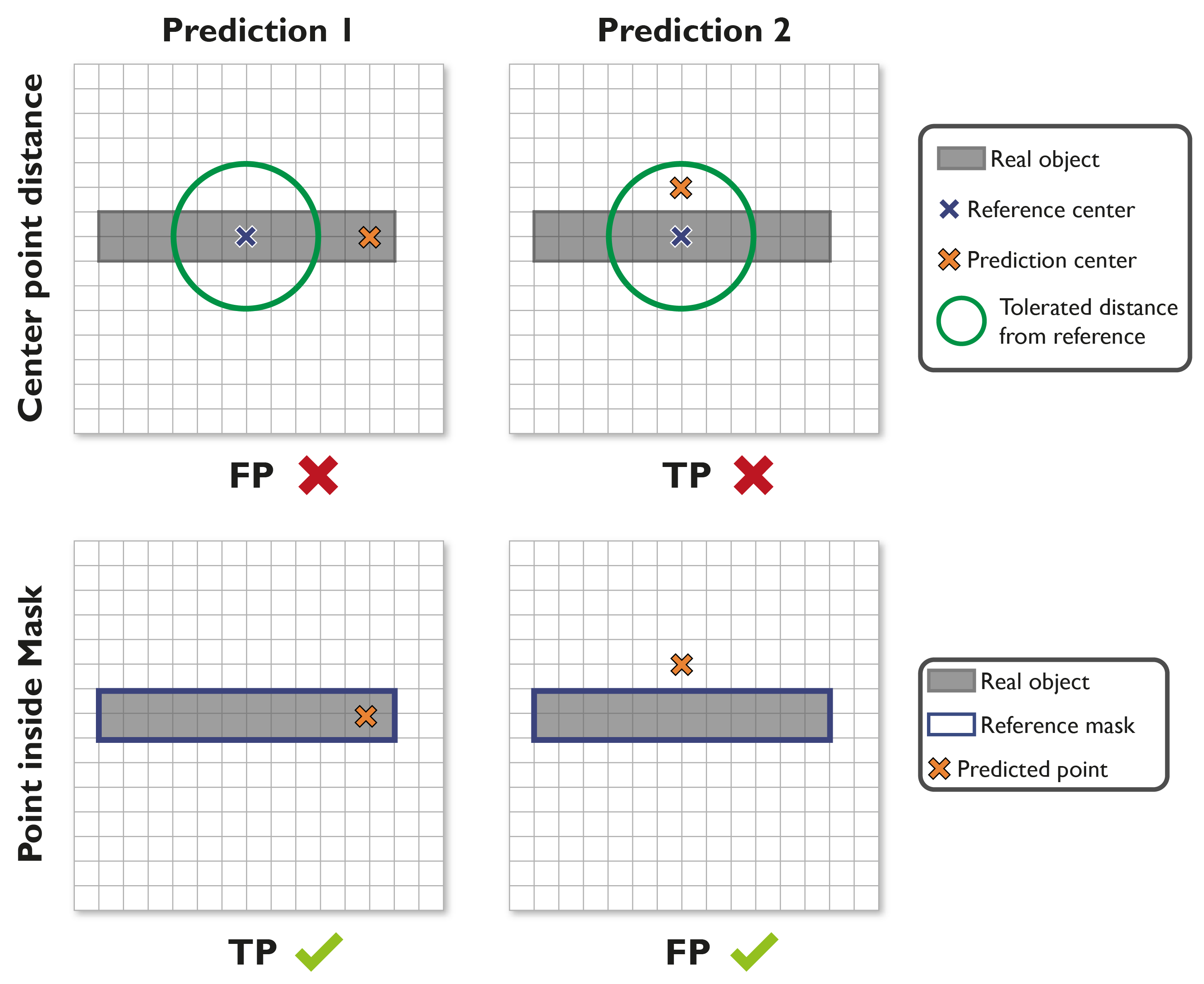}
    \caption{Effect of using center-based localization criteria for complex shapes (here: tubular objects). The center point distance is not an ideal criterion, identifying the \textit{Prediction 1} as a \acf{FP}, although it hits the elongated structure and identifying \textit{Prediction 2} as \acf{TP} although it is outside of the actual object. This is better handled by a binary Point inside Mask localization criterion, reversing \textit{Predictions 1} and \textit{2} into \ac{TP} and \ac{FP}, respectively.}
    \label{fig:center-vs-point}
\end{tcolorbox}
\end{figure}

\begin{figure}[H]
\begin{tcolorbox}[title= Pitfall: Mathematical implications of \textit{\ac{IoU}}-based localization criteria, colback=white]
    \centering
    \includegraphics[width=0.7\linewidth]{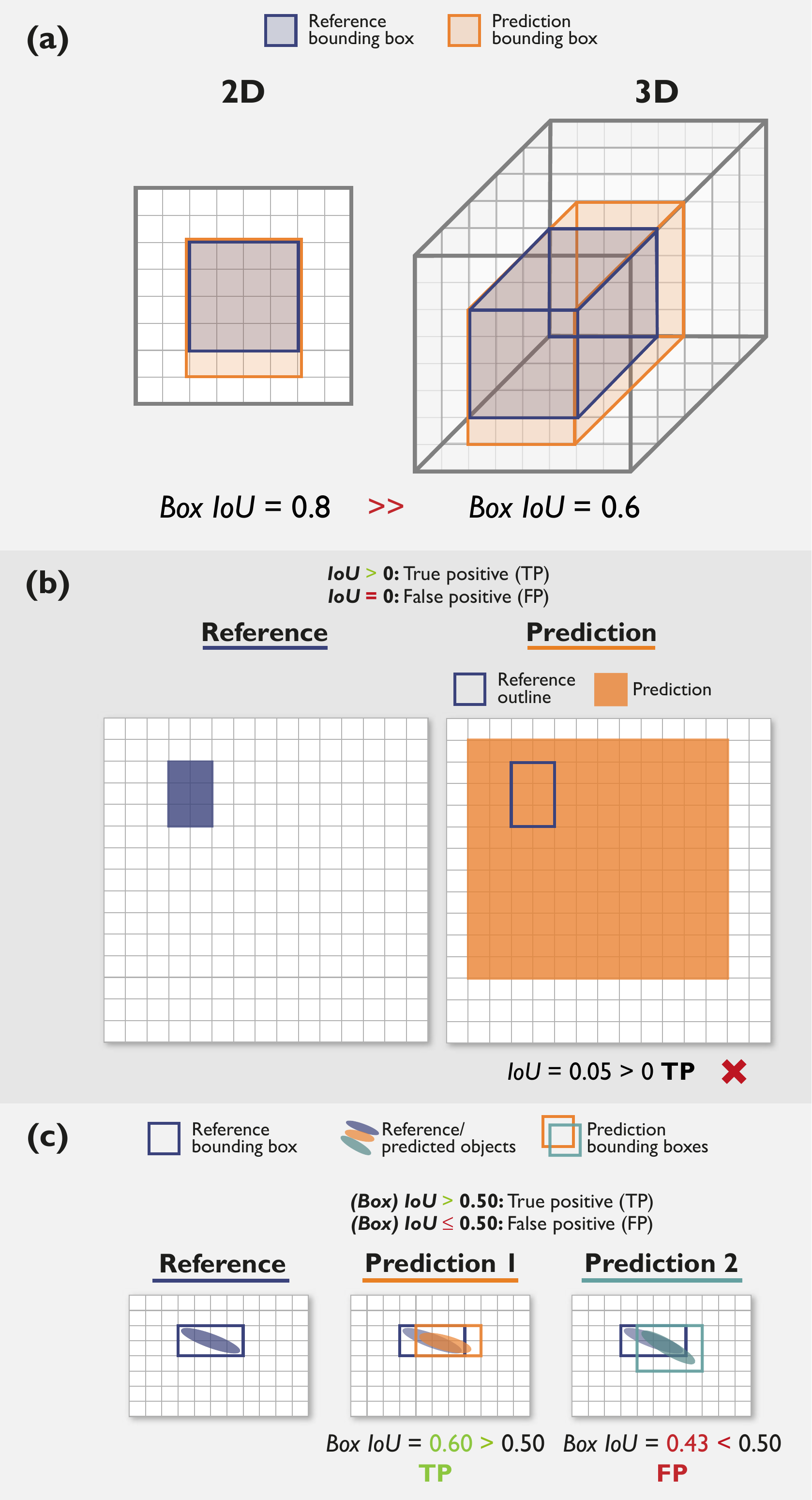}
    \caption{The \textit{\acf{IoU}} is a commonly used localization criterion in object detection. It comes with several limitations: 
    \textbf{(a)} The image dimension should be considered when setting the \textit{\ac{IoU}} (here: \textit{Box \ac{IoU}}) threshold for object detection. In 3D settings, the additional z-dimension results in a cubical increase in erroneous pixels. \textbf{(b)} Effect of a loose \textit{\ac{IoU}} criterion for object detection. When defining a \acf{TP} by an \textit{\ac{IoU}} > 0, the resulting localizations may be fooled by very large predictions.
    \textbf{(c)} Effect of defining \ac{TP} based on the \textit{\ac{IoU}} (here: \textit{Box \ac{IoU}}) threshold of the reference and predicted bounding boxes. Especially for diagonal, narrow objects, the number of bounding box pixels may change quadratically. Although \textit{Predictions 1} and \textit{2} are very similar, their bounding boxes diverge and lead to one of them being defined as \ac{TP}, the other as \acf{FP}.}
     \label{fig:bb-2d-3d}
\end{tcolorbox}
\end{figure}

\begin{figure}[H]
\begin{tcolorbox}[title= Pitfall: Mathematical implications of assignment strategies, colback=white]
    \centering
    \includegraphics[width=0.7\linewidth]{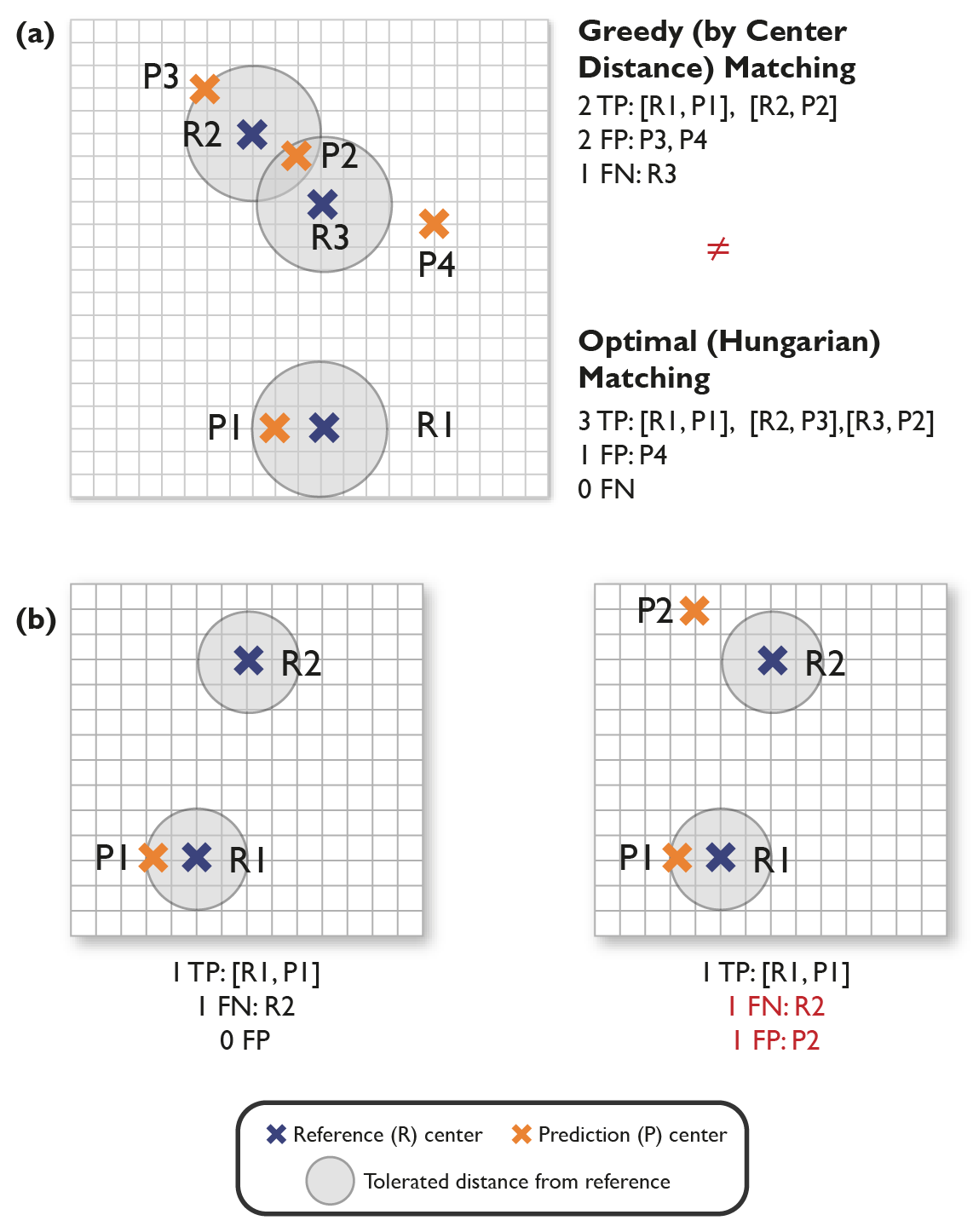}
    \caption{Effect of the choice of assignment strategies. \textbf{(a)} Conducting the assignment of predictions (P) and reference (R) objects sequentially for every reference object center (Greedy by Center Distance Matching), results in two \acf{TP} instances. Furthermore, P3 and P4 are considered \acf{FP} and R3 is a \acf{FN}. On the other hand, an optimal Hungarian matching yields three \ac{TP} instances, only one \ac{FP} and no \ac{FN} instances. \textbf{(b)} Both examples are identical except for the right example showing another prediction P2 for the reference center R2, which is completely missed in the left example. However, P2 is outside of the tolerated area and thus not considered a hit, although it might provide a better approximation than the left example.}
    \label{fig:od-assignment}
\end{tcolorbox}
\end{figure}

\begin{figure}[H]
\begin{tcolorbox}[title= Pitfall: One prediction assigned to multiple reference objects, colback=white]
    \centering
    \includegraphics[width=0.7\linewidth]{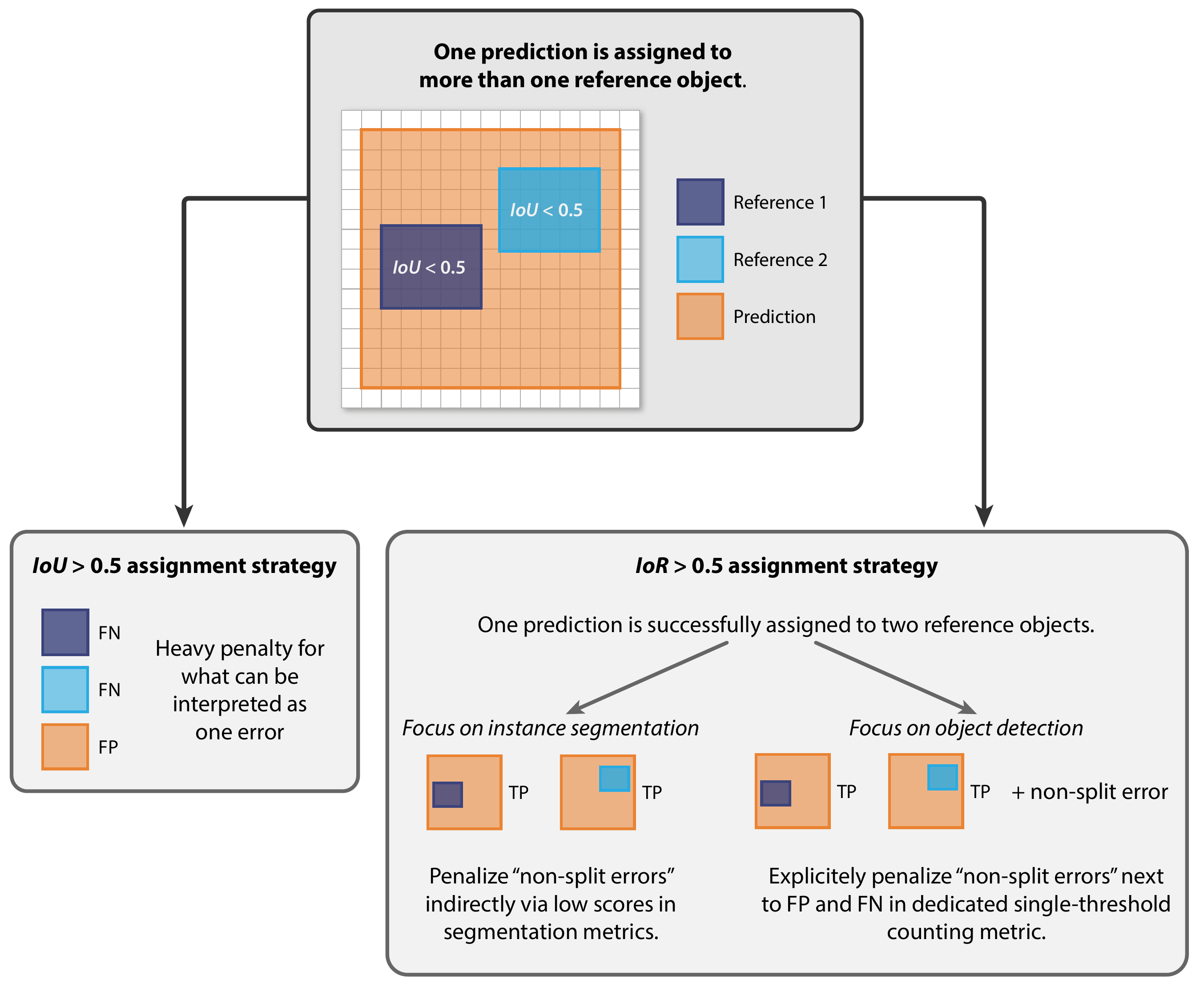}
    \caption{In case of a prediction assigned to multiple reference objects, an assignment strategy needs to be chosen. This may be based, for example, on the \textit{\acf{IoU}}>0.5 strategy, which may end up with a heavy penalty (two \acf{FN} and one \acf{FP}. Another option is to use the \textit{\acf{IoR}}>0.5 strategy, which examines whether the prediction was successfully assigned to the reference objects. In an additional step, the "non-split errors" will be penalized. Used abbreviations: \acf{FN}, \acf{FP} and \acf{TP}.}
    \label{fig:od-ior}
\end{tcolorbox}
\end{figure}

\begin{figure}[H]
\begin{tcolorbox}[title= Pitfall: Effect of annotation type, colback=white]
    \centering
    \includegraphics[width=1\linewidth]{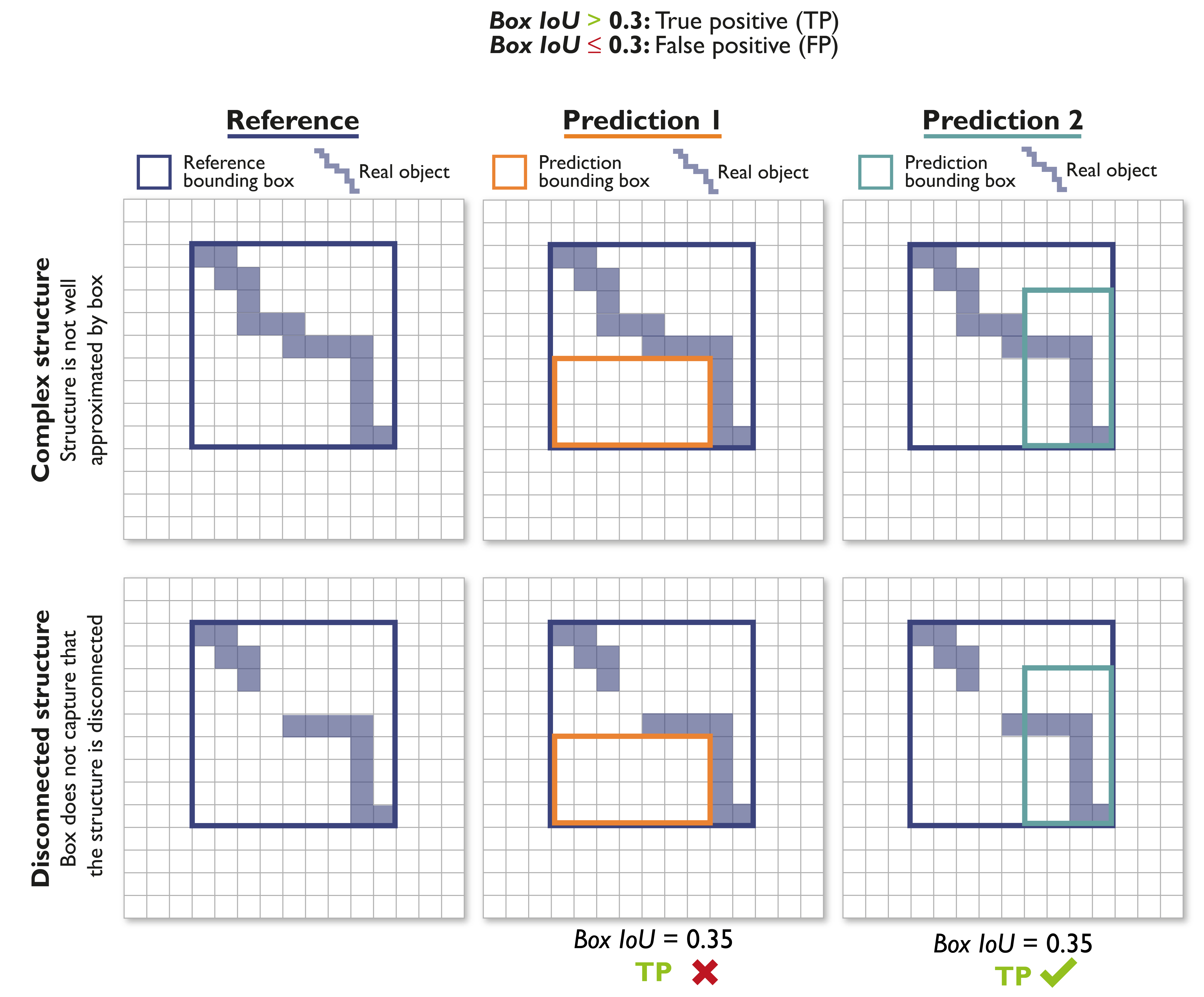}
    \caption{Bounding boxes are not well-suited for representing complex (top) and disconnected (bottom) shapes. Specifically, they are not well-suited for capturing multi-component structures. \textit{Predictions 1} and \textit{2} would both end up in a \acf{TP} detection, as the \textit{Box \acf{IoU}} is larger than the threshold 0.3. However, \textit{Prediction 1} is not hitting the real objects at all, as the given annotation does not represent them well.}
    \label{fig:disconnected}
\end{tcolorbox}
\end{figure}

\begin{figure}[H]
\begin{tcolorbox}[title= Pitfall: Effect of small structures on localization criterion, colback=white]
    \centering
    \includegraphics[width=0.9\linewidth]{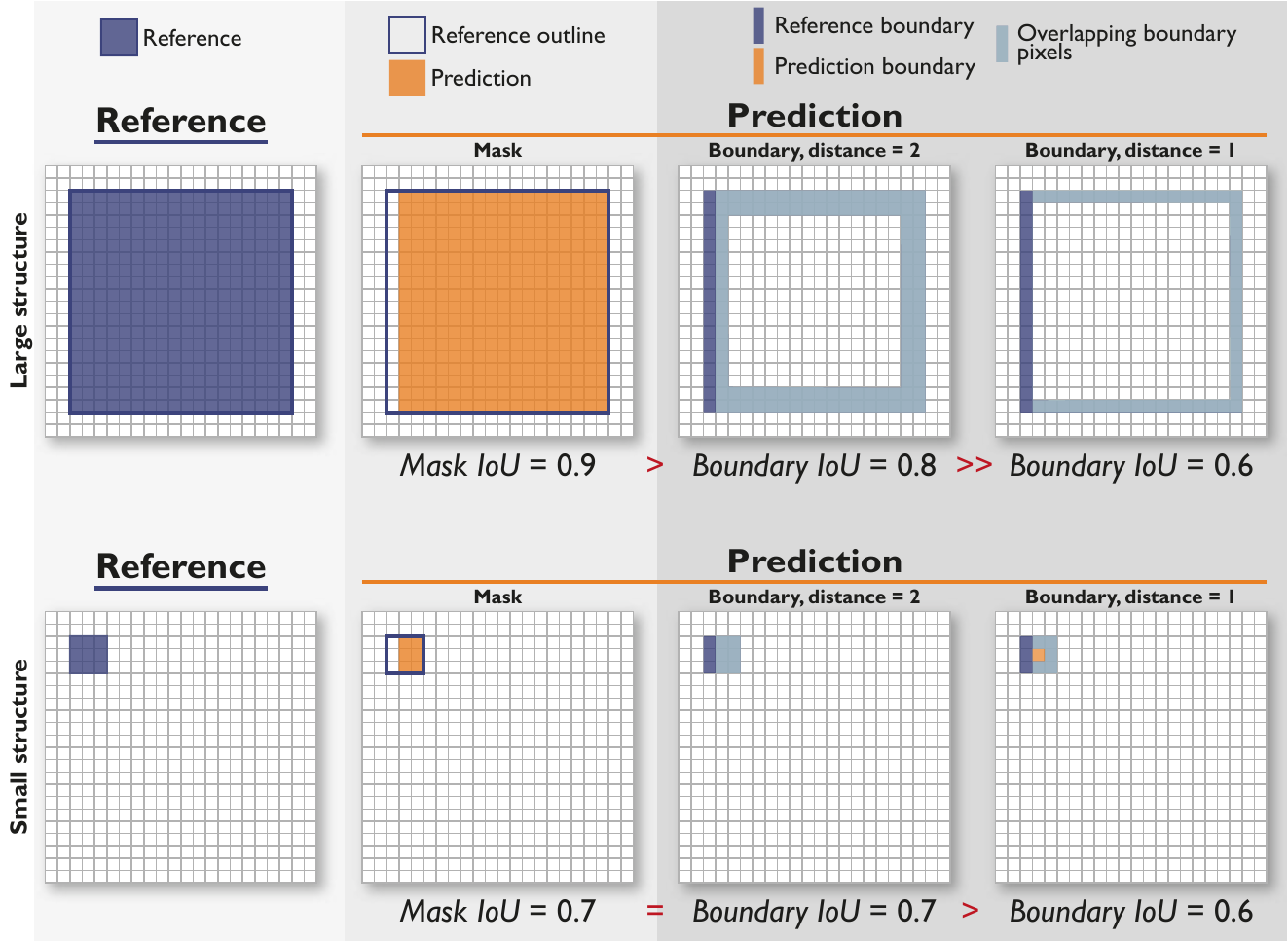}
    \caption{Comparison of \textit{Mask} and \textit{Boundary Intersection over Union (Boundary IoU)} localization criteria in the case of particular importance of structure boundaries. Overlapping pixels from the reference and prediction are shown in light blue. The \textit{Mask IoU} (second column) is less sensitive to boundary errors for large objects. The \textit{Boundary IoU} (third and fourth column) especially considers contours, (1) yields smaller metric scores, thus penalizing errors in the boundaries and (2) is more invariant to structure sizes, as it leads to very similar values for large and small structures (fourth column). }
     \label{fig:boundary-mask-iou} 
\end{tcolorbox}
\end{figure}

\begin{figure}[H]
\begin{tcolorbox}[title= Pitfall: Perfect \textit{Boundary IoU} for imperfect prediction, colback=white]
    \centering
    \includegraphics[width=\linewidth]{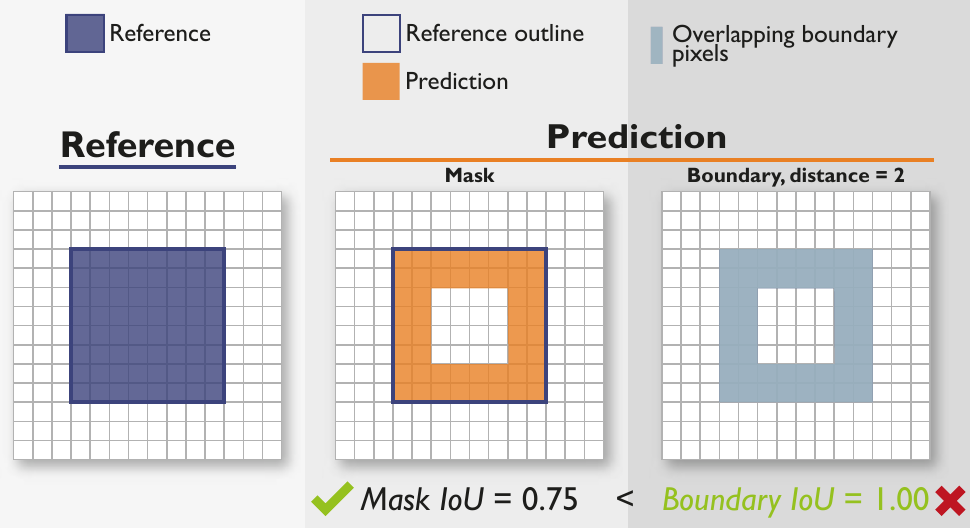}
    \caption{Effect of a perfect \textit{Boundary Intersection over Union (Boundary IoU)} score for an imperfect prediction. Overlapping pixels from the reference and prediction are shown in light blue. For a prediction with a hole in the middle, the \textit{Boundary \ac{IoU}} may result in a score of 1.00 if the distance to border contains all mask pixels (here: distance = 2). However, the \textit{Mask \ac{IoU}} spots the problem and results in a lower score.}
     \label{fig:boundary-iou} 
\end{tcolorbox}
\end{figure}

\begin{figure}[H]
\begin{tcolorbox}[title= Pitfall: \texttt{\acf{NaN}} handling for empty reference/prediction, colback=white]
    \centering
    \includegraphics[width=0.8\linewidth]{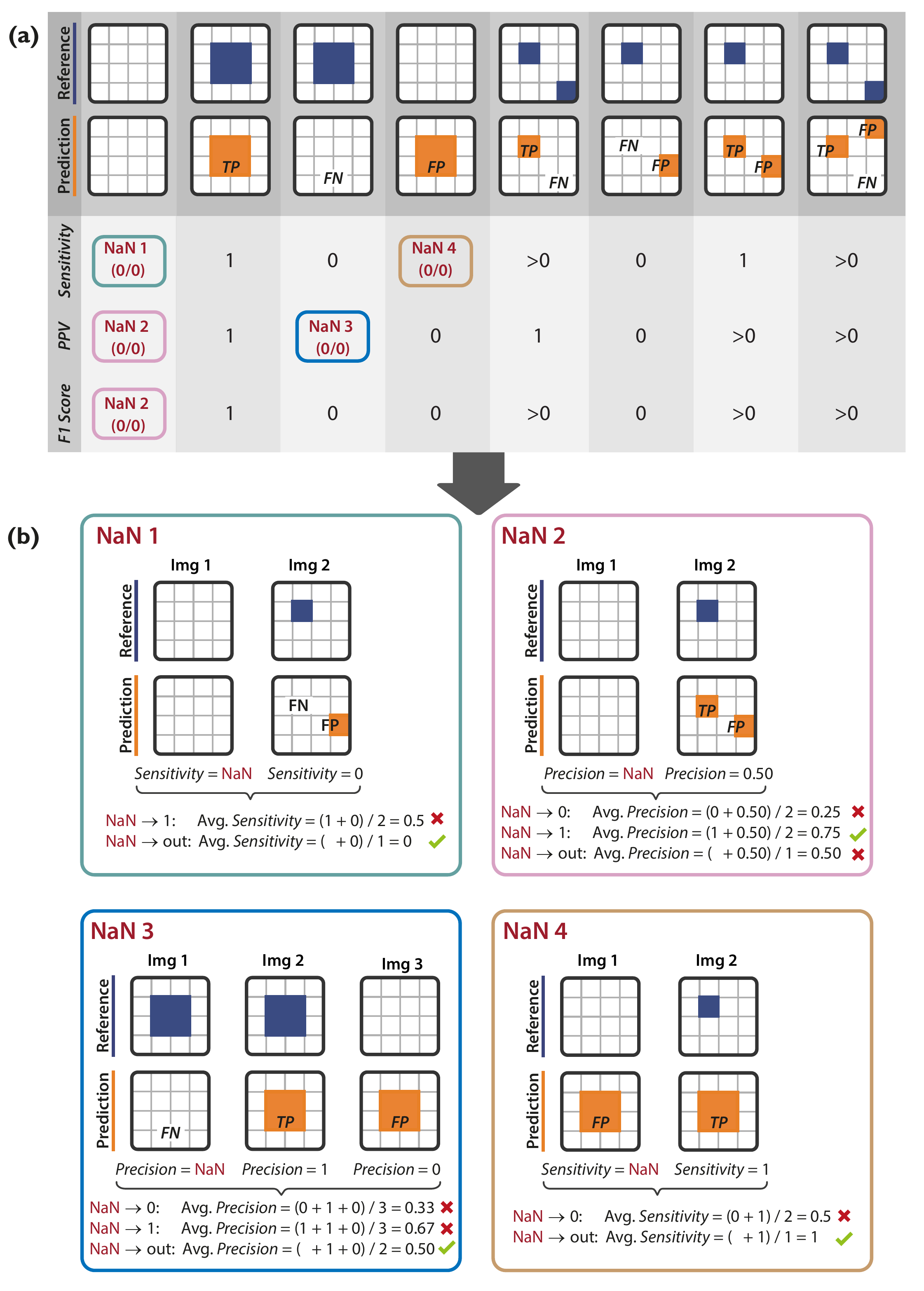}
    \caption{Effect of handling a \texttt{\acf{NaN}} caused by reference or prediction without target structure(s) in object detection/instance segmentation problems validated per image. \textbf{(a)} Demonstration of how and when \texttt{\ac{NaN}} can occur. Each column represents a potential scenario for per-image validation of objects, categorized by whether \acf{TP}, \acf{FN}, and \acf{FP} are present (n > 0) or not present (n = 0) after matching/assignment. The sketches on the top showcase each scenario when setting "n > 0" to "n = 1". For each scenario, \textit{Sensitivity}, \textit{\acf{PPV}} and \textit{F$_\text{1}$ Score} are calculated. Some scenarios yield undefined values (\acf{NaN}). \textbf{(b)} Effect of different \texttt{\ac{NaN}} handling strategies based on different conventions for the aggregation across multiple images. Four examples are shown for the \texttt{\ac{NaN}} scenarios from (a) (\textbf{\texttt{\ac{NaN}} 1-4}). \textbf{\texttt{\ac{NaN}} 1} and \textbf{4}: The intuitive penalization for FPs in "empty" images is already established by means of \textit{\ac{PPV}} scores (see \textbf{\texttt{\ac{NaN}} 4}) and further penalization by means of \textit{Sensitivity} is neither required nor appropriate. Instead, images without reference objects should be ignored when averaging \textit{Sensitivity} scores over images. \textbf{\texttt{\ac{NaN}} 2}: The intuitive penalization for \ac{FP} in "empty" images is established when assigning a \textit{\ac{PPV}} (and \textit{F$_\text{1}$} Score) of 1. \textbf{\texttt{\ac{NaN}} 3}: The intuitive penalization for \ac{FP} is established when removing images with \ac{FN} and no \ac{FP} from the aggregation of \textit{\ac{PPV}} (and \textit{F$_\text{1}$}) scores.}
     \label{fig:empty-ref-pred-od} 
\end{tcolorbox}
\end{figure}

\begin{figure}[H]
\begin{tcolorbox}[title= Pitfall: \acf{AP} disregards total number of images, colback=white]
    \centering
    \includegraphics[width=1\linewidth]{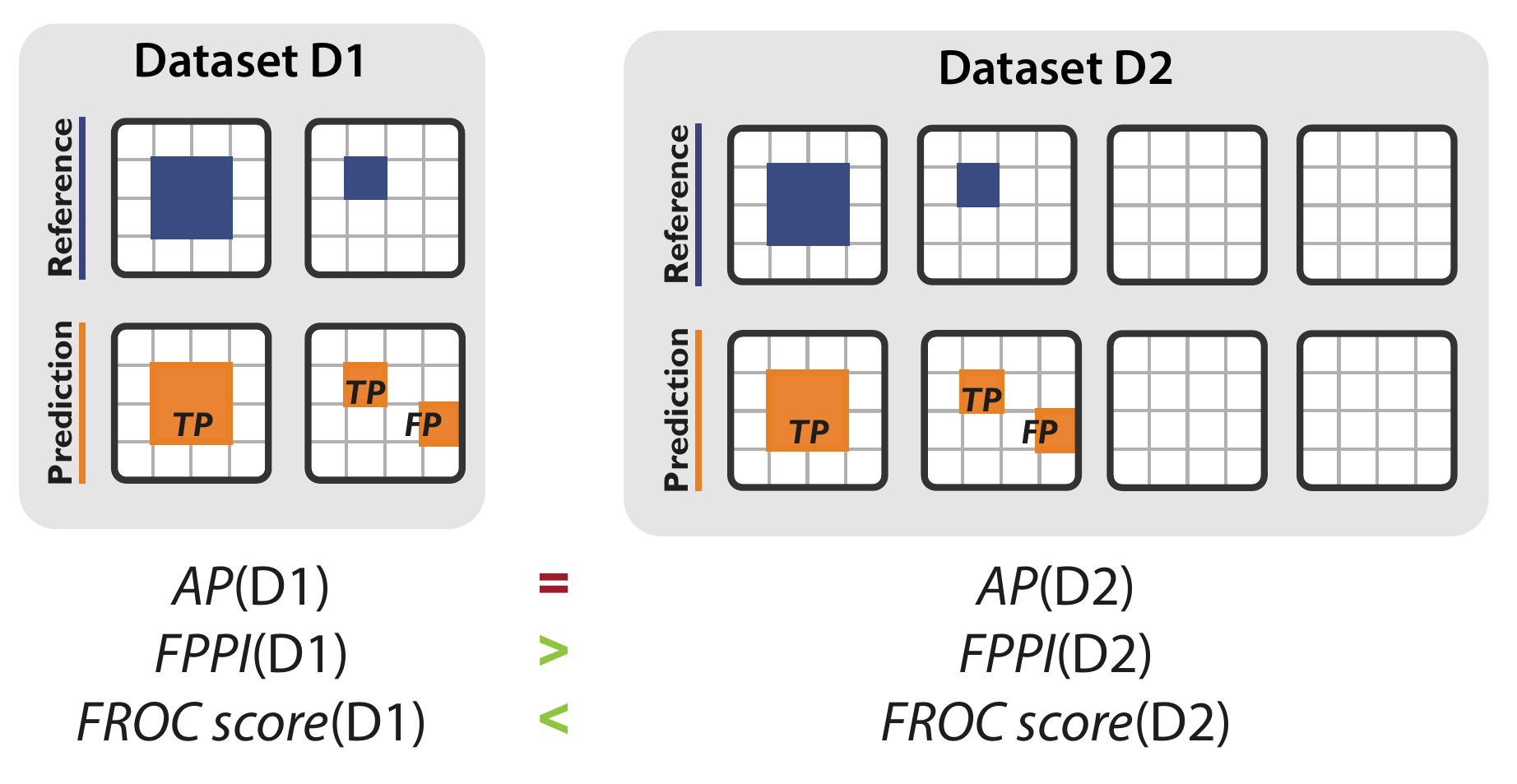}
    \caption{Effect of the number of images per data set on the metric scores. The \acf{AP} metric does not take into account the total number of images, yielding the same score for data sets D1 and D2. The \acf{FROC} curve plots the average number of \acf{FPPI} against the \textit{Sensitivity}, therefore accounting for the number of images. The \ac{FPPI} is lower for D2, yielding a higher \ac{FROC} score.}
     \label{fig:ap-froc} 
\end{tcolorbox}
\end{figure}

\begin{figure}[H]
\begin{tcolorbox}[title= Pitfall: \textit{\acf{FROC}} not standardized, colback=white]
    \centering
    \includegraphics[width=1\linewidth]{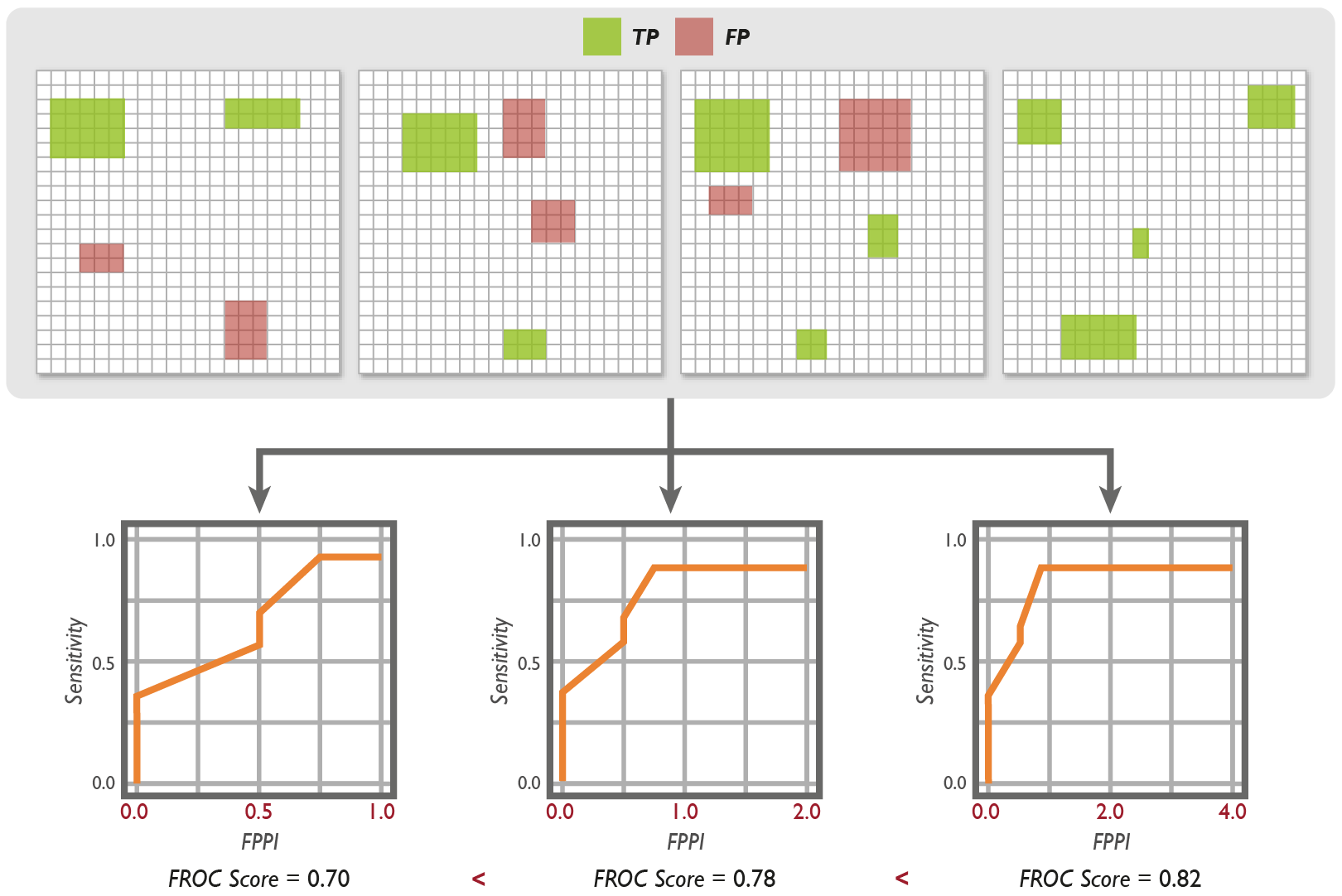}
    \caption{Effect of defining different ranges for the \textit{\acf{FPPI}} (which are unbounded to the top) used to draw the \textit{\acf{FROC}} curve for the same prediction (top). The resulting \textit{\ac{FROC}} Scores differ for different boundaries of the x-axis used for the \ac{FPPI} ([0, 1], [0, 2] and [0, 4]). Publications make use of different ranges for the x-axis, complicating comparison between works.}
    \label{fig:froc-no-standard}
\end{tcolorbox}
\end{figure}

\begin{figure}[H]
\begin{tcolorbox}[title= Pitfall: Large effects of small changes in predicted class probabilities, colback=white]
    \centering
    \includegraphics[width=1\linewidth]{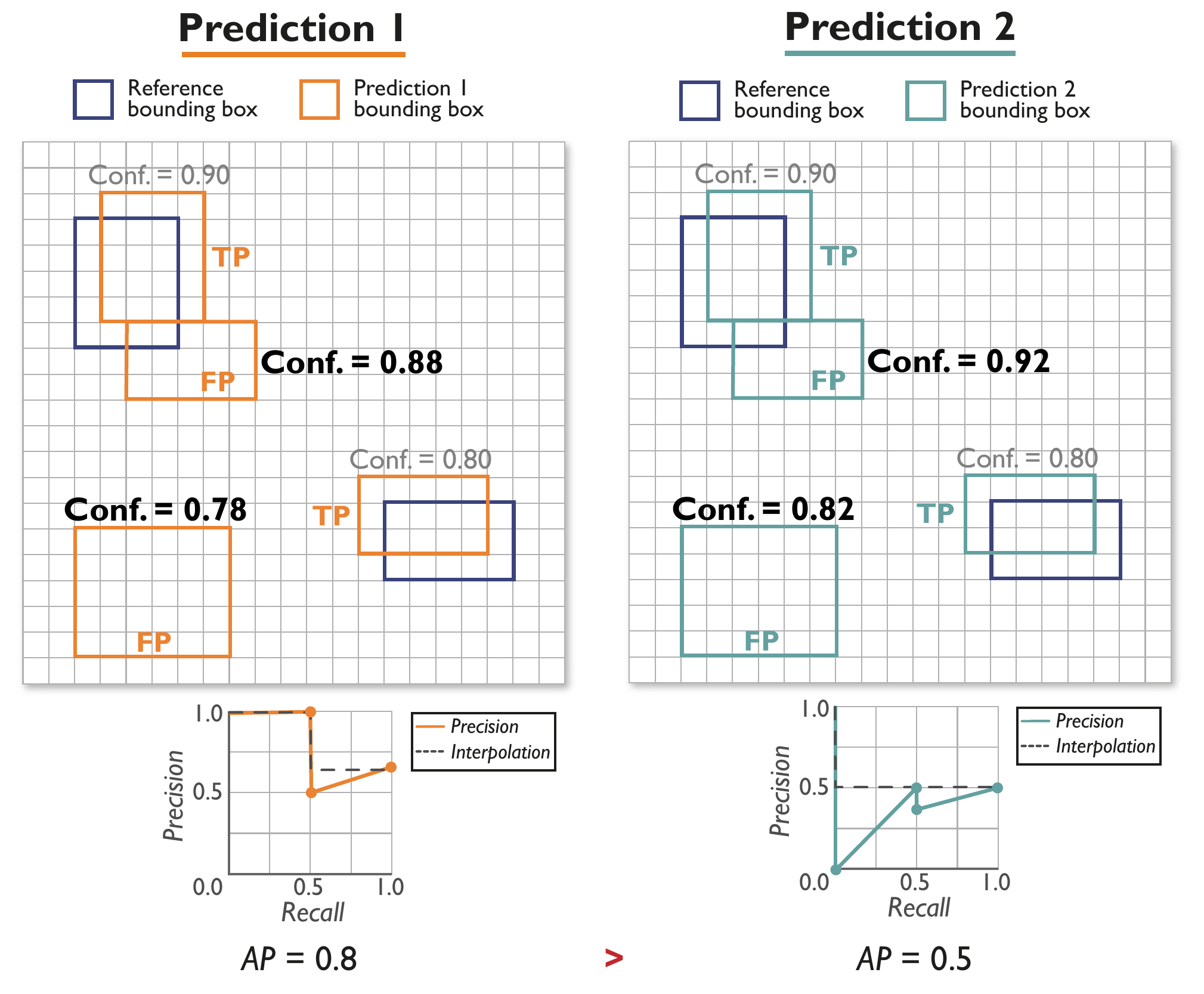}
    \caption{Effect of small changes in predicted class probabilities. Reference bounding boxes are shown in dark blue. \textit{Predictions 1} and \textit{2} detect the exact same bounding boxes with minor variations in their predicted class probabilities (represented by the confidence scores (conf.)). This leads to a different ranking and therefore varying \textit{\acf{PR}} curves, curve interpolations (dashed grey lines) and resulting \textit{\acf{AP}} scores.}
    \label{fig:AP-small-conf-changes}
\end{tcolorbox}
\end{figure}

\begin{figure}[H]
\begin{tcolorbox}[title= Pitfall: Predicted class probabilities neglected within ranking, colback=white]
    \centering
    \includegraphics[width=1\linewidth]{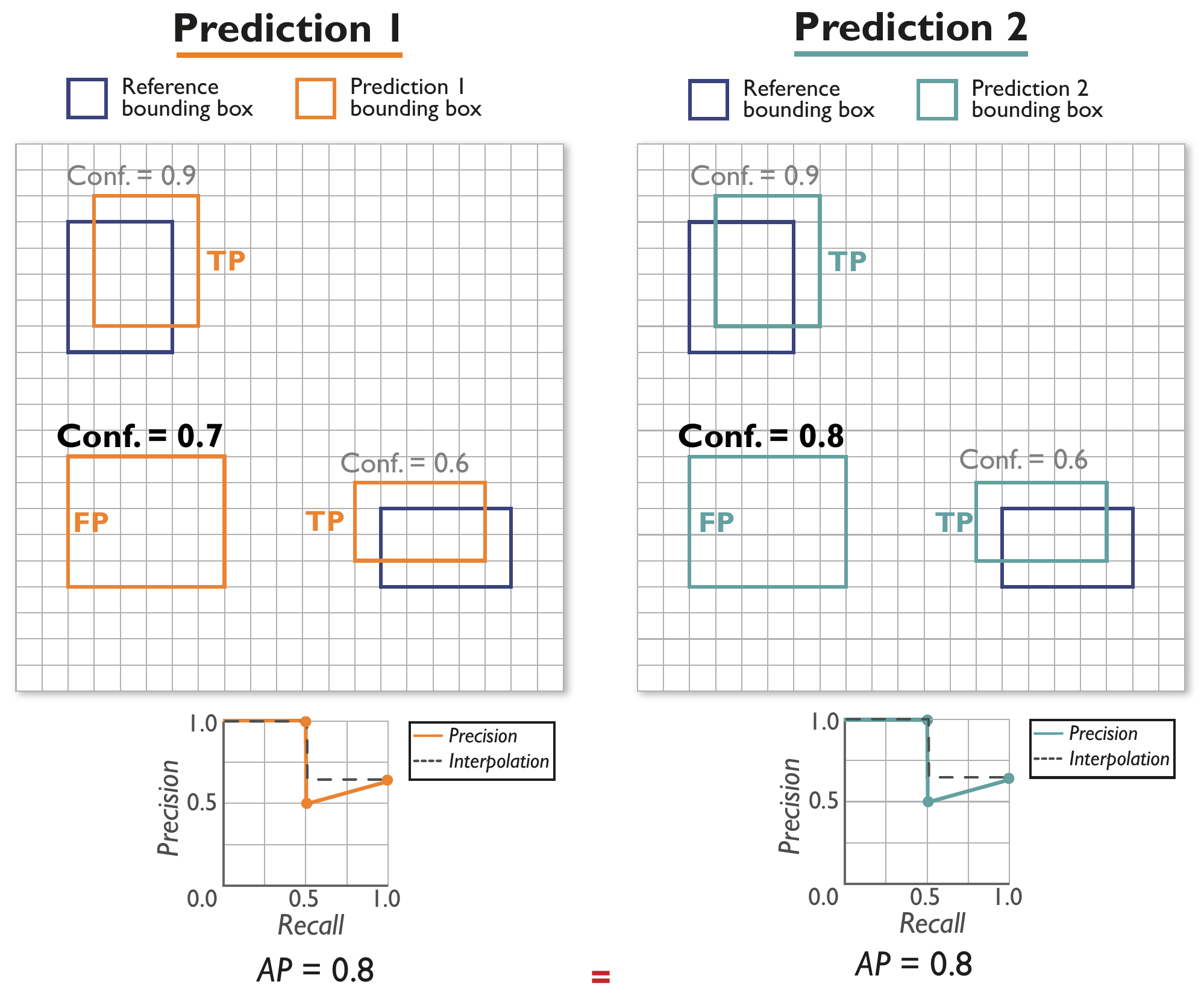}
    \caption{Effect of neglecting (the absolute values of) predicted class probabilities within the ranking. Reference bounding boxes are shown in dark blue. \textit{Predictions 1} and \textit{2} detect the exact same bounding boxes with variations in their predicted class probabilities (represented by the confidence scores (conf.)) that do not affect the ranking. Therefore, the \textit{\acf{PR}} curves, curve interpolations (dashed grey lines) and resulting \textit{\acf{AP}} scores are the same; the predicted class probabilities are hence unimportant within the ranking.}
    \label{fig:AP-conf-not-important}
\end{tcolorbox}
\end{figure}

\begin{figure}[H]
\begin{tcolorbox}[title= Pitfall: \ac{FP} with low predicted class probabilities, colback=white]
    \centering
    \includegraphics[width=1\linewidth]{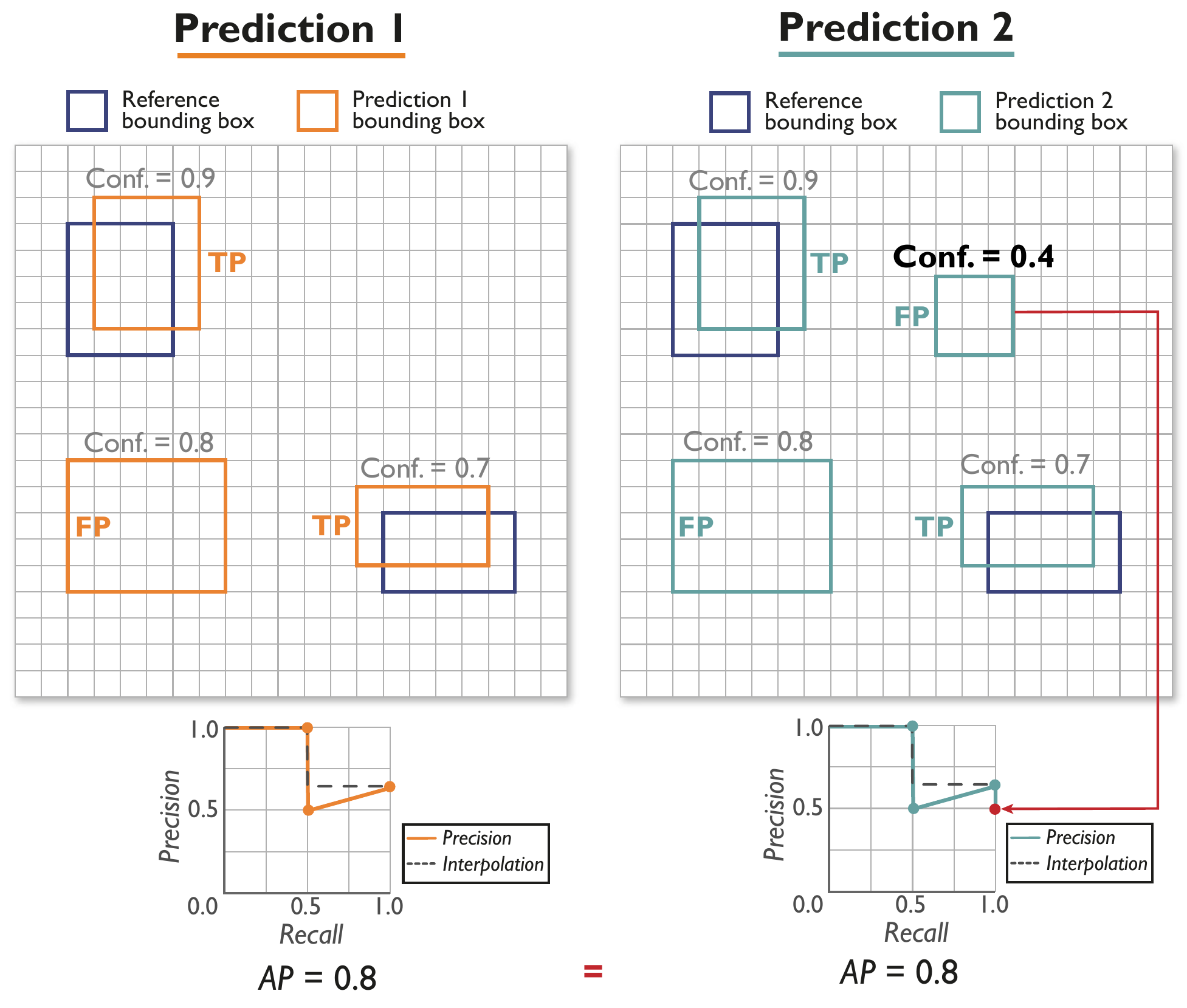}
    \caption{Effect of \acf{FP} predictions with low predicted class probabilities (represented by the confidence scores (conf.)). Reference bounding boxes are shown in dark blue. \textit{Predictions 1} and \textit{2} predict the exact same bounding boxes, but \textit{Prediction 2} shows one additional \ac{FP} detected box with low predicted class probabilities. This is not reflected in the \textit{\ac{AP}} score, as the \ac{FP} is located at the tail of the \textit{\acf{PR}} curve and does not change the curve interpolation (dashed grey lines).}
    \label{fig:AP-FP-tail}
\end{tcolorbox}
\end{figure}

\begin{figure}[H]
\begin{tcolorbox}[title= Undefined corner cases yield variations in metric scores, colback=white]
    \centering
    \includegraphics[width=1\linewidth]{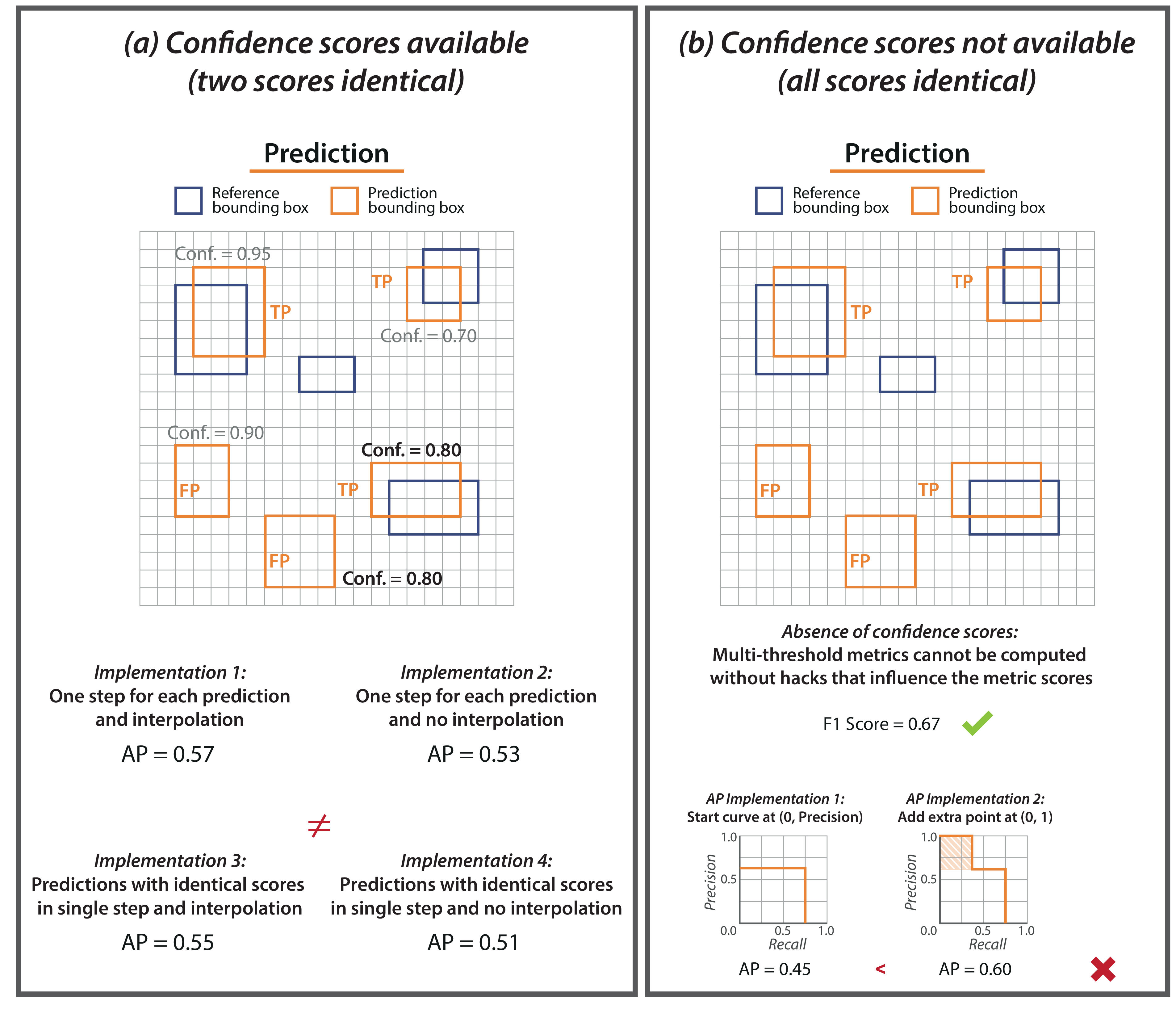}
    \caption{(a) Non-standardized metric implementation. In the case of the \textit{\acf{AP}} metric and the construction of the \textit{\acf{PR}}-curve, the strategy of how identical scores (here: confidence score of 0.80 is present twice) are treated has a substantial impact on the metric scores. Microsoft \acf{COCO} \citep{lin2014microsoft} and CityScapes \citep{Cordts2015Cvprw} are used as examples. (b) Multi-threshold metrics should only be computed if predicted class scores are available, although an increasing body of work  computes multi-threshold metrics such as \textit{\ac{AP}} in the absence of class scores (e.g. \citep{bai2017deep, de2017semantic, gao2019ssap, hirsch2020patchperpix, kulikov2020instance}). Otherwise, the strategy chosen for compensating the lack of class scores (here reflected by \textit{Implementations 1} and \textit{2}) leads to metric scores that are less well interpretable than those of established counting metrics working on a fixed confusion matrix (here: \textit{F$_\text{1}$ Score}).}
    \label{fig:corner-case-ap}
\end{tcolorbox}
\end{figure}

All pitfalls compiled for this work and relevant for object detection are summarized in Table~\ref{tab:properties-pitfalls}. Note that these pitfalls equally apply to instance segmentation problems. While most issues related to the actual metric selection have already been mentioned in the previous paragraphs, this section is primarily dedicated to technical peculiarities related to the localization and assignment criteria. These include: 

\begin{itemize}
    \item Mathematical implications of center-based localization criteria (Figures~\ref{fig:hit-criteria} -~\ref{fig:center-vs-point})
    \item Mathematical implications of \textit{\ac{IoU}}-based localization criteria (Figure~\ref{fig:bb-2d-3d})
    \item Mathematical implications of the choice of assignment strategies (Figure~\ref{fig:od-assignment} - \ref{fig:disconnected})
    \item Effect of small structures on localization criterion (Figure~\ref{fig:boundary-mask-iou})
    \item Perfect \textit{Boundary \ac{IoU}} for imperfect prediction (Figure~\ref{fig:boundary-iou})
    \item Possibility of reference or prediction without the target structure and \texttt{NaN} handling (Figure~\ref{fig:empty-ref-pred-od})
    \item \textit{Average Precision} \textit{vs.} \textit{Free-response ROC} score (Figure~\ref{fig:ap-froc})
    \item \textit{Free-response ROC score} is not standardized (Figure~\ref{fig:froc-no-standard})
    \item Effect of predicted class probabilities on multi-threshold metrics (Figures~\ref{fig:AP-small-conf-changes} - \ref{fig:AP-FP-tail})
    \item Non-standardized metric definition (Figure~\ref{fig:corner-case-ap})
\end{itemize}

\paragraph*{\textbf{Mathematical implications of center-based localization criteria}} Before calculating metrics for object detection tasks, it is necessary to define what qualifies a detection as a \textit{hit} (\ac{TP}) or \textit{miss} (\ac{FP}). There are multiple ways to define a hit, all of which come with their specific limitations. Below, the most commonly used center-based localization criteria are presented\footnote{For more details, please refer to the blogpost "Evaluation curves for object detection algorithms in medical images": \url{https://medium.com/lunit/evaluation-curves-for-object-detection-algorithms-in-medical-images-4b083fddce6e}.}\footnote{The presented pitfalls are also valid for approximations or bounding circles instead of the shown bounding boxes.}.
\begin{itemize}
    \item For the \textbf{center-cover criterion}, the reference object is considered a hit if the center of the reference object is inside the predicted detection. Figure~\ref{fig:hit-criteria}a shows how this criterion can be fooled by a model outputting very large boxes to maximize the chance of a correct detection. The same issue holds true for the Point inside Mask/Box criterion, which is fulfilled if a single predicted point lies inside the reference object.
    \item In the case of the \textbf{distance-based hit criterion}, a prediction is considered a hit if the distance $d$ between the center of the reference and the detected object is smaller than a certain threshold $\tau$. In Figure~\ref{fig:hit-criteria}b, both predictions have the same distance to their corresponding reference object centers. However, the prediction on the top right shows no overlap with the reference and should therefore not be counted as a hit.
    \item The \textbf{center-hit criterion} holds true if the center of the predicted object is inside the reference bounding box or contour. Given this definition, large reference objects are more likely to be hit, as shown in Figure~\ref{fig:hit-criteria}c. The left prediction is defined as a missed object (\ac{FP}), the right detection as a hit because of its larger size. 
\end{itemize}

\newpage
\begin{figure}[H]
\begin{tcolorbox}[title= Pitfall: Mathematical implications of center-based localization criteria, colback=white]
    \centering
    \includegraphics[width=1\linewidth]{images/Detection/Hit_criteria_3.png}
    \caption{Pitfalls for several center-based hit criteria in object detection. Reference objects are shown in dark blue, predictions in orange. The object centers are shown as blue/orange crosses. \textbf{(a)} The \textbf{center-cover criterion}, which requires the center of the reference to be inside the detection, can be fooled easily by predicting a very large bounding box/object. \textbf{(b)} Both predictions have the same distance to their corresponding reference center. The \textbf{distance-based criterion}, which requires the distance between center points not be exceeded, does not take into account the overlap between objects. However, the right prediction does not overlap with the reference and should, thus, not be considered a \acf{TP}. \textbf{(c)} The \textbf{center-hit criterion}, which requires the center of the prediction to be located inside the reference, favors large reference objects, as they are easier to detect. The left prediction is considered a \acf{FP}, as the reference center was not hit. The right prediction is considered a \ac{TP} because of the larger size of the object.}
    \label{fig:hit-criteria}
\end{tcolorbox}
\end{figure}

\newpage
The center point might not be a good reference for complex shapes such as tubular structures (Figure~\ref{fig:center-vs-point}). In those cases (and if annotations are provided in the form of masks), a binary “Point inside Mask” criterion might be the better choice.

\newpage
\paragraph*{\textbf{Mathematical implications of \textit{\ac{IoU}}-based localization criteria}} 
The most commonly used hit criterion is determined by computing the \textit{\ac{IoU}} between the predicted and the reference mask/bounding box/boundary. Pitfalls related to \textit{\ac{IoU}}-based criteria are mainly related to the setting of the threshold (Figure~\ref{fig:bb-2d-3d}). Many biomedical applications involve 3D rather than 2D images. When working with a higher dimension, it should be kept in mind that metrics may be affected. The additional dimension will lead to overlap errors being punished even more. Figure~\ref{fig:bb-2d-3d}a shows a comparison of the \textit{\ac{IoU}} for two rectangles (or bounding boxes) in 2D and 3D. Being mistaken by one voxel in the $z$-dimension will lead to a much lower \textit{\ac{IoU}} score in 3D compared to the 2D case. 

As \textit{\ac{IoU}}-based criteria take the overlap between regions into account, it is only possible to cheat with very large boxes if the \textit{\ac{IoU}} threshold is set to a very small value (here: 0), as shown in Figure~\ref{fig:bb-2d-3d}b. However, special care should be taken when applying the \textit{Box \ac{IoU}} in the presence of highly concave or elongated structures, as illustrated in Figure~\ref{fig:bb-2d-3d}c. This is because bounding boxes may quickly grow for narrow and diagonally placed objects, such as medical instruments, and result in \ac{FP} although visual inspection would indicate a correct prediction. 

\newpage

\newpage
\paragraph*{\textbf{Mathematical implications of the choice of assignment strategies}} Different assignment strategies may yield different proportions of \acp{TP}, \acp{FP} and \acp{FN}. This is shown in Figure~\ref{fig:od-assignment}a, in which two different assignment strategies based on the center distance are shown for the same situation with three reference and four predicted objects. Matching one reference object at a time (Greedy by Center Distance Matching) yields two \ac{TP} matches: [R1, P1] and [R2, P2]. This is due to the fact that the distance between R2 and P2 is smaller than the distance between R2 and P3 and that between R3 and P2. Thus, R2 is assigned to P2. The distance between R3 and P3 or P4 is greater than the defined tolerance, thus, P3 and P4 are seen as \ac{FP} and R3 is a \ac{FN}. If an Optimal (Hungarian) Matching \citep{kuhn1955hungarian} is applied, matching is performed differently by optimizing a cost function. This yields a different assignment with three \ac{TP} objects. Here, P3 is assigned to R2 and P2 is assigned to R3. As all reference objects are assigned, no \ac{FN} appears. Only P4 is counted as a \ac{FP}. 

In addition, researchers should think about whether \acp{FN} should be further distinguished into real \acp{FN}, i.e., reference objects for which no prediction exists (see left example in Figure~\ref{fig:od-assignment}b) and predictions that are outside a distance threshold or below an overlap threshold (see right example in Figure~\ref{fig:od-assignment}b). The latter case may be more useful in the present example, as R2 is not fully overlooked although P2 lies outside of the threshold area.

\newpage
\paragraph*{\textbf{Penalization of one prediction assigned to multiple reference objects requested}} Object detection and instance segmentation algorithms typically involve the step of assigning predicted objects to reference objects. This may result in one prediction being assigned to multiple reference objects or vice versa. Using the \textit{\ac{IoU}} > 0.5 (or similar threshold) as assignment strategy may end up in a very strict penalty of two \ac{FN} and one \ac{FP} if the \textit{\ac{IoU}} for both reference objects is smaller than the threshold (as shown for two reference objects in Figure~\ref{fig:od-ior}. Using the \textit{\ac{IoR}} may result in a less strict penalization for what could be interpreted as only a single error with an additional step of penalizing the non-split errors (either directly in object detection or indirectly in instance segmentation) \cite{matula2015cell}.

\newpage
\paragraph*{\textbf{Possibility of disconnected structures}} The \textit{Box \ac{IoU}} is sometimes employed despite access to pixel-mask annotations. A possible explanation is that researchers want to phrase their problem as an object detection problem and then apply the most commonly used validation methods. Such simplification might cause problems if structures are not well approximated by a box shape, or if structures yield multi-component masks, appearing to be disconnected. This may occur in the case of a tubular structure shown in a 2D tomographic image or a medical instrument occluded by tissue in an endoscopic image, for example. Figure~\ref{fig:disconnected} provides examples of a complex diagonal (top) and a disconnected structure (bottom). Both box predictions yield a \textit{Box \ac{IoU}} larger than 0.3, and are thus counted as \ac{TP} because of the chosen localization threshold. Nevertheless, \textit{Prediction 1} is not hitting the actual object at all. This is due to the fact that the target structures are not well approximated by the bounding box, leaving many empty pixels in the boxes.

\newpage
\paragraph{\textbf{Effect of small structures on localization criterion}}
\textit{Box IoU} and \textit{Mask IoU} are not sensitive to structure boundary quality in larger objects (cf. Section~\ref{sec:segmentation}). This is due to the fact that boundary pixels will increase linearly (quadratically  in 3D) while pixels inside the structure will increase quadratically (cubically in 3D) with an increase in structure size. In consequence, the \textit{IoU}-scores tend to be higher for large objects compared to small objects. For this reason, localization criteria such as the \textbf{\textit{Boundary IoU}} were designed. 

Figure~\ref{fig:boundary-mask-iou} shows an example of \textit{Mask IoU} and \textit{Boundary IoU} for a large (top) and a rather small structure (bottom). In the case of the \textit{Mask IoU}, the score drops substantially for the small structure, while the scores are more consistent for the \textit{Boundary IoU} when comparing small and large structures. This pitfall also applies to segmentation problems in which the \textit{(Mask) IoU} and \textit{Boundary IoU} are applied as overlap-based metrics.

It should further be noted that the \textit{Boundary IoU} is highly dependent on the chosen distance $d$, as illustrated in Figure~\ref{fig:boundary-mask-iou} (third vs fourth column). Similarly to the example provided in Figure~\ref{fig:outline}, the \textit{Boundary IoU} can be fooled to result in a perfect value of $1.0$. A prediction with a hole in the middle of the structure may result in a perfect metric score if the distance is chosen in a way that it incorporates all pixels of the predicted mask, as shown in Figure~\ref{fig:boundary-iou} \cite{cheng2021boundary}. The \textit{Mask IoU}, however, will be able to recognize the problem, as it completely measures the overlap between both structures. \cite{cheng2021boundary} propose to use the $\min(Boundary IoU, Mask IoU)$ to resolve this issue. Please note that the same limitations also affect other distance-based measures, such as the \textit{\ac{NSD}} or \textit{\ac{HD}} metrics. This pitfall also applies to segmentation tasks.

\newpage
\paragraph{\textbf{Possibility of reference/prediction without the target structure and \texttt{\ac{NaN}} handling}} When validating an object detection problem per image rather than per data set, a reference or prediction image without the target structure(s) may become problematic as some metric values will turn into \texttt{\ac{NaN}} due to division by zero errors (cf. Figure~\ref{fig:empty}). Figure~\ref{fig:empty-ref-pred-od}a shows potential scenarios for a validation per image categorized by the presence and absence of \ac{TP}, \ac{FP} and \ac{FN}. Four occurrences of \texttt{\ac{NaN}} are presented. To proceed with the validation, namely aggregating metric values for every image over the entire data set, a \texttt{\ac{NaN}} strategy needs to be defined for every use case. 

\newpage
\paragraph{\textbf{Average Precision \textit{vs.} Free-response ROC score}} While the \ac{AP} constitutes the standard metric for object detection and instance segmentation in the computer vision community, the \textit{\ac{FROC}} score is often favoured in the clinical context. In contrast to the \ac{AP}, the \ac{FROC} score takes into account the total number of images in the data set. As can be seen from Figure~\ref{fig:ap-froc}, both data sets D1 and D2 will yield the same \ac{AP} score, although data set D1 contains two images and D2 contains four images. The \ac{FROC} score, however, will reflect that the number of images is different for both data sets and that data set D2 contains two images that do not contain any \ac{FP}. Thus, the \ac{FPPI} will be lower in data set D2, yielding a higher \ac{FROC} score.

\newpage
\paragraph{\textit{\ac{FROC}} is not standardized} Although the \textit{\ac{FROC} Score} is often favored by clinicians over the \textit{\ac{AP}} given its simpler interpretation, it comes with a major drawback. While the values of the x-axis are clearly defined and bounded between 0 and 1 for the \textit{\ac{AP}}, there is no fixed or standardized definition for the \textit{\ac{FPPI}} used for the \textit{\ac{FROC}} curve. Depending on the defined range of the \textit{\ac{FPPI}}, the \textit{\ac{FROC} Score} will change. Figure~\ref{fig:froc-no-standard} shows three examples of different bounds for the x-axes (left: [0, 1], middle: [0, 2], right: [0, 4]) for the same prediction -- all of them yielding different \textit{\ac{FROC} Scores}.

\newpage
\paragraph{\textbf{Multi-threshold metric-related properties}} In the next paragraphs, we highlight some limitations of the multi-threshold metrics, exemplarily for the \textit{\ac{AP}} metric, which can be transferred to other multi-threshold metrics, such as the \textit{\ac{AUROC}} \cite{oksuz2018localization}. By definition, multi-threshold metrics are ranking metrics, which rank the predicted class probabilities or confidence scores (cf. Figure~\ref{fig:ap-example}). They are not designed to reflect the calibration of confidence or class scores, as shown in the following examples. Please note that we disregard the concrete choice of the localization criterion here for simplicity.

\paragraph{Predicted class probabilities} The \textit{\ac{PR}} curve and the resulting metric score \textit{\ac{AP}} highly depend on the ranking of predictions, based on their predicted class probabilities or confidence scores. Small changes in the scores can therefore significantly change the metric value, as shown in Figure~\ref{fig:AP-small-conf-changes}. On the other hand, as long as the ranking remains unchanged among predictions, the predicted class probabilities themselves are not important for the result, although they should be (see Figure~\ref{fig:AP-conf-not-important}). 
\newpage
\newpage

\newpage
\paragraph{\ac{FP} with low predicted class probabilities} False positive predictions with lower predicted class probabilities than the last correctly predicted reference, corresponding to the end of the \textit{PR} curve, do not affect the \textit{\ac{AP}} scores. Figure~\ref{fig:AP-FP-tail} shows two examples that are very similar, only differing in the number of wrongly predicted objects. \textit{Prediction 2}, with two \ac{FP}, performs worse than \textit{Prediction 1} with only one \ac{FP}. Nevertheless, the \textit{\ac{AP}} scores are the same for both models, given the low confidence of the second \ac{FP} of \textit{Prediction 2}. Thus, a prediction may contain numerous \ac{FP} objects with low confidence scores, but still yield a deceptively high \ac{AP} score, as this practice only increases the score but never decreases it by potentially hitting more reference objects.

\newpage
\paragraph{\textbf{Non-standardized metric definition}} Standard metric implementations do not always cover corner cases or handle them differently. For example, for the creation of the \ac{PR}-curve and the subsequent computation of the \textit{\ac{AP}} metric, predictions are typically ranked by their predicted class scores. Duplicate predicted class scores can either be computed in a single step per score or can all be treated together in a common step. Based on the chosen implementation, the \ac{PR}-curve will differ in appearance, causing a sometimes substantial change in the \textit{\ac{AP}} scores. Figure~\ref{fig:corner-case-ap}a provides such an example, in which the final scores differ by 0.06 solely because of two predictions having the same score. 

This difference in implementation is even more critical in the extreme case of all predictions having the same score (e.g., if no class scores are available; see Figure~\ref{fig:corner-case-ap}b). However, it should be noted that although this practice is not uncommon (e.g., \citep{bai2017deep, de2017semantic, gao2019ssap, hirsch2020patchperpix, kulikov2020instance}), multi-threshold metrics such as the \ac{AP} should generally be avoided if no class scores are available. Moreover, some implementations force the \ac{PR}-curve to start at the point (0, 1), i.e., to always start with a \textit{\ac{PPV}} value of 1 instead of the actual first \textit{\ac{PPV}} value that was computed. This may lead to a substantial overestimation if there are no confidence scores or if the number of instances is low.
 
\newpage
\section{Pitfalls related to analyses and post-processing}
\label{sec:aggregation-combination}
\begin{figure}[H]
\begin{tcolorbox}[title= Pitfall: Uninformative visualization, colback=white]
    \centering
    \includegraphics[width=1\linewidth]{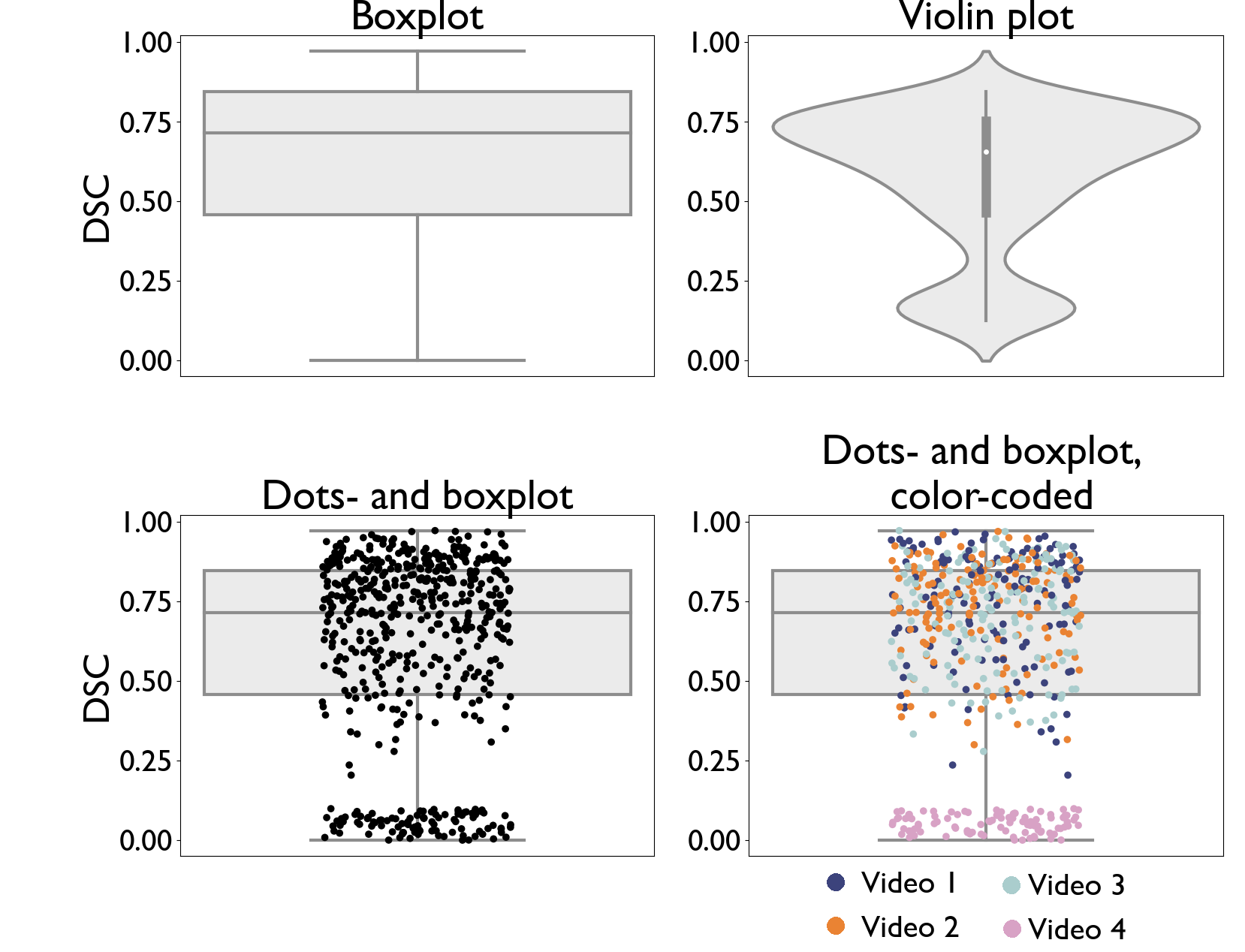}
    \caption{Effect of different types of visualization. A single boxplot (top left) does not give sufficient information about the raw metric value distribution (here: \textit{\acf{DSC}}). Using a violin plot (top right) or adding the raw metric values as jittered dots on top (bottom left) adds important information. In the case of non-independent validation data, color/shape-coding helps reveal data clusters (bottom right).}
    \label{fig:raw-metric-values-boxplot}
\end{tcolorbox}
\end{figure}

\begin{figure}[H]
\begin{tcolorbox}[title= Pitfall: Missing values for metrics with fixed upper/lower bounds, colback=white]
    \centering
    \includegraphics[width=0.9\linewidth]{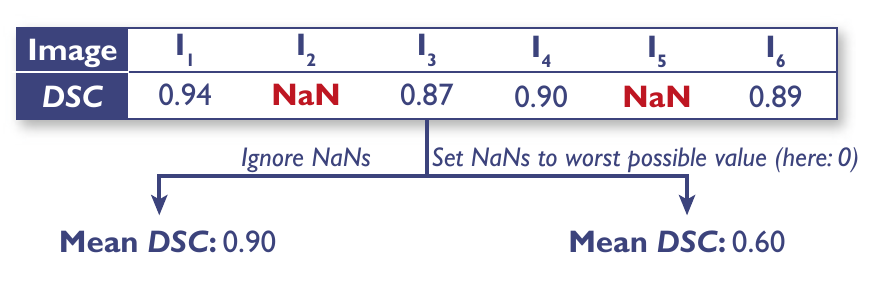}
    \caption{Effect of missing values when aggregating metric values. In this example, ignoring missing values leads to a substantially higher \textit{\acf{DSC}}) compared to setting missing values to the worst possible value (here: 0).}
    \label{fig:missings}
\end{tcolorbox}
\end{figure}

\begin{figure}[H]
\begin{tcolorbox}[title= Pitfall: Missing values for metrics without fixed upper/lower bounds, colback=white]
    \centering
    \includegraphics[width=0.9\linewidth]{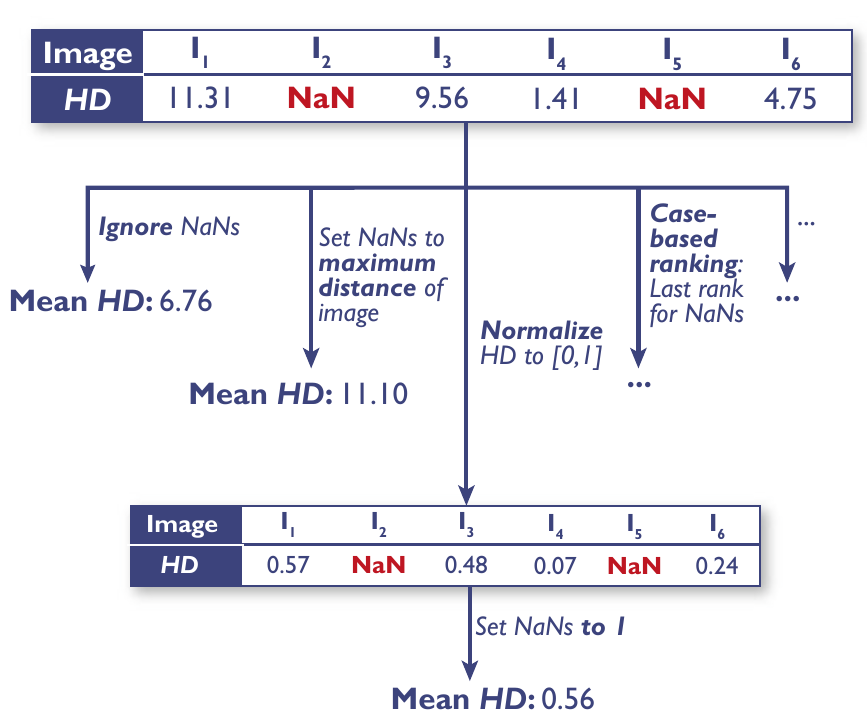}
    \caption{Effect of missing values when aggregating metric values for metrics without fixed boundaries (here: \textit{\acf{HD}}). In this example, ignoring or treating missing values in different ways leads to substantially different \textit{\ac{HD}} values.}
    \label{fig:missings-hd}
\end{tcolorbox}
\end{figure}

\begin{figure}[H]
\begin{tcolorbox}[title= Pitfall: Non-independence of validation data, colback=white]
    \centering
    \includegraphics[width=1\linewidth]{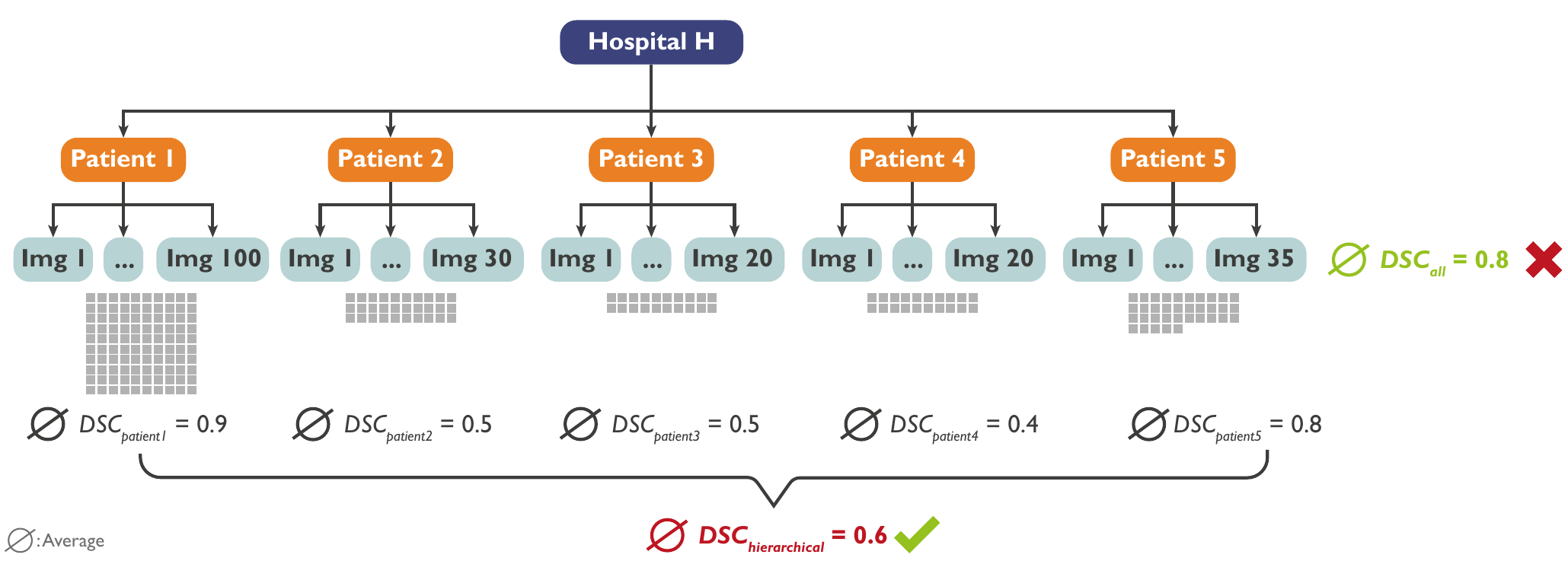}
    \caption{Effect of non-independence of validation data, caused here by an unequal number of data points per subject. The number of images taken from \textit{Patient 1} is much higher compared to that acquired from \textit{Patients 2-5}. Averaging over all \textit{\acf{DSC}} values results in a high averaged score. However, aggregating metric values per patient reveals much larger scores for \textit{Patient 1} compared to the others, which would have been hidden by simple aggregation. $\varnothing$ refers to the average \textit{\ac{DSC}} values.}
    \label{fig:hier-aggr}
\end{tcolorbox}
\end{figure}

\begin{figure}[H]
\begin{tcolorbox}[title= Pitfall: Ignoring multiple classes when aggregating, colback=white]
    \centering
    \includegraphics[width=1\linewidth]{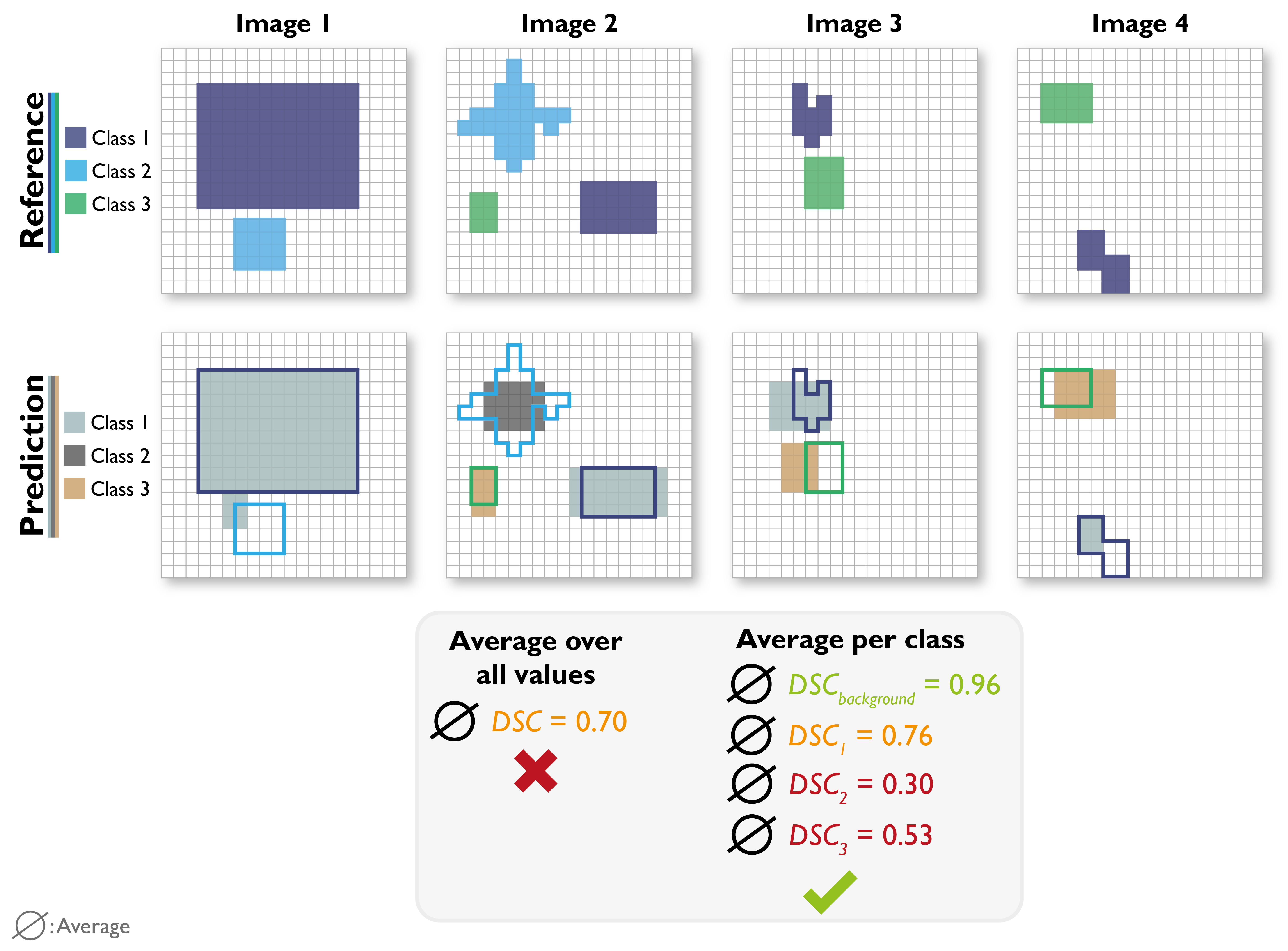}
    \caption{Effect of ignoring the presence of multiple classes when aggregating metric values. The overall average of all \textit{\acf{DSC}} scores for the four images results in a \textit{\ac{DSC}} score of 0.7. Averaging per class reveals a very low performance for classes 2 and 3. $\varnothing$ refers to the average \textit{\ac{DSC}} values.}
    \label{fig:aggr-per-class}
\end{tcolorbox}
\end{figure}

\begin{figure}[H]
\begin{tcolorbox}[title= Pitfall: Related metrics, colback=white]
    \centering
    \includegraphics[width=1\linewidth]{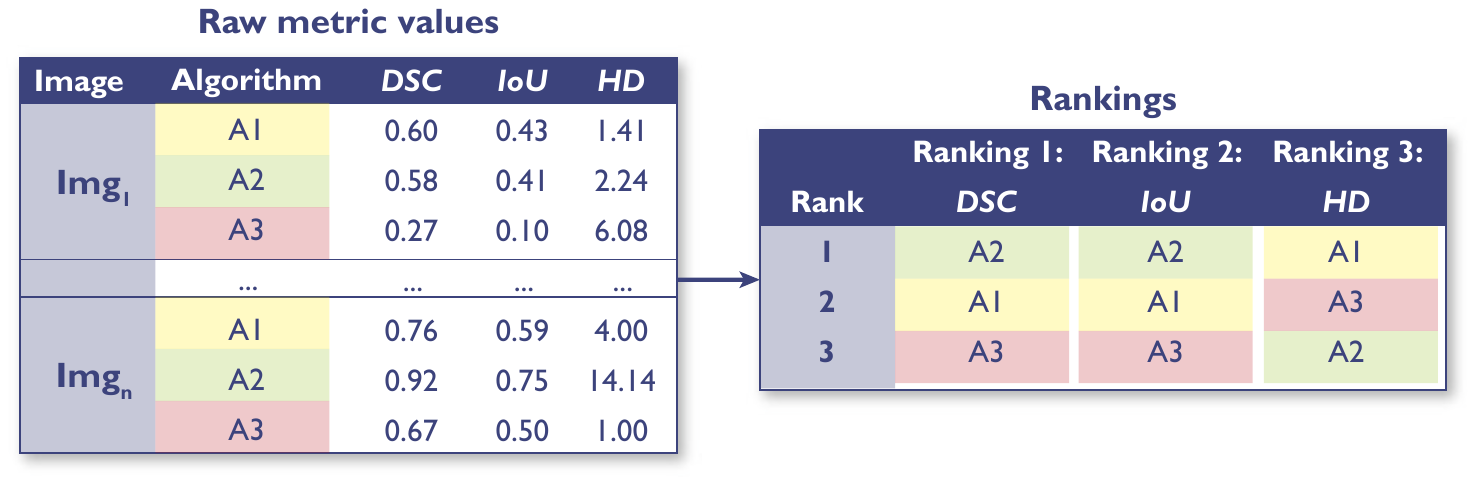}
    \caption{Effect of using mathematically closely related metrics. The \textit{\acf{DSC}} and \textit{\acf{IoU}} typically lead to the same ranking, whereas metrics from different families (here: \textit{\acf{HD}}) may lead to substantially different rankings.}
    \label{fig:combination}
\end{tcolorbox}
\end{figure}

\begin{figure}[H]
\begin{tcolorbox}[title= Pitfall: Ranking tables do not reflect ranking uncertainty, colback=white]
    \centering
    \includegraphics[width=1\linewidth]{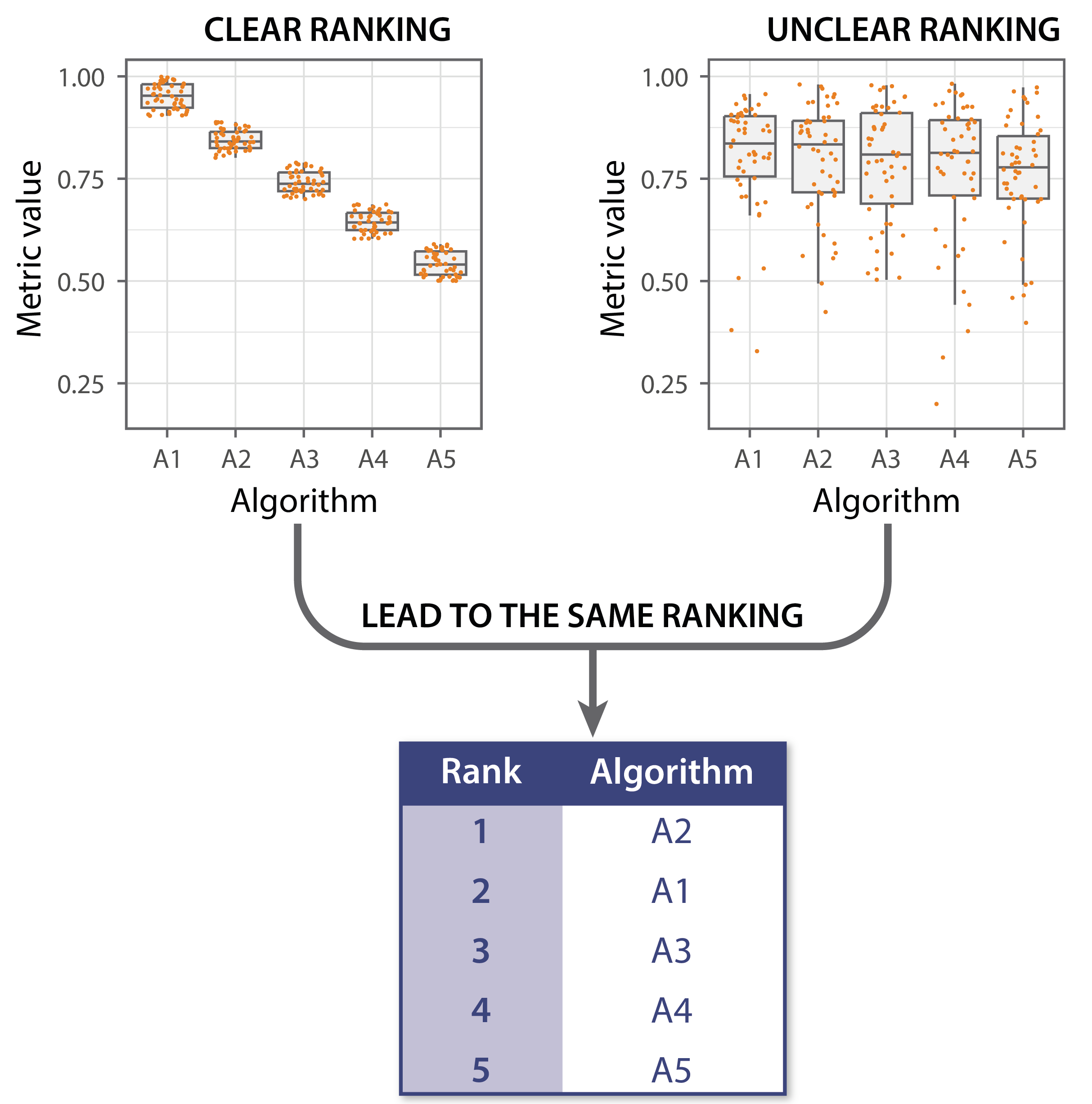}
    \caption{Effect of ranking uncertainty. The results of two benchmarking experiments with five algorithms $A1$-$A5$ differ substantially, as shown by the boxplots of the metric values for every algorithm. While the left case features a clear ranking visible from the boxplots, the right case does not, as performance is very similar across algorithms. However, both situations lead to the same ranking \citep{maier2018rankings,wiesenfarth2021methods}. Thus, solely providing ranking tables conceals information on ranking uncertainty.}
    \label{fig:ranking-uncertainty}
\end{tcolorbox}
\end{figure}

\begin{figure}[H]
\begin{tcolorbox}[title= Pitfall: Metric score differences leading to different rankings may be irrelevant, colback=white]
    \centering
    \includegraphics[width=1\linewidth]{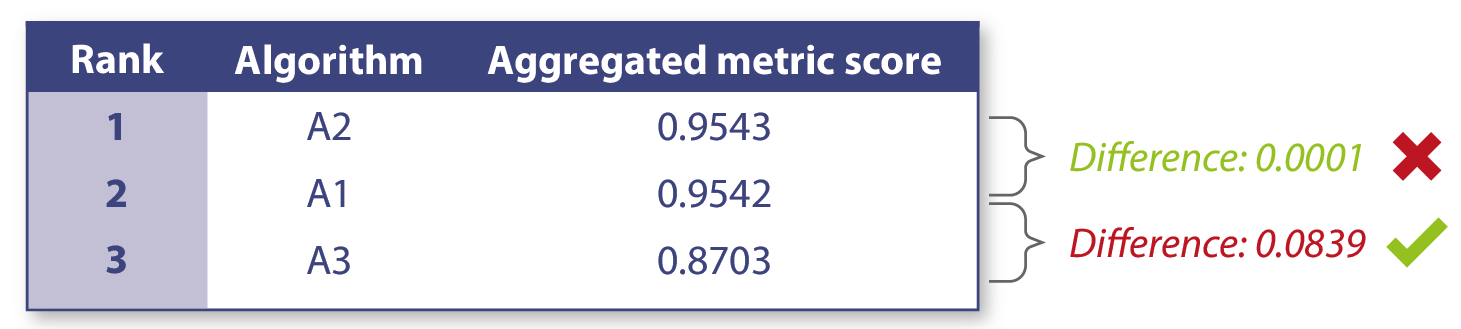}
    \caption{Effect of irrelevant metric score differences in rankings. The difference of the metric score aggregates of algorithms \textit{A1} and \textit{A2} is extremely low and not of biomedical relevance. However, going solely by numerical difference would assign them different ranks.}
    \label{fig:ranking-relevance}
\end{tcolorbox}
\end{figure}

\begin{figure}[H]
\begin{tcolorbox}[title= Pitfall: Non-determinism of algorithms, colback=white]
    \centering
    \includegraphics[width=0.7\linewidth]{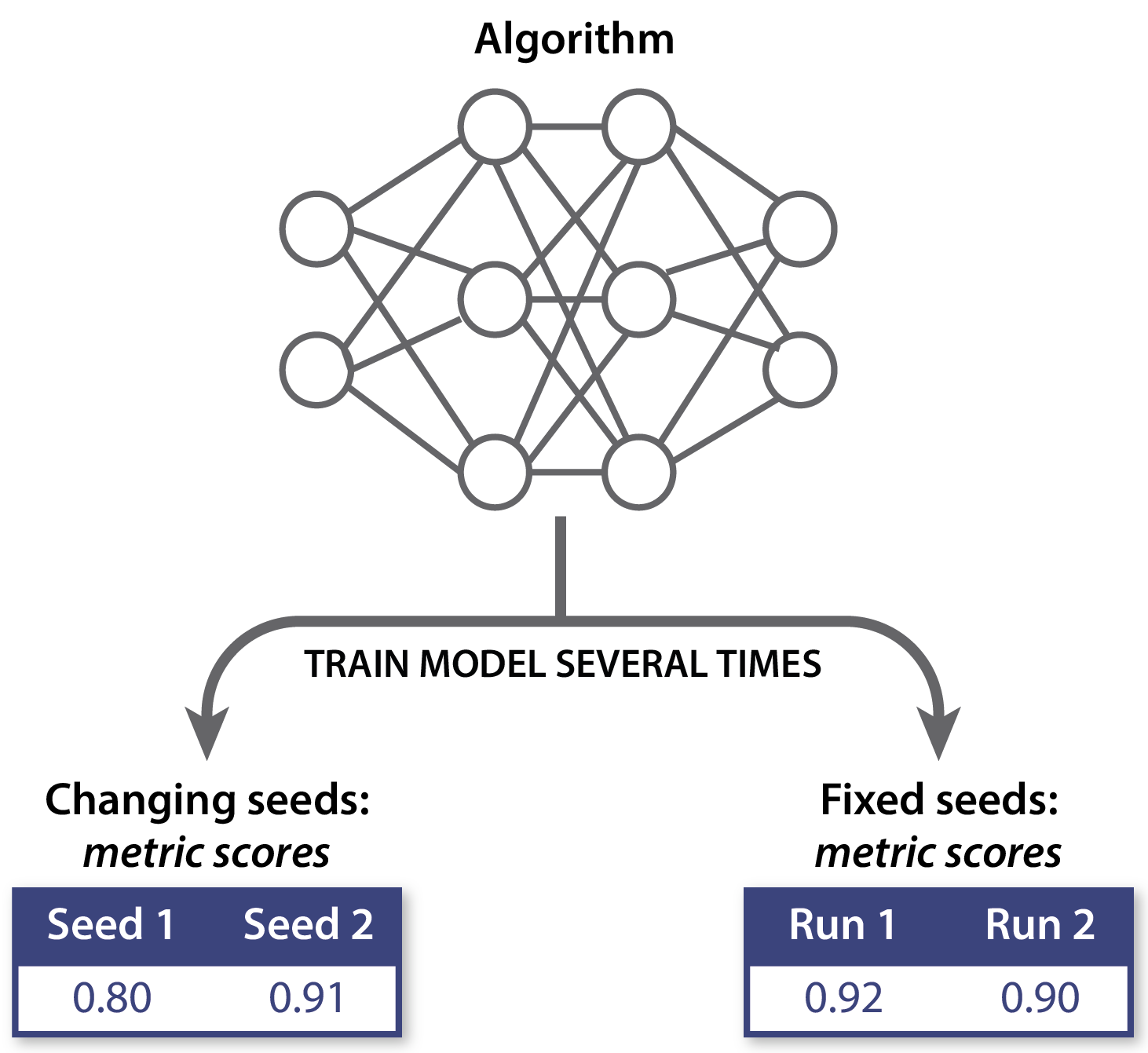}
    \caption[Effect of non-determinism of \acf{AI} algorithms. An algorithm trained under identical conditions may yield different results when changing seeds (left), but also with fixed seeds (right). The latter may, for example, be caused by parallel processes, the order of threads, auto-selection of primitive operations, and other factors. Fixing seeds does not guarantee reproducibility even for the same hardware/software configuration as many software libraries have a degree of randomness on their operations.]{Effect of non-determinism of \acf{AI} algorithms. An algorithm trained under identical conditions may yield different results when changing seeds (left), but also with fixed seeds (right). The latter may, for example, be caused by parallel processes, order of threads, auto-selection of primitive operations, and other factors \cite{pham2020problems}\protect\footnotemark. Fixing seeds does not guarantee reproducibility even for the same hardware/software configuration as many software libraries have a degree of randomness on their operations.}
    \label{fig:non-determinism}
\end{tcolorbox}
\end{figure}

A data set typically contains several hundreds or thousands of images. When analyzing, aggregating and combining metric values, a number of factors need to be taken into account. Pitfalls in this step are primarily related to the following aspects:

\begin{itemize}
    \item Uninformative visualization (Figure~\ref{fig:raw-metric-values-boxplot})
    \item Metric aggregation for invalid algorithm output (e.g. \texttt{NaN}) (Figures~\ref{fig:missings} - \ref{fig:missings-hd})
    \item Hierarchical data aggregation (Figure~\ref{fig:hier-aggr})
    \item Aggregation in the presence of multiple classes (Figure~\ref{fig:aggr-per-class})
    \item Combination of related metrics (Figure~\ref{fig:combination})
    \item Ranking uncertainty (Fig~\ref{fig:ranking-uncertainty})
    \item Insufficient biomedical relevance of metric score differences (Fig~\ref{fig:ranking-relevance})
    \item Non-determinism of algorithms (Fig~\ref{fig:non-determinism})
    \item Non-standardized metric implementation
\end{itemize}

\paragraph{\textbf{Uninformative visualization}} Relying on only reporting aggregated metric scores may result in missing essential information on algorithm performance. Therefore, raw metric values (e.g. per image) should always be shown, for example in the shape of boxplots, as depicted in the top left of Figure~\ref{fig:raw-metric-values-boxplot}. However, boxplots will only provide information on some key descriptive statistics, like median or 1st and 3rd quartiles. Another choice can be violin plots, which further visualize the raw data distribution. The top right of Figure~\ref{fig:raw-metric-values-boxplot} illustrates the multimodal distribution of the underlying data, invisible in the boxplot. Furthermore, using a violin plot and/or plotting the raw metric values for each data point on top (Figure~\ref{fig:raw-metric-values-boxplot}, top right and bottom left) will reveal the complete data distribution. In the example below, many values lie below the 3rd quartile, although the box looks tight. Nevertheless, even these two visualizations may hide important information. Assume a data set with metric values of four different videos. Color- or shape-coding the metric values by the video type (Figure~\ref{fig:raw-metric-values-boxplot} bottom right) reveals a huge cluster of extremely low \textit{\ac{DSC}} values only affecting Video 4 (pink), which would have been hidden by the other two types of visualization.
\newpage

\newpage
\paragraph{\textbf{Metric aggregation for invalid algorithm output (e.g. \texttt{\ac{NaN}})}} In challenges or benchmarking experiments, metric values are often aggregated over all test cases to produce a challenge ranking~\citep{maier2018rankings}. Missing data plays a crucial role when aggregating metric values and occurs primarily due to two reasons: invalid output of the algorithm or metric routine output resulting in \texttt{\ac{NaN}}, and non-submission of single cases (by accident or even for cheating~\cite{reinke2018exploit}). Figs.~\ref{fig:missings} and~\ref{fig:missings-hd} illustrate why a strategy on how to handle missing values may be crucial. 

In the case of metrics with fixed boundaries, such as the \textit{\ac{DSC}} or the \textit{\ac{IoU}}, missing values can easily be set to the worst possible value (here: 0). For spatial distance-based measures without lower/upper bounds, the strategy of how to treat missing values is not trivial. In the case of the \textit{\ac{HD}}, for example, one may choose the maximum distance of the image or normalize the metric values to $[0,1]$ and use the worst possible value (here: 1). Another possibility is to employ a case-based ranking scheme \cite{maier2018rankings} and assign the last rank for every missing submission. Furthermore, aggregating with the mean may not be a good choice as results are unlikely to be normally distributed. Crucially, however, every choice will produce a different aggregated value (Figure~\ref{fig:missings-hd}), thus potentially affecting the ranking. Another way of handling missing values would lie in rejecting the entire submission in a challenge.

However, metric values may also be undefined (\texttt{\ac{NaN}})) if either reference or prediction or both are empty. In the case of empty reference and prediction, an undefined metric value (e.g. \textit{\ac{DSC}}) may be a desirable outcome and should therefore  not necessarily be penalized.

\newpage
\newpage
\paragraph{\textbf{Hierarchical data aggregation}} Nowadays, most data sets are inherently hierarchically structured, meaning that the test cases are not independent. Data may, for example, come from several centers or hospitals, and for every center or even within one, different devices may be used for image acquisition, and images may be drawn from different subjects or patients. This should be kept in mind when visualizing and aggregating data points, especially if the individual tree nodes end in a large variation in the size of images. Figure~\ref{fig:hier-aggr} shows an example of five patients with an unequal number of images associated with them. Just averaging all metric values for every image would result in a high average \textit{\ac{DSC}} of 0.8. Averaging metric values per patient reveals that the \textit{\ac{DSC}} values are much higher for \textit{Patient 1}, overruling the other patients due to the high number of samples for this patient. Aggregating per patient first and averaging subsequently will resolve this issue.

\newpage
\paragraph*{\textbf{Aggregation per class}} Similar approaches should be chosen in the presence of multiple classes in a data set. The performance may differ significantly for the individual classes, as shown in Figure~\ref{fig:aggr-per-class}. The background class in particular will result in a nearly perfect averaged \textit{\ac{DSC}} value, whereas the average scores for classes 2 and 3 are much lower. Aggregating over all values, not considering the class, would hide this information. An alternative approach to the problem lies in the application of metrics that explicitly handle class balance, such as using the \textit{Generalized DSC} \cite{sudre2017generalised}.

\newpage
\paragraph{\textbf{Metric combination}} A single metric typically does not reflect all aspects that are essential for algorithm validation. Hence, multiple metrics with different properties are often combined. However, the selection of metrics should be well-considered as some metrics are mathematically related to each other~\citep{taha2014formal, taha2015metrics}. A prominent example is the \textit{\ac{IoU}} -- the most popular segmentation metric in computer vision -- which highly correlates with the \textit{\ac{DSC}} -- the most popular segmentation metric in medical image analysis. In fact, the \textit{\ac{IoU}} and the \textit{\ac{DSC}} are mathematically related (see Sec.~\ref{sec:fundamentals_ss}) \citep{taha2015metrics}.

Combining metrics that are related will not provide additional information for a ranking. Figure~\ref{fig:combination} illustrates how the ranking can change when adding a metric that measures different properties.

\newpage
\paragraph{\textbf{Ranking uncertainty}} Rankings themselves may be unstable. \citep{maier2018rankings} and \citep{wiesenfarth2021methods} demonstrated that rankings are highly sensitive to altering the metric aggregation operators, the underlying data set, or the general ranking method. Disregarding the robustness of rankings may thus lead to the winning algorithm ranking first solely by chance rather than true superiority. Such a case is illustrated in Fig.~\ref{fig:ranking-uncertainty}. Here, the boxplots of two different benchmarking experiments are presented, one of which represents a very clear ranking, the other showing a similar performance of all five algorithms. However, both scenarios would yield the same ranking table. The common practice of solely presenting ranking tables without further visualization \citep{wiesenfarth2021methods} may thus obscure important information.

\newpage
\paragraph{\textbf{Insufficient biomedical relevance of metric score differences}} Rankings may be uninformative in the domain context. For example, algorithms \textit{A1} and \textit{A2} in Fig.~\ref{fig:ranking-relevance} only differ by a very small amount (0.0001). While this difference would make one algorithm numerically superior, it may not be relevant for the actual application. Therefore, the algorithms should not receive different ranks, but rather share the same rank. On the other hand, algorithms \textit{A2} and \textit{A3} differ by a substantial amount, thus, ranking them differently is clinically relevant.

\newpage
\paragraph{\textbf{Non-determinism of algorithms}} \ac{AI} algorithms are subject to non-determinism. When training an algorithm under identical conditions (e.g., the same libraries and architecture), results will be subject to sometimes substantial changes if the random seeds are exchanged, as shown in Fig.~\ref{fig:non-determinism} (left). Although fixing random seeds will reduce the variability of results, there may still be differences in metric scores, for example caused by parallel processing \cite{pham2020problems} or usage of more than two \acp{GPU}.

\footnotetext{See for example: \url{https://pytorch.org/docs/stable/notes/randomness.html}}

\paragraph{\textbf{Non-standardized metric implementation}} Metric implementation is typically not standardized, as can for instance be seen in Figures~\ref{fig:froc-no-standard} and~\ref{fig:corner-case-ap}. This was further highlighted in a study presented in \cite{gooding2022multicenter}, in which the authors analyzed the differences between how metric implementations differ and their implications. The authors found, for example, that seven out of thirteen distinct methods were used for sampling the surface for distance-based metrics such as \textit{\ac{HD}} or that the thirteen participants of the study used five different definitions for the symmetric average distance \cite{gooding2022multicenter}. It was further shown that these differences result in a high variation in the final metric score, for instance of up to 9\% in the \textit{\ac{DSC}} and of up to 40\% for the \textit{\ac{HD}} metric \cite{gooding2022multicenter}.

\newpage
\section{Conclusion}
Choosing the right metric for a specific image processing task is a nontrivial undertaking. With this (dynamic) paper, we wish to raise awareness about some of the common flaws of the most frequently used reference-based validation metrics in the field of image processing and provide guidance of their use, encouraging researchers to reconsider common workflows. 

\section{Acknowledgements}

This work was initiated by the Helmholtz Association of German Research Centers in the scope of the Helmholtz Imaging Incubator (HI), the MICCAI Special Interest Group on biomedical image analysis challenges and the benchmarking working group of the MONAI initiative. It received funding from the European Research Council (ERC) under the European Union’s Horizon 2020 research and innovation programme (grant agreement No. [101002198], NEURAL SPICING). It was further supported in part by the Intramural Research Program of the National Institutes of Health (NIH) Clinical Center as well as by the National Cancer Institute (NCI) and the National Institute of Neurological Disorders and Stroke (NINDS) of the NIH, under award numbers NCI:U01CA242871 and NINDS:R01NS042645. The content of this publication is solely the responsibility of the authors and does not represent the official views of the NIH. T.A. acknowledges the Canada Institute for Advanced Research (CIFAR) AI Chairs program, the Natural Sciences and Engineering Research Council of Canada. F.B. was co-funded by the European Union (ERC, TAIPO, 101088594). Views and opinions expressed are however those of the authors only and do not necessarily reflect those of the European Union or the European Research Council. Neither the European Union nor the granting authority can be held responsible for them. V.C. acknowledges funding from NovoNordisk Foundation (NNF21OC0068816) and Independent Research Council Denmark (1134-00017B). B.A.C. was supported by NIH grant P41 GM135019 and grant 2020-225720 from the Chan Zuckerberg Initiative DAF, an advised fund of the Silicon Valley Community Foundation. G.S.C. was supported by Cancer Research UK (programme grant: C49297/A27294). M.M.H. is supported by the Natural Sciences and Engineering Research Council of Canada (RGPIN-2022-05134). A.Kara. is supported by French State Funds managed by the “Agence Nationale de la Recherche (ANR)” - “Investissements d’Avenir” (Investments for the Future), Grant ANR-10-IAHU-02 (IHU Strasbourg). M.K. was supported by the Ministry of Education, Youth and Sports of the Czech Republic (Project LM2018129). T.K. was supported in part by 4UH3-CA225021-03, 1U24CA180924-01A1, 3U24CA215109-02, and 1UG3-CA225-021-01 grants from the National Institutes of Health. G.L. receives research funding from the Dutch Research Council, the Dutch Cancer Association, HealthHolland, the European Research Council, the European Union, and the Innovative Medicine Initiative. S.M.R. wishes to acknowledge the Allen Institute for Cell Science founder Paul G. Allen for his vision, encouragement and support. C.H.S. is supported by an Alzheimer's Society Junior Fellowship (AS-JF-17-011). M.R is supported by Innosuisse grant number 31274.1 and Swiss National Science Foundation Grant Number 205320\_212939. R.M.S. is supported by the Intramural Research Program of the NIH Clinical Center. A.T. acknowledges support from Academy of Finland (Profi6 336449 funding program), University of Oulu strategic funding, Finnish Foundation for Cardiovascular Research, Wellbeing Services County of North Ostrobothnia (VTR project K62716), and Terttu foundation. S.A.T. acknowledges the support of Canon Medical and the Royal Academy of Engineering and the Research Chairs and Senior Research Fellowships scheme (grant RCSRF1819\textbackslash 8\textbackslash 25). B.V.C. was supported by Research Foundation Flanders (FWO grant G097322N) and Internal Funds KU Leuven (grant C24M/20/064).\\

We would like to thank Amine Yamlahi, Niklas Holzwarth, Marco Hübner, Amith Kamath, Dominik Michael, You Suhang, and Yannick Suter for proofreading the document and proposing pitfalls.

\addtocontents{toc}{\protect\setcounter{tocdepth}{-1}}
\bibliography{sample-base}

\addtocontents{toc}{\protect\setcounter{tocdepth}{1}}
\appendix
\section*{Appendix}

\section{Acronyms}
\begin{acronym}[]
\acro{AI}{artificial intelligence}
\acro{AP}{Average Precision}
\acro{ASSD}{Average Symmetric Surface Distance}
\acro{AUC}{Area under the curve}
\acro{AUROC}{Area under the Receiver Operating Characteristic curve}
\acro{BA}{Balanced Accuracy}
\acro{BM}{Bookmaker Informedness}
\acro{BS}{Brier Score}
\acro{CE}{Calibration Error}
\acro{clDice}{Centerline Dice Similarity Coefficient}
\acro{CI}{Confidence Interval}
\acro{COCO}{Common Objects in Context}
\acro{DSC}{Dice Similarity Coefficient}
\acro{EC}{Expected Cost}
\acro{ECE}{Expected Calibration Error}
\acro{FN}{False Negative}
\acro{FP}{False Positive}
\acro{FPPI}{False Positives per Image}
\acro{FPR}{False Positive Rate}
\acro{FROC}{Free-Response Receiver Operating Characteristic}
\acro{GPU}{Graphics Processing Unit}
\acro{HD}{Hausdorff Distance}
\acro{HD95}{Hausdorff Distance 95\% Percentile}
\acro{IoU}{Intersection over Union}
\acro{IoR}{Intersection over Reference}
\acro{LR+}{Positive Likelihood Ratio}
\acro{mAP}{mean Average Precision}
\acro{MASD}{Mean Average Surface Distance}
\acro{MCC}{Matthews Correlation Coefficient}
\acro{MCE}{Maximum Calibration Error}
\acro{MCP}{Maximum Class Probability}
\acro{NaN}{Not A Number}
\acro{NB}{Net Benefit}
\acro{NPV}{Negative Predictive Value}
\acro{NSD}{Normalized Surface Distance}
\acro{PPV}{Positive Predictive Value}
\acro{PR}{Precision-Recall}
\acro{PQ}{Panoptic Quality}
\acro{ROC}{Receiver Operating Characteristic}
\acro{TN}{True Negative}
\acro{TNR}{True Negative Rate}
\acro{TP}{True Positive}
\acro{TPR}{True Positive Rate}

\end{acronym}

\end{document}